\documentclass[12pt,a4paper]{book}
\usepackage[nottoc]{tocbibind}

\usepackage[square,numbers]{natbib}

\usepackage[utf8]{inputenc}
\usepackage[T1]{fontenc}
\usepackage[english]{babel}

\usepackage[oldstyle, scale=.95]{opensans} % police Open Sans
\usepackage{amsmath}
\usepackage{amsfonts}
\usepackage{fancyhdr}

\usepackage{amssymb} %Supposed to get the box with tilde to work

\usepackage{tensor}

\usepackage{xcolor} % où color selon l'installation
\definecolor{Prune}{RGB}{99,0,60} % l14-33 : couleurs de la charte graphique upsaclay
\definecolor{B1}{RGB}{49,62,72} 
\definecolor{C1}{RGB}{124,135,143}
\definecolor{D1}{RGB}{213,218,223}
\definecolor{A2}{RGB}{198,11,70}
\definecolor{B2}{RGB}{237,20,91}
\definecolor{C2}{RGB}{238,52,35}
\definecolor{D2}{RGB}{243,115,32}
\definecolor{A3}{RGB}{124,42,144}
\definecolor{B3}{RGB}{125,106,175}
\definecolor{C3}{RGB}{198,103,29}
\definecolor{D3}{RGB}{254,188,24}
\definecolor{A4}{RGB}{0,78,125}
\definecolor{B4}{RGB}{14,135,201}
\definecolor{C4}{RGB}{0,148,181}
\definecolor{D4}{RGB}{70,195,210}
\definecolor{A5}{RGB}{0,128,122}
\definecolor{B5}{RGB}{64,183,105}
\definecolor{C5}{RGB}{140,198,62}
\definecolor{D5}{RGB}{213,223,61}

\definecolor{HotCoral}{RGB}{240, 96, 92}
\definecolor{LivingCoral}{RGB}{252, 118, 106}
\definecolor{ViridianGreen}{RGB}{0, 150, 152}
\definecolor{PacificCoast}{RGB}{91, 132, 177}

\usepackage{mdframed}
\usepackage{multirow} %% Pour mettre un texte sur plusieurs rangées
\usepackage{multicol} %% Pour mettre un texte sur plusieurs colonnes
\usepackage{tikz}
\usepackage{graphicx}    % standard LaTeX graphics tool
\graphicspath{{template_files/}{chapters/figsquantumnesscosmo/}} % LateX will look for figures inside of this
\usepackage[absolute]{textpos} 
\usepackage{colortbl}
\usepackage{array}
\usepackage{geometry}
\usepackage{titlesec}

\renewcommand{\thesubsubsection}{\thesubsection-\alph{subsubsection}}

\usepackage[bottom]{footmisc}% places footnotes at page bottom
\usepackage{bm}
\usepackage{bbold}
\usepackage{dsfont} %For \mathds to write the identity operator

\usepackage{pdfpages}

\usepackage{subfiles} %To have multiple files

\usepackage[bookmarks=true,bookmarksdepth=4,backref]{hyperref} %for hyperlinks, bookmarksdepth=4 is there to allow the subsubsections
\hypersetup{ % paramétrage couleur des liens hypertextes, toujours garder colorlinks=true
    colorlinks=true,
    citecolor=LivingCoral, %Change the color of the citation
    linkcolor=ViridianGreen,
    urlcolor=purple}

\renewcommand*{\backref}[1]{}
\renewcommand*{\backrefalt}[4]{[{%
    \ifcase #1 Not cited.%
          \or Cited in~#2%
          \else Cited in~#2%
    \fi%
    }]}

\usepackage{cleveref} %Suppose to give better autocompletion

\newcommand{\ie}{{i.e.~}}

\let\oldsqrt\sqrt
\def\sqrt{\mathpalette\DHLhksqrt}
\def\DHLhksqrt#1#2{%
\setbox0=\hbox{$#1\oldsqrt{#2\,}$}\dimen0=\ht0
\advance\dimen0-0.2\ht0
\setbox2=\hbox{\vrule height\ht0 depth -\dimen0}%
{\box0\lower0.4pt\box2}}

\newcommand{\dd}{\mathrm{d}}
\newcommand{\sss}[1]{{\scriptscriptstyle{#1}}}

\newcommand{\uPl}{\mathrm{Pl}}

\newcommand{\usssPl}{\sss{\uPl}}

\newcommand{\setR}{\mathbb{R}}

\newcommand{\Hu}{\mathcal{H}}

\newcommand{\GN}{G_\text{N}}

\newcommand{\Mp}{M_\usssPl}

\newcommand{\beq}{\begin{equation}}
\newcommand{\eeq}{\end{equation}}
\newcommand{\bea}{\begin{equation}\begin{aligned}}
\newcommand{\eea}{\end{aligned}\end{equation}}

\newlength{\wsingfig}
\newlength{\wdblefig}
\newlength{\wquadfig}
\newlength{\wtriplefig}
\titleformat{\chapter}[hang]{\bfseries\LARGE\color{Prune}}{\thechapter\ -}{.1ex}
{\vspace{0.1ex}
}
[\vspace{1ex}]
\titlespacing{\chapter}{0pc}{0ex}{0.5pc}

\titleformat{\section}[hang]{\bfseries\Large}{\thesection\ }{0.5pt}
{\vspace{0.1ex}
}
[\vspace{0.1ex}]
\titlespacing{\section}{1.5pc}{4ex plus .1ex minus .2ex}{.8pc}

\titleformat{\subsection}[hang]{\bfseries\large}{\thesubsection\ }{1pt}
{\vspace{0.1ex} 
}
[\vspace{0.1ex}]
\titlespacing{\subsection}{3pc}{2ex plus .1ex minus .2ex}{.1pc}

\titleformat{\subsubsection}[hang]{\bfseries\normalsize}{\thesubsubsection\ }{1pt}
{\vspace{0.1ex}
}
[\vspace{0.1ex}]
\titlespacing{\subsubsection}{4.5pc}{2ex plus .1ex minus .2ex}{.1pc}
\begin{document}

%\def\biblio{}

%Title page

{\opensans

\begin{titlepage}

\newgeometry{left=6cm,bottom=2cm, top=1cm, right=1cm}

\tikz[remember picture,overlay] \node[opacity=1,inner sep=0pt] at (-13mm,-135mm){\includegraphics{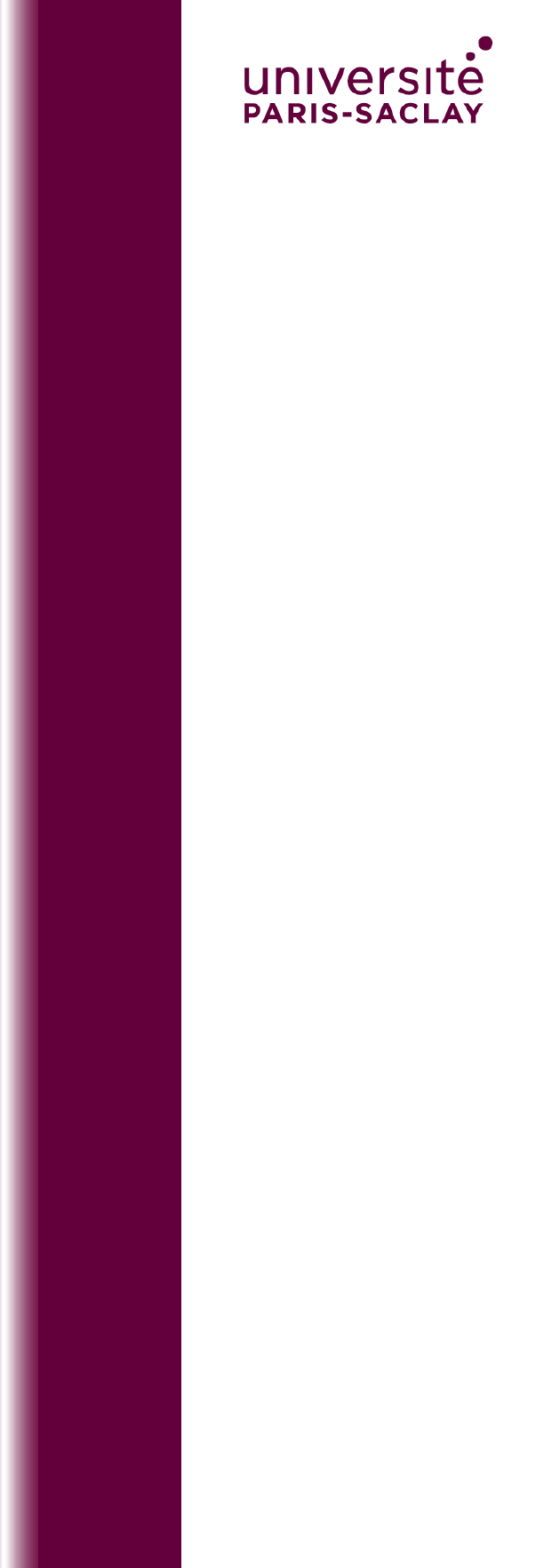}};

%*****************************************************
%******** NUMÉRO D'ORDRE DE LA THÈSE À COMPLÉTER *****
%******** POUR LE SECOND DÉPOT                   *****
%*****************************************************

\color{white}

\begin{picture}(0,0)
\put(-152,-743){\rotatebox{90}{\Large \textsc{THESE DE DOCTORAT}}} \\
\put(-120,-743){\rotatebox{90}{NNT : 2023UPASP078}}
\end{picture}

%*****************************************************
%******************** TITRE **************************
%*****************************************************

\flushright
\vspace{10mm} % à régler éventuellement
\color{Prune}
\fontfamily{cmss}\fontseries{m}\fontsize{22}{26}\selectfont
  \Huge 	Entanglement and decoherence in cosmology and in analogue gravity experiments  \\

\normalsize
\color{black}
\Large{\textit{Intrication et décohérence en cosmologie et dans les expériences de gravité analogue}} \\
%*****************************************************

\fontfamily{fvs}\fontseries{m}\fontsize{8}{12}\selectfont

\vspace{1.5cm}

\normalsize
\textbf{Thèse de doctorat de l'université Paris-Saclay} \\

\vspace{6mm}

\small École doctorale n°564 : physique en Île-de-France (PIF)\\ 
\small Spécialité de doctorat : Physique\\
\small Graduate School : Physique. Référent : Faculté des Sciences d’Orsay \\
\vspace{6mm}

{\footnotesize 
	Thèse préparée dans l'unité de recherche \textbf{IJCLab} (Université Paris-Saclay, CNRS), sous la direction de \textbf{Christos CHARMOUSIS}, Directeur de recherche, la co-direction de  \textbf{Jérôme MARTIN}, Directeur de recherche, et le co-encadrement de \textbf{Scott J. ROBERTSON}, Professeur Junior 
}
	\\
\vspace{15mm}

\textbf{Thèse soutenue à Paris, le 15 Septembre 2023, par}\\
\bigskip
\Large {\color{Prune} \textbf{Amaury MICHELI}} % Changer le Prénom et le NOM

%************************************
\vspace{\fill} % ALIGNER LE TABLEAU EN BAS DE PAGE
%************************************

\bigskip

\flushleft
\small {\color{Prune} \textbf{Composition du jury}}\\
{\color{Prune} \scriptsize {Membres du jury avec voix délibérative}} \\
\vspace{2mm}
\scriptsize
\begin{tabular}{|p{7cm}l}
\arrayrulecolor{Prune}
\bf{Chris WESTBROOK} &   \bf{Président} \\ 
Directeur de Recherche CNRS, Institut d'Optique Graduate School & \\
\bf{Ivan AGULLO} &  \bf{Rapporteur \& Examinateur } \\ 
Professeur associé (eq. HDR), Louisiana State University   &   \\ 
\bf{Silke WEINFURTNER} &  \bf{Rapportrice \& Examinatrice} \\ 
Professeur, University of Nottingham &   \\ 
\bf{Andreas ALBRECHT} &  \bf{Examinateur}  \\ 
Professeur Distingué, University of California, Davis    &   \\ 
\bf{David KAISER} &  \bf{Examinateur} \\ 
Professeur, Massachusetts Institute of Technology   &   \\

\end{tabular}

\end{titlepage}
}

%Abstract page
%Might have to modify the geometry of this page because if ends up being on the right hand-side then
%the frames will be very close to the edge of the page

%Limit the scope of the font modifications to this page
{\opensans

% page des résumés à garder en 2ème page. Si les résumés sont trop longs pour tenir sur une seule et même page, 
%on peut mettre un résumé par page
\thispagestyle{empty}
\newgeometry{top=1.5cm, bottom=1.25cm, left=2cm, right=2cm}

\fontfamily{rm}\selectfont

\lhead{}
\rhead{}
\rfoot{}
\cfoot{}
\lfoot{}

\noindent 
%*****************************************************
%***** LOGO DE L'ED À CHANGER IMPÉRATIVEMENT *********
%*****************************************************
\includegraphics[height=2.45cm]{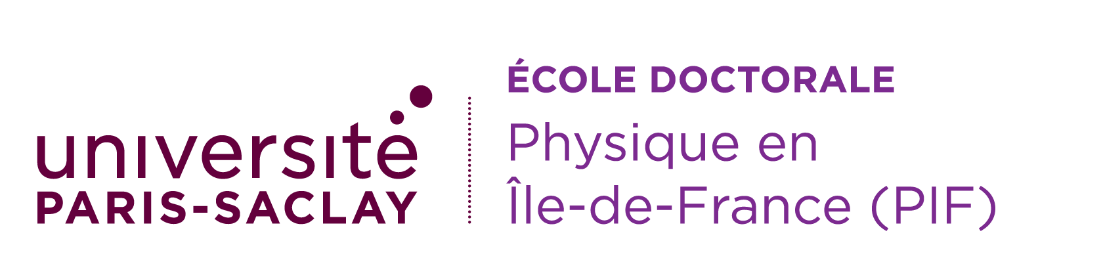}
\vspace{1cm}
%*****************************************************
\fontfamily{cmss}\fontseries{m}\selectfont

\small

\begin{mdframed}[linecolor=Prune,linewidth=1]

\textbf{Titre :} Intrication et décohérence en cosmologie et dans les expériences de gravité analogue

\noindent \textbf{Mots clés :} Gravité analogue, Cosmologie, Décohérence, Intrication, Espace-temps Courbe.

\vspace{-.5cm}
\begin{multicols}{2}
\noindent \textbf{Résumé :}

Cette thèse est consacrée à l'analyse de la création et destruction de corrélations quantiques dans le contexte de l'inflation cosmologique et d'une expérience analogue du préchauffage. 
L'inflation est une phase d'expansion accélérée de l'Univers, précédant le modèle dit standard de la cosmologie, introduite pour résoudre certaines lacunes du modèle.  L’inflation fournit également un mécanisme d'émergence des inhomogénéités primordiales par amplification de fluctuations quantiques initiales.
Elle est suivie d'une période de "réchauffement", durant laquelle on s'attend à ce que la plupart des particules soient générées et atteignent l'équilibre thermique, préparant ainsi le terrain pour le déroulement du modèle standard de la cosmologie. Pendant une période de "préchauffage", cette création procède en partie par excitation paramétrique de modes résonants des champs de matière initialement dans leur vide, un véritable processus quantique. La physique de l'inflation cosmologique et du préchauffage est celle d'un champ classique fort agissant sur un champ quantique pour produire des particules intriquées. Lorsque la source est la métrique de l'espace-temps elle-même, comme dans l'inflation, nous sommes dans le cadre de la théorie quantique des champs en espace-temps courbe (TQCEC). L'évolution des corrélations quantiques ainsi générées est le sujet de cette thèse. 
Dans la première partie du manuscrit, nous présentons le traitement quantique standard des perturbations cosmologiques durant l'inflation. Nous passons ensuite en revue les travaux antérieurs analysant la génération de corrélations quantiques entre des perturbations d'impulsions opposées à l'aide de mesures de «quanticité» telles que la non-séparabilité, la discorde quantique ou une inégalité de Bell. Partant de cette revue, nous présentons un calcul de l'évolution de la discorde quantique pour l'état des modes d'impulsions opposées lorsque la distillation des corrélations aux degrés de liberté environnementaux, appellée décohérence, est prise en compte à l'aide d'un modèle de Caldeira-Leggett. La décohérence place les perturbations dans un état comprimé mixte à deux modes, omniprésent dans le TQCEC et la physique quantique à basse énergie. Nous identifions les régimes dans lesquels les corrélations quantiques persistent malgré la décohérence et les régimes dans lesquels elles disparaissent. Enfin, nous procédons à une comparaison systématique des résultats de trois mesures différentes de quanticité appliquées au même état mixte comprimé à deux modes et démontrons un degré d'inéquivalence entre eux. 
La seconde partie du manuscrit est dédiée à l’analyse d’une expérience dite de "gravité analogue". La gravité analogue a émergé des travaux fondateurs de W. Unruh qui a proposé de concevoir des expériences de matière condensée pour tester les prédictions de la TQCEC dans un contexte où l'intrication peut, en principe, être mesurée. Depuis 2008, plusieurs groupes ont mené des expériences pour observer les propriétés de quasi-particules émises soit par un trou noir analogue, soit par l'analogue d'un univers en expansion. Nous nous concentrons ici sur une expérience imitant la dynamique du préchauffage à l'aide d'un gaz quasi unidimensionnel d'atomes d'hélium métastables, qui, lors de sa première réalisation, n'a pas pu mettre en évidence l’intrication. Il a ensuite été postulé qu'un degré suffisant d'interactions des quasi-particules pouvait expliquer cette absence. Nous commençons par passer en revue la génération de paires intriquées dans l'expérience et discutons l'absence d'intrication. Nous analysons ensuite les interactions du gaz de Bose unidimensionnel pour démontrer l’existence de nouveaux processus de dissipation pour les excitations générées au cours de l'expérience. Enfin, nous montrons l'effet de ces mêmes processus sur la corrélation. Nous concluons qu'ils pourraient être suffisants pour expliquer l'absence d'intrication dans l'expérience.

\end{multicols}

\end{mdframed}

\vspace{8mm}

\newpage

\includegraphics[height=2.45cm]{template_files/logo_usp_PIF.png}
\vspace{1cm}

\begin{mdframed}[linecolor=Prune,linewidth=1]

\textbf{Title:} Entanglement and decoherence in cosmology and in analogue gravity experiments


\begin{multicols}{2}
\noindent \textbf{Abstract:}

This thesis is dedicated to analysing the generation and destruction of quantum correlations in the context of inflationary cosmology and an experiment of 'analogue' preheating.
Inflation is a phase of accelerated expansion of the Universe, preceding the so-called Standard Model of Big Bang cosmology, introduced to solve some shortcomings of this model. It also provides a mechanism for the emergence of primordial inhomogeneities by amplification of initial quantum fluctuations. 
Inflation is followed by a 'reheating' period, in which 
most particles are expected to be generated and reach thermal equilibrium, setting the stage for the standard Big Bang of cosmology.
During a 'preheating' period, this creation proceeds partly by parametric excitation of resonant modes of the matter fields initially in their vacuum, a genuine quantum process.
The physics of both situations, inflation and preheating, is that of a strong classical field acting on a quantum field to produce entangled (quasi-)particles. When the classical source is the space-time metric itself, as in inflation, we are in the framework of Quantum Field Theory in Curved Space-time (QFTCS). The evolution of the generated quantum correlations is the topic of this PhD.

In the first part of the manuscript, we present the standard quantum treatment of cosmological perturbations during inflation.
We then review previous works analysing the generation of quantum correlations between opposite momenta perturbations using measures of 'quantumness' such as non-separability, quantum discord or Bell inequalities.
Building upon them, we present a computation of the evolution of quantum discord for the state of opposite momenta modes when the distillation of correlations to environmental degrees of freedom, i.e. decoherence, is taken into account using a Caldeira-Leggett model.
Decoherence places the perturbations in a mixed two-mode squeezed state, ubiquitous in QFTCS and low-energy quantum physics.
We identify regimes in which quantum correlations persist despite decoherence and regimes in which they disappear.
Finally, we systematically compare the results of three different measures of quantumness applied to the same mixed two-mode squeezed state and demonstrate a degree of inequivalence between them.

The second part of the manuscript is devoted to a so-called 'analogue gravity' experiment.
Analogue gravity ideas emerged from the seminal works of W. Unruh, who proposed designing condensed matter experiments to test the predictions of QFTCS in a context where entanglement can, in principle, be measured. Since 2008 several groups have performed experiments to observe the properties of quasi-particles emitted either by an analogue black hole or by the analogue of an expanding universe.
We here focus on an experiment mimicking the dynamics of preheating using a quasi-one dimensional gas of metastable Helium atoms, which in its first run failed to witness entanglement.
It was later postulated that a sufficient degree of quasi-particle interactions could explain its absence.
We start by reviewing the generation of entangled pairs in the experiment and the ensuing discussion on the absence of entanglement.
We then analyse the interactions of one-dimensional Bose gas and uncover new dissipation processes for the excitations generated during the experiment. Finally, we show the effect of the same processes on correlation. We conclude that they might be sufficient to explain the absence of entanglement in the experiment.

\end{multicols}
\end{mdframed}

}

\chapter*{Remerciements} 
%For the chapter to still appear in the table of contents despite
%not having a number
%see https://tex.stackexchange.com/questions/30122/generate-table-of-contents-when-section-sections-without-numbering-has-been
\addcontentsline{toc}{chapter}{\protect\numberline{}Remerciements}%

Je souhaiterais remercier toutes les personnes que j'ai côtoyées, rencontrées, croisées au cours des quatre années de préparation de cette thèse.
Sa forme et son contenu ne seraient pas les mêmes sans ces innombrables interactions.
\\

En premier lieu, je pense à double titre à Renaud Parentani, avec qui j'ai travaillé pendant les premiers mois de ma thèse.
Premièrement, parce que l'étude de la dissipation dans un système analogue réalisée dans la thèse est la poursuite directe d'un axe de recherche qu'il a initié et, il me semble, l'achèvement du sujet de stage que nous avions effectué ensemble avant le début de ma thèse.
Également, car lors de mes travaux sur la quanticité des perturbations cosmologiques, la seconde moitié de cette thèse, je n'ai cessé d'être impressionné par l'acuité et le caractère précurseur de ses contributions au domaine.
\\

Je tiens ensuite à remercier toutes les personnes qui m'ont permis de poursuivre ma thèse après une première année difficile.
Tout d'abord mes deux directeurs, Jérôme Martin et Scott Robertson, qui ont bien voulu assumer ce rôle alors qu'ils n'avaient rien demandé.
Ils ont été des directeurs géniaux d'un point de vue humain et ils sont probablement les personnes avec qui j'apprécie le plus de faire de la physique.
À ce titre, je remercie aussi Vincent Vennin, pour sa gentillesse, mais aussi parce que j'ai toujours un immense plaisir à travailler avec lui.
Je remercie également Christos Charmousis pour son soutien immédiat et ses conseils constants, ainsi que pour avoir signé tous les papiers administratifs pendant trois ans (sans rechigner).
\\

Évidemment, je ne pourrais remercier suffisamment Hsiao-Yin pour m'avoir soutenu (et supporté) durant des années éprouvantes, et m'avoir aidé à prendre les bonnes décisions.
Je remercie mes parents, mon frère et toute ma famille pour leur accompagnement depuis bien plus que quatre ans.
\\

Cette thèse doit beaucoup aux personnes qui ont contribué à ce que ses conditions de réalisation soient agréables. Je remercie énormément l'IAP (Institut
 d'Astrophysique de Paris) dans son ensemble pour son accueil, notamment le groupe  ${\cal G}\mathbb{R}\varepsilon\mathbb{C}{\cal O}$ et l'équipe informatique. 
Un immense merci à tou.te.s les doctorant.e.s de l'IAP qui ont joué un rôle essentiel dans ma motivation et mon épanouissement.
Merci à Aline, Louis et Nai qui ont partagé mon bureau.
Merci aux membres du club café de l'IAP Étienne, Marko et Nai (encore).
Merci à celleux avec qui je n'ai pas partagé un bureau ou une machine à café, mais avec qui j'ai passé des moments formidables, notamment Alix, Axel, Clément, Denis, Eduardo, Emma, Emilie, Ira, Julien, Louise, Matthieu, Marie, Marko, Pierre, Quentin.
\\

À l'extérieur de l'IAP, je remercie Thomas, un autre membre de la petite équipe intéressée par les aspects quantiques des perturbations cosmologiques et maintenant un bel ami, Victor, un autre amateur d'intrication qui subit mes délires de théoriciens dans la bonne humeur, ainsi que Chris, Denis et Quentin, le reste de la sympathique équipe collaborant sur COSQUA, Giulia, Florian et Tim, mes codoctorant.e.s de feu le LPT, Bartjan, Eugeny (in particular for explaining clearly the mapping between scalar-tensor theory and $f(R)$ gravity), Gatien, Karim, Marie et Sarah, toujours du LPT.
\\

Un grand merci à mes ami.e.s Alan, Benjy(s), Cyril, Daniel, Danilo, Dina, Étienne, JB, Léa, Lisheng, Louis, Ludo, Luisa, Matthieu(s), Naïla, Nilo, Virginie.
\\

Je suis également très reconnaissant envers mes professeurs de mathématiques du lycée Bernard Palissy, en particulier M. Akkouche, Mme et M. Puyou, qui, par la diversité de leurs enseignements, m'ont fait découvrir le plaisir de la résolution de problèmes.
\\

A big thank you to David Wands and Greg Kaplanek for their warm reception at ICG and Imperial College.
\\

Finally, I also want to thank the members of the websites Physics Forum and StackExchange, where I have many times find detailed and enlightening contributions, especially when preparing classes.
\\

This work was supported by the French National Research Agency via Grant No. ANR-20-CE47-0001 associated with the project COSQUA (Cosmology and Quantum Simulation).
\\

\chapter*{Synthèse en français} 

\addcontentsline{toc}{chapter}{\protect\numberline{}Synthèse en français}%

Cette thèse est consacrée à l'analyse de la création et destruction de corrélations quantiques dans le contexte de l'inflation cosmologique et d'une expérience analogue du préchauffage.

L'inflation est une phase d'expansion accélérée de l'Univers, précédant le modèle dit standard de la cosmologie, introduite pour résoudre certaines lacunes du modèle. L’inflation fournit également un mécanisme d'émergence des inhomogénéités primordiales par amplification de fluctuations quantiques initiales.
Elle est suivie d'une période de "réchauffement", durant laquelle on s'attend à ce que la plupart des particules soient générées et atteignent l'équilibre thermique, préparant ainsi le terrain pour le déroulement du modèle standard de la cosmologie. Pendant une période de "préchauffage", cette création procède en partie par excitation paramétrique de modes résonants des champs de matière initialement dans leur vide, un véritable processus quantique. La physique de l'inflation cosmologique et du préchauffage est celle d'un champ classique fort agissant sur un champ quantique pour produire des particules intriquées. Lorsque la source est la métrique de l'espace-temps elle-même, comme dans l'inflation, nous sommes dans le cadre de la théorie quantique des champs en espace-temps courbe (TQCEC). L'évolution des corrélations quantiques ainsi générées est le sujet de cette thèse.

Le premier chapitre du manuscrit~\ref{chapt:introduction} fait office d'introduction aux domaines dans lesquels les contributions de la thèse s'inscrivent : la cosmologie inflationnaire et les expériences dites de "gravité analogue".
La première partie du chapitre~\ref{sec:intro_cosmo} est une présentation de la description standard de l'évolution de l'Univers considéré en première approximation comme homogène. Ce traitement est basé sur la relativité générale dont les fondamentaux nécessaires sont tout d'abord rappelés en section~\ref{sec:GR}. Nous utilisons ensuite~\ref{sec:GRcosmo} cette théorie appliquée aux différentes composantes de l'Univers considérées comme homogènes et isotropes, hypothèse justifiée par les observations, pour dériver la géométrie de l'espace-temps (un Univers en expansion décrit par la métrique de Friedmann-Robertson-Lemaitre-Walker) et les équations régissant leur évolution (équations de Friedmann et équation d'état).
En se basant sur ces équations et sur les observations donnant la composition actuelle de l'Univers, nous présentons~\ref{sec:BBmodel} le modèle standard de la cosmologie décrivant l'histoire de celui-ci comme une succession d'ères. Chacune de ses ères est caractérisée par la domination de la composition de l'Univers par une de ses composantes, et un taux d'expansion qui lui est associé.
Cette première partie se conclut~\ref{sec:BBpuzzles} par une explication de certains problèmes associés au modèle standard de la cosmologie: problème de la platitude, de l'horizon et des monopoles.

Dans la seconde partie du chapitre~\ref{sec:inflation} nous présentons le modèle d'inflation cosmique, les prédictions associées pour l'état des inhomogénéités dans l'Univers, ainsi que comment cette période se connecte avec le modèle standard de la cosmologie via une période de (p)réchauffement.
L'inflation cosmique correspond à une période d'expansion \textit{accélérée} de l'Univers, par opposition aux périodes du modèle standard où le taux d'expansion est décroissant.
Nous commençons~\ref{sec:infl_solBB_puzzles} par montrer qu'une période suffisamment longue d'expansion accélérée permet de résoudre les trois problèmes du modèle standard détaillés précédemment.
Nous montrons ensuite comment une telle période d'inflation peut survenir si la composition de l'Univers est dominée par un champ scalaire homogène, l'inflaton, avec un potentiel suffisamment plat pour que son évolution soit lente (conditions de roulement lent).
Dans~\ref{sec:inhomogeneous_universe}, nous introduisons la description perturbative standard des inhomogénéités dans l'Univers, leur caractère perturbatif étant justifié par leur petitesse observée. Nous dérivons ensuite les équations d'évolution des perturbations pendant l'inflation qui prédisent une amplification de celles-ci. De plus, les grandes échelles d'énergie attendues pour l'inflation, suggèrent de décrire les perturbations par des champs quantiques. Nous montrons que ce caractère quantique fixe un niveau minimal pour ces  inhomogénéités si nous choisissons de considérer leur champ comme initialement sans particules. L'expansion de l'Univers amplifie ensuite ces fluctuations du vide, un phénomène de TQCEC. 
Le spectre de puissance des inhomogeneités prédit par ce scénario d'amplification des fluctuations du vide est en parfait accord avec les données du fond diffus cosmologique.
À l'issue d'une période d'inflation, l'Univers est essentiellement vide de particule. Afin de connecter cette période avec le modèle standard de la cosmologie, qui suppose un grand nombre de particules à l'équilibre, une période de réchauffement est nécessaire. Nous discutons du réchauffement dans la dernière section de ce chapitre~\ref{sec:preheating}, ainsi que du scénario de préchauffage, mentionné précédemment, dans lequel des particules sont créées par des amplifications paramétriques provoquées par les oscillations de l'inflaton au fond de son potentiel. Il s'agit là encore d'un effet de TQCEC.

La troisième et dernière partie~\ref{sec:introduction_AG} du chapitre d'introduction est consacrée à une présentation des expériences de gravité analogue.
Tout d'abord~\ref{sec:motivations_AG}, nous motivons l'utilité de telles expériences par quelques calculs d'ordre de grandeur montrant la difficulté d'observer directement un effet de TQCEC où la création de particules depuis le vide quantique médiée par la gravité.
Nous démontrons ensuite~\ref{sec:acoustic_metric} que les équations décrivant les perturbations de grande longueur d'onde d'un fluide peuvent être mises sous la forme de celle d'un champ scalaire évoluant dans un espace-temps courbe. La métrique de cet espace-temps est appelée métrique acoustique.
Dans~\ref{sec:progress_AG} nous passons rapidement en revue les progrès important effectué dans la conceptualisation et la réalisation d'expérience analogue en général.
Enfin, dans~\ref{sec:cosmo_AG}, nous procédons à une revue plus exhaustive des expériences faisant une analogie avec une situation cosmologique.

Le second chapitre~\ref{chapt:quantumness_cosmo} comprend les travaux effectués durant la thèse sur le caractère quantique ou classique des perturbations cosmologiques.
Comme détaillé dans le premier chapitre, dans le scénario standard de formation des structures, les inhomogénéités primordiales sont le résultat de fluctuations du vide quantique amplifiées par la gravité, notamment pendant l'inflation. Il est bien connu que l'état quantique qui en résulte est, pour les variables appropriées, un état comprimé  à deux modes (two-mode squeezed state) pour les modes de Fourier de direction opposés $\pm \bm{k}$. 
Ces aspects standards des perturbations cosmologiques quantiques sont décrits en détail dans l'article de revue reproduit dans la seconde partie du chapitre~\ref{sec:quantum_GW}.
Cet article passe également en revue l'état de l'art des discussions sur le caractère quantique ou classique des perturbations avant et jusqu'au milieu de la thèse. 
En particulier, nous revenons sur l'affirmation qu'un état très comprimé  serait classique en le présentant rigoureusement et en n'en montrant les limites: cela n'est vrai qu'en considérant la valeur de certains opérateurs.
Nous décrivons un certain nombre d'autres mesures de quanticité, basées sur les corrélations au sein de l'état comprimé, qui ont été appliquées aux perturbations cosmologiques (séparabilité, inégalités de Bell, discorde quantique, etc.) et qui montrent au contraire que l'état contient des corrélations quantiques.
Enfin, nous discutons les quelques articles prenant en compte l'effet que l'interaction des deux modes avec d'autres champs peut avoir sur les corrélations, phénomène dit de décohérence.
Il est attendu que la décohérence affaiblisse les corrélations et leur enlève tout caractère quantique. Les travaux effectués dans le contexte cosmologique tendent à confirmer cette intuition tout en montrant qu'il y a une compétition entre la décohérence et la compression de l'état pour déterminer le niveau de corrélation.
Dans la partie suivante~\ref{sec:discord_and_decoherence} nous reproduisons l'article rédigé pendant cette thèse consacré à l'évolution de la discorde quantique des perturbations cosmologiques en présence de décohérence. Les premières parties de l'article sont applicables au calcul de la discorde bipartite d'un champ scalaire dans un état gaussien. Nous commençons par montrer la dépendance de la discorde dans le choix de la "partition" du champ en groupe de degrés de libertés. Nous décrivons ensuite l'évolution de la discorde en l'absence de décohérence en utilisant plusieurs formalismes. Le cœur de l'analyse est contenu dans la troisième partie où l'évolution du système est suivie en présence de décohérence décrite par un modèle de Caldeira-Leggett qui préserve le caractère gaussien de l'état. Ceci nous permet de garder un formalisme simple pour le calcul de la discorde tout en autorisant une paramétrisation de l'interaction par sa dépendance temporelle dans le facteur d'échelle. Pour faciliter la compréhension de cette évolution, nous introduisons des paramètres de compression généralisés qui permettent une représentation géométrique simple de l'état, même en présence de décohérence.
Enfin, nous appliquons ces résultats au cas des perturbations cosmologiques qui fixe la dépendance temporelle du champ en absence de décohérence. Nous montrons que deux régimes existent : l'un où la décohérence est suffisamment forte pour effacer toute discorde, l'autre où la discorde, et donc les corrélations quantiques, restent larges à la fin de l'inflation. 
Dans la dernière partie~\ref{sec:comparing_quantumness_criteria}, nous reproduisons un article de la thèse comparant différentes mesures de quanticité pour les états comprimés à deux modes décohérés que nous avons étudiés dans l'article précédent. Ces états émergent dans de multiples contextes physique. L'article est en conséquence très générique et rédigé dans un langage de théorique quantique de l'information. Pour ces états, l'effet de la décohérence est paramétrisé simplement par la valeur de la pureté de l'état, qui décroit avec le degré d'interaction avec l'environnement. De tels états sont dits mixtes. Nous montrons que, même pour cette classe simple d'états mixtes, les critères sont inéquivalents, mais que pour tous les critères, la nature classique ou quantique de l'état est le résultat d'une compétition entre le niveau de pureté et le niveau de compression.

Le troisième chapitre décrit les progrès effectués dans la description théorique~\ref{chapt:analogue_preheating} d'une expérience de préchauffage analogue utilisant un gaz d'Hélium métastable piégé dans un piège magnétique quasi unidimensionnel.
Nous commençons~\ref{sec:presentation_exp} par décrire l'idée générale de l'expérience et les paramètres du gaz étudié.
Nous revenons ensuite~\ref{sec:model_exp} en détails sur la modélisation de l'expérience. Nous décrivons ce qu'est un état condensé, suivons la dynamique du gaz dans l'approximation d'une condensation complète puis étudions l'évolution de la partie non condensée via l'approche de Bogoliubov-de Gennes.
Le rôle des oscillations de l'inflaton dans le préchauffage qui produit une amplification paramétrique des autres champs dans l'Univers est joué dans l'expérience par l'oscillation radiale de la partie condensée du gaz qui génère des excitations longitudinales dans la partie non condensée.
L'oscillation radiale est produite par une modulation de la fréquence de piégeage radiale. Nous étudions la dynamique des excitations longitudinales en absence, puis en présence de modulation. Nous montrons dans ce dernier cas que la modulation produit une compression des modes normaux de direction opposées $\pm k$ en espace de Fourier. Cette compression se traduit par une création de paires de quasi-particules corrélées.
L'analogie avec le préchauffage et les moyens de mesurer cette production sont détaillés dans la partie suivante~\ref{sec:parametric_amplification}.
En particulier, nous insistons sur la nécessité de mesurer la non-séparabilité des paires produites afin de certifier qu'elles ont émergés du vide quantique et non d'une stimulation de la population thermique préexistante. 
Nous revenons également dans cette partie sur les résultats expérimentaux obtenus dans la première réalisation de l'expérience qui n'avait pas réussi à démontrer la non-séparabilité des paires.
Il avait alors été suggéré que des interactions entre les paires produites pouvaient détruire les corrélations quantiques. La suite de ce chapitre consiste précisément en l'étude de ces interactions afin d'évaluer leur capacité à faire disparaitre les corrélations quantiques générées dans l'expérience.
Nous commençons~\ref{sec:Madelung} par expliquer que le caractère unidimensionnel du gaz nous oblige à adopter un schéma perturbatif en termes de densité et de phase pour pouvoir utiliser l'approche de Bogoliubov-de Gennes.
Ce schéma perturbatif est ensuite utilisé dans l'article reproduit dans la partie suivante~\ref{sec:decay_nk} dans lequel il est montré que les interactions entre modes normaux peuvent mener au transfert des excitations produites vers d'autres modes, faisant effectivement décroitre le nombre et la corrélation des modes normaux produits.
Les interactions dominantes sont des processus dit de Beliaev et Landau stimulés par la population thermique de quasi-particules. Une quasi-particule créée dans un mode donné peut-être transférée dans un mode voisin par collision avec une quasi-particule thermique.
Nous calculons le temps de vie des quasi-particules dans un mode donné dû à ces processus.
L'article s'appuie largement sur des simulations numériques basées sur l'approximation de Wigner tronquée pour modéliser l'évolution du gaz. Ces simulations confirment nos  prédictions analytiques et en particulier le temps de vie calculé pour les modes normaux.
Nous donnons des détails supplémentaires sur ces simulations dans la partie~\ref{sec:TWA_simulation_1D_BEC}.
Enfin, dans la dernière partie~\ref{sec:decay_ck} nous présentons nos derniers résultats sur le temps de vie de la corrélation entre les modes normaux. Ce temps de vie apparait égal à celui de la population des modes. Cette égalité des temps de vie était une hypothèse du modèle effectif utilisé dans la littérature pour décrire l'effet des interactions sur la production de quasi-particules. En se basant sur un seuil estimé dans la littérature, nous montrons que les processus que nous avons identifiés semblent suffisants pour expliquer l'absence de corrélation quantique dans la première réalisation de l'expérience. Il s'agit du résultat majeur pour cette partie de la thèse.

La dernière partie revient sur les résultats obtenus durant la thèse et proposent quelques directions possibles de poursuite directe des travaux ainsi que des perspectives plus larges pour les domaines dans lesquels ils s'insèrent.

%Table of content
\newgeometry{top=4cm, bottom=4cm, left=2cm, right=2cm}

%%%%
%Before the Table of Content we update the parameter so that the links within the text are shown in black

\hypersetup{ linkcolor=black}

\tableofcontents

%Restore standard LaTeX values
%https://stackoverflow.com/questions/1670463/latex-change-margins-of-only-a-few-pages

\restoregeometry

%%%%
%After the Table of Content we update the parameter so that the links within the text are shown in color

\hypersetup{ linkcolor=ViridianGreen}

%Two lines added to be able to specifically change the back-ref to an un-numbered section
%https://tex.stackexchange.com/questions/62742/how-to-label-ref-an-un-numbered-section
\newcounter{structure}
\renewcommand{\thestructure}{Structure}

\chapter*{Structure of the manuscript} 

%Two other lines added to be able to specifically change the back-ref to an un-numbered section
%https://tex.stackexchange.com/questions/62742/how-to-label-ref-an-un-numbered-section
 \refstepcounter{structure}
\label{sec:structure}

%For the chapter to still appear in the table of contents despite
%not having a number
\addcontentsline{toc}{chapter}{\protect\numberline{}Structure of the manuscript}%

In this thesis, we have considered time-dependent scenarios relevant to cosmology and considered the amplification of quantum fluctuations in such scenarios. The analysis focused on the evolution of the `quantumness' of the amplified fluctuations. We conducted this study in the case of inflationary cosmology and an `analogue' preheating experiment.
In order to make this thesis by published works as self-contained as possible, we start in Chapt.~\ref{chapt:introduction} with a long introduction. It first goes over the necessary basics of relativistic cosmology in Sec.~\ref{sec:intro_cosmo}. In Sec.~\ref{sec:inflation}, we then review the motivations for cosmic inflation, its simplest implementation via single field slow-roll inflation and the consequences for the statistics of primordial inhomogeneities as well as, very briefly, for the production of particles in the early Universe.
In the latest part of the introduction, Sec.~\ref{sec:introduction_AG}, we briefly introduce analogue gravity experiments.
Following this introduction, in Sec.~\ref{chapt:quantumness_cosmo}, we reproduce the articles published during the PhD about quantum aspects of cosmological perturbations~\cite{Micheli:2022tld,Martin:2021znx,Martin:2022kph}.
The following chapter, Chapt.~\ref{chapt:analogue_preheating}, is then dedicated to the analogue preheating experiment studied during this PhD. The first sections explain in detail the set-up of the experiment. We also review the results of a series of publications of which the contribution of this PhD~\cite{Micheli:2022zet} is a continuation.
Finally, in the last chapter Chapt.~\ref{chapt:conclusion}, we draw some conclusions on the work done and give possible directions for what could be investigated next.

\chapter{Introduction}

\label{chapt:introduction}

\section{Homogeneous cosmology}

\label{sec:intro_cosmo}

We start this manuscript with a general presentation of cosmology. This presentation is relatively standard and inspired by several textbooks~\cite{Weinberg:2008zzc,Baumann:2022mni}, as well as lectures given at the ICFP master in Paris by Jérôme Martin, Sébastien Renaux-Petel and Marios Petropoulos.

\subsection{Gravity described by general relativity}
\label{sec:GR}

Cosmology is the study of the history of the universe on large scales, typically scales larger than the typical distance in between galaxies \ie $d \gg 1 \mathrm{Mpc} \approx 10^{22} \mathrm{m}$~\cite{Baumann:2022mni}. 
What are the relevant physical ingredients to take into account over these scales?
Of the four fundamental interactions (gravity, electromagnetic, weak and strong forces), gravity is the one that dominates
on cosmological scales.
First, the two latter have a very short range of interaction, while electromagnetic force and gravity
have an infinite range.
Of these two, gravity is by far the weakest. For instance the gravitational attraction between an electron and a proton is $40$ orders of magnitude smaller than the electromagnetic one. Still, electromagnetic charges can be positive and negative, leading to the phenomenon of screening. 
Therefore, on cosmological scales, assuming the universe to be charge neutral, the electromagnetic force plays no role, and gravity will be the main player
in defining the evolution of cosmological structures.
Currently, the most accurate theory describing the gravity force is General Relativity (GR) formulated by Einstein~\cite{Einstein:1915ca}.
We briefly recap the mathematical and physical concepts of GR necessary for the discussions in this manuscript
and refer to~\cite{Weinberg:1972kfs,Weinberg:2008zzc,Baumann:2022mni} for further details.
In GR, events, e.g. the emission/reception of a particle, are represented as points on a 4-dimensional Lorentzian manifold $\mathcal{M}$. A manifold is a space for which the neighbourhood of any event looks like that of Special Relativity: events $\mathcal{E}$ are represented by 4-vectors $x_{\mathcal{E}  } = \left( c t , \bm{x} \right)$ in $\setR\times\setR^3$ and the spacetime distance in between two infinitesimally close events $\mathcal{E}_1$ and $\mathcal{E}_2$ such that $x_{\mathcal{E}_2}^{\mu} = x_{\mathcal{E}_1}^{\mu} + \dd x^{\mu}$  is given by
\begin{equation}
    \dd s^2 = \eta_{\mu \nu} \dd x^{\mu} \dd x^{\nu} \, ,
\end{equation}
where the 2-tensor $\eta_{\mu \nu}$ is the Minkowski metric
\begin{equation}
     \eta_{\mu \nu} = \begin{pmatrix} -1 & 0 & 0 & 0 \\
     0 & 1 & 0 & 0 \\
     0 & 0 & 1 & 0 \\
     0 & 0 & 0 & 1
     \end{pmatrix} \, .
\end{equation}
We adopt for the rest of this manuscript the mostly-pluses convention $(-+++)$ for the signature of the metric~\cite{Baumann:2022mni}.
The Lorentzian manifolds used in GR are precisely equipped with a globally defined metric tensor $g_{\mu \nu}$ that
locally, by choosing an appropriate system of coordinates, reads like the Minkowski metric  $\eta_{\mu \nu}$. $g_{\mu \nu}$ dictates how distances and time duration between events are measured at a given manifold point. Generically the spacetime distance
between infinitesimally close events reads
\begin{equation}
    \dd s^2 = g_{\mu \nu} \left( x \right) \dd x^{\mu} \dd x^{\nu} \, .
\end{equation}

In GR the metric encodes all the information about the geometry of spacetime. 
It is a \textit{dynamical} quantity, which is affected by the matter-energy content of spacetime. 
Einstein's field equations give the relation in between the two
\begin{equation}
\label{eq:Einstein_eq}
G_{\mu \nu } = \frac{8 \pi \GN}{c^4} T_{\mu \nu} \, ,
\end{equation}
where $G_{\mu \nu } $ is the Einstein tensor, and $ T_{\mu \nu}$ is the stress-energy tensor characterising the matter-energy content.
$G_{\mu \nu } $ is related to the Ricci tensor $R_{\mu \nu}$, and the scalar curvature $R$, by
\begin{equation}
\label{def:einstein_tensor}
    G_{\mu \nu } = R_{\mu \nu} - \frac{1}{2} R g_{\mu \nu} \, .
\end{equation}
These two quantities are related to the Riemann tensor $R\indices{^{\mu}_{\nu}_{\rho}_{\sigma}}$ by
\begin{equation}
  R = R\indices{^{\mu}_{\mu}} ; \quad  R_{\mu \nu} =  R\indices{^{\alpha }_{\mu \alpha  \nu} } \,
\end{equation}
The Riemann tensor itself is completely built from second-derivatives of the metric\footnote{Different conventions exist for the expression of the components of the Riemann tensor. We follow the conventions of~\cite{Baumann:2022mni}.} 
\begin{equation}
    R\indices{^{\mu}_{ \nu \rho \sigma} } = \partial_{\rho} \Gamma^{\mu}_{\nu \sigma} - \partial_{\sigma} \Gamma^{\mu}_{\nu \rho} + \Gamma^{\mu}_{\rho \lambda} \Gamma^{\lambda}_{\nu \sigma} - \Gamma^{\mu}_{\sigma \lambda} \Gamma^{\lambda}_{\nu \rho}
\end{equation}
where
\begin{equation}
\label{def:levi_civita}
    \Gamma^{\mu}_{\nu \rho} = \frac{1}{2} g^{\mu \lambda} \left( \partial_{\nu} g_{\rho \lambda} + \partial_{\rho} g_{\nu \lambda} - \partial_{\lambda} g_{\nu \rho} \right) \, ,
\end{equation}
is the Levi-Civita connection.  Einstein's field equations~\eqref{eq:Einstein_eq} can be derived 
from an action principle. The evolution of the metric is described by the Einstein-Hilbert action 
\begin{equation}
\label{def:einstein_hilbert}
    S_{\mathrm{EH}} = \frac{c^3}{16 \pi G} \int \dd^4 x \sqrt{-g} R \, ,
\end{equation}
where $g$ is the determinant of the metric $g_{\mu \nu}$, 
while that of matter is described by the action
\begin{equation}
    S_{\mathrm{m}} = \frac{1}{c} \int \dd^4 x \sqrt{-g} \mathcal{L}_{\mathrm{m}} \, ,
\end{equation}
where $\mathcal{L}_{m}$ is a Lagrangian describing the behaviour of matter fields.
Defining
\begin{equation}
\label{def:stress_tensor}
    T_{\mu \nu} = \frac{-2}{\sqrt{-g}} \frac{\delta \sqrt{-g} \mathcal{L}_{\mathrm{m}}}{\delta g^{\mu \nu}} \, ,
\end{equation}
Eq.~\eqref{eq:Einstein_eq} is equivalent to the Euler-Lagrange equations for the total action
$S = S_{\mathrm{EH}} + S_{\mathrm{m}}$ varied with respect to $g^{\mu \nu}$.
Once the metric of spacetime is known, we can compute the trajectories of test bodies \ie 
bodies whose energy is small enough so that the modification they induce on the local metric can be neglected as a first approximation e.g. photons, dust grains or satellites. The spacetime trajectories $\underline{X} \left( u \right) = \left\{ x^{\mu} \left( u \right)\right\}$ of such test bodies, where $u$ is an affine parameter, are given by the \textit{geodesics} curves of this spacetime. Affinely parameterised geodesics are solutions of 
\begin{equation}
\label{def:geodesic_equation}
\frac{\dd^2 x^{\mu}}{\dd u^2} + \Gamma^{\mu}_{\nu \lambda} \frac{\dd x^{\nu}}{\dd u}  \frac{\dd x^{\lambda}}{\dd u}  = 0 \, .
\end{equation}
This equation is a generalisation to curved spacetime of Newton's second law (in a gravitational field), and there are several ways to derive it. Eq.~\eqref{def:geodesic_equation} can, for instance, be derived by requiring that test bodies going from $\underline{X} \left( u_1 \right)$ to $\underline{X} \left( u_2 \right)$ follow trajectories whose spacetime length is locally extremal in the sense that this length $\ell = \int_{u_1}^{u_2} \dd s \left( u \right)$ is stationary under infinitesimal changes around the trajectory~\cite{Weinberg:2008zzc}.
For massive particles, the 4-velocity $\dd \underline{X} \left( u \right) / \dd u$ is time-like, \ie $ g_{\mu \nu} \frac{\dd x^{\mu}}{\dd u}  \frac{\dd x^{\nu}}{\dd u} < 0$, and $u$ can be chosen to be the proper time $\tau$ associated to the particle rescaled by its mass $ u = c \tau / m$ (which changes the dimension of this parameter) so that
\begin{equation}
    g_{\mu \nu} \frac{\dd x^{\mu}}{\dd \tau}  \frac{\dd x^{\nu}}{\dd \tau}  = - m^2 \, .
\end{equation}
Then
\begin{equation}
\underline{P} = m \frac{\dd \underline{X} \left( \tau \right)}{\dd \tau} = \left( \frac{E}{c} , P^{i} \right) \, ,
\end{equation}
is the energy-momentum vector of the particle. 
$E$ is the energy of the particle, while $p = \sqrt{g_{ij}P^{i} P^{j}}$ is its physical three-momentum, such that $E^2 = p^2 c^2 + m^2 c^4$ by normalisation of the vector.
For massless particles, e.g. photons, there is no notion of proper time and $\dd \underline{X} \left( u \right) / \dd u$ is such that
\begin{equation}
    g_{\mu \nu} \frac{\dd x^{\mu}}{\dd \tau}  \frac{\dd x^{\nu}}{\dd \tau}  = 0 \, .
\end{equation}
The parameter $u$ can still be chosen so that $\underline{P} = \dd \underline{X} \left( u \right) /\dd u$ is the energy-momentum vector of the particle~\cite{Baumann:2022mni,Weinberg:2008zzc}. The normalisation of the 4-vector then gives $E = p c$. 
The geodesic equation Eq.~\eqref{def:geodesic_equation} can be conveniently rewritten as
\begin{equation}
\label{eq:geodesic_p}
    P^{\mu} \nabla_{\mu} P^{\nu} = 0 \, ,
\end{equation}
valid for massive and massless particles, where $\nabla_{\mu}$ is the covariant derivative with respect to the Levi-Civita connection.

\subsection{General relativistic cosmology}
\label{sec:GRcosmo}

Having reviewed the necessary basics of general relativity, we can apply the theory to the distributions of energy and matter in the Universe on large scales. 

\subsubsection{Geometry of the Universe: FLRW}

It is an observational fact that the distribution of matter in the (observable) Universe is isotropic on large scales, larger than $100 \mathrm{Mpc}$~\cite{Weinberg:2008zzc}.\footnote{We ignore the CMB dipole attributed to the peculiar velocity of our local group, or we work in a coordinate system boosted with respect to Earth so that this dipole is eliminated.}
We assume, in addition, that we, human observers, 
do not occupy a specific spatial position in the Universe and that the Universe, therefore, would appear as isotropic from any other location. It is the so-called cosmological principle~\cite{Weinberg:1972kfs}, and implies that the Universe is also homogeneous. The isotropy and homogeneity of the Universe translate in the mathematical framework of GR as the 
existence of a class of coordinate systems \ie of observers, in which the metric is invariant by translation and rotation. One can show that the metric in such coordinates is then of the Friedmann-Lemaître-Robertson-Walker (FLRW) form~\cite{Weinberg:1972kfs}. The line element reads
\begin{equation}
\label{def:FLRW}
    \dd s^2 = - c^2 \dd t^2 + a^2 (t) \left( \frac{\dd r^2}{1- \mathcal{K} r^2} + r^2 \dd \theta^2 + r^2 \sin^2 \theta \dd \varphi^2  \right) \, ,
\end{equation}
where the time coordinate $t$ is referred to as cosmic time, $\left( r , \theta , \varphi \right)$ are the spherical \textit{co-moving}
coordinates, $\mathcal{K}$ is called the spatial curvature, and $a(t)$ is the scale factor. The scale factor is the only dynamical quantity in Eq.~\eqref{def:FLRW}. The evolution of the homogeneous cosmological spacetime thus boils down to that of $a(t)$.

\subsubsection{Time and distances in expanding Universe}
\label{sec:physics_FLRW}

Let us briefly discuss how to describe distances and time intervals in cosmology with the FLRW metric.
We start with the spatial curvature $\mathcal{K}$. First, notice that if $\mathcal{K} = 0$ and $a(t)=1$, the FLRW metric reduces to the Minkowski metric in spherical coordinates.
For $\mathcal{K} \neq 0$, we can always set $\mathcal{K} = \pm 1$ by re-parameterising the comoving coordinates $\bm{r}^{\prime} = \sqrt{\mathcal{K}} \bm{r}$ and $a^{\prime} = a / \sqrt{\mathcal{K}}$. Notice that the density parameter for curvature $\Omega_{\mathcal{K}} = \mathcal{K} / a^2$, see Sec.~\ref{sec:Fried_eq} for the origin of this quantity, is invariant under such re-parameterisation.
Generically, the spatial metric induced by the metric \eqref{def:FLRW} on $t=\text{cst}$ hypersurfaces lead to the standard 3-dimensional Euclidian distance for $\mathcal{K}=0$, 
and to the geodesic distance over a unit sphere (respectively hyperbola) for $\mathcal{K} = +1$ (resp. $\mathcal{K} = -1$). $\mathcal{K}$ thus
encodes the geometry of the spatial sections of the Universe.
Let us now consider an arbitrary scale factor $a(t)$ and assumes $\mathcal{K} =0$ for simplicity. We compute the \textit{physical} distance in between two points located at fixed co-moving coordinates $\bm{r}_1 = \left( r_1 , 0 , 0 \right)$
and $\bm{r}_2 =\left( r_2 , 0 , 0 \right)$, as a function of cosmic time $t$. 
This physical distance is given by integrating the induced metric on $t=cst$ hypersurfaces in between these two points
\begin{equation}
   d = \int_{\bm{r}_1}^{\bm{r}_2} \sqrt{ g_{i j} \dd x^{i} \dd x^{j} } = a(t) \left| r_1 - r_2\right| \, ,
\end{equation}
where the summation using Latin indices, such as $i$, is limited to spatial indices, as opposed to summation using Greek indices, such as $\mu$, which covers all spacetime coordinates.
The scale factor $a(t)$, which evolves as a function of cosmic time, leads to dilation or contraction of the physical distances in the Universe. We come back to this point at the end of this part.
What are the trajectories of test bodies in such an expanding Universe? To answer, we have to solve the geodesics equation, Eq.~\eqref{eq:geodesic_p}. The Christoffel symbols for the FLRW metric and the computation details are given in Appendix~\ref{app:geodesics_FLRW}. Two special cases are worth mentioning. First, test bodies at rest in co-moving coordinates \ie with no peculiar velocity $\dd x^{r} / \dd u = 0$, follow geodesics curves. Such bodies are called \textit{co-moving} observers. Their proper time $\tau$ corresponds to cosmic time $t$. A second important case is that of photons. In Appendix~\ref{app:geodesics_FLRW}, it is shown that if we consider a photon received today at $t_0$ and trace back its evolution until an earlier time $t$, we find that
\begin{equation}
    E (t) = \frac{a_0}{a(t)} E_0 \, ,
\end{equation}
where $E_0$ is its energy, as seen by a co-moving observer, when received. The energy of a photon is directly related to its frequency via $E = h \nu$ where $h$ is the Planck constant. Therefore, independently of its reception frequency $\nu_0$, the photon was redshifted by an amount given by
\begin{equation}
\label{def:redshift_photons}
    \frac{\nu}{\nu_0} = \frac{a_0}{a} = 1 + z \, ,
\end{equation}
where $z$ is called the redshift. $z$ is a convenient proxy to parameterise the evolution of quantities in the FLRW metric. First, the redshift of a light source can be directly measured by spectroscopy \ie by comparing the frequencies of lines in the spectrum of the source to their tabulated values e.g.~\cite{Iye:2006mb}. Additionally, contrary to cosmic time $t$, the use of the redshift $z$ to label the time of occurrence of different events, e.g. emission of lights from different astrophysical objects, does not depend on the details of the evolution of $a \left( t \right)$.
Therefore, in the rest of this text, we often refer to the redshift of events rather than their time of occurrence.
We give a list of approximate redshifts used in the manuscript in Tab.~\ref{table:cosmo_redshifts}.

\begin{table}[]
    \centering
\begin{center}
\begin{tabular}{||c c c c c c||} 
 \hline
$ z_{\mathrm{GUT}}$ (without inflation) & $z_{\mathrm{dec \, \nu}}$  & $z_{\mathrm{BBN}}$  & $z_{\mathrm{LSS}}$ & $z_{\mathrm{eq}}$ & $z_{\mathrm{m}}$  \\ [0.5ex] 
 \hline\hline
 $10^{29}$ & $ 6 \times 10^9$ & $4 \times 10^8$ & 1090 & 3400 & 0.297  \\ [1ex] 
 \hline
\end{tabular}
\end{center}
\caption{Approximate values of redshift used in the manuscript. The approximation $z_{\mathrm{GUT}}$ is computed in Sec.~\ref{sec:validity_BBmodel}. $z_{\mathrm{dec \, \nu}}$ and
$z_{\mathrm{BBN}}$ are taken from Tab.~3.1 of~\cite{Baumann:2022mni}. 	 $z_{\mathrm{LSS}}$ is reported in \cite{Planck:2013pxb,Planck:2018vyg} where it is named $z_{\star}$.  The approximations $z_{\mathrm{eq}}$ and $z_{\mathrm{m}}$ are computed in Sec.~\ref{sec:compo_universe}. }
\label{table:cosmo_redshifts}
\end{table}

We close this discussion with a few words on the notion of distances in cosmology. Notably, neither the physical nor the co-moving distance between two objects can be directly measured in cosmology. The information we have about objects in astrophysics and cosmology comes from the light (and, in a few instances, the gravitational waves) that they emit. By measuring it, we try to infer the distance to the object. 
For instance, in a Newtonian spacetime, the perceived flux $L_{\mathrm{R}}$ of a static object located at a distance $d$ from us decays as
\begin{equation}
	L_{\mathrm{R}} = \frac{L_{\mathrm{E}}}{4 \pi d^2} \, ,
\end{equation}
where $L_{\mathrm{E}}$ is the luminosity emitted, and we assumed the emission to be isotropic. 
Therefore, if we know $L_{\mathrm{E}}$, by measuring $L_{\mathrm{R}}$, we can infer the distance to the object $d$. 
Because light propagation from the source is affected by the expansion of the Universe, this relation breaks down in FLRW Universe.
If we define the luminosity distance $d_{\mathrm{L}}$ by
\begin{equation}
	L_{\mathrm{R}}  = \frac{L_{\mathrm{E}}}{4 \pi d_{\mathrm{L}}^2} \, ,
\end{equation}
we find that~\cite{Weinberg:2008zzc}
\begin{equation}
	\label{def:luminosity_distance}
	d_{\mathrm{L}} (t_{\mathrm{R}}) = d_{\mathrm{E}/\mathrm{R}} (t_{\mathrm{E}}) \left( 1 + z_{\mathrm{E}/\mathrm{R}} \right)^2 \, ,
\end{equation}
where $d_{\mathrm{E}/\mathrm{R}} (t_{\mathrm{E}}) = a(t_{\mathrm{E}}) r_{\mathrm{E}/\mathrm{R}}$ is the physical distance 
to the source at the time of emission, and $z_{\mathrm{E}/\mathrm{R}} = a(t_{\mathrm{R}})/ a(t_{\mathrm{E}}) -1$ is the redshift to the source.
Therefore, if we trust the description of cosmology based on GR, in order to compute the physical distance to an object 
we also need an independent measure of its redshift $z_{\mathrm{E}/\mathrm{R}}$.
We can also view Eq.~\eqref{def:luminosity_distance} as a prediction of cosmological models with an expanding/contracting Universe which can be tested.
Generically, $z_{\mathrm{E}/\mathrm{R}}$ depends on the time of emission and reception of light, so on the distance to object $d_{\mathrm{E}/\mathrm{R}}$.
This relation depends on the details of the evolution of $a(t)$ \ie on the considered cosmological model.
Still, for sources close enough such that $z_{\mathrm{E}/\mathrm{R}} \ll 1$, we can derive the approximate relation
\begin{equation}
	\label{eq:Hubble_law}
	d_{\mathrm{L}} (t_{\mathrm{R}}) \approx \frac{z_{\mathrm{E}/\mathrm{R}}}{c H \left( t_{\mathrm{R}} \right)} \, ,
\end{equation}
where 
\begin{equation}
	\label{def:Hubble_parameter}
	H = \frac{\dot{a}}{a} \, ,
\end{equation}
is the rate of expansion of the Universe and is called the Hubble parameter. Eq.~\eqref{eq:Hubble_law} can be tested by plotting the luminosity distance of objects as a function of their measured redshifts (correcting for any Doppler effect). The first to put forward this linear relationship and to use observational data to extract the rate was Lemaître in 1927~\cite{Lemaitre:1927zz}. It was also identified by Hubble in~\cite{Hubble:1929ig}, whose name is now given to the relation of Eq~\eqref{eq:Hubble_law}. The analysis was later repeated with more accurate data and confirmed the result.
The expansion of the Universe is the first confirmed prediction of GR-based cosmology. We will present others later on. 

\subsubsection{Energy-matter content of the Universe}

Having detailed the basic features of the FLRW metric in full generality, we want to compute the precise dynamics of the scale factor $a(t)$ in cosmology.
It is given by Einstein's field equations \eqref{eq:Einstein_eq}. 
The Einstein tensor on the left-hand side is entirely determined by the metric and Eqs.~(\ref{def:einstein_tensor}-\ref{def:levi_civita}). 
The right-hand side depends on the matter-energy content of the Universe, which we now specify. The average distribution of matter, represented by the stress-energy tensor $T_{\mu \nu}$, has to be compatible with the homogeneity and isotropy observed. One can show~\cite{Baumann:2022mni} that the matter must then be described by a \textit{perfect fluid}. The stress-energy tensor of a generic perfect fluid reads
\begin{equation}
	T^{\mu \nu} = \left( \rho + p \right) \frac{u^{\mu} u^{\nu}}{c^2}  + p g^{\mu \nu} \, ,
\end{equation}
where $\rho$ is the energy density of the fluid, $p$ its pressure and $u^{\mu}$ its 4-velocity vector normalised to $g_{\mu \nu} u^{\mu} u^{\nu} = -c^2$. Notice that this expression is covariant. Imposing that the fluids composing the Universe appear isotropic and homogeneous to co-moving observers further requires that the fluid is co-moving \ie $u^{\mu } = \left( c , \bm{0} \right)$ in co-moving coordinates; the fluid is at rest and the energy density $\rho$ and pressure $p$ are a function of time $t$ but independent of $\bm{x}$.
Then $T\indices{^{\mu}^{\nu}}$ assumes the simple form
\begin{equation}
	\label{eq:tensor_perfect_fluid_up_down}
	T\indices{^{\mu}^{\nu}} = \begin{pmatrix}  \rho & 0 & 0 & 0 \\
		0 & p & 0 & 0 \\
		0 & 0 & p & 0 \\
		0 & 0 & 0 & p
	\end{pmatrix} \, .
\end{equation}

\subsubsection{Friedmann's equations}
\label{sec:Fried_eq}

In the presence of a collection of perfect fluids, Einstein's field equations \eqref{eq:Einstein_eq} reduce to
the two Friedmann equations~\cite{Weinberg:2008zzc,Baumann:2022mni} 
\begin{align}
	H^2 & = \frac{8 \pi \GN}{3 c^2 } \sum_{i} \rho_i - \frac{\mathcal{K} c^2}{a^2} \label{eq:friedmann_1} \, , \\
	\frac{\Ddot{a}}{a} & = - \frac{4 \pi \GN}{3 c^2} \sum_i \left( \rho_i  + 3 p_i \right) \label{eq:friedmann_2} \, .
\end{align}
where $H$ was defined in Eq~\eqref{def:Hubble_parameter}.
Additionally, we assume that each perfect fluid is separately covariantly conserved \ie we neglect conversion between the different species of fluid 
\begin{equation}
	\label{eq:cov_conserv}
	\nabla_{\mu} T^{\mu \nu}_i = 0 \iff \dot{\rho}_i + 3 H \left( \rho_i + p_i \right) =0 \, ,
\end{equation}
where $T^{\mu \nu}_i$ is the stress-energy tensor associated to the $i$th fluid.\footnote{The covariant conservation of the total stress-energy tensor is a consequence of Einstein's field equations and not an additional assumption. In the case of a single fluid Eq.~\eqref{eq:cov_conserv} can then be derived from Eqs.~(\ref{eq:friedmann_1})-(\ref{eq:friedmann_2}). } 
To close the system of equations, we have to specify the relation between the energy density of the fluid and its pressure. This relation is called the equation of state of the fluid and depends on the nature of each fluid.
The linear ansatz
\begin{equation}
	\label{def:lin_eos}
	p = w \rho \, ,
\end{equation}
where $w$ is called the equation of state parameter, covers most of the relevant cases, e.g. pressure-less matter corresponds to $w =0$,
and radiation to $w = 1/3$.
Combining Eq.~\eqref{eq:cov_conserv} and Eq.~\eqref{def:lin_eos}, we get the evolution of the energy density of the fluid
\begin{equation}
	\label{def:density_dilution}
	\rho \left( t \right) = \rho_0 \left[ \frac{a_0}{a\left( t\right)}\right]^{3 \left( 1+ w \right)} \, ,
\end{equation}
where $\rho_0$ and $a_0$ are the quantities evaluated at present time.

\subsection{Standard model of Big Bang cosmology}
\label{sec:BBmodel}

\subsubsection{Composition of the Universe}
\label{sec:compo_universe}

In the standard model of cosmology, we summarise the Universe's composition in four components: baryonic matter, cold dark matter, radiation and dark energy.
We will denote respectively $\rho_{\mathrm{b}}$, $\rho_{\mathrm{cdm}}$, $\rho_{\gamma}$ and $\rho_{\Lambda}$ their densities. The same subscripts are used for the pressure and all quantities related to a specific form of matter.
Baryonic and cold dark matters are taken to be pressure-less perfect fluids with $w_{\mathrm{m}}=0$.\footnote{The adjective `cold' precisely refers to the fact that this unknown form of matter behaves as a pressure-less fluid as opposed to a relativistic one $w \sim 1/3$~\cite{Baumann:2022mni}.} Because they dilute identically, we gather them in a single energy density $\rho_{\mathrm{m} }$. Radiation is a perfect fluid with $w_{\gamma}=1/3$. 
Dark energy is modelled by a \textit{cosmological constant}
$\Lambda$ added to the spatial curvature in the action \eqref{def:einstein_hilbert} such that
\begin{equation}
	S_{\Lambda} = \frac{c^3}{16 \pi \GN} \int \dd^4 x \sqrt{-g} \left( R - 2 \Lambda \right) \, .
\end{equation}
This results in the modified field equations
\begin{equation}
	G_{\mu \nu } + \Lambda g_{\mu \nu} = \frac{8 \pi \GN}{c^4} T_{\mu \nu} \, .
\end{equation}
While the cosmological constant in this presentation appears in the `geometry' part of the action, the $\Lambda g_{\mu \nu}$ term can be moved to the right-hand side of the field equations and understood as the stress-energy tensor of a perfect fluid with energy density $\rho_{\Lambda} = \Lambda c^4 / 8 \pi G$ and equation of state parameter
$w_{\Lambda} = -1$.
Dark energy is then seen as a type of matter with \textit{negative} pressure.
The standard model of Big Bang cosmology is often referred to 
as the Lambda-CDM model precisely because it features a cosmological constant $\Lambda$ as dark energy and cold dark matter in addition to ordinary matter. We will come back to the necessity of introducing dark matter and dark energy in Sec.~\ref{sec:validity_BBmodel}, 
and focus here on describing the physical content of the Lambda-CDM model.

\begin{table}
\label{table:cosmo_para}
\begin{center}
\begin{tabular}{||c c c||} 
 \hline
 $H_0$  & $\Omega^{(0)}_{\mathrm{b}} h^2$ & $\Omega^{(0)}_{\mathrm{cdm}} h^2$ \\ [0.5ex] 
 \hline
$67.36 \pm 0.54$ & $0.02237 \pm 0.00015$ & $0.1200 \pm 0.0012$  \\ [1ex]
 \hline
 $\Omega^{(0)}_{\Lambda}$ & $\Omega^{(0)}_{\mathcal{K}}$ & $T_{\gamma}^{(0)}$ \\ [1ex]
\hline
 $0.6847 \pm 0.0073$ & $ 0.0007 \pm 0.0019$ & $2.72548 \pm 0.00057$  \\ [1ex]
\hline
\end{tabular}
\caption{Values of the cosmological parameters given at $68\%$ confidence. The values for $H_0$, $\Omega^{(0)}_{\mathrm{b}} h^2$, $\Omega^{(0)}_{\mathrm{cdm}} h^2$,  $\Omega^{(0)}_{\Lambda}$ and $\Omega^{(0)}_{\mathcal{K}}$ are based on 
	the results of the mission \textit{Planck}~\cite{Planck:2018vyg} where $h = H_0 \, ( 100 \,  \text{km} \times \mathrm{s}^{-1} \times \mathrm{Mpc}^{-1} )^{-1}$.
	They are obtained by performing a Markov Chain Monte Carlo (MCMC) analysis on the temperature and
	polarisation power spectra of the CMB, see Sec.~\ref{sec:inhomogeneous_universe}, combined with lensing information. 
	The first four values are obtained by assuming a spatially-flat, $\Omega^{(0)}_{\mathcal{K}} =0$, $\Lambda$CDM model with adiabatic, Gaussian initial fluctuations. 
	$H_0$ and $\Omega^{(0)}_{\Lambda}$ are best-fit directly quoted from the paper. 
	The value of $\Omega^{(0)}_{\mathcal{K}}$ comes from an MCMC analysis performed over \textit{Planck} data combined with BAO data, where the curvature is included as an extra parameter.
	The value of the CMB photons' temperature $T_{\gamma}^{(0)}$ is reported from~\cite{Fixsen:2009ug}.
}
\end{center} 
\end{table}

The first Friedmann equation \eqref{eq:friedmann_1} can be rewritten by dividing by the critical energy density
\begin{equation}
	\label{def:crit_density}
	\rho_{\mathrm{c}}(t) = \frac{3 H^2(t) c^2}{8 \pi \GN} \, .
\end{equation}
It reads
\begin{equation}
	\label{eq:density_parameters}
	\Omega_{\mathrm{b}} + \Omega_{\mathrm{cdm}} + \Omega_{\gamma} + \Omega_{\Lambda} +\Omega_{\mathcal{K}} = 1   \, ,
\end{equation}
where we have defined the density parameters for each fluid $\Omega_{X} = \rho_X / \rho_{\mathrm{c}}$.
Each density parameter can then be understood as the contribution of the fluid to the total density of energy in the Universe. 
Let us say a few words about the contribution of the spatial curvature.
We can treat the spatial curvature as a fictitious perfect fluid at the level of Friedmann equations by taking its contribution to Eq.~\eqref{eq:friedmann_1} to be its energy density $\rho_{\mathcal{K}} = - 3 \Mp^2 \mathcal{K} / a^2$. Accordingly we take $w_{\mathcal{K}} = - 1 /3$ so that \eqref{def:density_dilution} is verified and that the contribution of curvature to  Eq.~\eqref{eq:friedmann_2} vanishes. Still, notice that the resulting density parameter can contribute positively or negatively to \eqref{eq:density_parameters} depending on whether the Universe is closed or open $\mathcal{K} = \mp 1$. The measured values of the cosmological parameters at present time\footnote{All quantities evaluated at present time will be shown with an exponent $(0)$ or an index $0$.}  are given in the table Tab.~\ref{table:cosmo_para}.
From these, we can compute the values of all density parameters at present time. From the values of $\Omega^{(0)}_{\mathrm{b}} h^2$, $\Omega^{(0)}_{\mathrm{cdm}} h^2$ and $H_0$, we find $\Omega^{(0)}_{\mathrm{b}} \approx 0.049$ and $\Omega^{(0)}_{\mathrm{cdm}} \approx 0.264$.
Finally, radiation encompasses all types of relativistic species. 
We assume that neutrinos are still relativistic today and we assume that their number is dominated by that in the Cosmic Neutrino Background, and similarly for photons, see Sec.~\ref{sec:inhomogeneous_universe}.
Then we have a simple relation between the density of photons and that of neutrinos~\cite{Planck:2018vyg}
\begin{equation}
	\rho_{\nu} = 3.046 \times \frac{7}{8} \times \left( \frac{7}{11} \right)^{4/3} \rho_{\gamma} \, .
\end{equation}
In general, the energy density of relativistic fluid at thermal equilibrium is directly related to its temperature
\begin{equation}
	\label{eq:density_relativistic_fluid}
	\rho = \frac{\pi^2}{30} g X \frac{\left( k_{\mathrm{B}} T \right)^4}{\left( \hbar c \right)^3}  \, .
\end{equation}
$g$ is the number of degrees of freedom of the particle, e.g. $g_{\gamma} =2$ for the two helicities of the photons, or $g_{e-} =2$ because
the electron has spin $1/2$ and so two spin states. The factor $X$ differentiates between bosons $X=1$ and fermions $X= 7/8$.
Using the relation of Eq.~\eqref{eq:density_relativistic_fluid} for the photons of the CMB at temperature $T^{(0)}_{\gamma}$ we can compute $\rho_{\gamma}$, and deduce $\rho_{\nu}$.
Combining the two and taking the value of $H_0$ given in Tab.~\ref{table:cosmo_para} we find the density parameter of radiation
$\Omega^{(0)}_{\gamma} = 9.21 \times 10^{-5}$; the contribution of radiation is negligible at present time.

\begin{figure}
    \centering
\centering
\includegraphics[width=0.95\textwidth]{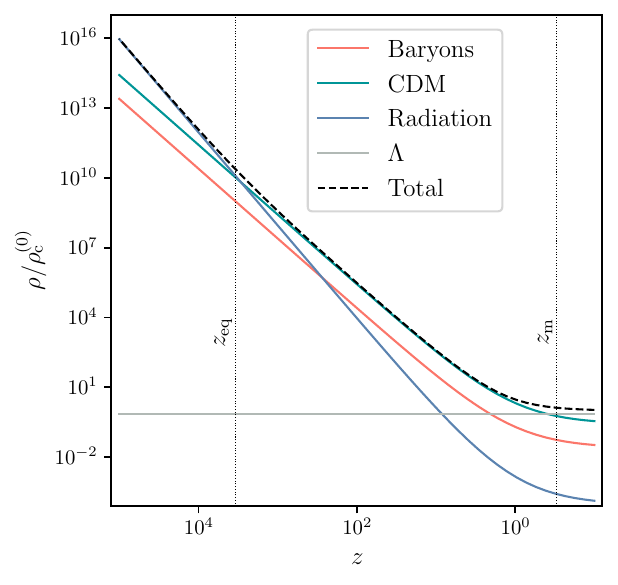} % first figure itself
\caption{Energy densities of the different constituents $\rho_i$ as a function of the redshift $z$ in logarithmic scale. The dominant constituent is seen to vary with redshift. The dotted lines show the approximate transition redshifts given in Tab.~\ref{table:cosmo_redshifts}.    }
\label{fig:compo_universe}
\end{figure}

\begin{figure}
\centering
		\includegraphics[width=0.95\textwidth]{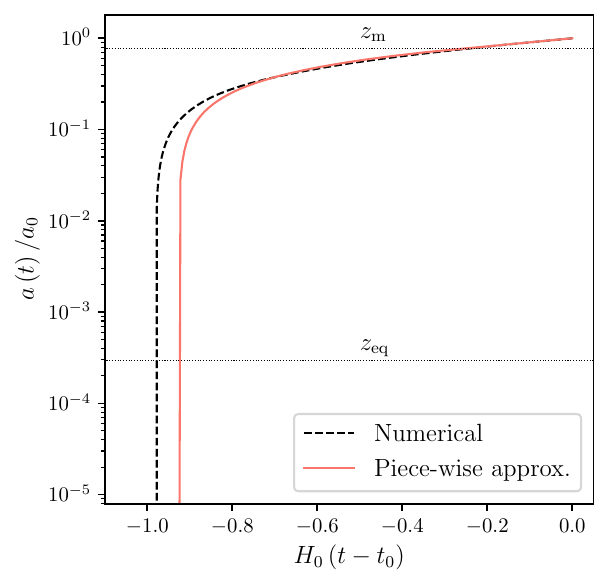} % first figure itself
		\caption{Evolution of the scale factor $a(t)$ as a function of cosmic time $t$. The dashed curve is obtained by numerically 
			solving Eq.~\eqref{eq:friedmann_1} backwards in time starting from present-day values given in Tab.~\ref{table:cosmo_para}. The red curve shows the piece-wise approximation computed in Eq.~\eqref{eq:evo_scale_factor_standard}. The dotted lines show the approximate redshifts of transition given in Tab.~\ref{table:cosmo_redshifts}.  }
		\label{fig:evol_scale_factor}
\end{figure}

From the present-day values of the density parameters, and knowing how each fluid dilutes \eqref{def:density_dilution}, we can infer the composition of the Universe at any past time as a function of the redshift $z$, see Fig.~\ref{fig:compo_universe}. 
We observe a succession of three different `eras', where the energy density of one fluid is orders of magnitude larger than that of the others. First, we have the present-day dark-energy-dominated era. Before that, we had a matter-dominated era and, even earlier, a radiation-dominated era. We can compute the redshifts of transition between these different eras by equating the density parameters of the relevant dominant contributions, e.g. between dark energy and matter, to find the redshift at which the matter domination era ended. We have
\begin{subequations}
\label{eq:transition_redshifts}
\begin{align}
\begin{split}
z_{\mathrm{m}} & = \left[ \frac{\Omega^{(0)}_{\Lambda}}{\Omega^{(0)}_{\mathrm{m}}} \right]^{1/3} - 1 \, ,
\end{split}	\\
\begin{split}
z_{\mathrm{eq}} & = \frac{\Omega^{(0)}_{\mathrm{m}}}{\Omega^{(0)}_{\gamma}}  - 1 \, ,
\end{split}
\end{align}
\end{subequations}
where  $z_{\mathrm{m}}$ (resp. $z_{\mathrm{eq}}$) is the redshift when the contributions of dark energy and matter (resp. matter and radiation) are equal. This last time is often referred to as `equality' time, hence its label. The numerical values obtained using the values of cosmological parameters given in Tab.~\ref{table:cosmo_para} are given in Tab.~\ref{table:cosmo_redshifts}
Notice that the curvature is measured to be negligible at present time.
In fact, since its value is currently small compared to that of matter and radiation, then it has always been negligible and will always be.
Indeed, as we go back in time to smaller values of $a(t)$, the contributions of matter and radiation grow faster than that of curvature, respectively, as $\rho_{\mathrm{m}} \approx a^{-3}(t)$ for matter, 
and $\rho_{\gamma} \approx a^{-4}(t)$ for radiation, versus $\rho_{\mathcal{K}} \approx a^{-2}(t)$ for curvature. 
In addition, since the budget is now dominated by dark energy, whose energy density is constant, while curvature's contribution dilutes, it will not dominate in the future as well.

One may wish to describe the evolution of the composition of the Universe as a function of cosmic time $t$, rather than as a function of the scale factor $a$. That requires solving the dynamics of the scale factor $a(t)$ as a function of $t$ given by Eq.~\eqref{eq:friedmann_1}. 
In the case where a single fluid $X$ dominates and we can neglect the other contributions in \eqref{eq:friedmann_1}, the equation is easily solved.
For a collection of fluids, it can either be solved numerically or by assuming instantaneous transitions between the different eras where a single fluid dominates the energy budget. Following the second route, and requiring the scale factor $a$ and the overall energy density $\rho$ to be continuous at the transitions\footnote{Notice that by Eq.~\eqref{eq:friedmann_1} it implies that $H$ is continuous as well.}, we get a piece-wise approximation of the dynamics. It reads
\begin{equation}
	\label{eq:evo_scale_factor_standard}
	\frac{a(t)}{a_0} =
	\begin{cases}
		e^{-H_0 \left(t_0 - t \right)} & \quad \text{for }  t > t_{\mathrm{m}} \, ,\\
		\frac{1}{1+z_{\mathrm{m}}} \left[ 1 + \frac{3}{2} H_0 \left( t - t_{\mathrm{m}} \right) \right]^{2/3}  & \quad \text{for } t_{\mathrm{m}} > t > t_{\mathrm{eq}} \, ,\\
		\frac{1}{1+z_{\mathrm{eq}}} \left[ 1 + 2 H_0 \left( \frac{1+z_{\mathrm{eq}}}{1+z_{\mathrm{m}}} \right)^{3/2} \left( t - t_{\mathrm{eq}} \right) \right]^{1/2} & \quad \text{for } t_{\mathrm{eq}} > t > t_{\mathrm{BB}} \, ,
	\end{cases}
\end{equation}
where $t_{\mathrm{m}}$ is the time when the contributions of dark energy and matter are equal, and $t_{\mathrm{eq}}$ is the time of equality. $t_{\mathrm{BB}}$ is the time of Big Bang defined by $a (t_{\mathrm{BB}}) =0$. $t_0 - t_{\mathrm{BB}}$ is then the age of the Universe.  We emphasise that the value of these times depends on the precise dynamics of the scale factor. In the piece-wise approximation above, their expressions in terms of the redshift of transition can be obtained by evaluating the above equations at transition time. 
Using the dynamics of the scale factor in Eq.~\eqref{eq:evo_scale_factor_standard} and the dilution equation, Eq.~\eqref{def:density_dilution}, we get the dynamics of the
total energy density
\begin{equation}
	\label{eq:evo_density_standard}
	\frac{\rho(t)}{\rho_{\mathrm{c}}^{(0)}} =
	\begin{cases}
		1 & \quad \text{for }  z_{\mathrm{m}} \geq z \, ,\\
		\left( \frac{1+z}{1+z_{\mathrm{m}}} \right)^3  & \quad \text{for } z_{\mathrm{eq}} \geq z \geq z_{\mathrm{m}}  \, ,\\
		\left( \frac{1+z_{\mathrm{eq}}}{1+z_{\mathrm{m}}} \right)^3 \left( \frac{1+z}{1+z_{\mathrm{eq}}} \right)^4 & \quad \text{for } z \geq z_{\mathrm{eq}}  \, .
	\end{cases}
\end{equation}
This evolution of the scale factor and density from present time is plotted in Fig.~\ref{fig:evol_scale_factor}.

\subsubsection{Thermal history of the Universe}

According to Eq.~\eqref{eq:evo_scale_factor_standard}, as we go back in time, the scale factor decreases, and the energy density of the fluids in the Universe increases, see Fig.~\ref{fig:evol_scale_factor}. 
Earlier times are therefore associated with larger energies.
Generically we can associate an energy scale $E$ to a given total energy density $\rho$ via~\cite{Baumann:2022mni}
\begin{equation}
	\label{def:energy_scale}
	E = \left( \hbar^3 c^3 \rho \right)^{1/4} = \sqrt{\hbar} c \left( 3 H^2 \Mp^2 \right)^{1/4} =  \left( 24 \pi H^2_{\mathrm{in}} t_{\mathrm{Pl}}^2 \right)^{1/4} \Mp c^2 \, ,
\end{equation}
where we have used Eq.~\eqref{eq:friedmann_1} to rewrite the energy scale in terms of the Hubble parameter and introduced the reduced Planck mass $\Mp = \sqrt{\hbar c / 8 \pi  \GN}$ and Planck time $ t_{\mathrm{Pl}}= \sqrt{\hbar \GN / c^5}$. 
In particular, notice that at a finite value of cosmic time $t_{\mathrm{BB}}$, the scale factor vanishes so that the energy density of the fluids in the Universe becomes infinite. This is the initial singularity in the Big Bang model. 
By analogy with the form of the energy density of a single relativistic fluid in Eq.~\eqref{eq:density_relativistic_fluid}, and neglecting the energy of curvature and of non-relativistic fluids, we can write~\cite{Baumann:2022mni}
\begin{equation}
	\label{def:temp_universe}
	\rho = \frac{\pi^2}{30} g_{\star} (T) \frac{\left( k_{\mathrm{B}} T\right)^4}{ \left(\hbar c \right)^3} \, ,
\end{equation}
where $T$ is a reference temperature (typically chosen to be that of photons $T_{\gamma}$), and
\begin{equation}
\label{def:effective_nbr_dof}
	g_{\star} (T) = \sum_{i} g_i X_i \left( \frac{T_i}{T} \right)^4   \, ,
\end{equation}
is called the effective number of degrees of freedom. The sum is over all the fluids making up the budget of the Universe.
Notice that here, different species of particles would be treated as a separate fluid, e.g. proton and neutrons, and not summarised in 
a common baryonic fluid.
At early enough time, when the energy is large enough, the constituents of the Universe were in thermal equilibrium in a primordial plasma at a common temperature $T$~\cite{Baumann:2022mni}. 
Then, $g_{\star}$ is just a weighted sum of the degrees of freedom of the different fluids.
Since the density of relativistic species dilutes as $a^{-4}$, Eq.~\eqref{def:temp_universe} gives an inverse proportionality relation between the temperature
and the scale factor $ T \propto a^{-1}$.\footnote{A more correct relation $ T \propto g_{\star}^{-1/3} (T)   a^{-1}$ can be obtained by considering the total entropy density~\cite{Baumann:2022mni}. The extra factor of $g_{\star}^{-1/3}$ accounts for species becoming non-relativistic, so that they drop out of the sum Eq.~\eqref{def:effective_nbr_dof}, and transferring their entropy to other species. } Note that the energy scale $E$ of Eq.~\eqref{def:energy_scale} is then proportional to the temperature.
This allows us to discuss the order of occurrence of the different events in the early Universe by referring directly to the temperature and energy of the Universe at that time rather than its redshift. 

We follow closely the account of these events made in Chapter 3 of~\cite{Baumann:2022mni}. Unless specified, the order of magnitude of the different quantities are also taken from this reference.
First, the proportionality coefficient in Eq.~\eqref{def:temp_universe} is not strictly a constant of time. The composition of the plasma changes, and so does $ g_{\star} (T)$. As the energy increases, atoms and nuclei are shattered by collisions with highly energetic particles. Even the protons and neutrons are eventually shattered into fundamental particles; the larger the energies of the collision, the larger the masses of the particles that can be produced.
Ultimately, when the temperature is around 100 GeV, the plasma contains all the fundamental particles of the Standard Model of particle physics. They are then relativistic and at thermal equilibrium.
If we consider an extension of the Standard Model, such as Grand Unification Theories (GUT)~\cite{Georgi:1974sy}, then at even larger temperature, around $T_{\mathrm{GUT}} \approx 10^{16}$ GeV for GUT~\cite{Weinberg:2008zzc}, 
other particles would also be present.

We here start our account of the events with Standard Model particles at equilibrium.
As temperature decreases, the heaviest particles stop being produced by particle-antiparticle pair creation, and they gradually decay until we are only left with electrons, positrons, protons, neutrons, photons, neutrinos and anti-neutrinos.
Initially, photons are trapped in the plasma due to Thompson scattering on electrons, while neutrinos are trapped due to interaction via weak nuclear force.
As temperature drops, $T_{\mathrm{dec \, \nu}} \approx 1 \, \text{MeV} $ the interaction rate of neutrinos with other species $\Gamma_{\nu}$ drops below the expansion rate $H^{-1}$ leading to a decoupling of the neutrinos from the plasma. Intuitively, the expansion is pulling away the reagents too fast compared to their typical interaction time
for them to have time to interact.
After decoupling, the neutrinos free-stream in the Universe. 
Therefore, the hot Big Bang model predicts a background radiation of neutrinos emitted in the early Universe. Since these neutrinos were at thermal equilibrium, this radiation should be that of a black body at temperature $ \left( 1 + z_{\mathrm{dec \, \nu}} \right)^{-1} T_{\mathrm{dec \, \nu}} \approx 1.9 \, \mathrm{K}$, where we have redshifted it to its value at present time.
Later, when the temperature is below the MeV, around $T_{\mathrm{BBN}} \sim 0.1 \, \text{Mev}$~\cite{Pitrou:2018cgg}, the so-called Big Bang Nucleosynthesis (BBN)~\cite{Alpher:1948ve} starts taking place \ie protons and neutrons fuse to
form nuclei of the lightest atoms, essentially up to the lithium. 
Only a little bit later, at $T_{\mathrm{rec.}} \approx 0.25-0.3 \, \text{eV}$, did the free electrons start to combine with protons to form
actual hydrogen atoms.
Then, since the photons were mainly interacting with free electrons via Thomson scattering, the decrease in the density of free electrons strongly reduced their interaction rate, eventually leading to their decoupling from the plasma at $T_{\mathrm{dec}} \approx 0.25 \, \mathrm{eV}$, and $z_{\mathrm{dec}} = 1100$. The hot Big Bang model thus also predicts a background black-body radiation of photons at temperature $\left( 1 + z_{\mathrm{dec}} \right)^{-1} T_{\mathrm{dec}} \approx T_{\gamma}^{(0)}$ today.

\subsubsection{Experimental confirmation and validity of the Lambda-CDM model}
\label{sec:validity_BBmodel}

Several predictions of the hot Big Bang model described above have been tested. We mention briefly a few of them in connection with the different aspects discussed previously.
First, we already mentioned in Sec.~\ref{sec:physics_FLRW} that
the predicted dynamical character of the distances in the Universe was observed as early as~\cite{Lemaitre:1927zz,Hubble:1929ig}.
Second, the model predicts the existence of two background thermal radiations, one made of neutrinos and one of photons, emitted when these particles decouple from the postulated primordial plasma.
The Cosmic Neutrino Background (C$\nu$B) has not been detected yet, as it is notoriously difficult to measure neutrinos, all the more with such small energies.
On the other hand, the Cosmic Microwave Background (CMB) of photons, where microwave refers to the current wavelength of these photons, was detected by Penzias and Wilson in 1965~\cite{Dicke:1965zz,Penzias:1965wn}. It was increasingly better studied by several missions, culminating with the Planck satellite~\cite{Planck:2013pxb}. The temperature of CMB photons is measured~\cite{Fixsen:2009ug} to be $T_{\gamma}^{(0)} \approx 2.7 K$, in accordance with the prediction of the hot Big Bang model.
The CMB is currently the best window we have in the conditions of the very early Universe, and we will come back to it repeatedly in the rest of this text.
Third, the hot Big Bang model, in combination with the standard model of particle physics, predicts the abundance of the different elements produced during BBN. These predictions are in excellent agreement with
the abundance measured (except for the abundance of Lithium, the so-called `cosmological lithium problem')~\cite{Pitrou:2018cgg}.
The hot Big Bang model thus allows us to explain most of the cosmological observations consistently. However,  it requires the introduction of dark energy 
and dark matter, the micro-physical nature of which remains elusive~\cite{Bertone:2004pz}. We briefly explain how observations lead to the introduction of dark energy and dark matter in the model.
In 1998, the authors of~\cite{SupernovaSearchTeam:1998fmf,SupernovaCosmologyProject:1998vns} extended the Hubble diagram to larger redshifts by measuring the luminosity distance and the redshift of distant supernovae.
Based on a more general version of Eq.~\eqref{eq:Hubble_law} valid for large redshifts and that parametrically depends on the density parameters of the different fluids $\Omega^{(0)}_{i}$,  
they demonstrated that expansion \textit{accelerates} and that some fluid behaving as a cosmological constant had to be introduced to account for it. This fluid is what we now refer to as dark energy.
On the other hand, the existence of dark matter can be seen as coming from observations at three different scales~\cite{Bertone:2004pz}. First, at the galactic scale, where the observed rotation curves of 
galaxies are in tension with Newtonian dynamics if we consider that the galaxies are only made up of luminous matter. 
This problem is solved by introducing supplementary matter, which is non-luminous, behaves like a pressure-less fluid and interacts only gravitationally with ordinary matter. Second, at the level of galaxy clusters, a comparison with total mass and luminous mass shows a deficit of gravitating mass that can be made up for by adding dark matter.
Finally, at cosmological scales, the measures of the CMB spectra allow us to differentiate between ordinary and dark matter and to constrain their abundances separately.

Despite its multiple successes, the hot Big Bang model cannot be valid until arbitrarily large energies. At the very least, when energies are of the order of the Planck Mass $\Mp c^2 \approx 10^{19} \, \text{GeV}$, we expect quantum gravitational effects to become relevant, and Einstein's equation should not be expected to be valid anymore~\cite{Guth:1980zm}.
Therefore, if we wish to discuss the `initial conditions' of the model, it is appropriate to set them at energies lower than this Planck scale, but the precise choice is arbitrary.
In this manuscript, we choose to set the initial conditions at a temperature close to these where the unification of the fundamental forces is postulated $T_{\mathrm{GUT}} \approx 10^{16} \, \text{GeV}$.
We can compute the associated redshift using the behaviour of the energy density in a radiation-domination era from Eq.~\eqref{eq:evo_density_standard} and equating it to the
energy density at GUT scale given by Eq.~\eqref{def:temp_universe} evaluated at $T_{\mathrm{GUT}}$. We take $g_{\star}\left( T_{\mathrm{GUT}} \right) \sim 160$, given for a GUT based on $\mathrm{SU}_5$ in~\cite{Guth:1980zm}, and find
\begin{equation}
	z_{\mathrm{GUT}} = \left( 1 + z_{\mathrm{eq}} \right)^{1/4} \left( 1 + z_{\mathrm{m}} \right)^{3/4} \left( \frac{\rho_{\mathrm{GUT}}}{\rho_{\mathrm{tot}}} \right)^{1/4} - 1 \approx 10^{29} \, .
\end{equation}

\subsection{The hot Big Bang puzzles}
\label{sec:BBpuzzles}

Additionally, the hot Big Bang model still leaves unexplained a few observational facts; these are often referred to as the Big Bang puzzles.

\subsubsection{Horizon problem}

\begin{figure}
    \centering
        \includegraphics[width=0.9\textwidth]{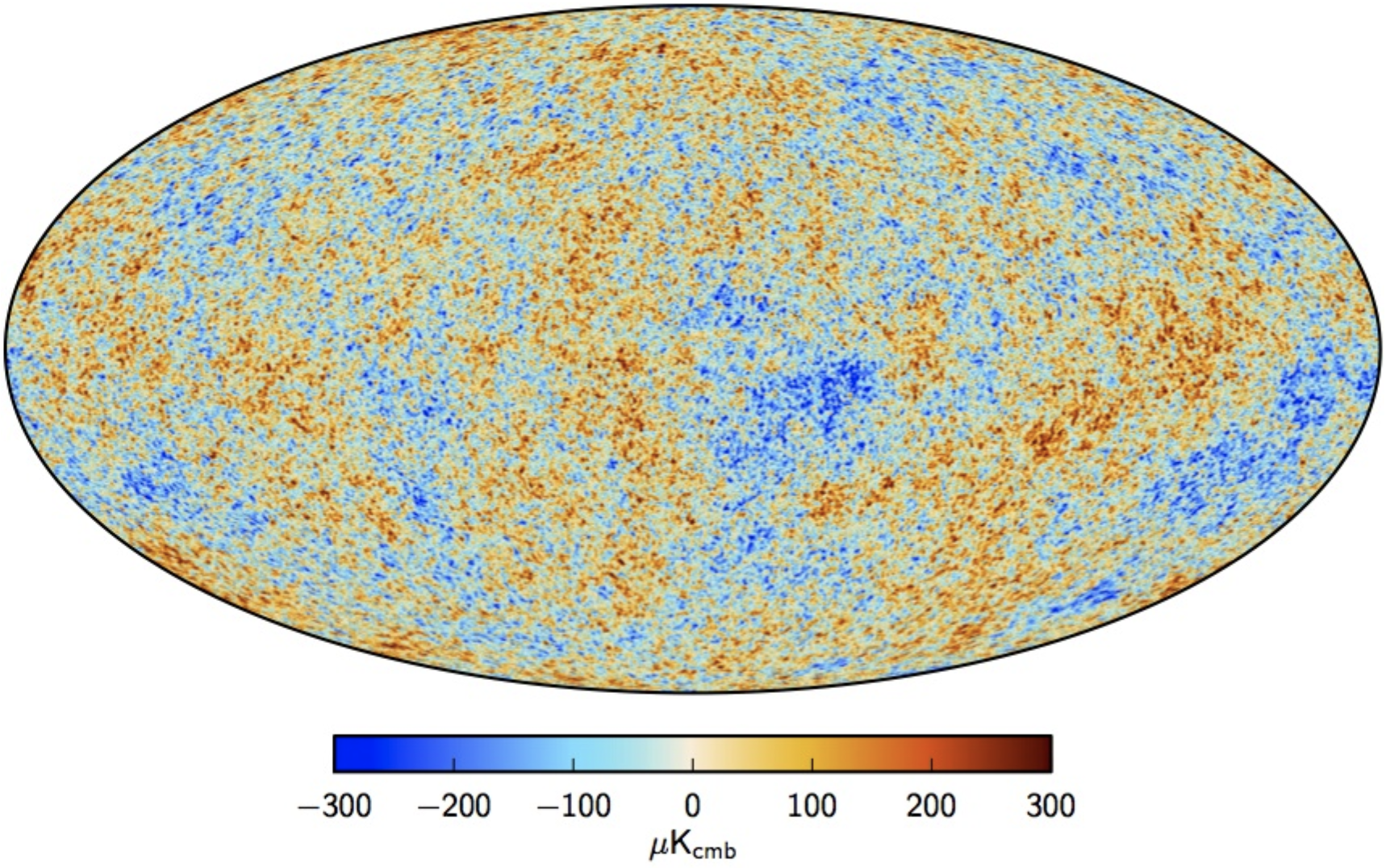} % second figure itself
        \caption{CMB map as measured by the satellite Planck~\cite{Planck:2015mrs}.  In this map the dipole, attributed to our peculiar velocity with respect to the CMB~\cite{COBE:1992syq}, and the monopole, corresponding to the average thermal emission at $T_{\gamma}^{(0)} = 2.72548 \pm 0.00057 \, \mathrm{K}$~\cite{Fixsen:2009ug}, have been subtracted. This map shows the anisotropies of the CMB which reflects the inhomogeneities in the underlying
        energy-matter content of the Universe at the time of emission. }
        \label{fig:CMB_background}
\end{figure}

First, the CMB is observed to be very isotropic $\Delta T / T \sim 10^{-5}$ once the dipole anisotropy of order $\Delta T_{\mathrm{dip}} / T \sim 10^{-3}$, attributed to the motion of the Solar system with respect ot the CMB, is removed ~\cite{COBE:1992syq,Baumann:2022mni}. A map of the CMB by Planck is given for illustration in Fig.~\ref{fig:CMB_background}. However, based on the Lambda-CDM model, any two points on this map separated by more than 1 degree should correspond to regions that were \textit{causally disconnected} at the time of emission.
Indeed, at any cosmic time $t$, two points are causally connected if they have been able to exchange a photon since Big Bang time. The distance travelled by such a photon is called the particle horizon $d_{\mathrm{H}}$. Its value is computed by following a photon (which we can always arrange to be radial) emitted at Big Bang time $t_{\mathrm{BB}}$ and received at time $t$ 
\begin{equation}
	d_{\mathrm{H}} (t) = c a(t) \int_{t_{\mathrm{BB}}}^t \frac{\dd t^{\prime}}{a \left( t^{\prime} \right)} \, .
\end{equation}
Assuming an instantaneous matter-radiation decoupling, all the photons received from the CMB come from the same 3D sphere called the Last Scattering Surface (LSS). 
If we consider the particle horizon at LSS time $t_{\mathrm{LSS}}$ we have
\begin{equation}
	\label{eq:horizon_LSS}
	d_{\mathrm{H}} \left( t_{\mathrm{LSS}} \right) = \frac{c \left( 1 + z_{\mathrm{m}} \right)^{3/2}}{H_0}  \left( \frac{2}{\sqrt{1+ z_{\mathrm{LSS}}}} - \frac{1}{\sqrt{1+ z_{\mathrm{eq}}}}   \right)  \, .
\end{equation}
We now compute the size of the particle horizon at LSS time using Eq.~\eqref{eq:horizon_LSS}. Given the current distance from us to LSS 
\begin{equation}
	\label{eq:distance_LSS}
	d_{\mathrm{LSS}} = c a_0 \int_{t_{\mathrm{LSS}}}^{t_0} \frac{\dd t^{\prime}}{a \left( t^{\prime} \right)} = \frac{c \left( 1 + z_{\mathrm{m}} \right)^{3/2}}{H_0}  \left[\frac{2 \left( 1+ z_{\mathrm{m}}\right)^2 + z_{\mathrm{m}}  }{ \left(1+ z_{\mathrm{m}}  \right)^{3/2}  } - \frac{2}{\sqrt{1+ z_{\mathrm{LSS}}}} \right]   \, . 
\end{equation}
Combining Eq.~\eqref{eq:horizon_LSS} and Eq.~\eqref{eq:distance_LSS}
we can compute the angular size separating these two points on the CMB
\begin{equation}
	\label{eq:angular_horizon}
	\delta \theta_{\mathrm{H}} = \arctan \left[ \frac{d_{\mathrm{H}} (t_{\mathrm{LSS}})}{ d_{\mathrm{LSS}}  } \right] \approx \frac{  \displaystyle  \frac{2}{\sqrt{1+ z_{\mathrm{LSS}}}} - \frac{1}{\sqrt{1+ z_{\mathrm{eq}}}}  }{ \displaystyle  \frac{2 \left( 1+ z_{\mathrm{m}}\right)^2 + z_{\mathrm{m}}  }{ \left(1+ z_{\mathrm{m}}  \right)^{3/2}  } - \frac{2}{\sqrt{1+ z_{\mathrm{LSS}}}}   } \, ,
\end{equation}
where we expanded the arctan at first order, because the angle is small, and computed all the other expressions using Eqs.~\eqref{eq:evo_scale_factor_standard}. Using the values of Tab~\ref{table:cosmo_redshifts}, we find $\delta \theta_{\mathrm{H}} \approx 0.017 \, \mathrm{rad.} \approx 1 \, \mathrm{deg.}$ as announced\footnote{For comparison, Moon's angular size seen from Earth is $0.5 \, \mathrm{deg.}$.}. This corresponds to a solid angle of $4 \pi \sin^2 \left( \delta \theta_{\mathrm{H}} / 4 \right)$, that is, we have $\sin^{-2} \left( \delta \theta_{\mathrm{H}} / 4 \right) \approx 50 000$ regions emitting photons at nearly identical temperatures, while any non-adjacent ones have not enjoyed any physical exchange allowing for equilibrium processes. We would then have to impose by hand that the initial conditions in these different regions were the same or accept that they have been set by processes violating causality. This is the horizon problem.

\subsubsection{Flatness problem}

A second puzzle comes from the low value of the spatial curvature density parameter $\Omega_{\mathcal{K}}$. Its value is constrained to be small today $\Omega_{\mathcal{K}}^{(0)} = 0.0007 \pm 0.0019$ and is compatible with $0$. Nevertheless, as detailed above, the contribution of spatial curvature to the energy budget is expected to grow with time compared to that of matter and radiation. Therefore, the fact that curvature is negligible today implies that it has to be fine-tuned to an even tinier value at earlier times. Using Eqs.~\eqref{eq:evo_scale_factor_standard}, we can compute the evolution of the Hubble parameter and so of the critical energy density as well as the evolution of the curvature energy density. We find that the density fraction of curvature reads
\begin{equation}
	\Omega_{\mathcal{K}}  = \Omega_{\mathcal{K}}^{(0)} \frac{\left( 1+ z_{\mathrm{eq}} \right) \left( 1+ z_{\mathrm{m}} \right)^3}{\left( 1+ z \right)^2} \quad \text{for } t_{\mathrm{eq}} > t > t_{\mathrm{BB}} \, .
\end{equation}
We can then, for instance, evaluate the curvature density parameter at GUT scale $z_{\mathrm{GUT}}$ and we find $\Omega_{\mathcal{K}} \left( z_{\mathrm{GUT}} \right) \approx  10^{-58}$. There is a large degree of arbitrariness in the choice of considering this time rather than a later one. In any case, even if evaluated at much lower energies, the fact still stands that the contribution of curvature is tiny.
One could argue that the curvature is simply vanishing $\mathcal{K}=0$ and that it is to be taken as another initial condition. However, as pointed out in \cite{Guth:1980zm}, the Universe is not \textit{exactly} described by an FLRW metric, and so the spatial curvature is not a parameter in the model that could be fixed to be vanishing by some symmetry principle justifying that the effect is just not actually realised in the physics we observe.
FLRW is an effective description of spacetime on large scales, and the value of the spatial curvature $\mathcal{K}$ is the result of some underlying physical processes. It would thus be more satisfactory to see the flatness of the Universe emerging as the result of a physical process. This is the flatness problem.

\subsubsection{Monopole problem}

Finally, the non-observation of magnetic monopoles is also considered a shortcoming of the model. Although absent in the standard model of particle physics, monopoles are present in high-energy extensions of it, e.g. Grand Unification Theories. They are expected to be produced in large numbers~\cite{Zeldovich:1978wj,Preskill:1979zi,Guth:1979bh,Einhorn:1980ym} in the early Universe when the energy densities get very large so that an observable number should persist until today. 
Monopoles form upon symmetry breaking when a field transition from a `fake' vacuum, which has become unstable, to a `true' vacuum, which is stable. When this vacuum has residual symmetry, the field can be found in different configurations of the new vacuum at different locations in space, forming bubbles of different configurations. Topological defects then appear at the boundaries between these bubbles, and monopoles are part of them.
The transition is a causal process, so the size of the region with a given field configuration cannot exceed the size of the particle horizon at that time $ d_{\mathrm{H}}  (t_{\mathrm{GUT}})$.
We can then put a lower bound on the number density of the monopoles forming on the boundaries of these regions $n_{\mathrm{mon}} > d_{\mathrm{H}} (t_{\mathrm{GUT}})^{-3}$.
Monopoles can only annihilate by combining with anti-monopoles, and these interactions are expected to be rapidly suppressed by the expansion of space~\cite{Zeldovich:1978wj,Preskill:1979zi} so that the number of monopoles is effectively conserved 
until today. We can then compute the present-time density parameter of monopoles 
\begin{equation}
	\Omega_{\mathrm{mon}}^{(0)} =  m_{\mathrm{mon}} c^2 \frac{H_0^3}{ \rho_{\mathrm{c}}} \frac{\left( 1 + z_{\mathrm{GUT}} \right)^3}{ \left( 1 + z_{\mathrm{eq}} \right)^{3/2} \left( 1 + z_{\mathrm{m}} \right)^{9/2} } \, ,
\end{equation}
where $m_{\mathrm{mon}}$ is the mass of a monopole. Taking  $m_{\mathrm{mon}} c^2 \approx 10^{18} \, \text{GeV}$ and $z_{\mathrm{GUT}} \approx 10^{29}$ we have
$\Omega_{\mathrm{mon}}^{(0)} \approx 10^{20}$  \ie monopoles would largely dominate the energy budget of the Universe today!
We know it is not the case and that the energy budget is dominated by dark energy.
Worse, we have yet to observe any such monopole. Searches for monopoles have constrained their number to $10^{-29}$ per nucleons~\cite{Jeon:1995rf,ParticleDataGroup:2022pth}. Taking all baryonic matter to be nucleons and assimilating all nucleons to protons of mass $m_{\mathrm{p}} c^2 = 938 \mathrm{MeV}$ we get a 
ratio of monopoles to nucleons of
\begin{equation}
	\eta_{\mathrm{mon}/\mathrm{b}} = \frac{\Omega_{\mathrm{mon}}^{(0)}}{\Omega_{\mathrm{b}}^{(0)}} \frac{m_{\mathrm{p}}}{m_{\mathrm{mon}}} \approx 2.5 \times 10^{3} \, ,
\end{equation}
more than $30$ orders of magnitude larger than the experimental upper limit. GUT models are then in clear tension with the standard hot Big-Bang model.

\section{Cosmic Inflation}
\label{sec:inflation}

The idea of cosmic inflation appeared in the study of phase transitions in the early Universe from GUT to the Standard Model~\cite{Guth:1980zm,Linde:1981mu,Albrecht:1982wi}. Guth~\cite{Guth:1980zm} recognised that a period of exponential expansion in the early Universe was a way to solve altogether the three puzzles of the Big Bang model that we detailed in the previous sub-section, Sec.~\ref{sec:BBpuzzles}. Shortly after, Linde~\cite{Linde:1981mu} and Albrecht and Steinhardt~\cite{Albrecht:1982wi} proposed a `new inflation' model to solve some problems related to the end of inflation in Guth's original formulation.
This section will present cosmic inflation as a generic possible cosmic era outside its initial study context. First, in Sec.~\ref{sec:infl_solBB_puzzles}, we show how a sufficiently long period of accelerated expansion solves the Big-Bang puzzles.
In Sec.~\ref{sec:SSR}, we then present a standard way to realise inflation using a single field slowly rolling on top of its potential. In Sec.~\ref{sec:inhomogeneous_universe}, we show, considering cosmological perturbations, that cosmic inflation gives testable and tested predictions for the statistics of inhomogeneities in the early Universe.
Finally, in Sec.~\ref{sec:preheating}, we say a few words about the connection between a period of inflation and the standard succession of eras of the hot Big Bang model.

\subsection{A solution to Big Bang puzzles}
\label{sec:infl_solBB_puzzles}

We had previously inferred from the present time values of the different density parameters $\Omega_{i}$ that the Universe was successively dominated by radiation, matter and finally, dark energy. Let us introduce, during radiation dominated era, a period when another perfect fluid $X$ dominates the energy budget of the Universe between $t_{\mathrm{in}}$ and $t_{\mathrm{f}}$ so that the scale factor dynamics is modified to 
\begin{equation}
	\begin{aligned}
		\frac{a(t)}{a_0} = \begin{cases}
			e^{H_0 (t_0 - t)} & \text{for } t \geq t_{\mathrm{m}}, \\
			\frac{1}{1+z_{\mathrm{m}}} \left[1 + \frac{3}{2} H_0 (t - t_{\mathrm{m}})\right]^{2/3} & \text{for } t_{\mathrm{m}} \geq t \geq t_{\mathrm{eq}}, \\
			\frac{1}{1+z_{\mathrm{eq}}} \left[1 + 2 H_0 \left(\frac{1+z_{\mathrm{eq}}}{1+z_{\mathrm{m}}}\right)^{3/2} (t - t_{\mathrm{eq}})\right]^{1/2} & \text{for } t_{\mathrm{eq}} \geq t \geq t_{\mathrm{f}}, \\
			\frac{1}{1+z_{\mathrm{f}}} \left[1 + \frac{3}{2} (w_{X} + 1) H_0 \phantom{ \left( \frac{1+z_{\mathrm{eq}}}{1+z_{\mathrm{m}}} \right)^{3/2} } \right. & \\
			 \left. \phantom{\frac{1}{1+z_{\mathrm{f}}} [ 1+ } \left(\frac{1+z_{\mathrm{eq}}}{1+z_{\mathrm{m}}}\right)^{3/2}   \left( \frac{1 + z_{\mathrm{f}} }{1 +  z_{\mathrm{eq}} } \right)^2 (t - t_{\mathrm{f}})\right]^{\frac{2}{3(w_{X} +1)}} & \text{for } t_{\mathrm{f}} \geq t \geq t_{\mathrm{in}}, \\
			\frac{1}{1+z_{\mathrm{in}}} \left[1 + 2 H_0 \left(\frac{1+z_{\mathrm{eq}}}{1+z_{\mathrm{m}}}\right)^{3/2} \right. & \\
			 \left. \phantom{\frac{1}{1+z_{\mathrm{f}}} [ 1+ } \left(\frac{1 + z_{\mathrm{f}}}{1 + z_{\mathrm{eq}}}\right)^2 \left(\frac{1+z_{\mathrm{in}}}{1+z_{\mathrm{f}}}\right)^{\frac{3}{2}(w_{X} + 1)} (t - t_{\mathrm{f}})\right]^{1/2} & \text{for } t_{\mathrm{in}} \geq t \geq t_{\mathrm{BB}}.
		\end{cases}
	\end{aligned}
\end{equation}

The evolution of the total energy density is accordingly changed to
\begin{equation}
	\label{eq:evo_density_infl}
	\frac{\rho(t)}{\rho_{\mathrm{c}}^{(0)}} =
	\begin{cases}
		1 & \quad \text{for }  z_{\mathrm{m}} \geq z \, ,\\
		\left( \frac{1+z}{1+z_{\mathrm{m}}} \right)^3  & \quad \text{for } z_{\mathrm{eq}} \geq z \geq z_{\mathrm{m}}  \, ,\\
		\left( \frac{1+z_{\mathrm{eq}}}{1+z_{\mathrm{m}}} \right)^3 \left( \frac{1+z}{1+z_{\mathrm{eq}}} \right)^4 & \quad \text{for } z_{\mathrm{f}} \geq z \geq z_{\mathrm{eq}}  \, , \\
		\left( \frac{1+z_{\mathrm{eq}}}{1+z_{\mathrm{m}}} \right)^3 \left( \frac{1+z_{\mathrm{f}} }{1+z_{\mathrm{eq}}} \right)^4 \left( \frac{1+z}{1+z_{\mathrm{f}} } \right)^{3(1+w_X)} & \quad \text{for }  z_{\mathrm{in}} \geq z \geq z_{\mathrm{f}}  \, , \\
		\left( \frac{1+z_{\mathrm{eq}}}{1+z_{\mathrm{m}}} \right)^3 \left( \frac{1+z_{\mathrm{f}} }{1+z_{\mathrm{eq}}} \right)^4 \left( \frac{1+z_{\mathrm{in}}}{1+z_{\mathrm{f}} } \right)^{3(1+w_X)} \left( \frac{1+z}{1+z_{\mathrm{in}} } \right)^{4} & \quad \text{for }   z \geq z_{\mathrm{in}}  \, .
	\end{cases}
\end{equation}

Notice that the redshift $z(E)$, associated with a certain energy scale $E$ larger than that of inflation, is now pushed to a larger value $z^{\mathrm{infl}}(E)$, to account for the extra degree of expansion during inflation. The link between the two can be obtained by equating the energy density of the Universe with and without the inclusion of inflation using Eq.~\eqref{eq:evo_density_standard} and Eq.~\eqref{eq:evo_density_infl}. We have 
\begin{equation}
	\label{eq:modif_redshfit_inflation}
	1 + z^{\mathrm{infl}} \left(E \right) = \left[1+  z \left(E \right) \right] e^{ \frac{N}{4} \left( 1 - 3 w_{X}\right)}  \, ,
\end{equation}
where we have introduced the number of $e$-folds
\begin{equation}
	\label{def:efolds}
	N = \ln \left( \frac{1+z_{\mathrm{f}}}{1 + z_{\mathrm{in}}} \right) = \ln \left( \frac{a_{\mathrm{f}}}{a_{\mathrm{in}}} \right) \, ,
\end{equation}
as a measure of the duration of inflation. We will have to take this change into account to keep fixing the initial condition at GUT scale, now corresponding to $z^{\mathrm{infl}}_{\mathrm{GUT}}$, to be consistent with our previous computations.
We will see that if, during this inflation period, the Universe undergoes an accelerated expansion, then we can solve the three puzzles of the hot Big Bang model. The intuitive reason behind it is that if we start from a locally homogeneous Universe over small regions and blow up any of these regions to be the size of our observable Universe, then the \textit{whole} Universe will appear homogeneous. Additionally, even if initially spatially curved, it will now appear flat as we have `zoomed in' on a relatively small section of it, making the curvature unnoticeable. Finally, suppose monopoles are produced before or during this period of inflation, and their production rate is small compared to that of expansion. In that case, their density is diluted down to tiny values.

\subsubsection{Horizon problem}

First, let us reconsider the particle horizon's angular size today in the CMB. Since the expansion history is unmodified until the radiation domination era, the expression Eq.~\eqref{eq:distance_LSS} of the co-moving distance to the LSS $r_{\mathrm{LSS}}$ is unchanged. Only the expression Eq.~\eqref{eq:horizon_LSS} of the particle horizon $r_{\mathrm{H}}$ is modified to
\begin{equation}
	\begin{alignedat}{2}
		\label{eq:particle_horizon_inflation}
		d^{\mathrm{infl}}_{\mathrm{H}} \left( t_{\mathrm{LSS}} \right) & = \frac{c \left( 1 + z_{\mathrm{m}} \right)^{3/2}}{H_0}  \Biggl\{  \frac{2}{\sqrt{1+ z_{\mathrm{LSS}}}} - \frac{1}{\sqrt{1+ z_{\mathrm{eq}}}} && \\
		& + \frac{\sqrt{1+ z_{\mathrm{eq}}}}{1+z_{\mathrm{f}}} \frac{1-3 w_{X}}{1+3 w_{X}} \left[ 1 - \left( \frac{1+z_{\mathrm{f}}}{1 + z_{\mathrm{in}}} \right)^{\frac{3 w_X + 1}{2}} \right]  \Biggl\}   \, , && \\
		& = d_{\mathrm{H}} \left( t_{\mathrm{LSS}} \right) +  \frac{c \left( 1 + z_{\mathrm{m}} \right)^{3/2}}{H_0} \frac{\sqrt{1+ z_{\mathrm{eq}}}}{1+z_{\mathrm{f}}} \frac{1-3 w_{X}}{1+3 w_{X}} \left( 1 - e^{- \frac{3 w_X + 1}{2} N } \right) &&  \, .
	\end{alignedat}
\end{equation}
First, notice that if the radiation domination era is uninterrupted \ie if the additional perfect fluid dominating from $t_{\mathrm{in,}}$ to $t_{\mathrm{f}}$ behaves as radiation \ie $w_X = 1/3$, or if the duration of inflation is negligible $N=0$, then the additional contribution vanishes as it should. 
Second, for the angular size of the horizon to grow, we need the fluid $X$ to be less stiff than radiation, \ie $w_{X} < 1/3$, or else the contribution of the term is negative. In addition, if $w_X > - 1/3$, then the term in the exponential is negative, and the term in brackets is bounded by one. The extra contribution will then be negligible because it is suppressed by $z_{\mathrm{f}}$ expected to be larger than $z_{\mathrm{BBN}} \sim 10^9$. On the other hand, if $w_X < - 1/3$, the contribution is exponential in $N$. Then, if inflation lasts long enough, we can make the angular horizon $\delta \theta_{\mathrm{H} }$ defined in Eq.~\eqref{eq:angular_horizon} arbitrarily large. For $w_X < - 1/3$, the second Friedmann equation Eq.~\eqref{eq:friedmann_2} shows that expansion accelerates. Thus, solving the horizon problem requires a phase of \textit{accelerated} expansion. How long should this period last? Let us compute the modification induced by inflation to the size of the angular horizon at present
\begin{align}
	\delta \theta_{\mathrm{H}}^{\mathrm{infl}} & = \arctan \left[ \frac{d_{\mathrm{H}}^{\mathrm{infl}} (t_{\mathrm{LSS}})}{ d_{\mathrm{LSS}}  } \right] \, ,\\
	& \approx \delta \theta_{\mathrm{H}} +  \frac{\sqrt{1+ z_{\mathrm{eq}}}}{1+z_{\mathrm{in}}} \frac{1-3 w_{X}}{1+3 w_{X}} e^{N} \left( 1 - e^{- \frac{3 w_X + 1}{2} N } \right) \nonumber \\
	& \times \left[ \frac{2 \left( 1+ z_{\mathrm{m}}\right)^2 + z_{\mathrm{m}}  }{ \left(1+ z_{\mathrm{m}}  \right)^{3/2}  } - \frac{2}{\sqrt{1+ z_{\mathrm{LSS}}}}  \right]^{-1}  \, .
\end{align}
We require that this size is larger than the celestial sphere \ie $\delta \theta_{\mathrm{H}}^{\mathrm{infl}} > 4 \pi$. 
Assuming that the inflation starts around GUT scale $z_{\mathrm{in}} = z^{\mathrm{infl}}_{\mathrm{GUT}}$, and using Eq.~\eqref{eq:modif_redshfit_inflation}, we find a lower bound on the number of e-folds that inflation should last. Neglecting 
$\delta \theta_{\mathrm{H}}$, and the $1$ in front of the exponential, the condition reads
\begin{align}
		\label{eq:efolds_horizon}
		N & > 1 + \frac{4}{1-3 w_{X}} \log \left\{  4\pi \left[ \frac{2 \left( 1+ z_{\mathrm{m}}\right)^2 + z_{\mathrm{m}}  }{ \left(1+ z_{\mathrm{m}}  \right)^{3/2}  } - \frac{2}{\sqrt{1+ z_{\mathrm{LSS}}}} \right]  \frac{1+z_{\mathrm{GUT}}}{\sqrt{1+ z_{\mathrm{eq}}}} \frac{1+3 w_{X}}{1-3 w_{X}} \right\}  \nonumber \, , \\
		& \gtrsim \frac{4}{1-3 w_{X}} \log \left( - 10 \pi \frac{1+3 w_{X}}{1-3 w_{X}}  \frac{z_{\mathrm{GUT}}}{\sqrt{ z_{\mathrm{eq}}}} \right)  \, ,
\end{align}
where, in the second line, we have used the values of the different redshifts in Tab.~\ref{table:cosmo_redshifts} to get a simpler estimate of the required number of $e$-folds. Although the precise number depends on the details of the phase of inflation, in particular on its energy scale, it only does so logarithmically so the order of magnitude will not be modified. Using our estimate $z_{\mathrm{GUT}} = 10^{29}$ and assuming that the phase of inflation is powered by a cosmological constant-like fluid $w_{X} = -1$, we get $N \gtrsim 65$. We will find that roughly the same number of e-folds is required to solve the other Big Bang puzzles.

\subsubsection{Flatness problem}

We move on to compute how the curvature density fraction at early times is modified when inflation is included. We had previously computed its value at a certain energy scale in the early Universe and found it to be un-naturally small. With the modified evolution, the curvature density parameter before inflation now reads 
\begin{equation}
	\Omega_{\mathcal{K}} \left( z \right)  = \Omega_{\mathcal{K}}^{(0)} e^{\left( 1 - 3 w_X \right) N } \frac{\left( 1+ z_{\mathrm{eq}} \right) \left( 1+ z_{\mathrm{m}} \right)^3}{\left( 1+ z \right)^2} \quad \text{for} \quad z \geq z_{\mathrm{in}} .
\end{equation}
We want to allow the curvature to be of order unity before the period of inflation starts and compute how much e-folds of inflation this requires.
Starting inflation around GUT scale again we take $\Omega_{\mathcal{K}} \left( z^{\mathrm{infl}}_{\mathrm{GUT}} \right) \approx 1$, which requires
\begin{equation}
	\label{eq:efolds_curvature}
	N >  \frac{2}{1-3w_X} \ln \left[  \frac{1  }{\Omega_{\mathcal{K}}^{(0)}} \frac{\left( 1 + z_{\mathrm{GUT} } \right)^2}{\left( 1 + z_{\mathrm{eq} } \right) \left( 1 + z_{\mathrm{m} } \right)^3} \right] \approx 66  \, ,
\end{equation}
which is almost the same duration as the one required to solve the horizon problem.

\subsubsection{Monopole problem}

Finally, we review the monopole problem as well. We assume that the monopole formed after a symmetry breaking before inflation proceeds. We can then repeat the computations of Sec.~\ref{sec:BBpuzzles} to have a lower bound on the density of monopoles, and we have
\begin{align}
	\begin{split}
		\Omega_{\mathrm{mon}}^{(0), \mathrm{infl}} & =  m_{\mathrm{mon}} c^2 \frac{H_0^3}{ \rho_{\mathrm{c}}} \frac{\left( 1 + z^{\mathrm{infl}}_{\mathrm{GUT}} \right)^3}{ \left( 1 + z_{\mathrm{eq}} \right)^{3/2} \left( 1 + z_{\mathrm{m}} \right)^{9/2} } e^{ - \frac{3}{2} N \left( 1 - 3 w_X \right)} \, , \\
		&  = \Omega_{\mathrm{mon}}^{(0)}  e^{- \frac{3}{4} N \left( 1 - 3 w_X\right) }  \, .
	\end{split}
\end{align}
The estimate of the ratio of the number of monopoles to that of baryons changes to
\begin{equation}
	\eta^{\mathrm{infl}} _{\mathrm{mon}/\mathrm{b}} = \eta _{\mathrm{mon}/\mathrm{b}} \, e^{- \frac{3}{4} N \left( 1 - 3 w_X\right) } \, .
\end{equation}
To agree with the current observations, we then have to require
\begin{equation}
	N > - \frac{4}{3} \frac{1}{1 - 3 w_X} \ln \left( \frac{10^{-29}}{\eta _{\mathrm{mon}/\mathrm{b}} }   \right)  \, .
\end{equation}
Evaluating this for a period of inflation powered by a cosmological constant-like fluid, \ie $w_X=-1$, we find that $N \gtrsim 25$ e-folds of inflation are required, which is the same order of magnitude found for the other problems. Again, changing the fluid parameter $w_X$, the energy scale of inflation by changing $z_{\mathrm{in}}$, or the scale at which we fix the initial condition for curvature, would only marginally change this result as the dependence is logarithmic.
The outcome of these computations is that a period of inflation lasting for roughly a $60$ e-folds can simultaneously solve the three Big-bang puzzles.
This initially constituted the main motivation to introduce such a period~\cite{Guth:1980zm}. In the next sub-section Sec.~\ref{sec:SSR} we demonstrate a simple way to realise such a period of inflation.

\subsection{Single field slow-roll inflation}
\label{sec:SSR}

We have effectively modelled inflation as the interruption of radiation domination era by a phase of domination by another fluid with an 
equation of state parameter $w_X$. We then showed that provided this domination is long-enough and leads to an accelerated expansion, 
\ie $w_X < -1/3$, we can solve the three Big-Bang puzzles. We now refer to this period as (cosmic) inflation. We stress that this requires the fluid to behave much differently from matter 
or radiation, mainly to have a negative pressure.
We now show a way to realise such a fluid.
It turns out to be sufficient to consider that this fluid is modelled by a scalar field $\varphi$ with a flat-enough potential $V\left(\varphi \right)$~\cite{Linde:1981mu,Albrecht:1982wi}, a scenario called single field slow-roll inflation. They are other ways to realise inflation using multiple scalar fields~\cite{Gong:2016qmq}, but single field slow-roll inflation is perfectly compatible with observations, so in the rest of this manuscript we will restrict to this simple scenario.  
%The effect of possible modifications to it on the quantumness of cosmological perturbations are briefly discussed in conclusion Sec.~\ref{chapt:conclusion}.

\subsubsection{Scalar field in FLRW}

We consider the action of a scalar field minimally coupled to gravity with a standard kinetic term
\begin{equation}
	\label{def:action_scalar}
	S_{\varphi} = - \frac{1}{c} \int \dd^4 x \sqrt{- g} \left[ \frac{1}{2} g^{\mu \nu} \partial_{\mu} \varphi \partial_{\nu} \varphi + V\left(\varphi \right)\right] \, .
\end{equation}
The field $\varphi$ will be referred to as the inflaton.
We can compute the stress-energy tensor of the field using the formula~\eqref{def:stress_tensor}, details are given in Appendix~\ref{app:tensor},
\begin{equation}
	\label{eq:Tmunu_scalar_field}
	T \indices{_{\mu \nu}} = \partial_{\mu} \varphi \partial_{\nu} \varphi - g \indices{_{\mu \nu}} \left[ \frac{1}{2}  g \indices{^{\alpha \beta}}  \partial_{\alpha} \varphi \partial_{\beta} \varphi+ V \left( \varphi \right)  \right] \, .
\end{equation}
Imposing that the scalar field distribution is homogeneous and isotropic, the spatial derivatives must vanish. We then find that the stress-energy tensor~\eqref{eq:Tmunu_scalar_field} assumes the perfect fluid form of Eq.~\eqref{eq:tensor_perfect_fluid_up_down} for
\begin{subequations}
	\begin{equation}
		\label{eq:energy_scalar}
		\rho = \frac{\dot{\varphi}^2}{2 c^2} + V \left( \varphi \right) \, ,
	\end{equation}
	\begin{equation}
		\label{eq:pressure_scalar}
		p = \frac{\dot{\varphi}^2}{2 c^2} - V \left( \varphi \right) \, .
	\end{equation}
\end{subequations}

There is \textit{a priori} no simple equation of state relating these two parameters, and we have to solve for the dynamics of the fluid to find out how
its energy and pressure evolve as the Universe expands.
Eq.~\eqref{eq:cov_conserv}, the conservation of the stress-energy tensor $T \indices{_{\mu \nu}}$, gives a first equation of evolution. For the scalar field, it gives the curved spacetime form of the Klein-Gordon equation~\cite{Klein:1926tv,gordonComptoneffektNachSchroedingerschen1926}
\begin{equation}
	\label{eq:klein_gordon}
	\Box \varphi - V^{\prime} \left( \varphi \right) c^2 = 0 
	\, ,
\end{equation}
where we introduced the d'Alembertian operator $\Box =  g \indices{_{\mu \nu}} \nabla^{\mu} \nabla^{\nu}$, and used the relation
\begin{equation}
	\Box \varphi  = - \frac{1}{\sqrt{-g}} \partial_{\mu} \left( \sqrt{-g} g^{\mu \nu} \partial_{\nu} \varphi \right) \, .
\end{equation}
valid for scalar fields. Notice the minus sign in Eq.~\eqref{eq:klein_gordon} in the mostly pluses signature convention.
For the FLRW metric we have
\begin{equation}
	\ddot{\varphi} + 3 H \dot{\varphi} + V^{\prime} \left( \varphi \right) c^2 = 0 \, .
\end{equation}
The effect of the expansion appears in the friction term proportional to the Hubble parameter. Assuming that the scalar field dominates the energy budget of the Universe, Eq.~\eqref{eq:friedmann_1}, which now reads
\begin{equation}
	\label{eq:friedmann_1_scalar_field}
	H^2 = \frac{8 \pi \GN}{3c^2} \left[  \frac{\dot{\varphi}^2}{2 c^2} + V \left( \varphi \right) \right] \, ,
\end{equation}
and Eq.~\eqref{eq:klein_gordon} form a closed system of equations completely describing the dynamics of the scalar field once the potential $V \left( \varphi \right)$ is fixed.
It only features two degrees of freedom so that initial conditions for the system are fixed by a choice of the initial field value and of its derivative $\left( \varphi_{\mathrm{in}} , \dot{\varphi}_{\mathrm{in}} \right)$,
and the subsequent evolution can be represented in a $\left( \varphi, \dot{\varphi} \right)$-plane.

\subsubsection{de Sitter and slow-roll}
\label{sec:dS_SR}

In order to get an accelerated expansion of the Universe, Eq.~\eqref{eq:friedmann_2} shows that we have to require $\rho+ 3 p < 0$, which translates for the scalar field in a domination of the potential energy over the kinetic one
\begin{equation}
	\dot{\varphi}^2 < V \left( \varphi \right) c^2 \, .
\end{equation}
Intuitively, we need the field to evolve slowly in a region of large potential energy. To solve the Big Bang puzzles, this regime should be sustained for at least $60$ e-folds.
An important limiting case is when the kinetic energy is negligible compared to the potential one $V(\varphi) c^2 \gg \dot{\varphi}^2$ in which case Eqs.~(\ref{eq:energy_scalar})-(\ref{eq:pressure_scalar}) give $p \approx - \rho$ \ie $w \approx -1$.
In such phase the Hubble parameter is a constant $H \approx H_{\mathrm{in}}$
and the expansion is exponential $a(t) = a_{\mathrm{in}} \exp [ H(t-t_{\mathrm{in}})   ]$; just as in the latest stage of the evolution \eqref{eq:evo_scale_factor_standard} when a cosmological constant dominate the energy-budget. The geometry of spacetime is then that of a de Sitter Universe.
We now consider the possibility to realise inflation by a phase of \textit{quasi-}de Sitter expansion.
Indeed, in a de Sitter spacetime, the exponential expansion never ends, while inflation ends eventually at $z_{\mathrm{f}}$, and the Universe becomes radiation-dominated.
Inflation can, therefore, only be close to a de Sitter expansion phase for a finite duration. We will require this phase to be long enough and parameterise the deviations from such a situation.
To that end, we introduce the hierarchy of Hubble flow functions $\{ \epsilon_{n}\} $ defined by
\begin{equation}
	\epsilon_{n+1} = \frac{\dd \ln \left| \epsilon_{n} \right|}{\dd N } \, ,
\end{equation}
where $\epsilon_0 = H_{\mathrm{in}} / H$ and the number of e-folds is computed from the start of inflation $N = \ln ( a / a_{\mathrm{in}} )$. In de Sitter $\epsilon_0 = 1$ and all the other flow functions vanish. 
Close to de Sitter we have $\epsilon_0 \approx 1$, and for this regime to persist over a few e-folds, we need this slow-roll parameter not to vary much \ie $\epsilon_{1} = \frac{\dd \ln \left| \epsilon_{0} \right|}{\dd N } \ll 1$. We can iterate this reasoning and require all flow functions $n \geq 1$ to be small $\epsilon_n \ll 1$. All flow functions in the hierarchy are of the same order. These are the slow-roll conditions, and the flow functions are often called the slow-roll parameters.
Satisfying the slow-roll conditions imposes constraints on the initial conditions and the potential's form. We illustrate that by computing the two first flow functions. The first one reads
\begin{equation}
	\label{eq:espilon1}
	\epsilon_1 = - \frac{\dot{H}}{H^2} = 1 - \frac{\ddot{a}}{a H^2} = 3 \frac{\frac{\dot{\varphi}^2}{2 c^2}}{\frac{\dot{\varphi}^2}{2 c^2} + V \left( \varphi \right)} \, ,
\end{equation}
where we have used Friedmann equations to express $\dot{H}$ in terms of the kinetic energy of the field
\begin{equation}
	\dot{H} = - \frac{4 \pi \GN}{c^4} \dot{\varphi}^2 \, .
\end{equation}
Notice that the Hubble parameter can only decrease $\dot{H}<0$, so $\epsilon_1 >0$. Since inflation happens when $\ddot{a} > 0$, the second equality shows that the condition for inflation is $\epsilon_1<1$. Inflation ends for $\epsilon_1 = 1$.
On the other hand, the slow-roll condition $\epsilon_1 \ll 1$, required to be close to de Sitter, is stronger than requiring inflation. The kinetic energy must be negligible compared to the potential one, not only smaller.
Similarly, the second flow function can be expressed as
\begin{equation}
	\label{eq:espilon2}
	\epsilon_2 = 6 \left[ \frac{\epsilon_1}{3} - 1 - \frac{V^{\prime} \left( \varphi \right) c^2}{3 H \dot{\varphi} } \right] \, .
\end{equation}
Since $\epsilon_1 \ll 1$ and $\epsilon_2 \ll 1$, we get the relation 
\begin{equation}
	\label{eq:SR_attractor_condition}
	\dot{\varphi} \approx - \frac{V^{\prime} \left( \varphi \right) c^2}{3 H } \, .
\end{equation}
This last relation corresponds to neglecting $\ddot{\varphi}$ in the Klein-Gordon equation \eqref{eq:klein_gordon}. It reduces the dynamics to a first-order equation, thereby reducing the space of allowed initial conditions to a choice of $\varphi_{\mathrm{in}}$
rather than a couple $(\varphi_{\mathrm{in}},\dot{\varphi}_{\mathrm{in}})$. In the slow-roll regime, Eq.~\eqref{eq:SR_attractor_condition} can then be used to
re-express the flow functions in terms of the derivative of the potential. We have
\begin{equation}
	\label{eq:SR_potential_1}
	\epsilon_1 \approx \frac{\dot{\varphi}^2}{2 V \left( \varphi \right) c^2} = \frac{c^4}{16 \pi \GN} \left( \frac{V^{\prime}}{V} \right)^2 \, , 
\end{equation}
where we have used the first Friedmann equation in the slow-roll limit
\begin{equation}
	\label{eq:friedmann_SR}
	H^2 \approx \frac{8 \pi \GN}{3 c^2} V \left( \varphi \right) \, .
\end{equation}
The second flow function can similarly be written in terms of the derivatives of the potential. We can approximate $\ddot{\varphi}$ by taking the time derivative of 
Eq.~\eqref{eq:SR_attractor_condition} and use Eq.~\eqref{eq:klein_gordon} and Eq.~\eqref{eq:friedmann_SR} to get
\begin{equation}
	\label{eq:SR_potential_2}
	\epsilon_2 \approx \frac{c^4}{4 \pi \GN} \left[ \left( \frac{V^{\prime}}{V}  \right)^2 - \frac{V^{\prime \prime}}{V} \right] \, .
\end{equation}
To summarise, Eqs.~(\ref{eq:SR_potential_1}-\ref{eq:SR_potential_2}) formalise our intuition that the potential should be flat enough in some
regions to sustain a slow-roll evolution, and Eq.~\eqref{eq:SR_attractor_condition} singles out a specific trajectory, \ie specific initial conditions in this region, for potential energy to dominate over kinetic energy.
The conjunction of these two conditions is a generic feature of the slow-roll solution~\cite{Liddle:1994dx} that we have illustrated at leading order in the slow-roll parameters. 
Generically~\cite{Liddle:1994dx}, for a given potential, one can construct order by order an analytic solution of the equations of motion that will satisfy the slow-roll conditions. The series converges towards a single trajectory in the phase-plane $(\varphi,\dot{\varphi})$, the slow-roll solution.
An essential feature is that, when the potential supports inflation, the different trajectories in this phase-space will get closer exponentially
fast in the number of e-folds. Therefore, provided we consider a scenario with more than a few e-folds of inflation, the slow-roll approximation will be a good 
approximation of the dynamics of the field \textit{irrespective} of the initial conditions at the start of inflation; it is an attractor solution.
This property makes slow-roll inflation partially immune to fine-tuning problems in the initial conditions.
The slow-roll approximation is then a powerful method to compute analytically, with an arbitrary level of precision, an inflationary trajectory with somewhat generic properties.

\begin{figure}
	\centering
	\includegraphics[width=0.9\textwidth]{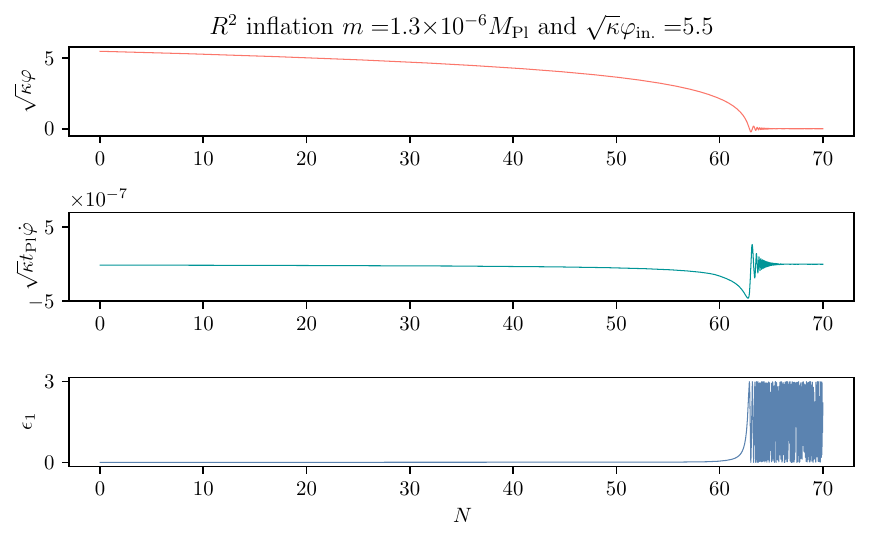} % first figure itself
	\caption{Evolution of the field $\varphi$, its time-derivative $\dot{\varphi}$ and the first flow function $\epsilon_1$,  as a function of the number of e-folds $N$
		in $R^2$-inflation with potential~\eqref{def:R2potential}. Inflation is taken to start
		at $N = 0$ where $\sqrt{\kappa}  \varphi_{\mathrm{in}} = 5.5$ while $\dot{\varphi}_{\mathrm{in}}$ is given by Eq.~\eqref{eq:SR_attractor_condition} evaluated at first order in slow-roll. For these values, the end of inflation $\epsilon_1 = 1$ is reached after  $N = 64$ e-folds,
		compatible with the estimate from the first-order slow-roll $N_{\mathrm{e}} - \frac{3}{4} - \sqrt{\frac{3}{4}} \approx 65$.}
	\label{fig:evolution_R2inflation}
\end{figure}

\begin{figure}
	\centering
	\includegraphics[width=0.9\textwidth]{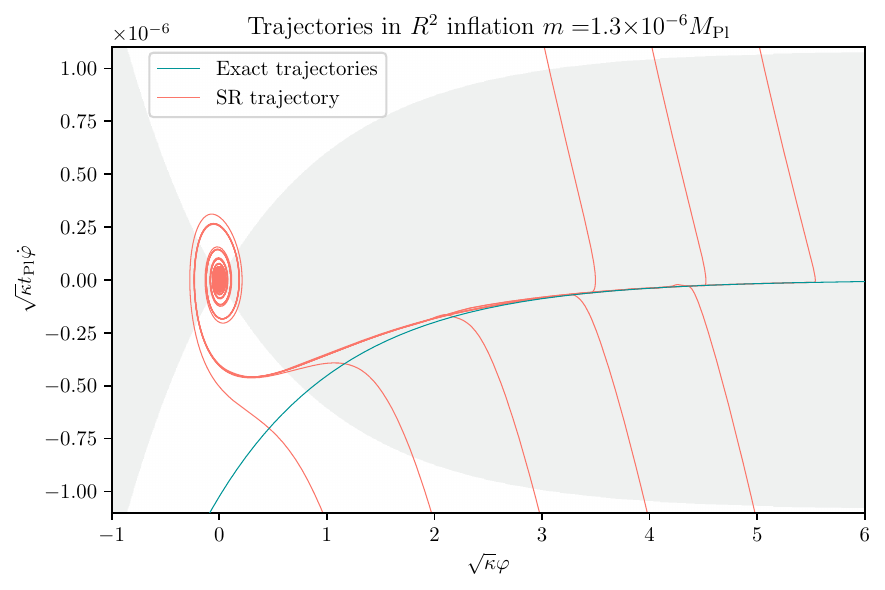} % second figure itself
	\caption{Trajectories in the phase-plane $(\varphi,\dot{\varphi})$ in $R^2$-inflation with potential~\eqref{def:R2potential}. 
		The red lines show the exact trajectories obtained by numerically solving Eq.~\eqref{eq:klein_gordon} and Eqs.~(\ref{eq:friedmann_1_scalar_field}) for different initial conditions.
		The green line shows the slow-roll trajectory evaluated at first-order, see Eq.~\eqref{eq:SR_trajectory_R2inflation}. The grey areas correspond to $\epsilon_1 \leq 1$ \ie values for which inflation proceeds. Once they leave the slow-roll attractor, the trajectories spiral towards the origin of the figure. This corresponds to the oscillations of the field, damped by Hubble friction, which can lead to preheating, see Sec.~\ref{sec:preheating}.  }
	\label{fig:trajectories_R2inflation}
\end{figure}

In order to illustrate the described inflationary dynamics and its slow-roll behaviour, we briefly study the case of $R^2$ inflation, also known as
Starobinsky inflation~\cite{Starobinsky:1980te}. Similar computations for other common inflationary models can be found in~\cite{Martin:2013tda}. The potential of this model reads
\begin{equation}
	\label{def:R2potential}
	V_{\mathrm{S}} \left( \varphi \right) = \frac{3}{4} \left( \frac{m c}{\sqrt{\kappa} \hbar} \right)^2 \left[ 1 - \exp \left( - \sqrt{\frac{2}{3}} \sqrt{\kappa} \varphi \right) \right]^2 \, ,
\end{equation}
where $\kappa = 8 \pi G / c^4$. The computation details for this case are given in Appendix~\ref{app:R2inflation}.
For the potential of Eq~\eqref{def:R2potential}, the slow-roll conditions are satisfied in the large field regime $\sqrt{\kappa} \varphi \gg 1$.
In Fig.~\ref{fig:trajectories_R2inflation}, the trajectories are shown to get close to the slow-roll trajectory in this regime, illustrating the 
attractor property.
Defining the number of e-folds to be $0$ at the start of inflation $N = \ln \left( a_{\mathrm{in}} / a \right)$ and the initial value
of the field to be $\varphi_{\mathrm{in}}$, the equation of motion is easily integrated to find the evolution of the field as a function of the number of e-folds
at first order in slow-roll 
\begin{equation}
	\label{eq:field_N_R2inflation}
	\varphi \left( N \right) = \sqrt{\frac{3}{2}} \frac{1}{\sqrt{\kappa}} \ln \left[ \frac{4}{3} \left( N_{e} - N \right) \right] \, ,
\end{equation}
where 
\begin{equation}
	N_{\mathrm{e}} = \frac{3}{4} e^{\sqrt{\frac{2}{3}} \sqrt{\kappa} \varphi_{\mathrm{in.}} } \, .
\end{equation}
Using the slow-roll approximation to find when $\epsilon_1$ reaches unity, we find that inflation lasts approximately $\Delta N  = N_{\mathrm{e}} - \frac{3}{4} - \sqrt{\frac{3}{4}}$, a value controlled by the field's initial value. Comparing with the exact evolution of $\epsilon_1$ shown in Fig.~\ref{fig:evolution_R2inflation}, it is seen to be a good approximation.

The energy scale of inflation is given by the value of the potential . 
Using Eq.~\eqref{def:energy_scale}, and evaluating the energy at the beginning of inflation, we find
\begin{equation}
	E_{\mathrm{infl}} \approx \sqrt{\frac{m}{\Mp}} \left( \frac{3}{4} \right)^{1/4} \Mp c^2  \, ,
\end{equation}
making it obvious that the mass scale $m$ in the potential controls the energy scale of inflation.
For Starobinsky inflation, the field's initial value is decoupled from the energy scale. We can tune the duration of inflation to be long enough to solve the Big Bang puzzles without requiring that the energy scale of inflation is super-Planckian $E_{\mathrm{infl}}>\Mp c^2$, a regime in which our classical description is not expected to be accurate anymore. 
In Figs.~\ref{fig:evolution_R2inflation} and \ref{fig:trajectories_R2inflation}, we chose $m$ such that $E_{\mathrm{infl}} = E_{\mathrm{GUT}}$, also used for illustration when considering Big-Bang puzzles, and chose the initial value of the field accordingly in order to have roughly $65$ e-folds of inflation.
Therefore, we have constructed an explicit example of a model of inflation able to solve the Big Bang puzzles without requiring super-Planckian energies or fine-tuning in the initial values of the field.
Many such models exist, but they are constrained by the observed properties of inhomogeneities in the early Universe. 
We review these aspects in the next sub-section Sec.~\ref{sec:inhomogeneous_universe}.

We close this sub-section by pointing out that the inflationary scenario we presented is not devoid of problems. For instance, although the attractor property removes some degree of fine-tuning in the initial conditions, we still have assumed that the inflaton field, and so Universe was homogeneous \textit{before} inflation starts. If for inflation to start, we need to require that the Universe is homogeneous across a-causal scales, then we have not solved the Horizon problem. The intuition here is that we only need to require homogeneity initially over a small patch, which is then inflated to our whole observable Universe. Numerical simulations have analysed the beginning of inflation in an inhomogeneous Universe, e.g.~\cite{East:2015ggf,Clough:2016ymm,Clough:2017efm,Aurrekoetxea:2019fhr,Bloomfield:2019rbs,Garfinkle:2023vzf}, showing that, under certain conditions, in large field models, inflation can start even in a very inhomogeneous environment, see however~\cite{Garfinkle:2023vzf}.
We refer to~\cite{Brandenberger:2016uzh,Baumann:2022mni} for a discussion of this initial condition problem and others.

\subsection{Inhomogeneous Universe}
\label{sec:inhomogeneous_universe}

In the previous sections, we have only discussed the physics of a homogeneous Universe.
Although on very large scales, when coarse-graining the distribution of matter, the Universe appears isotropic,
it is only an approximation. On smaller scales, the Universe is obviously inhomogeneous, being made up of galaxy clusters separated by large voids. Soon after the first papers on cosmic inflation as a solution to Big Bang puzzles, it was realised that inflation also offers a mechanism for generating primordial inhomogeneities. In~\cite{Mukhanov:1981xt}, Mukhanov and Chibisov proposed that quantum fluctuations amplified by an intermediate stage of de Sitter expansion could explain inhomogeneities in the Universe. Their work was independent of that of other authors working on inflation. They were building on the work of Starobinsky~\cite{Starobinsky:1980te}, who showed that quantum-corrected Einstein's field equation admitted a non-singular de Sitter Universe as a solution. This work provided the basis for $R^2$-inflation used as an example in the last sub-section, Sec.~\ref{sec:SSR}.
The analysis of matter inhomogeneities in the context of inflation was first done by~\cite{Mukhanov:1982nu,Hawking:1982cz,Starobinsky:1982ee,Guth:1982ec,Bardeen:1983qw}. The generation of gravitational waves in the context of inflation was first considered in~\cite{Rubakov:1982df}, see Sec.~\ref{sec:quantum_GW} for more details.
In this sub-section, we start by briefly explaining how to describe small inhomogeneities in cosmology. The CMB's temperature anisotropies are measured to be of order $\delta T/T \sim 10^{-5}$~\cite{Planck:2015mrs}, which directly reflects the small degree of inhomogeneities at LSS time~\cite{Sachs:1967er}.
We expect the inhomogeneities to be even smaller during inflation, which proceeded earlier, so the perturbative treatment should be perfectly valid then.
In this sub-section, we will show how a quantum theory of inflation can prescribe some initial conditions for structure formation, which were observationally
confirmed and allow us to constrain the specifics of the inflationary model.

\subsubsection{SVT decomposition}
\label{sec:SVT}

To account for the presence of inhomogeneities, we need to go beyond the description of the Universe in terms of 
the FLRW metric Eq.~\eqref{def:FLRW}.
In the standard structure formation scenario, small primordial inhomogeneities are gradually amplified by gravitational collapse.
In this work dealing with early Universe physics, we restrict attention to the early phase of this process when the inhomogeneities 
are still small compared to the background densities.
For this exposition, we follow the conventions and discussion of~\cite{Mukhanov:1990me}, except that we use the mostly-pluses convention for the signature.
We start by considering perturbations around the flat $\mathcal{K} =0$ FLRW metric
\begin{equation}
	\label{def:FLRW_perturbation}
	\dd s^2 = a^2 (\eta) \left[ - c^2 \left( 1+ 2 \phi \right) \dd \eta^2 + 2 B_{i} \dd x^{i} \dd \eta  + \left( \gamma\indices{_{ij}} + E\indices{_{ij}} \right) \dd x^{i} \dd x^{j}   \right] \, ,
\end{equation}
where we work with 3D Cartesian coordinates $x^{i}$, $\gamma\indices{_{ij}}$ is the Euclidian three-dimensional metric in Cartesian coordinates. We have introduced the conformal time $\eta$ related to cosmic time by $a^2 (\eta) \dd \eta^2 = \dd t^2$. During inflation the conformal time is negative $\eta < 0$, and the end of inflation corresponds to $\eta \to 0^{-}$.
There are $10$ perturbations parameterised by $\phi$, the spatial vector $B_i$ and the spatial 2-tensor $E\indices{_{ij}}$.
It is useful to perform a scalar-vector-tensor (SVT) decomposition of these perturbations.
We decompose each of these terms in objects having a well-defined transformation under the group of 3D spatial rotations, which is a symmetry of the (flat) background metric. To avoid confusion, we will refer to these objects as \textit{helicity} scalar-vector-tensors~\cite{Maggiore:2018sht,Baumann:2009ds}. Therefore, although $B_i$ transforms as a vector under a change of coordinates, we can decompose it in a helicity scalar and a helicity vector as
\begin{equation}
	\label{def:helmholtz_decompo}
	B_i = S_i -  \partial_{i} B \, ,
\end{equation}
where $B$ is an helicity scalar and $S_i$ is a divergence-free helicity vector
$\partial^{i} S_i = 0$. Eq.~\eqref{def:helmholtz_decompo} is nothing else than the Helmholtz decomposition of a vector.
We have a similar decomposition for the tensor 
\begin{equation}
	E\indices{_{ij}} = - 2 \psi \gamma\indices{_{ij}} + 2  \partial_{i}  \partial_{j} E + 2 \partial_{( i} F_{j)} + h\indices{_{ij}} \, ,
\end{equation}
where $\partial_{( i} F_{j)} = \left(\partial_{i} F_{j} + \partial_{j} F_{i} \right)/2$ is the symmetrised derivative. In this decomposition $\psi$ and $E$ are helicity scalars, $F_{j}$ is an helicity vector and $h\indices{_{ij}}$ an helicity tensor.
We have $\partial^{i} F_i = \partial^{i} h\indices{_{ij}} = 0$ and the tensor is trace-less $h\indices{^{i}_{i}} = 0$. We go to Fourier space to see the effect of rotations on these objects. For any wave-vector $\bm{k}$, the divergence-free condition reads $k^{i} S_i (\bm{k}) = 0$ \ie
$S_i (\bm{k})$ is transverse, while $\partial_{i} B (\bm{k}) \parallel \bm{k}$. Then under a rotation $R (\theta)$ around the direction $\bm{k}$, $\partial_{i} B (\bm{k})$ will be unaffected, while $S_i (\bm{k})$ will be rotated. In the helicity basis~\cite{Maggiore:2018sht,Baumann:2009ds} it would pick up a factor of $e^{\pm i \theta}$. This justifies the terminology of helicity SVT. The same reasoning applies to the components of $E_{ij}$ in Fourier space where $h\indices{_{ij}} (\bm{k})$ would instead pick up a factor of $e^{\pm 2 i \theta}$ in the helicity basis. 
In total we have four scalars $\psi$, $\phi$, $B$, $E$, two vectors $S_i$, $F_i$, and one tensor $h\indices{_{ij}}$.
No vector perturbations are produced during inflation, and any pre-existing ones would quickly decay~\cite{Grishchuk:1993ab}. They are thus conveniently ignored in the rest of this text.
We can repeat the SVT decomposition for the energy-momentum tensor by perturbing that of a perfect fluid.
The two sets of perturbations are related by the perturbed Einstein field equations, which, combined with the perturbed conservation equation, give the equations of motion of the perturbations.
We will not discuss the general case here and refer to~\cite{Mukhanov:1990me} for details. The main purpose of the SVT decomposition is that scalar, vector and tensors do not mix in the equations of motion at linear order in perturbation theory, see Appendix A of~\cite{Baumann:2009ds} for a proof.

\subsubsection{Gauge invariant variables}

There are some ambiguities in the definition of these perturbations.
General covariance is a basic property of General Relativity, and equations are generally formulated in a covariant way. This symmetry under coordinate transformations is broken in cosmology by working in coordinates where the distribution of matter is homogeneous and isotropic. Still, we could work with coordinates where the same Universe would no longer look homogeneous. For instance, starting from coordinates where the metric reads as in Eq.~\eqref{def:FLRW}, consider a small shift of the original spatial coordinates $\tilde{x}^i =  x^{i} + \xi^{i} (\eta , \bm{x} )$. The metric reads
\begin{align}
	\begin{split}
		\dd s^2 & = a^2 (\eta) \left( - c^2 \dd \eta^2 + \gamma\indices{_{ij}} \dd x^{i} \dd x^{j}  \right) \, , \\
		& = a^2 (\eta) \left\{ - c^2 \dd \eta^2 - 2 \xi^{\prime}_{i} \dd \tilde{x}^{i} \dd \eta  + \left[ \gamma\indices{_{ij}} - 2 \partial_{(i} \xi_{j)} \right] \dd \tilde{x}^{i} \dd \tilde{x}^{j}   \right\} \, .
	\end{split}
\end{align}
where $\prime$ denote derivatives with respect to conformal time $\eta$.
In the new coordinates, the metric looks like a perturbed FLRW metric, and a direct comparison with the ansatz~\eqref{def:FLRW_perturbation} gives non-vanishing values for the perturbation parameters. 
These perturbations are spurious since the Universe is still manifestly homogeneous by adopting the right system of coordinates (the old ones). This illustrates the \textit{gauge dependence} of the perturbations and of the quantities $\phi$,$B_i$,$\gamma_{ij}$ and $E_{ij}$. To avoid mistaking an effect of coordinate choice for a true deviation from FLRW, we can introduce gauge-invariant combinations of them. To build them, one should consider how each type of perturbation transforms under a small coordinate change 
$\tilde{x}^{\mu} =  x^{\mu} + \xi^{\mu} (\eta , \bm{x} )$.
In general, it is found that the transverse trace-less perturbations $h\indices{_{ij}}$ are gauge-invariant and that two gauge-invariant scalars, the so-called Bardeen variables~\cite{Bardeen:1980kt}, can be constructed
\begin{subequations}
\begin{align}
	\label{def:Bardeen_variables}
\begin{split}
	\Phi_{(\mathrm{B})} & = \phi - \Hu \left( B - E^{\prime} \right) + \left( B - E^{\prime} \right)^{\prime} \, ,
\end{split} \\
\begin{split}
	\Psi_{(\mathrm{B})} & = \psi - \Hu \left( B - E^{\prime} \right) \, ,
\end{split}
\end{align}
\end{subequations}
where $\Hu = a^{\prime}/a$.
We are interested specifically in the case of single field inflation where
the situation is simpler and these two variables are equal $\Psi_{(\mathrm{B})}=\Phi_{(\mathrm{B})}$~\cite{Mukhanov:1990me}.
Additionally, the perturbations around the background trajectory $\varphi_0 (\eta)$ of the field $\varphi = \varphi_0 (\eta) + \delta \varphi (\eta , \bm{x})$
only leads to scalar perturbations that can be gathered with metric perturbations in the gauge invariant combination~\cite{Mukhanov:1990me} 
\begin{equation}
	\label{def:invariant_scalar}
	\delta \varphi_{\mathrm{(gi)}} = \delta \varphi + \varphi_0^{\prime} \left( B - E^{\prime} \right) \, .
\end{equation}
The scalar field perturbation $\delta \varphi_{\mathrm{(gi)}}$ and the Bardeen variable $\Psi_{(\mathrm{B})}$ are related by the perturbed Einstein equation
\begin{equation}
	\label{eq:pert_einstein_bardeen_phi}
	\Hu  \Phi_{(\mathrm{B})} + \Phi_{(\mathrm{B})}^{  \prime} = \frac{\kappa}{2} \varphi_0^{\prime}   \delta \varphi_{\mathrm{(gi)} }
\end{equation}
It is found after lengthy computations~\cite{Mukhanov:1990me} that the perturbed actions of Einstein-Hilbert and of the scalar field
can be rewritten over a single scalar quantity $v$, called the Mukhanov-Sasaki variable~\cite{Mukhanov:1981xt,Kodama:1984ziu}, which is a combination of these two
\begin{equation}
	\label{def:MS_variable}
	v = a \left[  \delta \varphi_{\mathrm{(gi)}} + \mathfrak{z}  \Phi_{(\mathrm{B})} \right]   \, ,
\end{equation}
where 
\begin{equation}
	\mathfrak{z}  = \frac{a \varphi_0^{\prime}}{\Hu} = a \sqrt{\frac{2\epsilon_1}{\kappa}} \, .
\end{equation}
We use the fraktur font $\mathfrak{z}$ to distinguish this quantity from the redshift $z$.
The perturbed action reads ~\cite{Mukhanov:1990me}
\begin{align}
	\begin{split}
		\delta^{(2)} S_{\mathrm{S}} & = \frac{1}{2 c^3} \int \dd^4 x \left[ \left( v^{\prime} \right)^2 - \gamma^{ij} c^2 \partial_{i} v \partial_{j} v + \frac{ \mathfrak{z} ^{\prime \prime}}{ \mathfrak{z} } v^2 \right]  \, , \\
		& = \frac{1}{c^2} \int  \dd \eta \int_{\mathbb{R}^{3+}} \dd^3 \bm{k} \left[ v^{\prime}_{\bm{k}} v^{\prime}_{-\bm{k}} - c^2 k^2 v_{\bm{k}} v_{-\bm{k}} + \frac{ \mathfrak{z} ^{\prime \prime}}{ \mathfrak{z} }   v_{\bm{k}} v_{-\bm{k}} \right] \, ,
	\end{split}
	\label{def:pert_scalar_action}
\end{align}
where in the second line
we transformed the action to Fourier space 
\begin{equation}
	v \left( \bm{x} , \eta \right) = \int \frac{\dd^3 \bm{k}}{\left( 2 \pi \right)^{3/2}} e^{i \bm{k}.\bm{x}} v_{\bm{k}} \left( \eta \right) \, ,
\end{equation}
and folded the integration over $\mathbb{R}^{3+} = \left\{ \bm{k} | k_{\mathrm{z}} > 0 \right\}$. Since $v$ is real, we have 
$v^{\star}_{- \bm{k}} = v_{ \bm{k}}$.
It is more convenient to work in Fourier space because, at the linear level in perturbation theory, different pair of modes $\pm \bm{k}$
evolve independently as seen from Eq.~\eqref{def:pert_scalar_action}.
The case of tensor perturbations is studied in detail in \cite{Micheli:2022tld} reproduced in Sec.~\ref{sec:quantum_GW}, where the perturbed action is derived, and their evolution is followed during inflation and after.
We thus only quote relevant results for comparison with the scalar case.
First, the tensor perturbations $h_{ij}$ can be decomposed in two real scalar fields $\mu_{\lambda}$, representing its two polarisations $\pm$ in helicity basis, via
\begin{equation}
	\label{eq:hij_expanded_polarisation}
	h_{ij}(\bm{x},\eta) = \sqrt{32\pi\GN} \sum_{\lambda = \pm }
	\int \frac{\dd^3 \bm{k}}{(2\pi)^{3/2}a(\eta)}
	\varepsilon^{(\lambda)}_{ij} (\hat{\bm{k} }) \mu_{\lambda} (\bm{k},\eta)  e^{i\bm{k}\cdot\bm{x}} \, ,
\end{equation}
where $\varepsilon^{(\lambda)}_{ij}(\hat{\bm{k} })$ is the polarisation tensor for a wave in the
direction $\hat{\bm{k} }$ defined in Sec.~\ref{sec:quantum_GW}.
Using this expansion, the perturbed Einstein-Hilbert action for the tensor sector reads
\begin{align}
	\begin{split}
		\delta^{(2)}S_\textsc{t} & = \int\dd\eta \sum_{\lambda=\pm}
		\frac12\int\dd^3\bm{x}\, 
		\left[ \left(\mu_\lambda'\right)^2 - c^2
		\gamma^{ij}\partial_i\mu_\lambda \partial_j\mu_\lambda +
		\frac{a''}{a} \mu_\lambda^2 \right], \\
		&  = \int  \dd \eta  \sum_{\lambda=\pm} \int_{\mathbb{R}^{3+}} \dd^3 \bm{k} \left[ \mu^{\prime}_{\lambda} \left( \bm{k} \right) \mu^{\prime}_{\lambda} \left( -\bm{k} \right) - \left( c^2 k^2 - \frac{a^{\prime \prime}}{a} \right) \mu_{\lambda} \left( \bm{k} \right) \mu_{\lambda} \left( - \bm{k} \right)  \right] \, .
	\end{split}
	\label{def:pert_tensor_action}
\end{align}
It is made of two copies of the action of scalar perturbations, up to the substitution $\mathfrak{z} \to a$.
Finally, although the Mukhanov-Sasaki variable $v$ allows having the most compact form for the equations of motion, $\delta \varphi$ loses its meaning after inflation so that the definition Eq.~\eqref{def:MS_variable} is not valid anymore\footnote{Another definition valid for a perfect fluid can be given~\cite{Mukhanov:1990me}.}.
A convenient quantity to work with is $\zeta$ defined as
\begin{equation}
	\label{def:zeta}
	\zeta = \frac{2}{3} \frac{\Hu^{-1}  \Phi_{(\mathrm{B})}^{  \prime} +   \Phi_{(\mathrm{B})}  }{1 + w} +  \Phi_{(\mathrm{B})} \, ,
\end{equation}
where $w = \rho / p$ is the effective equation of state parameter for the total pressure and energy density.
The definition~\eqref{def:zeta} only contains geometrical quantities and is therefore valid in any background spacetime, in particular when entering radiation domination at the end of inflation.
During inflation, $\zeta$ can be related to the Mukhanov-Sasaki variable.
First, using the expression of energy and pressure density  Eqs.~(\ref{eq:energy_scalar})-(\ref{eq:pressure_scalar}) and the definition of the first slow-roll parameter  Eq.~\eqref{eq:espilon1}, we find
\begin{equation}
	w_{\mathrm{infl}} = \frac{2}{3} \epsilon_1 -1 \, .
\end{equation}
We then use the perturbed Einstein equation Eq.~\eqref{eq:pert_einstein_bardeen_phi} and the definition Eq.~\eqref{def:zeta} to get the simple relation
\begin{equation}
	\zeta = \frac{v}{ \mathfrak{z} } \, .
\end{equation}
A crucial property of $\zeta$ is that when a single perfect fluid dominates, e.g. during inflation or radiation domination,  it is conserved on large scales~\cite{Bardeen:1983qw,Mukhanov:1990me,Martin:1997zd}. It thus allows smoothly connecting the evolution of gravitational potentials from inflation to later times, for example at LSS, and to compute the CMB power spectrum. The fluctuations of temperature measured in the CMB photons are related to Bardeen potentials by the Sachs-Wolfe effect~\cite{Sachs:1967er}, which is directly connected to $\zeta$ by definition.
Similarly, the physical quantity to compute for tensor is $h_{ij}$ rather than $\mu_{\lambda}$. However, being intrinsically geometrical, they are always well-defined and differ merely by a factor of $a$.

\subsubsection{Evolution of inflationary perturbations}
\label{sec:evo_inf_pert}

We will now solve the equations of motion and describe the evolution of the perturbations during inflation.
The equation of motion of the scalar perturbation is straightforwardly derived from the action~\eqref{def:pert_scalar_action}
\begin{equation}
	\label{eq:MS_equation}
	v^{\prime \prime}_{\pm \bm{k}} + \left( c^2 k^2 - \frac{ \mathfrak{z} ^{\prime \prime}}{ \mathfrak{z} }  \right) v_{\pm \bm{k}} = 0 \, .
\end{equation}
It is the equation of an oscillator with a time-dependent frequency
\begin{equation}
	\omega_{k \, \mathrm{S}}^2 = c^2 k^2 - \frac{ \mathfrak{z} ^{\prime \prime}}{ \mathfrak{z} } \, .
\end{equation}
The behaviour of the mode $\bm{k}$ will be different when $\omega_{k \, \mathrm{S}}^2$ is positive or negative. When $\omega_{k \, \mathrm{S}}^2 > 0$, we expect the mode to oscillate, and when $\omega_{k \, \mathrm{S}}^2 < 0$, we expect the mode to be amplified.
To understand the evolution of $\mathfrak{z}^{\prime \prime}/\mathfrak{z}$, we will work again in the slow-roll approximation where the background is a
quasi-de Sitter expansion. We express the time-dependent part of the frequency in slow-roll parameters
\begin{equation}
	\label{eq:exp_zprimeprime}
	\frac{ \mathfrak{z} ^{\prime \prime}}{ \mathfrak{z} }  = \frac{a^{\prime \prime}}{a} +  \frac{\epsilon_{2}}{2} \left( \frac{a^{\prime \prime}}{a} + \Hu^2 \right) +  \frac{\Hu^2}{2} \left( \epsilon_2 \epsilon_3 + \frac{\epsilon_{2}^2}{2}  \right) \, .
\end{equation}
The terms $a^{\prime \prime} / a$ and $\Hu$ must also be expanded in slow-roll parameters to be consistent. To understand the mode evolution qualitatively, let us first consider the de Sitter case, where the flow functions vanish. We have $a = - 1 / H \eta$, then $\mathfrak{z}^{\prime \prime} / \mathfrak{z} = a^{\prime \prime} / a = 2 / \eta^2 $. First, notice that, in de Sitter, the frequency for scalar and tensor perturbations coincide\footnote{This is a general feature of power law inflation models of which de Sitter is a special case~\cite{Martin:2013tda}}.
Second, the sign of $\omega_{k \, \mathrm{S}}^2$ is then directly obtained by a comparison between the physical size of the Hubble radius $c / a H = - c \eta$ and the wavelength of the mode $k^{-1}$. When the mode is \textit{super-Hubble} \ie its wavelength is larger than the Hubble radius,
then $k \left( a H \right)^{-1} = - c k \eta \ll 1$ and $\omega_{k \, \mathrm{S}}^2$ is negative: the mode is amplified.
On the other hand, when the mode is \textit{sub-Hubble} then $k \left( a H \right)^{-1} = - c k \eta \gg 1$ and the mode oscillates $\omega_{k \, \mathrm{S}}^2 > 0$. 
Since the physical size of the Hubble radius decreases as time passes, more modes $\bm{k}$ will become super-Hubble and be amplified during inflation.
The picture is similar in slow-roll, but the amplification condition must be corrected to include the slow-roll parameters and have a precise estimate of the amount of amplification undergone by a mode.
Let us follow the evolution of one mode $\bm{k}$ which is initially sub-Hubble.
Starting in the regime $c^2 k^2 \gg \mathfrak{z}^{\prime \prime}/ \mathfrak{z}$, since $\omega_{k \, \mathrm{S}} \approx c k $, we have
\begin{equation}
	\label{eq:subHubble_vk} 
	v_{\bm{k}} \left( \eta \right) \approx A_{\bm{k}}  e^{i c k \eta} + B_{\bm{k}}  e^{- i c k \eta} \, ,
\end{equation}
where $A_{\bm{k}}$ and $B_{\bm{k}}$ are two constants fixing the initial conditions for the evolution of the mode.
Then in the regime $c^2 k^2 \ll \mathfrak{z} ^{\prime \prime}/\mathfrak{z}$ where $\omega_{k \, \mathrm{S}} \approx \mathfrak{z}^{\prime \prime}/ \mathfrak{z} $, we have the following general solution
\begin{equation}
	\label{eq:superHubble_vk} 
	v_{\bm{k}} \left( \eta \right) \approx \alpha_{\bm{k}} \mathfrak{z} \left( \eta \right) + \beta_{\bm{k}} \mathfrak{z} \left( \eta \right) \int_{\eta}^{0} \frac{\dd \eta^{\prime}}{\mathfrak{z}^2 \left( \eta^{\prime} \right)} \, ,
\end{equation}
where $\alpha_{\bm{k}}$ and $\beta_{\bm{k}}$ are two complex coefficients. Dividing by $\mathfrak{z}$, we find the corresponding super-Hubble expression for $\zeta$
\begin{equation}
	\label{eq:superHubble_zetak} 
	\zeta_{\bm{k}} \approx \alpha_{\bm{k}}   + \beta_{\bm{k}}  \int_{\eta}^{0} \frac{\dd \eta^{\prime}}{\mathfrak{z}^2 \left( \eta^{\prime} \right)} \, .
\end{equation}
Since $\mathfrak{z}^2>0$, the second term will decay; it is called the decaying mode. In contrast, the first term is a constant and is called the growing mode.
As announced when introducing it, we find that $\zeta$ goes to a constant $\zeta_{\bm{k}} \approx \alpha_{\bm{k}} $ when $\eta \to 0^{-}$ on large scales.
Therefore, to connect computations made for $v$ during inflation to CMB observations, we only need to extract the value $\alpha_{\bm{k}}$ from the evolution of the Mukhanov-Sasaki variable.
This is done by continuously joining the two solutions given in Eq.~\eqref{eq:subHubble_vk} and Eq.~\eqref{eq:superHubble_vk} in the regime $c^2 k^2 \approx \mathfrak{z}^{\prime \prime}/\mathfrak{z}$.
The number of e-folds corresponding to this regime depends on the scales considered. We here are interested in the range of modes $\bm{k}$ corresponding to that measured by CMB experiments like Planck. By tracing back the evolution of their physical size from today to the inflationary period, we can estimate (assuming certain values for energy scale of inflation and reheating) that they crossed out the Hubble radius from 
60 to 53 e-folds before the end of inflation~\cite{Martin:2004um}. In the next part Sec.~\ref{sec:quantum_IC}, we first fix the initial conditions to derive the coefficients $A_{\bm{k}}$ and $B_{\bm{k}}$. We then compute an approximate solution for the evolution of the Mukhanov-Sasaki variable valid during this period of $7$ e-folds and extract from it the $\alpha_{\bm{k}}$ coefficient of Eq.~\eqref{eq:superHubble_zetak}.

\subsubsection{Quantum initial conditions for structure formation}
\label{sec:quantum_IC}

A good way to characterise the properties of the observed CMB temperature fluctuations $\delta T /T$ is to view them as the result of a random process.
Intuitively, each patch of a large enough size in a given direction $\hat{\bm{n}}$ is considered as an independent realisation of the process $\delta T /T (\hat{\bm{n}})$~\cite{Grishchuk:1997pk}.
We can then deduce the statistical properties of the underlying process by computing $n$-point correlation functions.
Following this strategy, it is found~\cite{Planck:2019kim} that the fluctuations are Gaussian within an excellent approximation \ie all the information is contained in the $2$-point correlation function also known as the power spectrum $\langle \delta T /T (\hat{\bm{n}}) \delta T /T (\hat{\bm{n}}^{\prime}) \rangle$. 
These correlation functions are usually decomposed in spherical harmonics to be analysed and are related to that of curvature perturbations $\langle \zeta_{\bm{k}} \zeta_{\bm{k}^{\prime}} \rangle$ by the Sachs-Wolfe effect~\cite{Sachs:1967er}.
In this picture, the curvature perturbations are also seen as a result of an initial random process, the characteristics of which are fixed by the 
initial conditions.
These conditions are fixed in the very early Universe at the beginning of inflation, where the energy scale at play, e.g. around the GUT scale, makes
the description of matter in terms of \textit{quantum} fields relevant.
We can therefore consider a quantum version of the scalar field theory developed so far.
The quantisation procedure is reviewed in detail in~\cite{Micheli:2022tld}, reproduced in Sec.~\ref{sec:quantum_GW}, for the tensor perturbations. The procedure for scalar perturbations proceeds in the same fashion;  we do not detail it here. However, for completeness, we present the conclusions of the quantisation procedure for both scalar and tensor.
Our goal in this part is to derive the standard results for the spectra of the tensor and scalar perturbations, which can be compared to cosmological observations. 
The quantum aspects of the perturbations are one of the two topics of this manuscript, and we relegate in-depth discussions of the quantum aspects to later sections.

First, the Mukhanov-Sasaki variable is promoted to an operator $\hat{v}_{\bm{k}}$ whose canonically conjugated operator
\begin{equation}
	\label{def:conjugate_MS}
	\hat{p}_{ \pm \bm{k}} = \hat{v}_{ \pm \bm{k}}^{\prime} \, ,
\end{equation}
is computed from the action  Eq.~\eqref{def:pert_scalar_action}, and we impose canonical commutation relation
\begin{equation}
	\left[\hat{v}_{ \bm{k}} , \hat{p}_{\bm{k}^{\prime}}\right] = i \hbar \delta \left( \bm{k} + \bm{k}^{\prime} \right) \, .
\end{equation}
Using these operators, we can compute quantum expectation values for the different quantities, e.g. $\langle \hat{\zeta} (\bm{x} , \eta ) \hat{\zeta} (\bm{x}^{\prime} , \eta ) \rangle$. They would match the average value of the quantity if we measured it repeatedly after having prepared the state multiple times with the same initial conditions. However, we have only access to a single realisation of our Universe. Still, under an ergodicity assumption~\cite{Grishchuk:1997pk}, these quantum expectation values are assumed to match the measured statistical expectation values computed over different patches of the Universe at that time. To compare them with the correlation functions measured in the CMB, the expectation values at the end of inflation have to be evolved to LLS time. This evolution is summarised in transfer functions~\cite{Baumann:2022mni}.
Using this correspondence, we have to compute the quantum evolution of the operators and evaluate their expectation values at the end of inflation.
For that, we can easily compute from Eq.~\eqref{def:pert_scalar_action} the Hamiltonian of the theory. As the action is quadratic, and only mixes the modes $\pm k$, we restrict to a pair of modes
\begin{equation}
	\label{def:H_scalar_pert} 
	\hat{H}_{\pm \bm{k} ,  \mathrm{S}} = \hat{p}_{\bm{k}} \hat{p}_{-\bm{k}} + \left( c^2 k^2 - \frac{ \mathfrak{z} ^{\prime \prime}}{ \mathfrak{z} } \right) \hat{v}_{\bm{k}} \hat{v}_{-\bm{k}}  \, ,
\end{equation}
and write the associated Schrödinger equation. To solve this equation, we are then back to the problem of specifying the initial conditions.
The simplest and minimal choice we can make is that initially, there were only \textit{vacuum} fluctuations. 
When a mode $\bm{k}$ is very sub-Hubble $c k \eta \to  - \infty$ it does not feel the expansion of spacetime, and the quantisation procedure is that of a simple harmonic oscillator of frequencies $k$. We can introduce the standard creation/annihilation operators
\begin{subequations}
	\begin{align}
		\hat{v}_{\bm{k}} \left(- \infty \right) & = \sqrt{ \frac{\hbar c}{2k} }
		\left[ \hat{a}_{\bm{k}} \left( - \infty  \right) + \hat{a}_{-
			\bm{k}}^{\dagger} \left( - \infty  \right) \right] \, ,
		\\ \hat{p}_{\bm{k}} \left( - \infty \right) & = - i \sqrt{\frac{\hbar 
				k}{2 c }} \left[ \hat{a}_{\bm{k}} \left( - \infty  \right) - \hat{a}_{-
			\bm{k}}^{\dagger} \left( - \infty \right) \right].
	\end{align}
	\label{def:scalar_creation_operators}
\end{subequations}
The vacuum state in the sub-Hubble limit is then defined as the state annihilated by both operators $\hat{a}_{ \pm\bm{k}} \left(  - \infty \right) \left| 0 \right\rangle_{ \pm\bm{k}}$, and is a well-know Gaussian wave-function, see Sec.~\ref{sec:quantum_GW}.
Evolving the mode under the quadratic Hamiltonian~\eqref{def:H_scalar_pert} from this initial Gaussian vacuum state preserves the Gaussianity of the wave function. The expectation values will then satisfy the observed Gaussian character of the correlation functions.
The choice of initial vacuum state for all modes $\bm{k}$ in the sub-Hubble regime is called the Bunch-Davies vacuum~\cite{Bunch:1978yq}. Working in the Heisenberg picture, the evolution can be summarised using a mode function $u_{k}$, such that
\begin{align}
	\hat{v} \left( \bm{x} , \eta \right) & = \int \frac{\mathrm{d}^3
		\bm{k}}{(2 \pi)^{3/2}} \left[ e^{i \bm{k}.\bm{x}} u_{k} \left( \eta
	\right) \hat{a}_{\bm{k}}\left(  - \infty \right) + e^{- i
		\bm{k}.\bm{x}} u_{k}^{\star} \left( \eta \right)
	\hat{a}_{\bm{k}}^{\dagger} \left( - \infty \right) \right] \, .
	\label{def:v_mode_expansion}
\end{align}
Using Eq.~\eqref{def:v_mode_expansion}, one can show that $\hat{v}$ is a solution of the Heisenberg equation if and only if the mode function $u_{k}$ is a solution of the Mukhanov-Sasaki equation, Eq.~\eqref{eq:MS_equation}.
We have then reduced the quantum evolution to the classical one. Note that this reduction is a general feature of the evolution under a quadratic Hamiltonians, see Sec.~\ref{sec:quantum_GW}. The considerations of Sec.~\ref{sec:evo_inf_pert} on the evolution of the classical solution $v_{\bm{k}}$ are then mapped completely to the mode function $u_k$. Recall that, to solve the dynamics, we have to first find the coefficients in the sub-Hubble limit Eq.~\eqref{eq:subHubble_vk}, fixing the initial conditions, and then join the obtained sub-Hubble solution to the super-Hubble one in the regime 
$k^2 \approx \mathfrak{z}^{\prime \prime}/\mathfrak{z}$. 
First, the initial conditions are uniquely fixed by choice of the Bunch-Davies vacuum as an initial state. It imposes that the mode function $u_k$ matches in the asymptotic past $c k \eta \to  - \infty$ the Minkowski mode function \ie
\begin{equation}
	\label{def:BunchDavies_condition}
	u_{k} \xrightarrow[ c k \eta \to - \infty]{}  \sqrt{\frac{\hbar c}{2k}} e^{- i c k \eta} \, .
\end{equation}
Comparing with Eq.~\eqref{eq:subHubble_vk} it fixes the coefficients to be $A_{\bm{k}} = \sqrt{\hbar c / 2 k}$ and $B_{\bm{k}}=0$. The quantum origin of the perturbations is then reflected in the $k$-dependence, which is that of the quantum vacuum.
Second, to connect with the super-Hubble solution, we have to solve the Mukhanov-Sasaki equation Eq.~\eqref{eq:MS_equation} in the regime 
$k^2 \approx \mathfrak{z}^{\prime \prime}/\mathfrak{z}$ with these initial conditions in the past.
In general, the expression of $\mathfrak{z}^{\prime \prime}/\mathfrak{z}$ given by Eq.~\eqref{eq:exp_zprimeprime} is too complicated to solve the equation when the flow functions are time-dependent.
However, their evolution is itself slow-roll suppressed. 
Therefore, if we focus on a narrow enough time window around a reference time $\eta_{\star}$, we can Taylor expand the flow functions.
Since we focus on CMB scales, we consider a typical pivot scale $k_{\star}$ and define its Hubble crossing time by $\eta_{\star}$ such that $a_{\star} H_{\star} = c k_{\star} $. We denote by a $\star$ quantities evaluated at $\eta_{\star}$. 
We Taylor expand the flow-function about $\eta_{\star}$
\begin{equation}
	\label{eq:expansion_flow_functions_SR}
	\epsilon_{n} \left( \eta \right) = \epsilon_{n \star} \left\{ 1 + \epsilon_{n+1 \star} \left( N - N_{\star} \right)  + \mathcal{O} \left[  \left( N - N_{\star} \right)^2 \epsilon_{n}^2 \right] \right\} \, .
\end{equation}
All other CMB scales $k \neq k_{\star}$ will exit the Hubble radius at times $\eta$ such that $|N-N_{\star}| \approx 3.5$~\cite{Martin:2004um}. Since we expect $\epsilon_{1} ( \eta_{\star} ) \ll 1 $, the expansion is expected to be valid for a sufficiently large time window to describe all CMB scales. To solve the dynamics, we first compute the dynamics of the scale factor at first order in slow-roll. By definition of the first slow-roll parameter, we have
\begin{equation}
	\Hu \approx - \frac{1}{\eta} \left( 1 + \epsilon_{1 \star} \right) \, ,
\end{equation}
which, by integration, gives 
\begin{equation}
	\ln \left( \frac{a}{a_{\star}} \right) = -  \left( 1 + \epsilon_{1 \star} \right) \ln \left( \frac{\eta}{\eta_{\star}} \right) \, .
\end{equation}
Using $\Hu_{\star} \approx - (1 + \epsilon_{1 \star})/ \eta_{\star}$ we get
\begin{equation}
	\label{eq:expansion_scale_factor_SR}
	a \left( \eta \right)  \approx \frac{1}{H_{\star} \eta} \left[ 1 + \epsilon_{1 \star} - \epsilon_{1 \star} \ln \left( \frac{\eta}{\eta_{\star}} \right)  \right]  \, .
\end{equation}
We can then use Eq.~\eqref{eq:expansion_scale_factor_SR}, and Eq.~\eqref{eq:expansion_flow_functions_SR} for $\epsilon_1$, to get the expansion of $\mathfrak{z}$ around $\eta_{\star}$. We have
\begin{equation}
	\mathfrak{z} \left( \eta \right) \approx \sqrt{\frac{2 \epsilon_{1 \star}}{\kappa}} \frac{1}{H_{\star} \eta} \left[ 1 + \epsilon_{1 \star} - \left( \epsilon_{1 \star} + \frac{\epsilon_{2 \star}}{2} \right) \ln \left( \frac{\eta}{\eta_{\star}} \right)  \right] \, ,
	\label{eq:expansion_z_SR}
\end{equation}
where we have used $N-N_{\star} \approx - \ln ( \eta / \eta_{\star})$.
We can then compute the time-dependent parts of the frequencies driving the evolution of scalar and tensor perturbations
\begin{equation}
	\label{eq:SRexpansion_frequency_tensor}
	\frac{a^{\prime \prime}}{a} = \frac{2 + 3 \epsilon_{1 \star}}{\eta^2} \, ,
\end{equation}
and
\begin{equation}
	\label{eq:SRexpansion_frequency_scalar}
	\frac{ \mathfrak{z} ^{\prime \prime}}{ \mathfrak{z} } = \frac{2 + 3 \epsilon_{1 \star} + \frac{3}{2} \epsilon_{2 \star} }{\eta^2} \, ,
\end{equation}
where the terms in the logarithm of the conformal time have cancelled out.
Due to these cancellations, and the simple remaining time-dependence, the equation of motions, Eq.~\eqref{eq:MS_equation}, and the equivalent for tensor perturbations, can be integrated in terms of Hankel functions. For Eq.~\eqref{eq:MS_equation} we have
\begin{equation}
	\label{eq:scalar_mode_function}
	u_{k} \left( \eta \right) = C_k \sqrt{\frac{- c^2 \hbar \eta \pi}{4}} H^{(1)}_{\frac{3}{2} +
		\epsilon_{1 \star} + \frac{\epsilon_{2 \star}}{2}  } \left( - c k \eta \right) + D_k  \sqrt{\frac{- c^2 \hbar \eta \pi}{4}}   H^{(2)}_{\frac{3}{2} +
		\epsilon_{1 \star}+ \frac{\epsilon_{2 \star}}{2}  } \left( - c k \eta \right)    \, ,
\end{equation}
where $C_k$ and $D_k$ fix the initial conditions. Considering the limit of the Hankel functions in the asymptotic past\footnote{Implicitly, we assume that this asymptotic behaviour is reached sufficiently quickly to be within the range of validity of the expansion around $\eta_{\star}$. The same assumption is made when looking at the asymptotic future limit below.}, we have
\begin{equation}
	H^{(1/2)}_{\nu} \left( - c k \eta \right) \xrightarrow[ k \eta \to - \infty]{} \sqrt{\frac{2}{  - c k \eta \pi  }}  e^{  \pm i \left( c k \eta - \nu \pi /2 - \pi /4 \right)} \, ,
\end{equation}
so that the Bunch-Davies vacuum corresponds to $C_k = - e^{i \frac{\pi}{2} \left( \epsilon_1 + \frac{\epsilon_2}{2} \right)}$, and $D_k =0$. 
The mode function is now completely specified. We want to extract from it the value of the coefficient $\alpha_{\bm{k}}$ in front of the growing mode in Eq.~\eqref{eq:superHubble_vk}. To do so, we consider  $u_k/\mathfrak{z}(\eta)$, effectively the mode function of $\zeta$, and we take the asymptotic future limit. 
Generically
\begin{equation}
	H^{(1)}_{\nu} \left( - c k \eta \right) \xrightarrow[ c k \eta \to 0^+]{} - \frac{i}{\pi} \Gamma \left( \nu \right) \left( \frac{2}{- c k \eta} \right)^{\nu} \, ,
\end{equation}
where we have introduced the Gamma function $\Gamma (\nu)$.
Using the expansion of $\mathfrak{z} (\eta)$ of Eq.~\eqref{eq:expansion_z_SR} we get
\begin{align}
\label{eq:super_hubble_mode_function_zeta}
	\begin{split}
		\frac{u_k}{\mathfrak{z}} \underset{c k \eta \to 0^+}{\sim}   & i \frac{C_k}{2} \sqrt{\frac{\hbar \kappa H_{\star}^2 }{k^3 c \epsilon_{1 \star}}} \left[  1 +  \left( 1 - \ln 2 - \gamma_{\mathrm{E}}   \right) \epsilon_{1 \star} \phantom{\ln \left( \frac{k}{k_{\star}} \right)} \right. \\
		& \left. + \left( 2 - \ln 2 - \gamma_{\mathrm{E}}   \right) \frac{\epsilon_{2 \star}}{2}    - \left( 2 \epsilon_{1 \star} + \epsilon_{2 \star}  \right) \ln \left( \frac{k}{k_{\star}} \right)\right] \, ,
	\end{split}
\end{align}
where we have used $\Gamma \left( 3/2 + \nu  \right) \approx \sqrt{\pi} [1 + (2  - 2 \ln 2 - \gamma_{\mathrm{E}}) \nu ]/2$, $\gamma_{\mathrm{E}}$ being the Euler constant, and used $c k_{\star} = - \left( 1 + \epsilon_{1 \star} \right)/\eta_{\star}$ to replace $\eta_{\star}$.
We find that $u_k/ \mathfrak{z}$ is constant in the super-Hubble limit so that the matching to the super-Hubble solution is trivial in this case: we set $\alpha_{\bm{k}}$ equal to the right of  Eq.~\eqref{eq:super_hubble_mode_function_zeta}. At first order in slow-roll, we thus found that for scales close to $k_{\star}$, $u_k \approx \alpha_{\bm{k}} \mathfrak{z}(\eta)$ in the super-Hubble limit. 

\subsubsection{Connection to observations}

Using Eq.~\eqref{def:v_mode_expansion} we can compute the 2-point function related to the CMB temperature fluctuations.
We have
\begin{equation}
	\left \langle \hat{v} \left( \bm{x} , \eta \right) \hat{v} \left( \bm{x} + \bm{r} , \eta \right) \right \rangle = \int_0^{+\infty} \dd \ln k \,  \frac{\sin \left( k r \right)}{kr} \frac{k^3}{2 \pi^2} \left| u_k \right|^2 \, ,
\end{equation}
where we have used homogeneity and isotropy of the mode function to obtain the expression on the right-hand side. The power spectrum is then conventionally 
defined as
\begin{equation}
	\label{def:powerspectrum_v}
	\mathcal{P}_{v}  \left( k , \eta \right)  = \frac{k^3}{2 \pi^2}  \left| u_k \right|^2 \, .
\end{equation}
From Eq.~\eqref{def:powerspectrum_v}, we deduce the power spectrum of $\zeta$
\begin{equation}
	\label{def:powerspectrum_zeta}
	\mathcal{P}_{\zeta} \left( k , \eta \right)  = \frac{k^3}{2 \pi^2} \frac{ \left| u_k \right|^2 }{\mathfrak{z}^2} \, ,
\end{equation}
which is a dimensionless quantity.
These expressions are true at any time. We are interested in the value of $\mathcal{P}_{\zeta}(k)$ for CMB scales at the end of inflation. In the previous part, we have derived the asymptotic value of $u_k/\mathfrak{z} $, given in Eq.~\eqref{eq:super_hubble_mode_function_zeta}, which combined with Eq.~\eqref{def:powerspectrum_zeta} gives
\begin{align}
	\label{eq:curvature_power_SR}
\begin{split}
\mathcal{P}_{\zeta} \left( k \right)  & = \frac{H_{\star}^2 \kappa \hbar} {8 \pi^2 \epsilon_{1 \star} c} \left[ 1 + 2 \left( 1 - \ln 2 - \gamma_{\mathrm{E}}   \right) \epsilon_{1 \star}  \phantom{\ln \left( \frac{k}{k_{\star}} \right)    } \right. \\
& \left. + \left( 2 - \ln 2 - \gamma_{\mathrm{E}}   \right) \epsilon_{2 \star}  - \left( 2 \epsilon_{1 \star} + \epsilon_{2 \star}  \right) \ln \left( \frac{k}{k_{\star}} \right)     \right]  \, .
\end{split}
\end{align}

It is the standard formula for the power spectrum of scalar perturbations on CMB scales at the end of inflation~\cite{Mukhanov:1990me}. We can perform 
similar computations for tensor perturbations, see Sec.~\ref{sec:quantum_GW}. 
We introduce a mode function $u_{k, \lambda}^{\mathrm{T} }$ for each scalar polarisation which will satisfy Eq.~\eqref{eq:MS_equation} where $\mathfrak{z}^{\prime \prime}/\mathfrak{z}$ is replaced by $a^{\prime \prime}/a$. The power spectrum for tensor perturbations then reads
\begin{equation}
	\mathcal{P}_{\mathrm{T}} \left( k , \eta \right) = \frac{4 \pi \kappa }{\pi^2 a^2 c} k^3 \left| u_{k, \lambda}^{\mathrm{T} } \right|^2 \, .
\end{equation}
A direct comparison of Eq.~\eqref{eq:SRexpansion_frequency_tensor} and Eq.~\eqref{eq:SRexpansion_frequency_scalar} then shows that the equation are equal under the substitution $\epsilon_{1 \star} \to \epsilon_{1 \star} + \epsilon_{2 \star} / 2$. Therefore, we can repeat the same procedure of matching sub and super-Hubble limits using an intermediate slow-roll approximation.
For CMB scales at the end of inflation, we then have
\begin{equation}
	\label{eq:tensor_power_SR}
	\mathcal{P}_{\mathrm{T}} \left( k \right)  =  2 \frac{H_{\star}^2 \kappa \hbar}{\pi^2 c} \left[ 1 + 2 \left( 1 - \ln 2 - \gamma_{\mathrm{E}}   \right) \epsilon_{1 \star} -  2 \epsilon_{1 \star} \ln \left( \frac{k}{k_{\star}} \right)     \right]  \, .
\end{equation}
The spectra Eq.~\eqref{eq:curvature_power_SR} and Eq.~\eqref{eq:tensor_power_SR} are the key predictions of the theory of inflationary perturbations. We briefly discuss how we can extract information about inflation by measuring them. First,
if we neglect the slow-roll corrections, both power spectra Eq.~\eqref{eq:curvature_power_SR} and Eq.~\eqref{eq:tensor_power_SR} are scale invariant. The deviations from scale invariance appear in the first order in the slow-roll parameter due to departure from an exact de Sitter expansion. 
These deviations are customarily parameterised by introducing the spectral indices
\begin{equation}
	\label{def:ns}
	n_{\mathrm{S}} = 1 + \frac{\dd \ln \mathcal{P}_{\zeta}}{\dd \ln k} \, ,
\end{equation}
and 
\begin{equation}
	\label{def:nt}
	n_{\mathrm{T}} = \frac{\dd \ln \mathcal{P}_{\mathrm{T}}}{\dd \ln k} \, .
\end{equation}
Further deviations could be parameterised by higher derivatives of the power spectra, e.g. the so-called running of the spectral index. 
Truncating at the spectral index we can write $\mathcal{P}_{\zeta} = A_{\mathrm{s}} \left( k / k_{\star} \right)^{n_{\mathrm{S}} -1}$ and $\mathcal{P}_{\mathrm{T}} = A_{\mathrm{T}} \left( k / k_{\star} \right)^{n_{\mathrm{T}} }$.
Scale invariance corresponds to $n_{\mathrm{S}} = 1$ and $n_{\mathrm{T}} = 0$. Eq.~\eqref{eq:curvature_power_SR} and Eq.~\eqref{eq:tensor_power_SR} gives in the first order in slow-roll $n_{\mathrm{S}} = 1 - 2 \epsilon_{1 \star} - \epsilon_{2 \star}$ and $n_{\mathrm{T}} = - 2 \epsilon_{1 \star}$.
The behaviour of the spectra thus depends on the values of the slow-roll parameters, which are model-dependent quantities. Nevertheless, since $\epsilon_1 > 0$ for any inflation model, inflation generically predicts a slightly \textit{red-tilted} spectrum for tensor perturbations $n_{\mathrm{T}}<0$, \ie a spectrum with an amplitude decreasing with $k$. On the other hand, the scalar spectrum can be blue-tilted $n_{\mathrm{S}}>1$ in specific inflationary models where $\epsilon_2$ becomes negative~\cite{Martin:2013tda}.
The experimental measures of the power spectra constrain the values of the spectral indices $n_{\mathrm{S}}$ and $n_{\mathrm{T}}$, and so constrain the space of allowed inflationary models to that which can give rise to these values.
Another important quantity is the tensor-to-scalar ratio
\begin{equation}
	\label{def:tensor_to_scalar}
	r = \frac{A_{\mathrm{T}}}{A_{\mathrm{S} }} \approx 16 \epsilon_{1 \star} \, ,
\end{equation}
where we have evaluated the expression in the first order in slow-roll parameters. The level of tensor perturbations is slow-roll suppressed compared to scalar perturbations. 

So far, only the dominant scalar perturbations have been identified in the CMB, and only an upper bound for the ratio is known $r < 0.036$ at $95 \%$ confidence level~\cite{BICEP:2021xfz}.
The latest measurement of the CMB by Planck has provided significant support for the prediction of single field inflation. It found no trace of non-Gaussianity in the inhomogeneities at its level of precision~\cite{Planck:2019kim}, and the spectrum of scalar perturbations which was found to be slightly red-tilted with a spectral index measured~\cite{Planck:2018vyg} to $n_{\mathrm{S}} = 0.9649 \pm 0.0042$. The overall amplitude was measured to be $A_{\mathrm{S}} =  2.101^{+0.031}_{-0.034} \times 10^{-9}$.
The measure of the index constrains the value of the slow-roll parameters at the Hubble crossing for the cosmological scale.
A measure of the tensor perturbations would allow direct access to the 
value of the first slow-roll parameter via the index $n_{\mathrm{T}}$, and then to the value of the Hubble parameter $H_{\star}$. We recall that its value is directly related to the energy scale of inflation via Eq.~\eqref{def:energy_scale}. We get
\begin{equation}
	E_{\star} = \left( \frac{3 \pi^2}{2} r A_{\mathrm{S}} \right)^{1/4} \Mp c^2 \, ,
\end{equation}
so the observational upper bound on $r$ gives an upper bound on the energy scale of inflation. We find roughly $E_{\star} \leq 10^{16} \, \mathrm{GeV}$.

To close this sub-section, we want to emphasise that the tremendous practical success of the inflationary mechanism to predict the CMB observations rests on the assumption that these cosmological-scale inhomogeneities emerged from initial vacuum fluctuations, a purely quantum phenomenon.
Therefore the CMB temperature fluctuations would be a trace of quantum physics playing a role on cosmological scales, an observation at odds with the common lore that quantum physics is only relevant on very short scales.
This, albeit commonly accepted, surprising feature of the model is worth a deeper analysis.
First, in the treatment we presented previously, it was unclear when the physics described stopped being quantum and became completely classical. This is the quantum-to-classical transition problem.
Second, although the measure of the power spectrum matched the one predicted from amplified vacuum fluctuations, this only constitutes indirect proof of the quantum origin of the perturbations. We could provide an \textit{ad-hoc} initial stochastic Gaussian distribution that would reproduce the final measured power spectrum, see Sec.~\ref{sec:quantum_GW}. One might thus want to seek other features in the final distribution that initial classical inhomogeneities would not be able to reproduce.
Finally, finding such a feature would demonstrate that a quantum treatment of the fluctuations of geometry is required in the early Universe,
so that gravity (at least at a linear level) has to acquire the status of a quantum theory.
These are the motivations for the analysis conducted in Chapt.~\ref{chapt:quantumness_cosmo}.

\subsection{(P)reheating: connecting inflation to the standard model of cosmology}
\label{sec:preheating}

\subsubsection{Reheating processes}

In Sec.~\ref{sec:infl_solBB_puzzles}, we introduced a transitory phase of inflation within the radiation domination era.
We have explained that such a phase allows us to solve the hot Big Bang puzzles while giving a generation mechanism
for inhomogeneities in the early Universe, the seeds for structure formation.
During inflation, the Universe's energy budget is entirely dominated by the inflaton field $\varphi$ (and its perturbations $\delta \hat{\varphi}$). Any pre-existing form of matter is expected to be highly diluted by the expansion so that most of the Universe's matter content must be produced after inflation~\cite{Dolgov:1982th,Traschen:1990sw}.
At the end of inflation, the energy of the inflaton must be somehow transferred to particles in other fields and eventually into the particles of the Standard Model.
In addition, to allow BBN to proceed as predicted and observed, these particles must thermalise and do so at a temperature $T_{\mathrm{reheat}}$ at least as large as the temperature then: $T_{\mathrm{reheat}} \geq  T_{\mathrm{BBN}} \approx 1 \, \mathrm{MeV}$.
The processes leading to the initial radiation domination of the standard Hot Big Bang model, and the period during which they unfold, are commonly referred to as \textit{reheating}\footnote{Since we expect inflation to break any pre-existing thermal equilibrium and dilute the existing forms of matter, the Universe at the end of inflation would be empty and cold. Hence, we must (re)heat it.}~\cite{Albrecht:1982mp}. We give a brief overview of reheating to connect with the work on an `analogue preheating experiment' presented in Chapt.~\ref{chapt:analogue_preheating}. We refer to the review~\cite{Amin:2014eta} for details.

The generation of particles in reheating is expected to proceed in two ways. First, by perturbative decays of inflaton particles $\varphi$, to particles of other matter fields $\chi$. An example of such a process is a 3-body interaction $\varphi  \to \chi \chi$, corresponding to the annihilation of $\varphi$-particles in pairs of $\chi$ particles . In the first papers on reheating, such perturbative decay processes were considered by adding an effective decay rate in the Klein-Gordon equation~\eqref{eq:klein_gordon} of the inflaton. 
Additionally, it was later realised~\cite{Dolgov:1989us,Traschen:1990sw,Kofman:1994rk} that the oscillations of the inflaton at the end of inflation, see Fig.~\ref{fig:evolution_R2inflation}, effectively acting as a strong classical field, could trigger an exponential production of particles in modes of other fields $\chi$, within some energy range related to the frequency of the oscillations.
This is the phenomenon of parametric amplification. The parametric processes can initially be much more efficient than the perturbative ones~\cite{Kofman:1997yn}. 
This phase of parametric growth is referred to as \textit{preheating}~\cite{Kofman:1994rk}.
However, reheating never completes at this stage~\cite{Kofman:1997yn}. First, parametric resonances produce particles in a very non-thermal state. Interactions between the produced particles are thus necessary to drive the distribution to a thermal one. Second, the parametric resonance process is going to become inefficient as the oscillations of the inflaton are damped by the energy transfer to other particles. Perturbative channels are necessary to ensure in the late stages a total decay of the inflaton and completion of reheating~\cite{Kofman:1997yn,Amin:2014eta}.
In~\cite{Micheli:2022zet}, reproduced in Sec.~\ref{sec:decay_nk}, we consider some interactions between quasi-particles produced in an analogue preheating experiment.
However, this analysis focuses on how these interactions affect the production of quantum coherence in the early stage of parametric resonance rather than on how they could lead to the thermalisation of the excitations.

\subsubsection{Illustration of preheating}

To close this sub-section, we illustrate the phenomenology of parametric resonance by considering a simple model of preheating used in~\cite{Kofman:1997yn}. 
Assume that the potential of the inflaton field has a minimum at $\varphi =0$. We approximate the potential to be quadratic around its minimum
\begin{equation}
	V_{\varphi} \left(\varphi \right) \approx \frac{ m^2 c^2}{2 \hbar^2} \varphi^2 \, ,
\end{equation}
and assume that this approximation represents well the part probed by the inflaton in its late-time oscillations. 
Assume now that the inflaton is coupled to a quantum scalar field $\hat{\chi}$. We assume $\hat{\chi}$ to be massless for simplicity.
We take the coupling to the inflaton of the form $\lambda^2 \varphi^2 \hat{\chi}^2$. The action for $\chi$ reads
\begin{equation}
	\label{def:action_chi}
	S_{\chi} = - \frac{1}{c} \int \dd^4 x \sqrt{- g} \left(\frac{1}{2} g^{\mu \nu} \partial_{\mu} \hat{\chi} \partial_{\nu} \hat{\chi} + \frac{1}{2} \lambda^2 \varphi^2 \hat{\chi}^2    \right)  \, .
\end{equation}
As we did previously for $\hat{v}$, we can perform a mode expansion of the scalar field
\begin{align}
	\hat{\chi} \left( \bm{x} , t \right) & = \int \frac{\mathrm{d}^3
		\bm{k}}{(2 \pi)^{3/2}} \left[ e^{i \bm{k}.\bm{x}} \chi_{k} \left( t
	\right) \hat{c}_{\bm{k}} + e^{- i
		\bm{k}.\bm{x}} \chi_{k}^{\star} \left( t \right)
	\hat{c}_{\bm{k}}^{\dagger}  \right] \, .
	\label{def:chi_mode_expansion}
\end{align}
The mode functions $\chi_k$ are going to satisfy the equation of motion of the Fourier modes of the classical version of the field
\begin{equation}
	\label{eq:EOM_chi_k}
	\ddot{\chi}_{\bm{k}} + 3 \frac{\dot{a}}{a} \dot{\chi}_{\bm{k}}  + \left[ \frac{c^2 k^2}{a^2(t)}  + \lambda^2 \varphi^2(t) \right] \chi_{\bm{k}} = 0
\end{equation}
We make the simplifying assumption that, over the characteristic time scale of the expected growth, we can neglect the expansion of space and take $a=1$ in Eq~\eqref{eq:EOM_chi_k}. As a second simplifying assumption, we assume that the inflaton $\varphi$ oscillates at constant amplitude fixed by its mass $\varphi(t) = \varphi \sin (m c^2 t/\hbar)$. This assumption eventually breaks down because the oscillations are damped by the expansion of space and the transfer of energy to $\hat{\chi}$. Under these assumptions Eq.~\eqref{eq:EOM_chi_k} can be recast in the form of a Mathieu equation~\cite{Kofman:1997yn}
\begin{equation}
	\label{eq:mathieu_eq}
	\frac{\partial^2 \chi_{\bm{k}} }{\partial z^2} + \left[ \mathcal{A}_k + 2 q \cos \left( 2 \theta \right) \right] \chi_{\bm{k}} = 0 \, ,
\end{equation}
where $\mathcal{A}_k = 2 q +  c^2 k^2 /m^2$, $q = \lambda^2 \varphi^2 /2 m^2$ and $\theta = m c^2 t/\hbar$.
Eq.~\eqref{eq:mathieu_eq} is a well-known equation. Its analysis~\cite{kovacicMathieuEquationIts2018} shows that for some $q$-dependent ranges of $\mathcal{A}_k$ around specific values $\mathcal{A}_k^{(l)}$,
the solution will exhibit exponential growth. Outside of these instability bands, the modes oscillate. In the narrow resonance regime, $q \ll 1$, the instability bands are located around $\mathcal{A}_k^{(l)} \approx l^2$ and of size $\sim q^{l}$. Therefore, the primary band $\mathcal{A}_k^{(1)} \approx 1$ is the largest one, and the resonant modes have frequencies $c k \approx m c^2/\hbar$ \ie half of the driving frequency. We refer to~\cite{kovacicMathieuEquationIts2018} for more on the instability regime.
Notice that in a classical setting, where the Fourier modes of the classical field satisfy Eq.~\eqref{eq:EOM_chi_k}, an initial vacuum of $\hat{\chi}$ excitation corresponds to $\chi_{\bm{k}} =0$. In this case, the solution of Eq.~\eqref{eq:EOM_chi_k} gives  $\chi_{\bm{k}} =0$ at all times. In the quantum case, however, the vacuum corresponds to the normalisation Eq.~\eqref{def:BunchDavies_condition}, which leads to an amplification of vacuum fluctuations, a pure quantum phenomenon.
As we will detail in Sec.~\ref{sec:analogue_cosmo_exp}, a signature of this quantum origin is that the generated pairs are entangled. The entanglement can be generated even if a small incoherent population is initially present.

Note the similarity of this process with the vacuum amplification of scalar and tensor perturbations during the slow-roll part of the evolution. The different dynamics of the background, growing for inflation and oscillating for preheating, lead to different excitation spectra. While any super-Hubble mode is amplified during inflation, only modes in well-defined resonant modes are produced during preheating.
Both of these effects fit in a large class of phenomena where a quantum field is excited by a strong classical field such as the Schwinger effect~\cite{Schwinger:1951nm} or the Hawking effect~\cite{Hawking:1974rv,Hawking:1975vcx}.
These effects are described using the formalism of quantum field theories in curved spacetimes (QFTCS).
We see that such QFTCS effects have significant cosmological consequences. Yet, they are hard to demonstrate experimentally.
These effects are small, and the field strengths accessible in laboratories are often insufficient to generate a signal strong enough to
be detected.
Therefore, a first avenue to explore is to look for signatures of such quantum effects directly in cosmology. This is the topic of Chapt.~\ref{chapt:quantumness_cosmo} for the
generation of inflationary perturbations.
A second possible avenue is to rely on analogue systems, which are described by the same equations of motion as a QFTCS situation while being experimentally realisable.
This perspective led to the development of the  \textit{analogue gravity}  community, which we present in the next section, Sec.~\ref{sec:introduction_AG}.

%%%%%%%
%To add the bibliography when compiling the subfile only
%\bibliographystyle{plain}
%\biliography{bibmanuscript}

\section{Analogue gravity}

\label{sec:introduction_AG}

\subsection{Motivations}
\label{sec:motivations_AG}

The idea of analogue gravity first emerged in the study of Hawking radiation~\cite{Hawking:1974rv,Hawking:1975vcx,Unruh:1976db}.
By studying the evolution of modes of a field in the geometry of a star collapsing to a black hole, Hawking predicted that the black hole would emit a thermal black-body spectrum of the particles associated with that field.
The temperature of such radiation was predicted to be inversely proportional to $M$, the mass of the black-hole
\begin{equation}
	\label{def:Hawking_radiation}
	T_{\mathrm{H}} = \frac{\hbar c^3}{8 \pi G M k_{\mathrm{B}}} \approx \frac{M_{\odot}}{M}  60 \mathrm{nK} \, .
\end{equation}
For astrophysical black holes of mass typically larger than a few solar masses~\cite{KAGRA:2021duu}, this temperature is much smaller 
than that of the cosmic microwave background $T_{\mathrm{CMB}} \approx 3 \, \mathrm{K}$~\cite{Fixsen:2009ug}, so that the hole effectively absorbs energy rather than emits. The temperature is even smaller than the CMB temperature anisotropies $\delta T / T_{\mathrm{CMB}} \approx 10^{-5}$~\cite{COBE:1992syq}.
Thus, even if the prediction of Hawking is correct, the radiation of astrophysical black holes seems too small to be detected~\cite{Hawking:1975vcx}.\footnote{As already mentioned in~\cite{Hawking:1975vcx}, smaller black-holes could have formed as \textit{primordial black-holes} from the collapse of large density fluctuations at the end inflation~\cite{Zeldovich:1967lct,Carr:1974nx}. If they exist, such black holes can be of all masses, and the smallest ones would emit at a high rate. }

The Hawking effect illustrates the difficulty of observing QFTCS effects directly.
A similar illustration can be found in the Unruh effect~\cite{Unruh:1976db}. It predicts that a uniformly accelerating observer in the Minkowski vacuum would observe thermal radiations of particles at a temperature
\begin{equation}
	\label{def:Unruh_temperature}
	T_{\mathrm{U}} = \frac{\hbar}{c} \frac{\alpha}{2 \pi k_{\mathrm{B}}} \, ,
\end{equation}
where $\alpha$ is the proper acceleration of the observer.
Getting a temperature of 1 K in Eq.~\eqref{def:Unruh_temperature}, with peak radiation in the microwave, requires accelerating the detector
to $\alpha \approx 10^{20} \mathrm{m}/\mathrm{s}^2$. To get a reference, we consider the initial linear accelerator Linac4 at the LHC~\cite{Vretenar:2020quc} that takes $\mathrm{H}^{-}$ ions, that we will assume to be protons, from rest to $160 \, \mathrm{MeV}$ of kinetic energy over $80 \, \mathrm{m}$. To get a very rough estimate, we assume the acceleration to be uniform throughout and compute the proper acceleration of the particle (which, for these velocities, does not differ much from the acceleration in the laboratory frame). We get $a \approx 5.70 \times 10^{14} \mathrm{m}/\mathrm{s}^2$ \ie six orders of magnitude lower than the acceleration necessary even to get microwave radiation. Other estimates and a proposition to indirectly observe the Unruh effect can be found in~\cite{Bell:1964kc}. 

Still, a few years after the predictions of Hawking radiation, Unruh realised that the equations describing the propagation of linear perturbations (\ie sound waves) in a trans-sonic fluid could be recast in the form of the equation of motions of a scalar field in a black-hole metric~\cite{Unruh:1980cg}.
The intuition behind this result is that in a trans-sonic fluid, there is a sonic horizon. Past this horizon, the flow velocity is larger than the speed of sound, and sound waves can only propagate downstream, like particles beyond the horizon of a black hole. Unruh coined the word `dumb'-holes for these geometries where sound is trapped.
Following the same reasoning leading to Hawking radiation, a quantum fluid with such a sonic horizon would emit thermal radiation of sound waves. Unruh found the temperature to be typically small 
\begin{equation}
	T \approx 3 \times 10^{-7} \,  \frac{c_{\mathrm{s}}}{c} \left(  \frac{ R}{1 \mathrm{mm} } \right)^{-1}  \mathrm{K}    \, ,
\end{equation}
where $c_{\mathrm{s}}$ is the speed of sound in the fluid and $R$ the size of the horizon. Nevertheless, to quote his words, observing the radiation in this fluid is still `a much simpler experimental task than creating a $10^{-8} \, \mathrm{cm}$ black hole'.

In addition to opening up the possibility for the experimental observation of an effect akin to Hawking radiation, the study of this \textit{analogue} system was
also acknowledged as a means to assess the robustness to UV modifications of QFT above the Planck scale~\cite{Unruh:1980cg}. Indeed, the derivation assumes no back-reaction of the produced particles on the background metric, and also that there are similar quantum fluctuations available until arbitrarily larger energy, even beyond the Planck scale, to fuel the radiation indefinitely~\cite{Parentani:2002bd}. This is the so-called trans-planckian problem~\cite{Martin:2000xs,Niemeyer:2000eh,Niemeyer:2001qe}. While the physics beyond the Planck scale \ie a theory of quantum gravity, is unknown, we know the fluid theory breakdowns when considering scales close to the inter-particle separation. Therefore, the analogue fluid can be used to investigate how a modified UV theory affects the predicted radiation. Ref.~\cite{Unruh:1980cg} marks the beginning of the analogue gravity programme, initially focused on the analogue of the Hawking effect.
Before reviewing the field's main achievements, we demonstrate the advertised correspondence between the propagation of sound waves on a fluid and that of a scalar field in curved spacetime.

\subsection{Sound-waves on a fluid as scalar field in curved spacetime}
\label{sec:acoustic_metric}

We closely follow the demonstration made in~\cite{Barcelo:2005fc}. Consider a fluid not submitted to any external force described by a pressure field $p$, a mass density field $\rho$ and a velocity field $\bm{v}$. We work in the laboratory frame $(ct, \bm{x})$ and the motion of the fluid is described by the continuity and Euler's equations
\begin{subequations}
	\begin{equation}
		\label{eq:Euler}
		\partial_{t} \rho + \vec{\nabla} \cdot \left( \rho \bm{v} \right) = 0 \, ,
	\end{equation}
	\begin{equation}
		\label{eq:continuity}
		\partial_{t} \bm{v} + \bm{v} \left( \vec{\nabla} \cdot \bm{v} \right) = - \frac{1}{\rho} \vec{\nabla} p \, .
	\end{equation}
\end{subequations}
The continuity equation expresses the local conservation of the mass density, while Euler's equation comes from applying Newton's third law to a mesoscopic volume element.
Assuming the fluid to be irrotational and making a Helmholtz decomposition of the velocity field, we have $\bm{v} = - \vec{\nabla}  \phi $, where $\phi$ is the velocity potential. Then 
$ \vec{\nabla} \cdot \bm{v} =  - \bm{v} \times ( \vec{\nabla} \times \bm{v}) + \vec{\nabla} (v^2)/2 =  \vec{\nabla} (v^2)/2 $.
Assuming now the fluid to be barotropic \ie the energy density field is a function of the pressure field $\rho = \rho (p)$, we can define the specific enthalpy of the fluid
\begin{equation}
	\label{def:enthalpy}
	h \left( p \right) = \int_0^p \frac{\dd p^{\prime}}{\rho \left( p^{\prime} \right)} \, ,
\end{equation}
such that $\vec{\nabla} h = \vec{\nabla} ( p ) / \rho$.\footnote{The enthalpy of a thermodynamic system of volume $V$ is defined as the sum of its internal energy $U$ and the energy due to external pressure $P$ exerted on it $H = U + P V$. Consider the enthalpy variation in the system to see how Eq.~\eqref{def:enthalpy} is related to the enthalpy. It reads $\dd H = T \dd S + V \dd P$. If we consider an isentropic system $\dd S = 0$, then $\dd H = V \dd P$. The specific enthalpy $h$ is the enthalpy per unit of mass, so $\dd h =  \dd P / \rho$ since the volume divided by the mass is the inverse mass density. Integrating this equation gives back Eq.~\eqref{def:enthalpy}.} Euler equation~\eqref{eq:Euler} then reads
\begin{equation}
	\vec{\nabla} \left( - \partial_{t} \phi + h + \frac{v^2}{2} \right) = 0 \, ,
\end{equation}
which can be integrated to 
\begin{equation}
	\label{eq:Euleur_baro_irrot}
	- \partial_{t} \phi + h + \frac{v^2}{2} = 0 \, .
\end{equation}
The integration constant $C$ can always be absorbed by redefining the velocity potential $\phi \to \phi + C t $, which does not affect the velocity field. We set it to $0$ in Eq.~\eqref{eq:Euleur_baro_irrot}. Once the functional link between the density $\rho$ and pressure field $p$ is fixed, the equations Eq.~\eqref{eq:continuity} and Eq.~\eqref{eq:Euleur_baro_irrot} form a closed system for two dynamical degrees of freedom, the velocity potential $\phi$ and the pressure field $p$. We now consider the propagation of perturbations on top of a background solution $\phi_0$ and $p_0$ of these equations; this is the definition of sound waves.
We have
\begin{subequations}
	\begin{align}
		\begin{split}
			p & = p_0 + \delta p \, , 
		\end{split} \\
		\begin{split}
			\phi & = \phi_0 + \delta \phi \,  , 
		\end{split} \\
		\begin{split}
			\bm{v} & = - \vec{\nabla} \phi_0 - \vec{\nabla} \left(  \delta \phi \right) = \bm{v}_0 + \delta \bm{v} \,   , 
		\end{split}
	\end{align}
\end{subequations}
where we assume the perturbations to be of small amplitude $|\phi_0| \gg |\delta \phi |$ and $|p_0| \gg |\delta p|$. Varying the specific enthalpy Eq.~\eqref{def:enthalpy} we have
\begin{equation}
	h \left( p \right) \approx h \left( p_0 \right) + \frac{\delta p}{\rho_0} \, .
\end{equation}
The perturbations of Eq.~\eqref{eq:continuity} and Eq.~\eqref{eq:Euleur_baro_irrot}
at linear order, then reads
\begin{subequations}
	\begin{align}
		\begin{split}
			\label{eq:pert_continuity}
			- \partial_t \delta \phi + \frac{\delta p}{\rho_0} + \bm{v}_0 \cdot \delta \bm{v} & = 0 \, ,
		\end{split} \\
		\begin{split}
			\label{eq:pert_Euler}
			\partial_t \delta \rho + \vec{\nabla} \left( \delta p \bm{v}_0 \right) + \vec{\nabla} \cdot \left( \rho_0  \delta \bm{v} \right) & = 0 \, ,
		\end{split} 
	\end{align}
\end{subequations}
Using Eq.~\eqref{eq:pert_continuity} we get
\begin{equation}
	\label{eq:pressure_as_fn_velocity_pot}
	\delta p = \rho_0 \left[ \partial_t \delta \phi + \bm{v}_0 \cdot \vec{\nabla} \left( \delta \phi \right)  \right] \, ,
\end{equation}
that completely expresses the pressure perturbation as a function of the velocity potential perturbation. We define the local speed of sound $c_0$ by
\begin{equation}
	c_0^{-2} =  \left. \frac{\partial \rho}{\partial p}  \right| _{\rho_0} \, .
\end{equation}
Notice that the speed of sound is \textit{a priori} a field that depends on time and position. Using Eq.~\eqref{eq:pressure_as_fn_velocity_pot}, the density perturbation can then be expressed as
\begin{equation}
	\label{eq:density_as_fn_velocity_pot}
	\delta \rho = \frac{\rho_0 }{c_0^2} \left[ \partial_t \delta \phi + \bm{v}_0 \cdot \vec{\nabla} \left( \delta \phi \right)  \right] \, .
\end{equation}
Combining this equation with Eq.~\eqref{eq:pert_Euler} we get an equation for the perturbation of the velocity potential alone
\begin{equation}
	- \partial_t \left\{  \frac{\rho_0 }{c_0^2} \left[ \partial_t \delta \phi + \bm{v}_0 \cdot \vec{\nabla} \left( \delta \phi \right)  \right] \right\} + \vec{\nabla} \cdot \left\{ \partial_t \delta \phi -  \frac{\rho_0 }{c_0^2} \left[ \partial_t \delta \phi + \bm{v}_0 \cdot \vec{\nabla} \left( \delta \phi \right)  \right]\right\} \, .
\end{equation}
This last equation can then be rewritten as a Klein-Gordon equation, see Eq.~\eqref{eq:klein_gordon}, for $\delta \phi$ as if it were a minimally coupled scalar field on curved spacetime without a potential~\cite{Barcelo:2005fc}
\begin{equation}
	\label{eq:analogue_EOM}
	\Box \delta \phi = - \frac{1}{\sqrt{-g}} \partial_{\mu} \left( \sqrt{-g} g^{\mu \nu} \partial_{\nu} \delta \phi \right) = 0 \, ,
\end{equation}
where the \textit{acoustic} metric is defined in the laboratory coordinates $(t , \bm{x})$ by
\begin{equation}
	\label{def:acoustic_metric}
	g\indices{_{\mu} _{\nu}} = \frac{\rho_0}{c_0} 
	\begin{pmatrix}
		- \left( c_0^2 - v_0^2 \right) & - \bm{v}_0 \\
		- \bm{v}_0 & \mathds{1}_3
	\end{pmatrix}
	\, ,
\end{equation}
where $\mathds{1}_3$ is the three-dimensional identity matrix.
Eq.~\eqref{eq:analogue_EOM} is the basis of the analogy. We give the associated acoustic line-element\footnote{Notice that the metric is not dimensionless and the line-element does not have the dimension of a length squared. However, since the field obeys a linear equation of motion, the metric can always be rescaled by an arbitrary factor to be made dimensionless.}
\begin{equation}
	\label{eq:acoustic_line_element}
	\dd s^2 = \frac{\rho_0}{c_0} \left[ - \left( c_0^2 - v_0^2 \right) \dd t^2 - 2 \dd t \bm{v}_0\cdot \dd \bm{x} + \dd \bm{x} \cdot \dd \bm{x} \right] \, .
\end{equation}
Although we used a classical framework, since Eq.~\eqref{eq:analogue_EOM} is a linear equation, we would obtain the same result by considering a quantum fluid with a large coherent background treated classically and some perturbations on top, e.g. phonons on a Bose-Einstein condensate $\hat{\phi} \approx \phi_0 \hat{\mathds{1}} + \delta \hat{\phi}$, see Sec.~\ref{chapt:analogue_preheating}.
Notice~\cite{Barcelo:2005fc} that the analogy has limits. First~\cite{Barcelo:2005fc}, we only reproduce the kinematics of a field on a given spacetime metric and, although this metric could be dynamical, e.g. if $\phi_0$ is made time-dependent, its dynamics is \textit{not} described by analogue Einstein's field equations. Second, we can change coordinates, e.g., redefining time $t \to \tau$ to match the form that a curved spacetime metric would have in a specific set of coordinates; see below for an example with Schwarzchild metric. However, measuring devices will ultimately experience the time of the laboratory frame $t$ and not the transformed one $\tau$, see however~\cite{Fedichev:2003dj}. Finally, 
not every metric can be reproduced, at least not in the simple treatment we presented. In particular, a generic $3+1$ spacetime metric is defined by $6$ independent functions: the metric can be represented by a symmetric $4\times4$ matrix in coordinates which has $10$ independent coefficients, but these coefficients depend on a choice of coordinate system which we are free to choose, adding $4$ constraints.
On the other hand, the acoustic metric features three functions $p_0$, $\phi_0$ and $c_0$. The continuity equation relates the first two; we are only left with $2$ functions to fix.
Still, tuning these two functions allows us to reproduce many physically relevant spacetimes, see~\cite{Barcelo:2005fc} for examples. 

To close this discussion, we show, following~\cite{Unruh:1980cg} (while keeping the mostly pluses convention used in this manuscript), how to reproduce the near horizon part of a black-hole metric.
First, let us assume that the speed of sound $c_0$ is everywhere constant and that the background flow is spherically symmetric, stationary and convergent so that $\rho (r)$ and $\bm{v} = v^r(r) \mathbf{u}_r$, with $ v^r(r) <0$. Going to spherical coordinates the line element~\eqref{eq:acoustic_line_element} then reads
\begin{equation}
	\dd s^2 = \frac{\rho_0}{c_0} \left\{ - \left( c_0^2 - v_0^2 \right) \dd t^2 - 2 v_0^r \dd t \dd r + \dd r^2 + r^2 \left[ \dd \theta^2 + \sin^2 \theta \dd \varphi^2 \right] \right\} \, .
\end{equation}
We define a new time coordinate
\begin{equation}
	\tau = t + \int^{r}_0 \frac{v_0^r}{c_0^2 - \left( v_0^r \right)^2} \dd r^{\prime} \, ,
\end{equation}
and assume that the fluid becomes trans-sonic at a radius $R$ such that $v^r(r) \approx -c_0 + 2 \alpha (r-R)$ close to the sonic horizon.
The line element close to the horizon $r \approx R$ then reads
\begin{equation}
	\dd s^2= \frac{\rho_0}{c_0} \left[ - 2 \alpha \left( r - R \right) c_0 \dd \tau^2 + \frac{c_0}{2 \alpha \left( r - R \right)} \dd r^2 + r^2 \left( \dd \theta^2 + \sin^2 \theta \dd \varphi^2 \right) \right] \, .
\end{equation}
Comparing with Schwarzchild metric in usual spherical coordinates close to the horizon $r \approx r_{\mathrm{s}} = 2 G M / c^2$, where
\begin{equation}
	\dd s^2_{\mathrm{Sch.}} =  - \frac{c^2}{r_{\mathrm{s}}^2} \left( r - r_{\mathrm{s}} \right) \dd t^2 + \frac{r_{\mathrm{s}}}{ r - r_{\mathrm{s}}} \dd r^2 + r^2 \left( \dd \theta^2 + \sin^2 \theta \dd \varphi^2 \right)  \, ,
\end{equation}
we have the mapping $c_0 = c$ and $\alpha = c^3 / 4 G M $ up to the factor $\rho_0 / c_0$ in front. In the analogy, the speed of sound plays the role of the speed of light, and the factor $\alpha$ is related to the mass of the black hole.
This analogy at the linear level, with a conformal factor, is sufficient to get the Hawking effect~\cite{Unruh:1980cg,Barcelo:2005fc}. Here we only use the formula~\eqref{def:Hawking_radiation} to find what the analogue of Hawking's temperature would be
\begin{equation}
	T_{\mathrm{H}} = \frac{\hbar}{2 \pi k_{\mathrm{B}}} \left(   \left.  \frac{\partial v^r}{\partial r} \right| _{r= R} \right)^{-1} \, .
\end{equation}

Theoretical refinements of this idea and experimental progress towards observing an analogue Hawking effect have been the central focus in the analogue gravity community until recently.

\subsection{Progress and achievements in Analogue Gravity}
\label{sec:progress_AG}

Several accounts of the progress in the study of analogue systems exist in the literature~\cite {Parentani:2002bd,Barcelo:2005fc,Jacquet:2020bar,Almeida:2022otk}. 
We limit ourselves here to a short account mainly following~\cite{Parentani:2002bd,Almeida:2022otk} to illustrate the diversity of platforms used and highlight the successes of the analogue gravity programme so far. 
The initial focus was on analogue black holes and the Hawking effect. A first theoretical success was in understanding that the break-down of the fluid model for large frequencies did not preclude the existence of Hawking radiation, and, provided dispersion was weak enough, preserved approximately its thermality at the expected temperature given in Eq.~\eqref{def:Hawking_radiation}  ~\cite{Unruh:1994je,Brout:1995wp,Jacobson:1999zk,Parentani:2002bd}. 
On the experimental side, many systems were considered as possible platforms to realise the analogy.
For instance, the possibility of using superfluid He-3~\cite{Volovik:1997xi,Kopnin:1997jy,Jacobson:1998he}, 
or Bose-Einstein condensates~\cite{Garay:1999sk}, two typically quantum fluids, was quickly suggested.
The opportunity of using electromagnetic radiation in a medium was also analysed~\cite{Leonhardt:2000fd,Schutzhold:2004tv}.
Since, as detailed in Sec.~\ref{sec:acoustic_metric}, the analogy stands both for quantum and classical fields depending on the nature of the fluid considered, a classical analogue black-hole in water waves system was also considered~\cite{Schutzhold:2002rf}. 

The mechanism behind Hawking radiation is that of a scattering of modes at the horizon of black-hole. The Hawking radiation corresponds to one of the scattered modes of this process. When vacuum fluctuations source the process, it is called \textit{spontaneous}. The same type of scattering leading to Hawking radiation can also be triggered by sending on the horizon a classical wave or by letting external noise impinge on it. We get a \textit{stimulated} Hawking radiation, the analogue of which can be studied in classical systems. 

The first experimental results started to be reported in 2007 with experiments in electromagnetic optical fibres~\cite{Philbin:2007ji} and water waves~\cite{Rousseaux:2007is}. In the following years, experiments were improved in these platforms, e.g. see~\cite{Drori:2018ivu} in fibres and~\cite{Weinfurtner:2010nu,Euve:2015vml} for water waves, and performed in a growing diversity of platform, e.g. optical crystals~\cite{Faccio:2009yw}, polaritons~\cite{nguyenAcousticBlackHole2015}, superfluid He-3~\cite{Clovecko:2018qnj} or Bose-Einstein condensates (BEC)~\cite{Steinhauer:2015saa,MunozdeNova:2018fxv}. New theoretical tools were introduced to analyse them, e.g. 2-point density correlation functions~\cite{Balbinot:2007de}.
To conclude this sketch of the landscape of analogue black-hole experiments and analysis, we point out that, as of June 2023, only one group~\cite{Steinhauer:2015saa,MunozdeNova:2018fxv} claims to have observed spontaneous analogue Hawking radiation \ie seeded by vacuum fluctuations.

\subsection{Cosmological analogues}
\label{sec:cosmo_AG}

Despite the initial dominance of analogue black holes in the analogue gravity field, analogies with other systems have also been considered.
The analogues of cosmological situations are of particular relevance for this manuscript that presents results on the analysis of an analogue preheating experiment, see Sec.~\ref{sec:preheating} and Sec.~\ref{chapt:analogue_preheating}. Before moving on to this specific set-up, we review in this sub-section the first analogue cosmology systems that have been considered or realised.

\subsubsection{Early theoretical investigations}

Ref~\cite{Volovik:1997xi}  is an early example of analogy with cosmology. 
It investigates analogies between defects in superfluid He-3 and topological defects in the early Universe, such as cosmic strings.

While the Schwarzchild black hole is a time-independent and stationary spacetime, the most relevant cases for cosmology are rather time-dependent and homogeneous spacetime, as represented by the FLRW metric of Eq.~\eqref{def:FLRW}. Analogues of time-dependent spacetimes were also quickly considered by authors.

First, although not an analogue gravity experiment, the `Bose-nova' experiment~\cite{donleyDynamicsCollapsingExploding2001} was analysed by~\cite{Calzetta:2002jz,Calzetta:2002zz,Calzetta:2005yk} drawing inspiration from inflationary cosmology and the analogue gravity endeavour. 
In the experiment, the interactions between atoms in a Bose gas are changed from repulsive to attractive, leading to a contracting of the BEC.
They drew a parallel between the dynamics of certain excitations on top of the condensate and that of perturbations in inflation, which are `frozen' and amplified when super-Hubble, before oscillating at later times when re-entering the Hubble radius. They attribute some oscillations seen in~\cite{donleyDynamicsCollapsingExploding2001} to these perturbations. 

The possibility of simulating time-dependent spacetimes with such Bose gas with tunable interactions was first explored in~\cite{Barcelo:2003et,Barcelo:2003wu}. The authors showed that in a constant density condensate, with an appropriately modulated speed of sound, phonons \ie quantised sound waves, behave as a scalar field in an effective expanding FLRW metric. The dynamics then lead to the creation of phonons from the vacuum~\cite{Barcelo:2003wu}.
Another way to realise an effective FLRW metric is to consider an expanding (or contracting) condensate or to modulate the sound speed and the trapping frequency simultaneously, as suggested by~\cite{Fedichev:2003bv}. The authors also generalised the computations of~\cite{Barcelo:2003wu} for the creation of phonons and suggested using density-density correlations, described in Sec.~\ref{chapt:analogue_preheating} of this manuscript, to demonstrate the presence of quasi-particles.
Some authors have also considered the physical status of the coordinates in which the analogue metric is derived. In ~\cite{Fedichev:2003id}, the authors considered an FLRW spacetime in different coordinates and studied the possibility of observing an analogue of the Gibbons-Hawking effect. 
In~\cite{Fedichev:2003dj}, the authors suggested building a detector sensitive to different time coordinates to test their different responses to quasi-particle creation.
Other authors have considered the analogy of an expanding Universe with a purely expanding Bose gas without modulating the speed of sound. For instance
\cite{Weinfurtner:2004mu} considered a three-dimensional Bose gas, while \cite{Uhlmann:2005hf} considered lower dimensional trapped gases with non-standard self-interactions. 
The authors of~\cite{Fischer:2004bf} argued that the proposed expansion of the gas, or change of its scattering length, comes with difficulties that can be circumvented by considering a two-component gas.

Numerical simulations of the generation of quasi-particles in analogue cosmology were also performed. In~\cite{Jain:2007gg}, the authors performed numerical simulations of a two-dimensional Bose gas with time-dependent scattering length to evaluate particle production for inflationary or cyclic universe analogues. They used the Truncated Wigner Approximation (TWA) method that became quite standard in analogue gravity, see below Sec.~\ref{sec:TWA_simulation_1D_BEC} and e.g.~\cite{vanregemortelSpontaneousBeliaevLandauScattering2017}.
In a follow-up work~\cite{Weinfurtner:2008if}, the authors study the implications of breaking down the analogue gravity picture beyond the hydrodynamical regime where we have an acoustic metric see Sec.~\ref{sec:acoustic_metric}.

Finally,  in~\cite{Alsing:2005dno,Schutzhold:2007mx}, the authors proposed to use ion traps rather than Bose gases to build analogue cosmology set-ups.

\subsubsection{Analogue cosmology experiments}
\label{sec:analogue_cosmo_exp}

The first analogue cosmology experiments were performed in the 2010s. 
We list a few examples of them to give a panorama.
In~\cite{Jaskula:2012ab}, the authors report on the observations of the production of sound waves following the modulation of the trapping frequency of an elongated cigar-shaped gas. The creation proceeds by a parametric transfer of the energy of the condensate to pairs of phonons in resonant modes at half the driving frequency. Initially designed as an analogue dynamical Casimir effect experiment proposed in~\cite{Carusotto:2010mei},
the results of the experiment were later re-interpreted as an analogue of 
preheating~\cite{Busch:2014vda}. In~\cite{Robertson:2018gwi}, the analogy was extended to the late time dynamics where a redistribution of resonant phonons to other modes, akin to preheating, see Sec.~\ref{sec:preheating}, is conjectured to have happened. The authors of~\cite{Jaskula:2012ab} showed that the produced phonons were correlated but not entangled.

Soon after, in~\cite{Hung:2012nc}, the authors performed a quench \ie a sudden variation, of the scattering length of a two-dimensional BEC.
They observed the amplification of density fluctuations attributed to the creation of phonon pairs and the formation of coherent density oscillations, a signature of the (classical) correlation of the pairs, similar to Sakharov oscillations~\cite{Sakharov:1966aja,Grishchuk:2011wk} in primordial inhomogeneities.
In a recently updated run of the experiment~\cite{Chen:2021xhd}, they reported the observation of the quantum entanglement of the pairs. This arguably makes~\cite{Chen:2021xhd} a successful DCE analogue experiment.

Another analogue cosmology experiment using a Bose gas was reported in~\cite{Eckel:2017uqx}. 
There the authors designed a ring-shaped two-dimensional BEC on which they imprinted an azimuthal phonon pattern before expanding the radius of the BEC.
They observed a redshifting of the phonons' wavelength akin to that of cosmological perturbations in FLRW and a damping of their amplitude that they partially attribute to an analogue of Hubble friction. At the end of the expansion, they witness the generation of transverse oscillations of the condensate, whose energy then transfers to topological defects (dark solitons, then vortices). They point out that this transfer is analogue to some reheating models. Refined analyses were performed in~\cite{Eckel:2020qee,Banik:2021xjn}.

Very recently, in~\cite{Viermann:2022wgw}, the authors report a precise realisation of the suggestion of~\cite{Barcelo:2003wu,Fedichev:2003bv} in a BEC. The authors first observed the propagation of a localised wave packet of phonons in real space. They showed that, by tuning the geometry of the condensate, they could mimic the propagation in positive, flat or negative curvature spacetimes. Then they perform a co-ramp of the scattering length and the trapping frequency to generate phonons in pairs. They report the observation of analogue Sakharov oscillations, as in~\cite{Chen:2015bcg,Chen:2021xhd} following a quench of the scattering length but not that of entanglement of the pairs.

Finally, other experimental platforms were also used for analogue cosmology experiments.
The authors of~\cite{Wittemer:2019agm} considered the relative motion of two trapped ions. The relative motion is seen as the analogue of a mode of a quantum field. By changing the trap frequency, they managed to squeeze its state without directly evidencing the entanglement of the two ions. 
Sakharov-like oscillations were observed in a quantum fluid of light~\cite{Steinhauer:2021fhb}, but no direct evidence of entanglement was shown. 

Lastly, a second analogue preheating experiment was designed and realised using a fluid-fluid interface~\cite{Barroso:2022vxg}.
The authors predicted how interactions between the different modes of perturbations generically lead to a transfer of energy from the primary resonant modes to other modes. They were able to observe experimentally different signatures of this transfer, distinguishing them from, for instance, extra Floquet resonances.

\subsubsection{Experiments beyond QFTCS?}

In the earliest analysis and experiments, the analogue gravity community focused on simulating pure QFTCS effects. There, quantum perturbations are acted upon by a classical background field, produced particles are assumed to be non-interacting and not back-react on the classical background.
In contrast, analogues of (p)reheating, during which interactions and backreaction are essential, have recently received much attention.
In addition to the experiment~\cite{Barroso:2022vxg}, several theoretical studies have been published~\cite{Zache:2017dnz,Chatrchyan:2020cxs,Butera:2022kwi,Wang:2023wld}.
The interactions between the produced quasi-particles and the backreaction of these quasi-particles on the condensate are both analysed.
For instance, in~\cite{Butera:2022kwi}, the authors were able to numerically follow the gradual fragmentation of the initially homogeneous condensate into patches of different densities due to the backreaction of the perturbations, and the decay of the initially excited field, the analogue of the inflaton.

In a similar spirit, we study in this manuscript a possible dissipation channel for the excitations produced by the initial parametric amplification in~\cite{Jaskula:2012ab}.
Contrary to~\cite{Barroso:2022vxg}, the excitations of the system are quantum.
As detailed in Sec.~\ref{sec:preheating}, we expect that (quasi-)particles will be generated from the quantum vacuum, and a signature of this creation out of the vacuum is that they appear in entangled pairs~\cite{Busch:2014vda}. 
Unfortunately, the measured correlations in the original experiment were insufficient to demonstrate the entanglement of the pairs. In~\cite{Busch:2014vda}, the authors showed that a small degree of dissipation was sufficient to explain this absence of entanglement.
However, a precise microphysical mechanism to explain this dissipation was missing.
In Sec.~\ref{chapt:analogue_preheating}, we present the progress made during this PhD in this direction.
In particular, we compute the decay rate of the quasi-particle number and pair correlation induced by the first-order interaction between them. We verify the validity of our predictions by comparing them with numerical simulations.
These findings could be helpful in optimising the visibility of entanglement in a new run of the experiment~\cite{Jaskula:2012ab}.

%%%%%%
%To add the bibliography when compiling the subfile only
%\bibliogr%aphystyle{plain}
%\bibliography{bibmanuscript}

\chapter{Quantumness and decoherence of cosmological perturbations}
\label{chapt:quantumness_cosmo}

\section{Content of this chapter}

In this chapter, we reproduce the references~\cite{Micheli:2022tld,Martin:2021znx, Martin:2022kph}. They are the original works produced during the PhD devoted to finding features in the state of cosmological perturbations that could reveal their quantum origin.
The three next sections are made up of a published article which is briefly introduced. In these introductions we give some context, point out possible mismatches between existing conventions, and sometimes add details skipped in the paper.

\section{Review: `Quantum cosmological gravitational waves?'}
\label{sec:quantum_GW}

We start this discussion of the search for quantum features in cosmological perturbations with the review article~\cite{Micheli:2022tld}.
Although published later than Ref.~\cite{Martin:2021znx} and Ref.~\cite{Martin:2022kph}, it exposes in detail the formalism used in other references while skipping the technicalities of these specific works. We, therefore, reproduce it first as an introduction to the two other works.
The detailed exposition is contained in Sec.~3 of~\cite{Micheli:2022tld}, except for Sec.~3.7, written by Patrick Peter, to be considered separately.
Although the focus of the review is on tensor perturbations, most of this exposition is equally applicable to scalar perturbations. 
Still, note that the discussion of the evolution of the perturbations in a simplified cosmological model made in Sec.~3.8 cannot be applied to scalar perturbations. The latter are described by the Mukhanov-Sasaki variable $v$ and the equation Eq.~\eqref{eq:MS_equation} \textit{only} during inflation. The considerations on the extreme squeezing of perturbations in de Sitter apply to scalar and tensor.
Finally, the core of the analysis in Sec.~4 and the conclusion in Sec.~5 cover both the scalar and tensor cases.
We also point out that the Lagrangian considered in Eq.~(22) of~\cite{Micheli:2022tld}, reproduced below, differs from that in Eq.~\eqref{def:pert_tensor_action} by a total derivative $\dd [ \Hu ( \mu_{\lambda}^{\star}  \mu_{\lambda} )^{\prime} ] / \dd \eta$. The form used in~\cite{Micheli:2022tld} is more standard in the cosmology literature, e.g.~\cite{Grishchuk:1990bj}. A canonical transformation relates them, and they give the same results when computing the expectation values of the same operators. 
Nevertheless, these different conventions result in different definitions for the conjugated field $\pi_{\lambda} = \partial L / \partial \mu_{\lambda}^{\prime}$: in the convention of Eq.~\eqref{def:pert_tensor_action} we have $\pi_{\lambda} = \mu_{\lambda}^{\prime}$, while in that of Eq.~(22) of \cite{Micheli:2022tld} we have $\pi_{\lambda} = \mu_{\lambda}^{\prime} - \Hu  \mu_{\lambda}$.
An important consequence, as explained in~\cite{Agullo:2022ttg}, is that the creation/annihilation operators, Bogoliubov coefficients, and so the squeezing parameters, see Eq.~(42) of~\cite{Micheli:2022tld}, defined from these two different fields will in general be different.
This difference is manifest when comparing Fig.~5 of~\cite{Micheli:2022tld} and Fig.~3 of~\cite{Martin:2021znx}, where the direction of squeezing is horizontal in the first case and vertical in the second.
This mismatch translates the difference between the fluctuations of $\hat{v}^{\prime}$, which are asymptotically of order unity, and those of $\hat{v}^{\prime} - \Hu  \hat{v}$, which, due to the second term, grow faster than those of $\hat{v}$.
However, note that they agree when the expansion of the background encoded by $\Hu$ can be neglected, e.g. in radiation domination for tensor perturbations.

Two additional references~\cite{Prokopec:2006fc,Adamek:2013vw} would have deserved citation in the literature review, but we were unaware of them at the time of writing. In~\cite{Prokopec:2006fc}, the authors considered the decoherence induced by isocurvature perturbations on adiabatic perturbations in a two-field model, where the fields are coupled via the gravitational perturbations. They show that the adiabatic perturbations can be efficiently decohered, in the sense that the off-diagonal matrix elements are suppressed, and compute the associated entanglement entropy.
In~\cite{Adamek:2013vw}, the authors also considered the decoherence caused by one field on another in de Sitter spacetime. In addition, they considered the effect of a modified dispersion relation for the `measured' scalar field. They showed that there is a competition between squeezing and decoherence to determine whether the final state of the scalar field particles is separable or not.

The observability of the features, in particular using the measures introduced in Refs.~\cite{Martin:2021znx, Martin:2022kph} reproduced in Secs.~\ref{sec:discord_and_decoherence} and \ref{sec:comparing_quantumness_criteria}, is discussed at the very end of Sec.~4.
The discussion concludes that no viable measurement protocol has been proposed to demonstrate the quantum origin of cosmological perturbations experimentally.

\includepdf[pages=-]{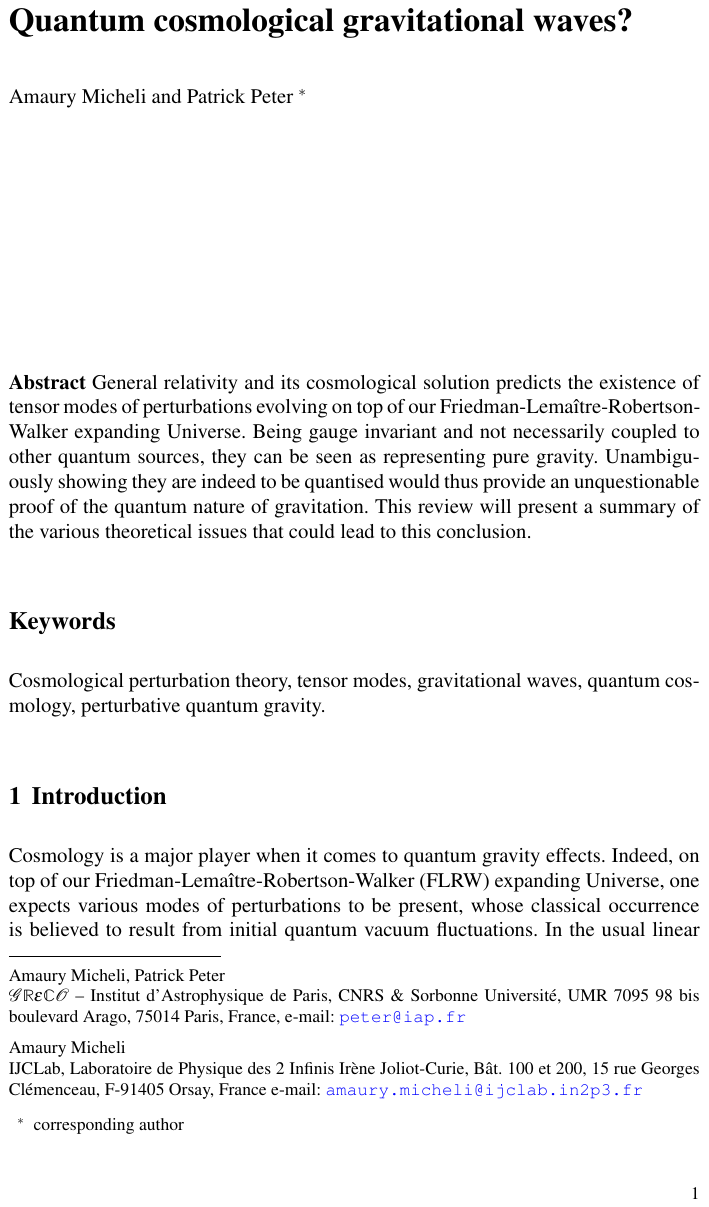}

\section{Article: `Discord and decoherence'}
\label{sec:discord_and_decoherence}

Our general strategy to evaluate the quantumness of cosmological perturbations is to consider correlations between different perturbations and see if these can be labelled quantum because they exhibit features that cannot be accounted for by a classical theory, e.g. they violate
a Bell inequality~\cite{Martin:2022kph}.
The non-classicality of these correlations is expected to be affected by the 
interaction of the perturbations with other degrees of freedom and themselves, a phenomenon known as decoherence.
In~\cite{Martin:2021znx} reproduced below, we analysed one specific measure of the non-classicality of correlations, the quantum discord, in the cosmological context. 
The main result of this paper, derived in Sec.~4, is the computation of the quantum discord in the presence of a minimal decoherence model, extending known results~\cite{Martin:2015qta}.
Similarly to,~\cite{Adamek:2013vw} we show that there is a competition between the dynamics of inflation generating quantum correlations and decoherence that destroys them: the rate of interaction with the external degrees of freedom leading to decoherence must be large enough for the quantum discord to have disappeared by the end of inflation.
Again, we point out that the convention used for $\hat{p}$, the field conjugated to the Mukhanov-Sasaki field $\hat{v}$ matches Eq.~\eqref{def:pert_scalar_action}, and so differs from that used in~\cite{Micheli:2022tld}, see Sec.~\ref{sec:quantum_GW} for details.

\includepdf[pages=-]{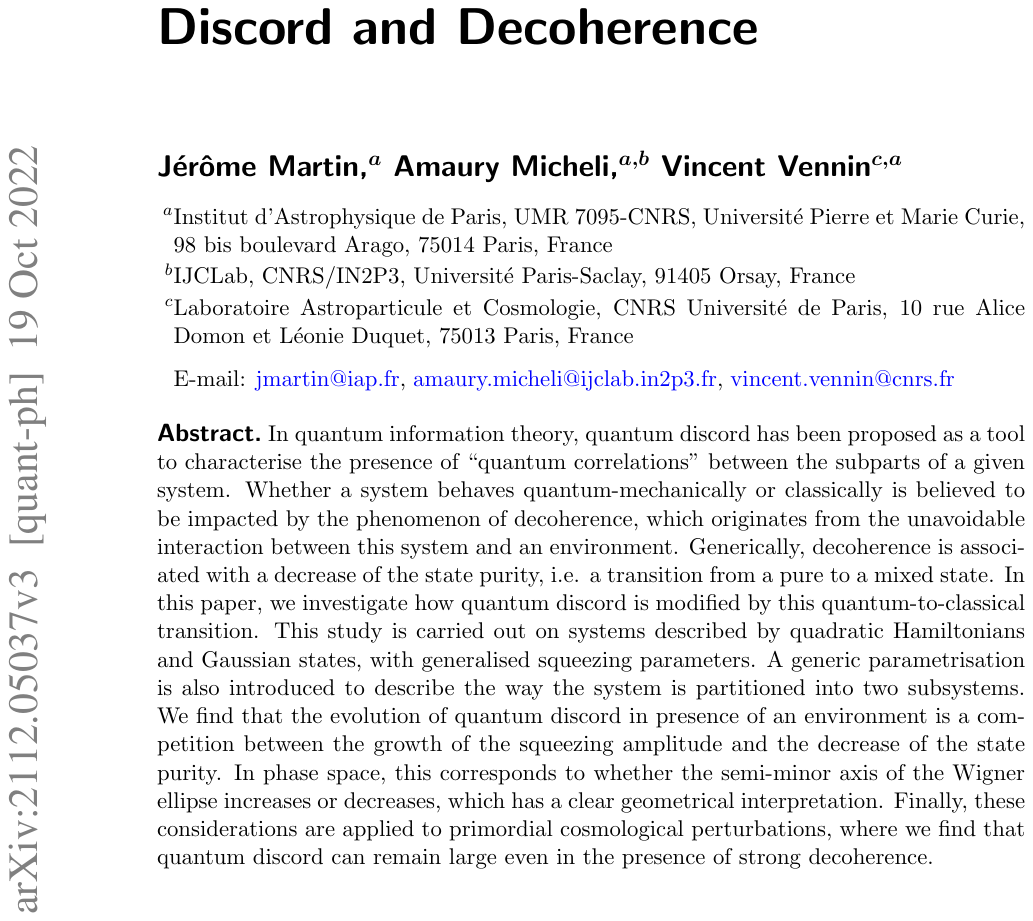}

\section{Article: `Comparing quantumness criteria'}
\label{sec:comparing_quantumness_criteria}

As shown in the review~\cite{Micheli:2022tld}, many quantumness criteria have been used to assess the degree of non-classicality of cosmological perturbations in slightly different situations.
In~\cite{Martin:2022kph} reproduced in this section, we thoroughly compare three such criteria over a general class of states, the two-mode squeezed thermal states, which cover most of the decohered states considered in the literature on inflationary perturbations and beyond.
The analysis is kept at a very general level, and the paper is presented as one of quantum information. We supplement our general analysis with the study of the effect on an initial two-mode squeezed states of two simple Gaussian channels \ie transformations of the state that needs not be unitary, modelling losses and noise in a measurement process.
Therefore, although the initial motivation behind this work is the study of cosmological perturbations, the comparison also applies to the study of the experiment~\cite{Jaskula:2012ab} in which we expect the state of correlated quasi-particles to be mixed and quasi-Gaussian, see Sec~\ref{sec:decay_ck}. 

\includepdf[pages=-]{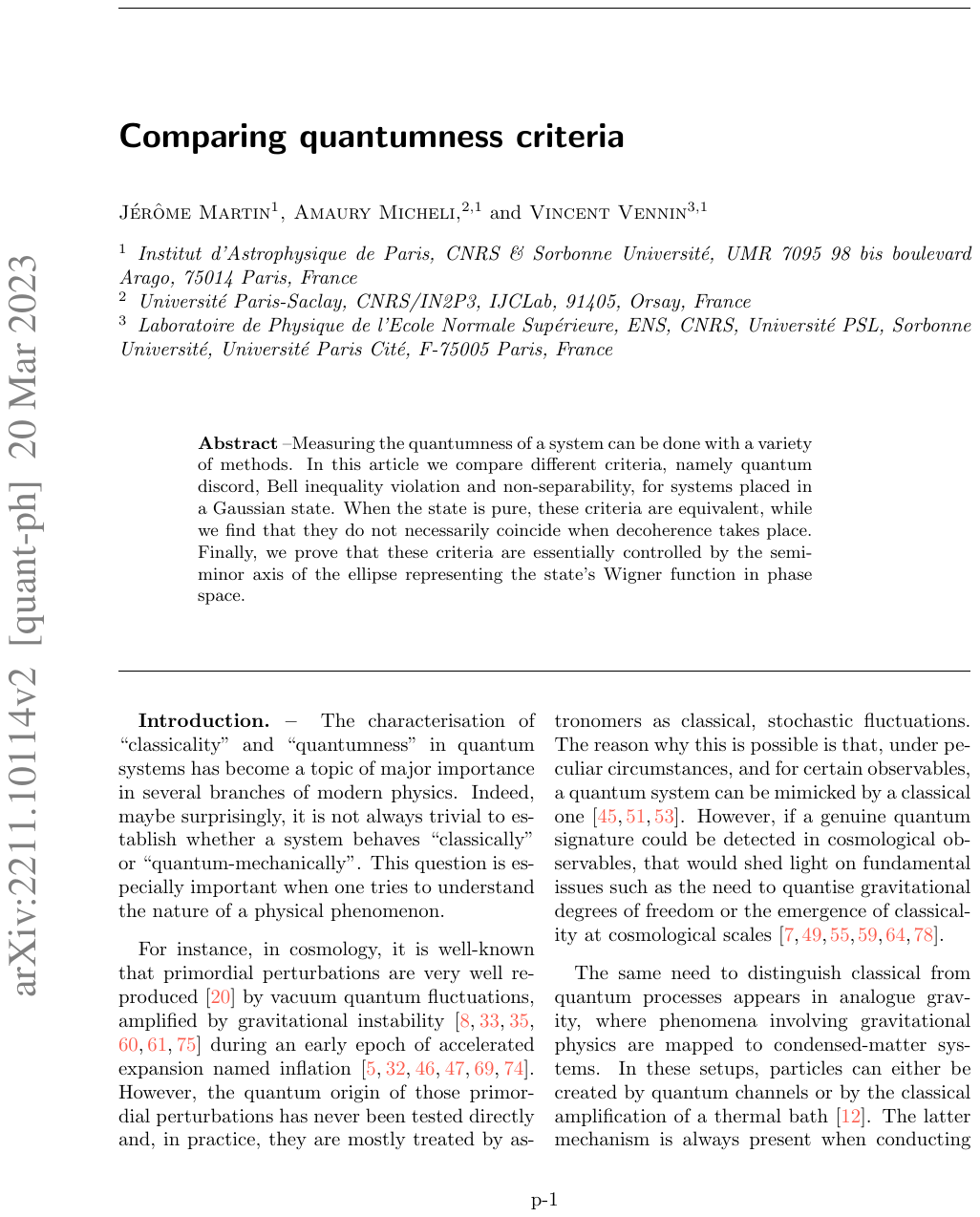}

\chapter{An analogue preheating experiment with BEC}
\label{chapt:analogue_preheating}

\section{Content of this chapter}

This chapter is devoted to analysing the analogue preheating experiment~\cite{Jaskula:2012ab}, which is the second topic of the PhD.
In Sec.~\ref{sec:presentation_exp}, we present the original experimental set-up, its goal and what was (not) observed in its first run.
Then, in Sec.~\ref{sec:model_exp}, we present the model we use to describe it and review the results obtained in previous publications on the re-analysis of the experiment~\cite{Busch:2013sma,Busch:2014vda,Robertson:2016evj,Robertson:2017ysi,Robertson:2018gwi}.
The latest publication~\cite{Micheli:2022zet} is reproduced in Sec.~\ref{sec:decay_nk}. It contains part of the original results obtained during this PhD on the analysis of the experiment. It presents a general decay mechanism for phonons in a 1D Bose gas via Beliaev-Landau channels and its application to resonant phonons in a parametric amplification set-up akin to that of~\cite{Jaskula:2012ab}.
Given the importance of numerical simulations in this latest analysis, we devote Sec.~\ref{sec:TWA_simulation_1D_BEC} to a more detailed explanation of the algorithm used. 
Finally, Sec.~\ref{sec:decay_ck} presents preliminary results on the decay, via similar Beliaev-Landau processes, of phononic pair-correlation of the type generated during parametric amplification. These results will be the object of a future publication.
We conclude that these processes are sufficient to explain the absence of entanglement in~\cite{Jaskula:2012ab}, which is the main result of the second part of this PhD.

\section{Presentation of the experiment}
\label{sec:presentation_exp}

\begin{figure}
	\centering
	\begin{minipage}{0.3\textwidth}
		\centering
		\includegraphics[height=0.4\textheight]{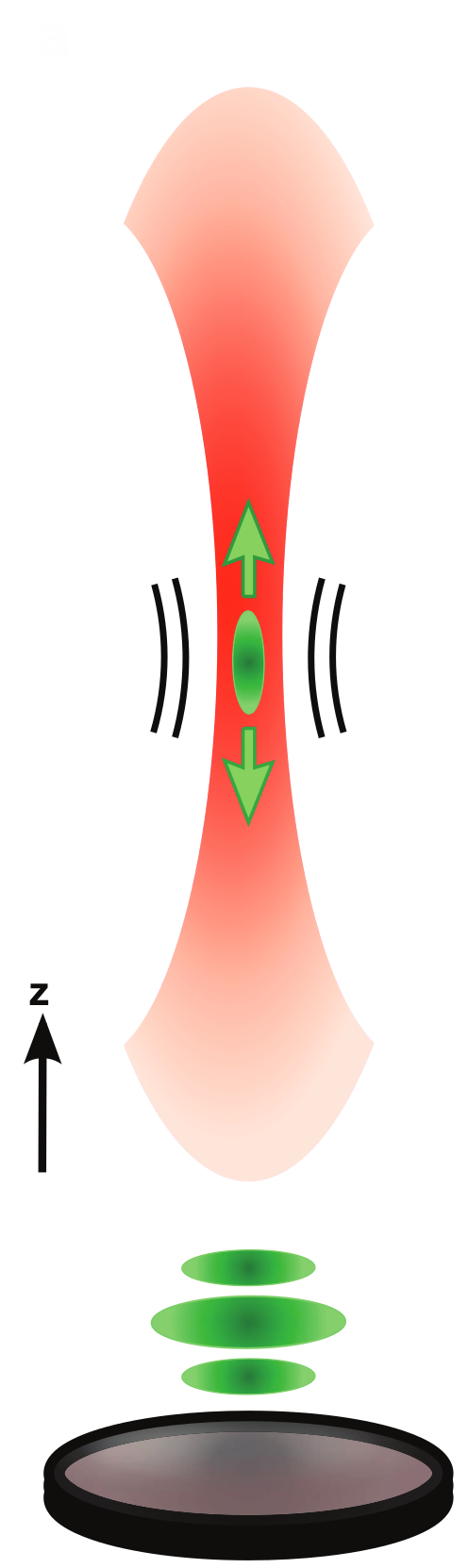}
		\caption{Representation of the trapped cold atoms used for the analogue Dynamical Casimir effect. This figure is adapted from~\cite{Jaskula:2012ab} with only a change of colors. }
		\label{fig:trapped_atoms}
	\end{minipage}\hfill
	\begin{minipage}{0.6\textwidth}
		\centering
		\includegraphics[width=\textwidth]{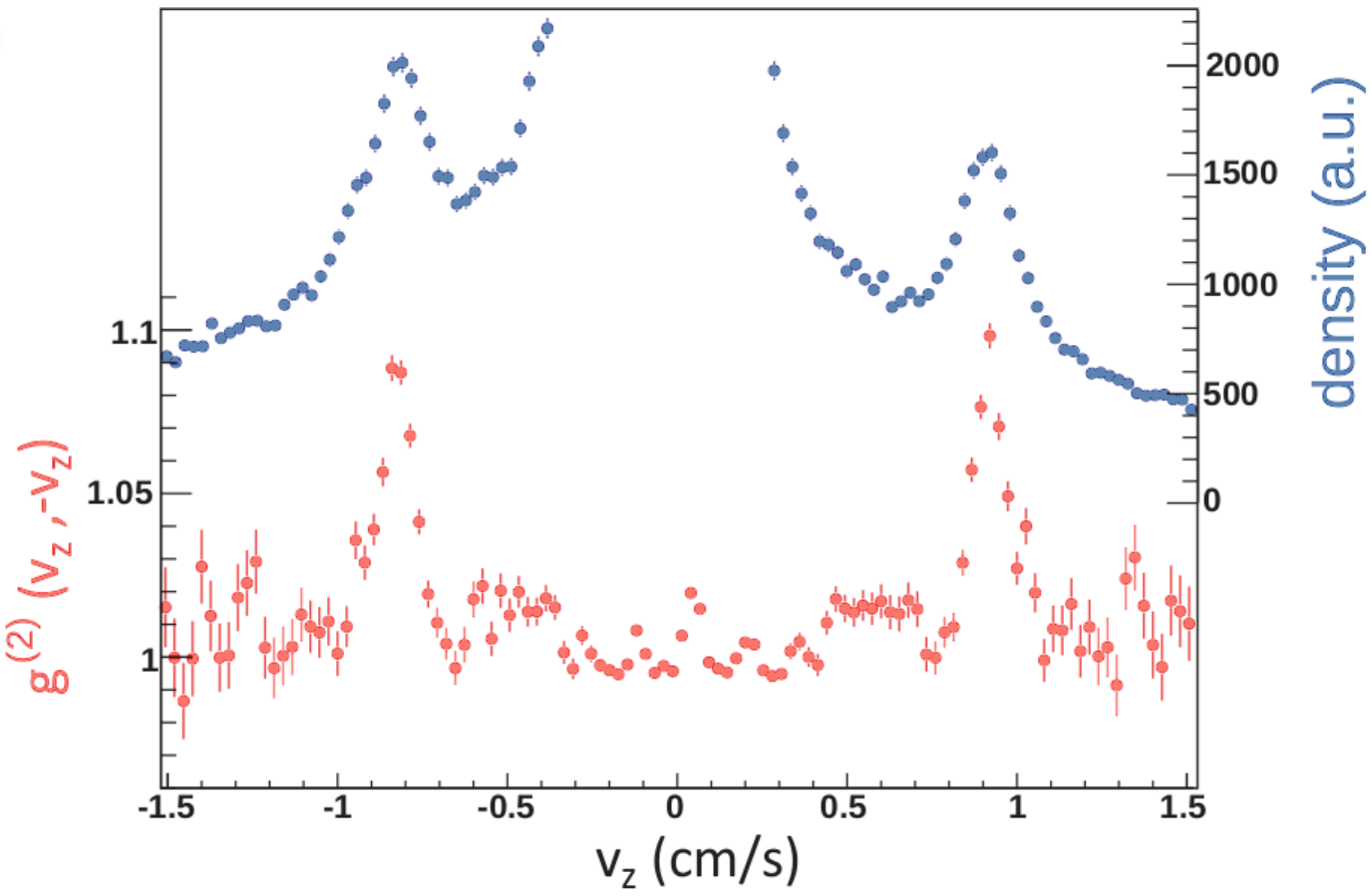}
		\caption{In red, $2$-point correlation function $g^{(2)}_k$ of Eq.~\eqref{def:small_g2} as measured after opening the trap in the experiment~\cite{Jaskula:2012ab}. This figure is adapted from~\cite{Jaskula:2012ab} with only a change of colors. }
		\label{fig:measured_small_g2}
	\end{minipage}
\end{figure}

In~\cite{Jaskula:2012ab}, the authors report on the results of their experiment of an analogue Dynamical Casimir Effect (DCE)~\cite{mooreQuantumTheoryElectromagnetic1970,DeWitt:1975ys,Davies:1976hi}. 
The dynamical Casimir effect refers to the production of particles expected for a quantum field confined in a space region when the boundaries of this region are varied in time. This creation can be understood on the same basis as the cosmological pair production~\cite{Parker:1968mv} discussed in Sec.~\ref{sec:quantum_GW}: an `in' vacuum before the expansion, and an `out' vacuum after the expansion, are related by a Bogoliubov transformation~\cite{mooreQuantumTheoryElectromagnetic1970}.
Analogues of the DCE were experimentally observed in super-conducting circuits~\cite{wilsonObservationDynamicalCasimir2011,Lahteenmaki:2011cwo}, where the index of the medium is modulated to mimic changing boundary conditions, and in an optical system~\cite{vezzoliOpticalAnalogueDynamical2019}.
In~\cite{Jaskula:2012ab}, the authors prepared a gas of approximately $10^5$ meta-stable Helium atoms\footnote{This number of atoms could be inaccurate due to the detector's saturation when the condensed part of the gas reaches it. Compared with the current experiment run, it is likely over-estimated, and we will take the value of $1.5\times10^4$ for numerical applications below. This number is obtained by inferring the one-dimensional density of the condensed gas $n_{0}$ from the speed of sound and trap frequency reported in~\cite{Jaskula:2012ab}.
	Since most atoms are initially in the condensate, we assimilate this density to the total density of the gas $n_{\mathrm{1D}}$. The number of atoms is then obtained by assuming that the longitudinal size of the gas is given by twice the Thomas-Fermi radius, see below. We thank Victor Gondret for the discussions on this point.} $\mathrm{ \indices{^4} He}^{\star}$ cooled down to $200 \mathrm{nK}$. The atoms are placed in a magnetic trap of an elongated cigar shape in the vertical direction, see Fig.~\ref{fig:trapped_atoms}.
The trapping potential is divided between a radial and vertical (longitudinal) part
\begin{equation}
	\label{def:trap_potential}
	V_{\mathrm{ext}} \left( \bm{x} \right) = \frac{1}{2} m \omega_{\perp}^2 r^2 + \frac{1}{2} m \omega_{z}^2 z^2 \, ,
\end{equation}
where $m$ is the mass of the Helium atoms, $r^2 = x^2 + y^2$, $\omega_{\perp}$ is the radial trapping frequency and $\omega_{z}$ is the vertical
trapping frequency. For the anisotropic trap of this experiment the authors report $\omega_{\perp} / 2 \pi = 1500$ Hz while $\omega_{z} / 2 \pi = 7$ Hz, so that $\omega_{\perp} \gg \omega_{z}$ and the gas is effectively one-dimensional. 

The experimenter can change these trapping frequencies during the experiment, and two types of time dependence were investigated in~\cite{Jaskula:2012ab}. The first is a quench of the frequencies by a factor $\sqrt{2}$ \ie the trap is suddenly made tighter. In the second experiment, the trap frequencies are modulated according to $\omega^2 = \omega_{0}^2 [   1 + A \sin ( \omega_{\mathrm{m}} t)   ] $ for frequencies $\omega_{\mathrm{m}} / 2 \pi \in [ 900 , 5000]$ Hz and $A=0.1$, which is the peak-to-peak modulation amplitude of $\omega$. In both cases, due to the already strong trapping in the radial direction, the frequency change does not excite the atoms radially but leads to longitudinal excitations. 
These excitations are sound waves on top of the condensed atoms for long enough wavelengths. This mechanism of longitudinal waves generation via parametric resonance was also used in~\cite{Engels:2007zz}.
As we will show in detail in Sec.~\ref{sec:dynamics_no_external_drive} the excitations can generally be interpreted as quasi-particles.
A modulation at frequency $\omega_{\mathrm{m}}$ is expected to lead to a parametric creation of them in pairs of frequencies $\omega_{1/2}$ such that $\omega_{\mathrm{m}} = \omega_{1} + \omega_{2}$. To preserve the isotropy of the gas, they have to be of opposite momentum $\pm \bm{k}$, and so are of the same frequency $\omega_{\bm{k}}$. Even in the case where no quasi-particles are present in the initial state, owing to the quantum nature of the gas, pairs are going to be produced out of the vacuum, a mechanism very reminiscent of preheating.
Note that, in agreement with a remark made in Sec.~\ref{sec:preheating}, since any time dependence would lead to quasi-particle creation, an analogy could also be made with inflation by expanding the trap rather than modulating it.
A distinctive feature of this creation out of the vacuum is that we expect the pairs to be entangled~\cite{Busch:2013sma}.

In~\cite{Jaskula:2012ab}, the authors showed that the modulation produced quasi-particles in pairs of opposite momentum $\pm \bm{k}$. In addition, they showed that these pairs are correlated, yet, the correlations were not strong enough to demonstrate that the pairs were entangled.
In contrast, in other analogue DCE experiments~\cite{Lahteenmaki:2011cwo,vezzoliOpticalAnalogueDynamical2019}, the authors explicitly demonstrated that the generated pairs of photons were in an entangled state, thereby proving that they had been created out of the vacuum; a genuine quantum effect.
Subsequent analyses~\cite{Busch:2013sma,Busch:2014vda,Robertson:2016evj,Robertson:2017ysi,Robertson:2018gwi} tried to account for the absence of entanglement and the non-linear processes at play in~\cite{Jaskula:2012ab}. The emergence of entanglement in the early-time dynamics was studied in~\cite{Busch:2013sma,Busch:2014vda} using simple quadratic models. The authors pointed out the analogy with the parametric creation during preheating. Then, in~\cite{Robertson:2016evj}, the authors accounted more precisely for the interplay between the change in the trapping frequency and the production of excitations, with a  focus on entanglement in~\cite{Robertson:2017ysi}.
In the latest paper of the series~\cite{Robertson:2018gwi}, the authors numerically analysed the redistribution of energy of the generated pairs via interactions between quasi-particles and their backreaction on the condensate. They also dealt with the late-time processes drawing an analogy with the processes following preheating in the early Universe.
We point out that the focus of~\cite{Robertson:2016evj,Robertson:2017ysi,Robertson:2018gwi} was on the quench experiment rather than the modulation. Still, following a quench, the gas would exhibit `breathing' oscillations in the radial direction, and these oscillations would, in turn, lead to pair creation in well-defined resonant modes. We come back to this point in Sec.~\ref{sec:dynamics_condensate}. 

In the next section Sec.~\ref{sec:model_exp}, we present the mathematical formalism that they used to describe the experiment, which is independent of the type of modulation considered.
Similarly, the computations presented in Sec.~\ref{sec:decay_nk} apply generically to any quasi-particles in a 1D Bose gas, irrespective of their production mechanism.

\section{Modelling the experiment}
\label{sec:model_exp}

In Sec~\ref{sec:condensed_state}, we start by defining a condensed state in the standard description of a gas using quantum field theory.
Then, in Sec.~\ref{sec:dynamics_gas}, we describe the general dynamics of a weakly interacting gas in a (quasi-)condensed state, including when the radial trapping frequency is varied.
In Sec.~\ref{sec:parametric_amplification}, we briefly examine the specific case of a modulation of the radial trapping frequency, draw the link with preheating and explain how relevant information about the created quasi-particles is experimentally accessed.
Finally, in Sec.~\ref{sec:Madelung}, we amend the standard BdG approximation presented in  Sec.~\ref{sec:dynamics_gas} for the specific case of a one-dimensional gas.

\subsection{Condensed state}
\label{sec:condensed_state}

The standard mathematical description of a mono-atomic Bose gas of Helium atoms uses a quantum field theory for a complex scalar field $\hat{\Psi} (\bm{x})$ dubbed the atomic field. An extensive presentation of this formalism can be found in~\cite{pitaevskiiBoseEinsteinCondensation2003}, and we briefly recap essential aspects.
The atomic field $\hat{\Psi} (\bm{x})$ is defined as the operator whose action on any quantum state describing the system $ | \Psi \rangle$ removes one atom at position $\bm{x}$ \ie it is a destruction operator in real space. Similarly, its hermitian conjugate $\hat{\Psi}^{\dagger}(\bm{x})$ is a creation operator in real space. We impose that they satisfy the canonical commutation relation
\begin{equation}
	\label{def:AtomicFieldCCR}
	\left[ \hat{\Psi}\left( \bm{x} \right) , \hat{\Psi}^{\dagger}  \left( \bm{x}^{\prime} \right)\right] = \delta \left( \bm{x} - \bm{x}^{\prime} \right) \, ,
\end{equation}
while the atomic field commutes with itself at any point $\left[ \hat{\Psi}\left( \bm{x} \right) , \hat{\Psi}\left( \bm{x}^{\prime} \right)\right] = 0$.
The atomic density at $\bm{x}$ is given by
\begin{equation}
	\label{def:atomic_density}
	\hat{\rho} \left(\bm{x} \right) =  \hat{\Psi}^{\dagger} \left(\bm{x} \right)   \hat{\Psi} \left(\bm{x} \right) \, ,
\end{equation}
which is a hermitian operator. The total number of atoms in the system $N$ is obtained by integrating the density field over space.
We can also define the phase $\hat{\theta}$ of the gas by writing the atomic field
\begin{equation}
	\label{def:madelung_form}
	\hat{\Psi} \left(\bm{x} \right) =  e^{i \hat{\theta}} \sqrt{\hat{\rho} \left(\bm{x} \right)} \, .
\end{equation}
Mathematical problems are attached to the definition of a phase operator~\cite{Haldane:1981zz,moraExtensionBogoliubovTheory2003}. We briefly discuss them in Sec.~\ref{sec:quasi_condensate}. The decomposition in density and phase of Eq.~\eqref{def:madelung_form} is called the Madelung form~\cite{madelungAnschaulicheDeutungGleichung1926}. It will prove helpful later when dealing with a one-dimensional gas for which no true condensed state can emerge, and we can only have a  \textit{quasicondensate}~\cite{petrovLowdimensionalTrappedGases2004}.
For three, and even two-dimensional, systems, there exists a temperature $T_{\mathrm{cond.}}$ below which the gas can be placed in a condensed state. In a condensed state, a single-particle state, typically the ground state of the gas, is macroscopically occupied, \ie the number of particles in this state $N_0$ is such that when the total number of particles $N$ diverges in the thermodynamical limit, the fraction $n_0 = N_0 / N$ remains finite~\cite{pitaevskiiBoseEinsteinCondensation2003}. Such a state is characterised by long-range order for some correlations~\cite{pitaevskiiBoseEinsteinCondensation2003}, typically the one-body correlation function $g_{1}$ defined by
\begin{equation} 
	\label{def:g1}
	g_{1} \left( \bm{x} \, , \, \bm{x}^{\prime} \right) = \left \langle \hat{\Psi}^{\dagger} \left( \bm{x} \right) \hat{\Psi} \left( \bm{x}^{\prime} \right)  \right \rangle \, .
\end{equation}
If we consider a homogeneous system, $g_1$ only depends on the difference $ \bm{x} - \bm{x}^{\prime}$ and for a condensate state is expected in the thermodynamical limit to behave as
\begin{equation}
	g_{1} \left( \bm{x} \, , \, \bm{x}^{\prime} \right) \xrightarrow[\left| \bm{x} - \bm{x}^{\prime} \right| \rightarrow + \infty ]{} n_{0} \neq 0 \, .
\end{equation}
For a condensed gas, given that a macroscopic amount of particles are in the same state, we consider them as a separate gas for which we define the operator $\hat{\Psi}_{0}$ that destroys a particle of the macroscopically occupied state. The density of particles in this state is then given by
\begin{equation}
	n_0 = \left \langle \hat{\Psi}^{\dagger}_0 \hat{\Psi}_0 \right \rangle \, .
\end{equation}
For condensed states, where the number of particles $N_0$ is huge, removing or adding a particle in this macroscopically occupied state should keep the system's state roughly the same. The action of $\hat{\Psi}_{0}$ can then be approximated as a multiplication by a complex function $\Psi_0$ such that
\begin{equation}
	\hat{\Psi}_{0} \left| \Psi \right \rangle \approx \Psi_{0} \left| \Psi \right \rangle \, ,
\end{equation}
and similarly for its hermitian conjugate. Therefore we treat this operator as a complex function
\begin{equation}
	\hat{\Psi}_{0} \left( \bm{x} \right) \approx \Psi_0 \left( \bm{x} \right) \hat{\mathds{1}} \, ,
\end{equation}
where $\hat{\mathds{1}}$ is the identity operator.
If we consider all particles to be in the macroscopically occupied state, then removing an atom from the gas corresponds to removing an atom from the macroscopically occupied state hence
\begin{equation} \label{eq:fullycondensed}
	\hat{\Psi} \left( \bm{x} \right) \approx \Psi_0 \left( \bm{x} \right) \hat{\mathds{1}} \, ,
\end{equation}
and the total density $n$ then equals the condensed density $n_0$.
Assuming that the condensed state is homogeneous, it trivially exhibits non-diagonal long-range order
\begin{equation}
	g_{1} \left( \bm{x} \, , \, \bm{x}^{\prime} \right) \approx n_{0} \neq 0 \, .
\end{equation}
Still, we need to refine the above prescription as only part of the atoms are in the macroscopically occupied state. The standard approach is to split the atomic field between a condensed part and a perturbation~\cite{pitaevskiiBoseEinsteinCondensation2003}
\begin{equation}
	\label{eq:condensate_pert}
	\hat{\Psi} = \Psi_{0} \left( \hat{\mathds{1}} + \delta \hat{\Psi} \right)  \, ,
\end{equation}
where $\delta \hat{\Psi}$ is the atomic field for the non-condensed part. Notice that here we have defined the perturbations in a relative manner using the condensed wavefunction as a pre-factor. It is more common to use absolute perturbation $\hat{\Psi} = \Psi_0 \hat{\mathds{1}} + \delta \hat{\Psi}$.
The Bogoliubov-de Gennes approximation (BdG) consists in assuming the contribution of the non-condensed part to be a perturbation on top of the condensate $|\delta \hat{\Psi}| \ll 1$.
Note the similarity with the analysis that lead to the introduction of the acoustic metric in Sec.~\ref{sec:acoustic_metric}, where we considered the perturbations of a fluid around a background flow.
We now write a Hamiltonian describing the dynamics of this atomic field.

\subsection{Dynamics of the gas}
\label{sec:dynamics_gas}

For dilute Bose gas as in~\cite{Jaskula:2012ab}, the interactions are dominated by two-body contact interactions, which are described by the pseudo-potential~\cite{pitaevskiiBoseEinsteinCondensation2003}
\begin{equation}
	\hat{V} \left( \bm{x} \right) = \frac{g}{2} \hat{\Psi}^{\dagger} \left( \bm{x} \right) \hat{\Psi}^{\dagger} \left( \bm{x} \right) \hat{\Psi} \left( \bm{x} \right) \hat{\Psi} \left( \bm{x}\right) \, ,
\end{equation}
where the interaction constant can be related to the $s$-wave scattering length $g = 4 \pi \hbar^2 a_s / m$.
The trapping potential is given by Eq.~\eqref{def:trap_potential}, and we can write down the Hamiltonian for the gas
\begin{equation}
	\label{def:H_3Dgas}
	\hat{H} = \int \dd {\bm{x} } \left[ \frac{\hbar^2}{2 m} \vec{\nabla} \hat{\Psi}^{\dagger} \cdot \vec{\nabla} \hat{\Psi} +  V_{\mathrm{ext}} \left( \bm{x} \right) \hat{\Psi}^{\dagger} \hat{\Psi}  + \frac{g}{2} \hat{\Psi}^{\dagger} \hat{\Psi}^{\dagger} \hat{\Psi} \hat{\Psi}  \right]  \, .
\end{equation}
Its Heisenberg equation of motion derived using Eq.~\eqref{def:AtomicFieldCCR} reads
\begin{equation}
	\label{eq:EOM_3D_Heinsenberg}
	i \hbar \partial_{\rm t} \hat{\Psi} = - \frac{\hbar^2}{2 m} \Delta \hat{\Psi} +  V_{\mathrm{ext}} \left( \bm{x} \right) \hat{\Psi} + g  \hat{\Psi}^{\dagger}  \hat{\Psi} \hat{\Psi} \, .
\end{equation}
One can check that the number of atoms given by
\begin{equation}
	N \hat{\mathds{1}} = \int \hat{\Psi}^{\dagger}  \hat{\Psi} \dd \bm{x}
\end{equation}
is conserved for an evolution via Eq.~\eqref{eq:EOM_3D_Heinsenberg}. 
We solve Eq.~\eqref{eq:EOM_3D_Heinsenberg} by splitting the atomic field into condensed and perturbation parts as in Eq.~\eqref{eq:condensate_pert}, and use the BdG approximation to write first an equation for the condensed part neglecting the non-condensed part which is the classical version of Eq.~\eqref{eq:EOM_3D_Heinsenberg}
\begin{equation}
	\label{eq:3D_GPE}
	i \hbar \partial_{\rm t} \Psi_0 = - \frac{\hbar^2}{2 m} \Delta \Psi_0 +  V_{\mathrm{ext}} \left( \bm{x} \right) \Psi_0 + g  \left| \Psi_0 \right|^2 \Psi_0 \, .
\end{equation}
This equation is called the Gross-Pitaevskii equation (GPE).

The BdG approximation is only valid for weakly interacting gas, so before moving on to the computation of the dynamics of the gas, we do
some order of magnitude computations for the gas in~\cite{Jaskula:2012ab} to ensure that the approximations we applied are valid.
First, we approximate the gas to have the geometry of a cylinder. The radius of the cylinder is directly given by the trapping frequency 
$a_{\perp} = \sqrt{\hbar/m \omega_{\perp}}$, see Sec.~\ref{sec:dynamics_condensate}. The length of the gas is more complicated to compute.
Applying the analysis of~\cite{petrovLowdimensionalTrappedGases2004}, see in particular Fig.~5, given the number of atoms $N \approx 10^4$, and the radial extension $a_{\perp}$ of~\cite{Jaskula:2012ab}, shows the gas is in the Thomas-Fermi regime. Then, the vertical condensed density profile has an extension of twice the Thomas-Fermi radius given by
\begin{equation}
	R_{\mathrm{TF}} = a_z \left(  3 \frac{a_z a_s}{a_{\perp}^2} N \right)^{1/3} \, ,
\end{equation}
where $a_z = \sqrt{\hbar/m \omega_{z}}$. Due to the second term in brackets, this extension is much larger than the naive expectation $a_z$. While in the Thomas-Fermi regime the density profile is not homogeneous, in the rest, we neglect the vertical trapping and consider a homogeneous profile of length $L$. The length is estimated using the above to $L \approx 2  R_{\mathrm{TF}}$. We then compute the gas parameter $n a_s^3$, where $n \approx N/(2 R_{\mathrm{TF}} \pi a_{\perp}^2)$ is the number density of atoms, and $a_s$ is the $s$-wave scattering length for $\indices{^4}\mathrm{He}$ atoms in $2^3S_1$ meta-stable state. It is measured~\cite{moalAccurateDeterminationScattering2006} to be $a_s = 7.5 \, \mathrm{nm}$. The condition for a weakly interacting 3D gas is that the gas parameter is much smaller than unity, \ie that the typical distance between atoms $n^{-1/3}$ is larger than the $s$-wave scattering length. We find $n a_s^3 \approx 2 \times 10^{-6}$, so we are well within this regime.
It is also useful to compute the gas parameter in 1D $ n_{\mathrm{1D}} a_s = N a_s/(2 R_{\mathrm{TF}} ) \approx 0.19$, which is used as a parameter in the papers analysing the experiment e.g.~\cite{Robertson:2018gwi}. Following~\cite{petrovLowdimensionalTrappedGases2004},
the gas is in the weakly interacting regime in one-dimension when the interaction energy per particle is much less than the characteristic kinetic energy of particles. This condition translates in the requirement that the Lieb-Liniger interaction constant~\cite{Lieb:1963rt} $\gamma = m g_1/ (\hbar n_{\mathrm{1D}})$, where $g_1$ is the one-dimensional interaction constant, is much less than one $\gamma \ll 1$.
For our radially confined gas we find~\cite{olshaniiAtomicScatteringPresence1998,petrovLowdimensionalTrappedGases2004} $g_1 =  g /  (2 a_{\perp}^2)$, which is derived in Eq.~\eqref{eq:quadratic_H_pert_BdG_longitidunal} below. We then have
$\gamma = 2 a_s/ ( n_{\mathrm{1D}}  a_{\perp}^2) \approx  4 \times 10^{-4}$. Again, we are in a weakly interacting regime.
Having checked the consistency of our treatment, we describe the behaviour of the condensed part of the gas.

\subsubsection{Condensate}
\label{sec:dynamics_condensate}

\begin{figure}
	\centering
		\centering
		\includegraphics[width=0.9\textwidth]{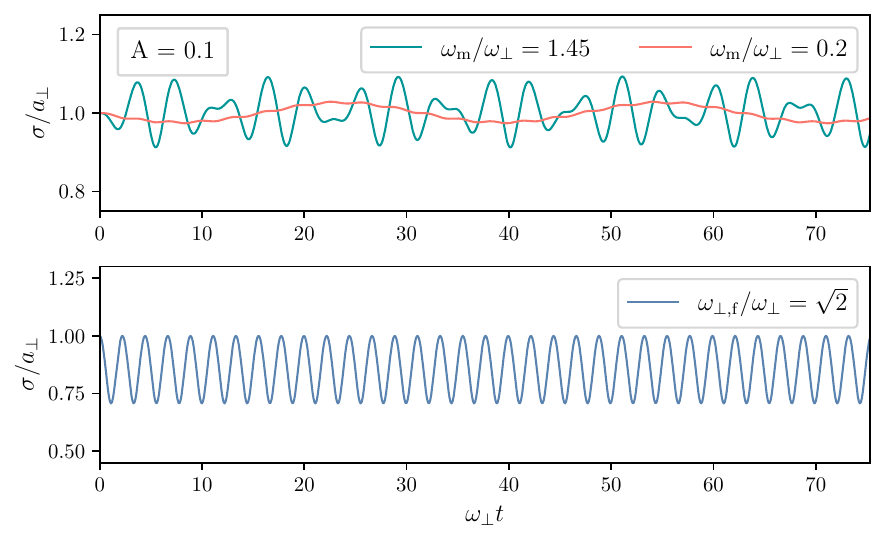}
		\caption{(Upper panel) Behaviour of $\sigma$, the radial extension of the condensate, under a modulation of the transverse trapping frequency $\omega_{\perp}$ according to $\omega_{\perp}^2 = \omega_{\perp}^2 \left[ 1 + A \sin \left( \omega_{\mathrm{m}} t \right) \right]$. $\sigma$ is adimensionalised by the transverse size $a_{\perp}$ associated to $\omega_{\perp}$, and the time is adimensionalised by $ \omega_{\perp}$. The green curve matches the benchmark values used in~\cite{Jaskula:2012ab}, $A= 0.1$ and $ \omega_{\mathrm{m}} = 1.45 \omega_{\perp}$. $\sigma$ features oscillations at two frequencies $2 \omega_{\perp}$ and $\omega_{\mathrm{m}}$.  The red curve corresponds to $\omega_{\mathrm{m}} \ll \omega_{\perp}$, $\sigma$ then predominantly features oscillations at $\omega_{\mathrm{m}}$, while the breathing oscillations at $2 \omega_{\perp}$ are suppressed. 
		(Lower panel) Behaviour of $\sigma$, the radial extension of the condensate, under a quench of the transverse trapping frequency $\omega_{\perp} \to \omega_{\perp,\mathrm{f}}$. Following the quench $\sigma$ oscillates at a frequency $2 \omega_{\perp,\mathrm{f}}$.}
    \label{fig:modulation_and_quench_sigma}
\end{figure}

The wave function of the condensed part, which will act as a background for the excitations on top, is shaped by the trap's effect and the interaction between atoms. We want to account for the effect of the change in the trap size on the solution of this equation. We follow the method and the approximations made in~\cite{Robertson:2016evj,Robertson:2017ysi,Robertson:2018gwi}. First, we expect the physics to be mainly one-dimensional in the vertical direction due to the anisotropy of the trap $\omega_{\perp} \gg \omega_{z}$. We thus separate the condensate wavefunction in a vertical part $\phi_0(z ,t)$ and radial part $\psi_0 (r , \theta , t)$ such that
\begin{equation}
	\label{def:factorisation_ansatz}
	\Psi_0 = \frac{1}{\sqrt{2 \pi}} \psi_0 \left(r , \theta , t \right) \phi_0 \left(z ,t\right) \, ,
\end{equation}
where we choose the normalisation of these fields such that $\phi_0$ is normalised as the wavefunction of 1D condensate
\begin{subequations}
	\begin{align}
		\begin{split}
			\int_{- \infty}^{+ \infty} \left| \phi_0 \left(z ,t\right) \right|^2 \dd z & = N_0 \, ,
		\end{split} \\
		\begin{split}
			\int_0^{+ \infty} \left| \psi_0 \left(r , \theta , t \right) \right|^2 r \dd r & = 1 \, ,
		\end{split}
	\end{align}
\end{subequations}
where $N_0$ is the number of atoms in the condensate. 
The factorisation ansatz.~\eqref{def:factorisation_ansatz} can only describe cylindrically symmetric condensates whose radial profile 
is the same along their vertical extension. It can still provide a good approximation of the ground state of the gas, as shown in~\cite{Robertson:2016evj} by comparing its profile to a numerically computed one. Again, we appeal to the strong anisotropy of the trap $\omega_{\perp} \gg \omega_{z}$ to neglect the effect of the vertical trapping. To still give the gas a finite extension, we assume it is confined in a box of length $L$ in the vertical direction. In addition, we assume the condensed part of the gas to be homogeneous in the vertical direction \ie $\phi_0 = n_0 = N_0 /L$ where
$n_0$ is the linear density of the condensate along the vertical direction. Finally, homogeneity for a finite-size gas requires working with periodic boundary conditions. The GPE over the radial direction then reads
\begin{equation}
	\label{eq:radial_GPE}
	i \hbar \partial_{\rm t} \psi = - \frac{\hbar^2}{2 m} \frac{1}{r} \frac{\partial}{\partial r} \left( r \frac{\partial \psi}{\partial r} \right)  +  \frac{\hbar^2}{2 m a_{\perp}^4}  r^2 \psi + \frac{\hbar^2}{m} 2 n_0 a_s  \left| \psi \right|^2 \psi \, ,
\end{equation}
where the factor of $n_0$ is inherited from the vertical homogeneous profile entering the non-linear interaction term, and we recall that $a_{\perp} = \sqrt{\hbar / m \omega_{\perp}}$. Assuming that the gas is stationary before any change in the trap frequency, we have
\begin{equation}
	\psi_0  \left(r , \theta , t \right) = \tilde{\psi} \left( r \right) e^{-  i \frac{ \mu }{\hbar} t } \, ,
\end{equation}
where $\mu$ is the chemical potential \ie the energy per atom. The stationary GPE now reads
\begin{equation}
	\label{eq:stationnary_radial_GPE}
	\mu \tilde{\psi}  = - \frac{\hbar^2}{2 m} \frac{1}{r} \frac{\partial}{\partial r} \left( r \frac{\partial \tilde{\psi}}{\partial r} \right)  +  \frac{\hbar^2}{2 m a_{\perp}^4}  r^2 \tilde{\psi} + \frac{\hbar^2}{m} 2 n_0 a_s  \left| \tilde{\psi} \right|^2 \tilde{\psi} \, .
\end{equation}
To find the ground state's wave function in which the gas condenses, we have to minimise the right-hand side of this wave function. It can be done approximately by using a Gaussian ansatz for the radial profile
\begin{equation}
	\label{eq:radial_profile}
	\tilde{\psi} = \frac{\sqrt{2}}{\sigma} e^{- \frac{r^2}{2 \sigma^2}} \, ,
\end{equation}
where $\sigma$ controls the extension of the gas. If we neglect the interaction of the gas $g = 0$, then the Gaussian profile is an exact solution
The extension is then $\sigma = a_{\perp}$, and the chemical potential is $\mu = \hbar \omega_{\perp}$, completely controlled by the trap frequency.
This justifies the intuition that $a_{\perp}$ is an estimate of the radial size of the condensate in the trap.
To compute the modifications to these values due to interactions, we insert the ansatz~\eqref{eq:radial_profile} in Eq.~\eqref{eq:stationnary_radial_GPE} and integrate over the radial direction
\begin{equation}
	\mu = \frac{m \omega_{\perp}^2}{2 } \sigma^2 + \frac{\hbar^2}{2 m \sigma^2} \left( 1 + 4 n_0 a_s \right) \, .
\end{equation}
The first term is due to the trap. It favours smaller extensions, where the atoms lie in the centre of the trap with minimal potential energy, while the second term, due both to quantum pressure and the interactions, favours larger extensions, where the atoms are far apart from each other. 
Minimising the right-hand side over $\sigma$ we find~\cite{gerbierQuasi1DBoseEinsteinCondensates2004,Robertson:2016evj} the extension of the ground state $\sigma_0 = a_{\perp} (1 + 4 n_0 a_s )^{1/4}$ and the associated energy $\mu_0 = \hbar \omega_{\perp} (1 + 4 n_0 a_s )^{1/2}$. We found an approximate wavefunction for the condensate when the trap frequency is constant
\begin{equation}
	\label{eq:static_condensate}
	\Psi_0 = \sqrt{\frac{n_0}{2 \pi \sigma_0^2}} e^{- \frac{r^2}{2 \sigma_0^2}} e^{- i \frac{ \mu_0 }{\hbar} t} \, .
\end{equation}
We now have to consider the change in the radial trapping frequency. In~\cite {Jaskula:2012ab}, both the vertical and radial trapping frequencies are quenched (or modulated). Nevertheless, since the radial trapping frequency is much tighter, the change of the radial frequency injects much more energy for the same ratio of final and initial frequencies. Therefore, we neglect changes in the vertical size. For a time-dependent harmonic radial trap $\omega_{\perp} (t)$ one can build an exact solution of the equation of motion for a time-varying trap from a stationary solution at fixed trap frequency $\omega_{\perp,0}$~\cite{kaganEvolutionBosecondensedGas1996}. Starting from our approximate stationary solution of Eq.~\eqref{eq:static_condensate}, we can then build the approximate time-dependent solution
\begin{equation}
	\label{eq:timedependent_condensate}
	\Psi_0 \left( r, t \right) = \sqrt{\frac{n_0}{2 \pi \sigma^2(t)}} e^{- \frac{r^2}{2 \sigma^2(t)}} e^{i \left[ \frac{m r^2}{2 \hbar} \frac{\dot{\sigma}}{\sigma} - \frac{\mu_0}{\hbar} \int^t_0 \frac{\sigma^2_0}{\sigma^2(t^{\prime})} \dd t^{\prime}   \right] } \, ,
\end{equation}
where the radial extension of the gas $\sigma$ is now time-dependent and satisfies the Ermakov-Pinney~\cite{Lidsey:2003ze,leachErmakovEquationCommentary2008} differential equation
\begin{equation}
	\label{eq:EOM_sigma}
	\ddot{\sigma} + \omega_{\perp}^2 (t) \sigma = \frac{\sigma_0^4 \omega_{\perp,0}^2}{\sigma^3} \, .
\end{equation}
The trap modulation's effect has been accounted for by a change in the geometry of the condensate. Notice that when the trap is held 
at $\omega_{\perp , 0}$, then $\sigma = \sigma_0$.
It is instructive to solve Eq.~\eqref{eq:EOM_sigma} for the two types of change considered in~\cite{Jaskula:2012ab}: a quench and a modulation.
A quench of the radial trapping frequency from $\omega_{\perp}$  to $\omega_{\perp, \mathrm{f} }$ will lead to oscillations of the radial size of the gas at $2 \omega_{\perp, \mathrm{f}}$, see lower panel of Fig.~\ref{fig:modulation_and_quench_sigma}.
In the case of a modulation at $\omega_{\mathrm{m}}$, if it is slow enough \ie $\omega_{\mathrm{m}} \ll \omega_{\perp}$, the radial size will dominantly oscillate at $\omega_{\mathrm{m}}$. However if the modulation is fast $\omega_{\mathrm{m}} \gg \omega_{\perp}$ then the behaviour of $\sigma$ has two characteristic frequencies $\omega_{\mathrm{m}}$ and $2 \omega_{\perp}$, see upper panel of Fig.~\ref{fig:modulation_and_quench_sigma}. The oscillations at $2 \omega_{\perp}$ are present in the slow case, but suppressed.
Notice that the equation also has resonances, for instance, at $\omega_{\mathrm{m}} = 2 \omega_{\perp}$ where $\sigma$ grows exponentially. For the benchmark value in~\cite{Jaskula:2012ab} $\omega_{\mathrm{m}} \approx 1.5  \omega_{\perp}$, shown in the upper panel of Fig.~\ref{fig:modulation_and_quench_sigma}, we are in an intermediate regime where we do expect to have oscillations at $\omega_{\mathrm{m}}$ and $2 \omega_{\perp}$, two similar frequencies.

We now compute the dynamics of the perturbation and show that it only depends on that of the trap indirectly, through that
of the condensate. 
In particular, the perturbations undergo parametric amplification at half the frequency of oscillations of the condensate.
Here we will neglect the backreaction of perturbations on the condensate and refer to~\cite{Robertson:2018gwi} for treatment of that point.

\subsubsection{Perturbations}
\label{sec:dynamics_pert_bdg}

First, we will neglect excitations with nodes in the radial direction so that 
\begin{equation} 
	\label{eq:BdG_longitidunal}
	\hat{\Psi} = \Psi_0 \left( r, t \right) \left[ \hat{\mathds{1}} + \delta \hat{\Psi} (z,t) \right] \, .
\end{equation}
Notice that we are dealing with a relative perturbation here, so the absolute perturbations have an $r$ dependence given by the profile of the condensate.
Following~\cite{Robertson:2018gwi}, this approximation can be justified by computing the dispersion relation $\omega (k , n)$ of the three-dimensional excitations, perturbing Eq.~\eqref{eq:EOM_3D_Heinsenberg} around Eq.~\eqref{eq:static_condensate}, where $k$ is the wavenumber in the vertical direction and $n$ is a quantum number labelling excitations in the radial direction.
This procedure was done analytically in the non-interacting limit and numerically for excitations with no azimuthal dependence in~\cite{Robertson:2016evj}. The authors found that azimuthally symmetric excitations with a non-trivial radial part typically have an energy $2 \omega_{\perp}$ larger than the longitudinal ones.
Therefore, at the initial time, the thermal radial excitations will be negligible in numbers compared to the longitudinal ones. Additionally, anticipating the rest of the section, for a modulation at $\omega_{\mathrm{m}}$
we expect parametric resonance to occur in modes with a frequency half that of driving $\omega_{\mathrm{m}}/2$, and possibly at the trap frequency $\omega_{\perp}$. On the other hand, for a quench, we expect oscillation at the trap frequency $\omega_{\perp}$. For the benchmark value $\omega_{\mathrm{m}} = 1.5 \omega_{\perp}$, both of these frequencies are lower than the frequency of the first radial mode, which will thus not be excited. We will neglect the radial excitations from now on. We proceed by inserting Eq.~\eqref{eq:BdG_longitidunal} in Eq.~\eqref{def:H_3Dgas} and collect the terms of second order in $\delta \hat{\Psi}$ to form a Hamiltonian at first non-trivial order for the perturbations $\hat{H}^{(2)}$. At this stage, a slight technical complication comes from using relative perturbations in Eq.~\eqref{eq:BdG_longitidunal}. $ \hat{\Psi}^{(\dagger)}$ are canonically conjugate operators, see Eq.~\eqref{def:AtomicFieldCCR}, and $\sqrt{n_0} \delta \hat{\Psi}^{(\dagger)}$ inherit this property (after integration over the radial profile) so that
\begin{equation}
	\left[ \delta \hat{\Psi} \left( \bm{x} \right) , \delta \hat{\Psi}^{(\dagger)} \left( \bm{x}^{\prime} \right) \right] = \frac{1}{n_0} \delta \left( \bm{x} - \bm{x}^{\prime} \right) \, .
\end{equation}
However, the time-dependent factor of $\Psi_0$ in front of $\delta \hat{\Psi}$ in the expansion of $\hat{\Psi}$ will lead to an additional term $i \hbar \dot{\Psi_0} \delta \hat{\Psi}$ in the equation of motion at the first order of $\delta \hat{\Psi}$ which is therefore not directly the Heisenberg equation derived from $\hat{H}^{(2)}$. This extra factor corresponds to implementing a canonical transformation $\delta \hat{\Psi} \to \Psi_0 \delta \hat{\Psi}$ with a time-dependent factor which, as well-known, leads to a modified Hamiltonian. The computations are performed in Appendix.~\ref{app:canonical_transfo_BdG}, and we give the resulting Hamiltonian (that we still call $\hat{H}^{(2)}$ for simplicity)
\begin{equation}
	\label{eq:quadratic_H_pert_BdG_longitidunal}
	\hat{H}^{(2)} = \int_0^L n_0 \left[ \frac{\hbar^2}{2 m} \frac{\partial \delta \hat{\Psi}^{\dagger}}{\partial z} \frac{\partial \delta \hat{\Psi}}{\partial z} + g_1 n_0 \delta \hat{\Psi}^{\dagger} \delta \hat{\Psi} + \frac{g_1 n_0}{2} \left( \delta \hat{\Psi}^{\dagger \, 2} + \delta \hat{\Psi}^2 \right) \right] \dd z \, ,
\end{equation}
where we have integrated over the radial profile, and $g_1(t) = g  / 2 \pi \sigma^2(t)$ is the effective 1D interaction constant.
Eq.~\eqref{eq:quadratic_H_pert_BdG_longitidunal} is exactly the BdG Hamiltonian for a one-dimensional gas whose interaction 
constant is made time-dependent. In general, the interaction constant of a Bose gas can be controlled using a Fesbach resonance, as in~\cite{Viermann:2022wgw}. However, here the interaction constant is an effective one coming from the dimensional reduction, and the time dependence is due to the trap.
We can compute the associated Heisenberg equation of motion, which could have directly computed by perturbing Eq.~\eqref{eq:EOM_3D_Heinsenberg} as in~\cite{Robertson:2016evj}
\begin{equation}
	\label{eq:EOM_BdG_1D}
	i \hbar \partial_{t} \delta \hat{\Psi}  = -\frac{ \hbar^2}{2 m} \partial_{z}^2 \delta \hat{\Psi} + g_1(t) n_0 \left( \delta \hat{\Psi} + \delta \hat{\Psi}^{\dagger}   \right) \, .
\end{equation}
This equation is solved by decomposing the perturbation in Fourier modes
\begin{equation}
	\label{eq:fourier_modes}
	\delta \hat{\Psi} (z,t) = \frac{1}{\sqrt{N_0}} \sum_{k \in 2 \pi \mathbb{Z}^{\star} / L} e^{i k z} \hat{a}_{k} \, .
\end{equation}
where $\hat{a}_{k}$ is the adimensional annihilation operator for atoms with momentum $\hbar k$. The $k=0$ term is removed to make the number of condensed atoms conserved, and the total number of atoms conserved at first order in perturbations\footnote{Notice that thanks to this, the Hamiltonian~\eqref{def:H_3Dgas} vanishes at first order in perturbation, and the quadratic Hamiltonian~\eqref{eq:quadratic_H_pert_BdG_longitidunal} is the first non-trivial contribution.}. 
The creation/annihilation operators satisfy the canonical commutation relation
\begin{subequations}
	\label{eq:CCR_fourier}
	\begin{align}
		\begin{split}
			\left[ \hat{a}_{k} , \hat{a}_{k^{\prime}}^{\dagger} \right] & = \delta_{k , k^{\prime}} \, ,
		\end{split} \\
		\begin{split}
			\left[\hat{a}_{k} , \hat{a}_{k^{\prime}} \right] & = 0   \, ,
		\end{split}
	\end{align}
\end{subequations}
where $\delta_{k , k^{\prime}}$ is the Kronecker delta.
We have
\begin{equation}
	\label{eq:nbr_atoms}
	N = \int \dd z \left \langle \hat{\rho} \right \rangle = N_0 + \sum_{k \in 2 \pi \mathbb{Z}^{\star} / L} \left \langle \hat{a}_{k}^{\dagger} \hat{a}_{k} \right \rangle \, ,
\end{equation}
where the second piece gives the number of excited atoms not in the condensate. This piece is referred to as the \textit{depletion}.
We then write the Hamiltonian~\eqref{eq:quadratic_H_pert_BdG_longitidunal} in Fourier modes
\begin{equation}
	\label{eq:quadratic_H_pert_BdG_Fourier}
	\hat{H}^{(2)} = \sum_{k \in 2 \pi \mathbb{Z}^{\star} / L} \left( \frac{\hbar^2 k^2}{2 m} + g_1 n_0 \right) \hat{a}^{\dagger}_{k} \hat{a}_{k} + \frac{g_1 n_0}{2} \left( \hat{a}^{\dagger}_{ k} \hat{a}^{\dagger}_{- k} + \hat{a}_{k} \hat{a}_{- k}\right)  \, .
\end{equation}
The linear Heisenberg equation of motion~\eqref{eq:EOM_BdG_1D} can also be recast as an equation over $\hat{a}_{k}$ and $\hat{a}_{-k}^{\dagger}$. We have
\begin{align}
	\label{eq:BdG_EOM_ak}
	i \hbar \, \partial_{t}
	\begin{pmatrix}
		\hat{a}_{k} \\
		\hat{a}^{\dagger}_{- k}
	\end{pmatrix}
	=
	\underbrace{\begin{pmatrix}
			\frac{\hbar^2 k^2}{2 m} + g_1 n_0 &  g_1 n_0 \\
			-g_1 n_0  & -  \frac{\hbar^2 k^2}{2 m} - g_1 n_0
	\end{pmatrix}}_{= M_{k}}
	\begin{pmatrix}
		\hat{a}_{k} \\
		\hat{a}^{\dagger}_{- k}
	\end{pmatrix}
	\, .
\end{align}
Notice that, just as for cosmological perturbations, see Sec.~\ref{sec:evo_inf_pert}, as a consequence of homogeneity of the Hamiltonian~\eqref{def:H_3Dgas} these equations preserve the total momentum only mixing operators having the same action, raising or lowering the total momentum by $k$. Hence, only the mode $\pm k$ are coupled at quadratic order. 

\subsubsection{Dynamics in the absence of external drive}
\label{sec:dynamics_no_external_drive}

Let us first consider the time-independent case where the trap frequency is held at a fixed value $\omega_{\perp , 0}$ so that $\sigma = \sigma_0$. For a free system  $g = 0$, $M_k$ is diagonal and atoms of different momenta evolve independently. In the presence of interactions, $g \neq 0$, there is a non-trivial mixing between $\pm \hbar k$ atoms. 
At the level of the Hamiltonian, this manifests in the fact that Eq.~\eqref{eq:quadratic_H_pert_BdG_Fourier} is not diagonal in the atom number basis.
Therefore, the ground state of the interacting gas does not only contain atoms at rest, a part of them will be moving; this is the so-called quantum depletion~\cite{pitaevskiiBoseEinsteinCondensation2003}. To quantify this depletion we need to diagonalise the Hamiltonian or equivalently 
the matrix $M_k$ of Eq.~\eqref{eq:BdG_EOM_ak}. As for cosmological perturbations, see Sec.~\ref{sec:quantum_GW}, this is done via a Bogoliubov 
transformation
\begin{equation}
	\label{def:phonons}
	\begin{pmatrix}
		\hat{a}_{k} \\
		\hat{a}^{\dagger}_{- k}
	\end{pmatrix}
	= 
	\underbrace{
		\begin{pmatrix}
			u_k & v_k \\
			v_{-k}^* &  u_{-k}^*
	\end{pmatrix}}_{= P_{k}}
	\begin{pmatrix}
		\hat{b}_{k} \\
		\hat{b}^{\dagger}_{- k}
	\end{pmatrix}
	\, ,
\end{equation}
where we have defined the Bogoliubov coefficients that satisfy the standard relations $|u_k|^2  - |v_k|^2 = |u_{-k}|^2 - |v_{-k}|^2 = 1$ and $u_k v_{-k} = u_{-k} v_{k}$ for the pairs $(\hat{b}_{\pm k}, \hat{b}_{\pm k}^{\dagger})$ to be canonically conjugated. 
These operators define the quasi-particles of the system, often called \textit{phonons}, although this term is sometimes reserved only for low-lying excitations. We here use these words interchangeably.
We then pick $P_k$ to diagonalise $M_k$, and we find the expressions of the two Bogoliubov coefficients~\cite{pitaevskiiBoseEinsteinCondensation2003}
\begin{align}
	u_{k} &  = \sqrt{\frac{ \left( \frac{\hbar^2 k^2}{2 m} + g_1 n_0 \right) + \hbar \omega_k }{2 \hbar \omega_k}} =  \frac{ \sqrt{\frac{\hbar^2 k^2}{4 m} + g_1 n_0} + \sqrt{\frac{\hbar^2 k^2}{4 m}}   }{\sqrt{2 \hbar \omega_k } } \, , \\
	v_k & = - \sqrt{\frac{ \left( \frac{\hbar^2 k^2}{2 m} + g_1 n_0 \right) - \hbar \omega_k }{2 \hbar \omega_k }} = - \frac{\sqrt{\frac{\hbar^2 k^2}{4 m} + g_1 n_0} - \sqrt{\frac{\hbar^2 k^2}{4 m}} }{\sqrt{ 2 \hbar \omega_k } } \, ,
\end{align}
where we have defined the frequency 
\begin{equation}
	\label{def:BdG_dispersion}
	\omega_{k} = c |k| \sqrt{ 1 + \frac{\xi^2 k^2}{4} } \, ,
\end{equation}
the speed of sound $c^2 = g_1 n_0 / m $ and the healing length $\xi = \hbar /mc$.
Eq.~\eqref{def:BdG_dispersion} is the standard Bogoliubov-de Gennes dispersion relation~\cite{pitaevskiiBoseEinsteinCondensation2003}.
The diagonalised dynamics then reads
\begin{align}
	\partial_{t}
	\begin{pmatrix}
		\hat{b}_k \\
		\hat{b}_{-k}^{\dagger}
	\end{pmatrix}
	=
	\begin{pmatrix}
		-i \omega_{k} & 0 \\
		0 & i \omega_{k}
	\end{pmatrix}
	\begin{pmatrix}
		\hat{b}_k \\
		\hat{b}_{-k}^{\dagger}
	\end{pmatrix} \, ,
\end{align}{}
and so
\begin{equation}
	\hat{b}_k(t) = e^{-i \omega_{k} t} \hat{b}_k(0) \, . 
\end{equation}
Using the phononic creation and annihilation operators, we rewrite the Hamiltonian, which is now diagonal
\begin{equation}
	\label{eq:quadratic_H_pert_BdG_phonons}
	\hat{H}^{(2)} = \sum_{k \in 2 \pi \mathbb{Z}^{\star} / L}  \hbar \omega_k \hat{b}^{\dagger}_{k} \hat{b}_{k} \, .
\end{equation}
The phonons describe the gas (at the first order in perturbation) as a collection of free quasi-particles of energy $\hbar \omega_k$.
We can get a physical understanding of the nature of these excitations by considering two limits.
In the large wavenumber limit $k \xi \gg 1$, we have $\omega_k \sim \hbar k^2 / 2 m$, the dispersion of free atoms, and the Bogoliubov coefficients give $u_k \to 1$ and $v_k\to 0$: the quasi-particles are close to free atoms. 
On the other hand, for $k \xi \ll 1$, we have $\omega_k \sim c |k|$, the dispersion of sound waves, and the Bogoliubov coefficients becomes equal  $u_k \sim v_k \sim \sqrt{m c / (2 \hbar |k|)} $: the low wavenumber excitations are that of a fluid. It is in this hydrodynamic regime that the analogy of analogue gravity is usually formulated, see Sec.~\ref{sec:introduction_AG}.

\subsubsection{Dynamics in the presence of an external drive}
\label{sec:dynamics_external_drive}

We now consider the case where the interaction constant $g_1(t)$ is varied, or equivalently as considered in~\cite{Bruschi:2013tza,Busch:2013sma,Busch:2014vda} the speed of sound $c(t)$ is varied. 
We follow the analysis of \cite{Busch:2014vda}. The passage matrix $P_k$ defined in Eq.~\eqref{def:phonons} is time-dependent, and then the equation of motion for the phononic operators reads
\begin{equation}
	\label{eq:EOM_BdG_timedependent}
	\partial_{t}
	\begin{pmatrix}
		\hat{b}_k \\
		\hat{b}_{-k}^{\dagger}
	\end{pmatrix}
	=
	\begin{pmatrix}
		-i \omega_{k}(t) &  \frac{\dot{\omega_k}}{2 \omega_k} \\
		\frac{\dot{\omega_k}}{2 \omega_k} & i \omega_{k}(t)
	\end{pmatrix}
	\begin{pmatrix}
		\hat{b}_k \\
		\hat{b}_{-k}^{\dagger}
	\end{pmatrix} \, ,
\end{equation}
where we have used $|u_k|^2-|v_k|^2=1$, and the fact that only $c(t)$ is varied here while the mass is kept constant. The anti-diagonal term comes from the time dependence of the passage matrix. It shows that, in addition to the adiabatic change of frequency $\omega_{k}(t)$ in the phase, the evolution leads to a non-trivial mixing between the phononic modes $\pm k$, and so a quasi-particle creation. Notice the similarity with Eq.~(32) of~\cite{Micheli:2022tld}, reproduced in Sec.~\ref{sec:quantum_GW}, for cosmological perturbations. This equation is
then similarly solved by introducing a Bogoliubov transformation with coefficients $\alpha_k(t)$ and $\beta_k(t)$ such that
\begin{equation}
	\label{eq:time_evo_phonons}
	\begin{pmatrix}
		\hat{b}_{k}(t) \\
		\hat{b}^{\dagger}_{- k}(t)
	\end{pmatrix}
	= 
	\begin{pmatrix}
		\alpha_k(t) e^{- i \int^t_{t_{\mathrm{in}}} \omega_{k} \dd t^{\prime}  } & \beta^{\star}_k(t) e^{- i \int^t_{t_{\mathrm{in}}} \omega_{k} \dd t^{\prime}  } \\
		\beta_k(t) e^{ i \int^t_{t_{\mathrm{in}}} \omega_{k} \dd t^{\prime}  } &  \alpha_k^{\star}(t) e^{i \int^t_{t_{\mathrm{in}}} \omega_{k} \dd t^{\prime}  }
	\end{pmatrix}
	\begin{pmatrix}
		\hat{b}_{k} (t_{\mathrm{in}}) \\
		\hat{b}^{\dagger}_{- k} (t_{\mathrm{in}})
	\end{pmatrix}
	\, ,
\end{equation}
where $\alpha_k(t_{\mathrm{in}})=1$, $\beta_k(t_{\mathrm{in}})=0$ and $\hat{b}_k^{ (\dagger) }(t_{\mathrm{in}})$ corresponds to the operators of Eq.~\eqref{def:phonons} evaluated at some time $t_{\mathrm{in}}$.
In Eq.~\eqref{eq:time_evo_phonons}, we have factored out the adiabatic evolution of the phase, and Eq.~\eqref{eq:EOM_BdG_timedependent} is then equivalent to
\begin{align}
	\label{eq:EOM_bogo_phonon}
	\dot{\alpha_k} & = \frac{\dot{\omega_k}}{2 \omega_k} \beta_k e^{ 2 i \int^t_{t_{\mathrm{in}}} \omega_{k} \dd t^{\prime}  } \, , \\
	\dot{\beta_k} & = \frac{\dot{\omega_k}}{2 \omega_k} \alpha_k e^{ - 2 i \int^t_{t_{\mathrm{in}}} \omega_{k} \dd t^{\prime}  } \, .
\end{align}

We consider a situation where the frequency of the trap is varied during a finite duration such that $\omega_{\perp}(t) \xrightarrow[t \to - \infty]{} \omega_{\perp , \mathrm{in}}$ and $\omega_{\perp}(t) \xrightarrow[t \to + \infty]{} \omega_{\perp , \mathrm{out}}$. In these two asymptotic regions, the Bogoliubov coefficients $u_k^{\mathrm{in , out}}$ and $v_k^{\mathrm{in , out}}$ are time-independent, so the phonons defined by the creation and annihilation operators $\hat{b}_{\pm k}^{ (\dagger) \, \mathrm{in , out}}$ of Eq.~\eqref{def:phonons}, evaluated in the asymptotic regions, have a well-defined number. These pairs of in and out operators are related by a Bogoliubov transformation. This can be seen by fixing $t_{\mathrm{in}} \to - \infty$ in Eq.~\eqref{eq:time_evo_phonons}. Then the operators $\hat{b}^{(\dagger)}_{\pm k}(t)$ in the asymptotic future will correspond to $\hat{b}_{\pm k}^{ (\dagger) \, \mathrm{out}}$ (up to the running phase that we factorised out) so
\begin{equation}
	\label{eq:in_out_phonon_operator}
	\hat{b}_{\pm k}^{ \mathrm{out}} = \alpha_k(+\infty)  \hat{b}_{\pm k}^{\mathrm{in}}  + \beta_k^{\star}(+\infty) \hat{b}^{\dagger \, \mathrm{in}}_{\mp k} \, .
\end{equation}
The dynamics is then completely fixed by solving Eq.~\eqref{eq:EOM_bogo_phonon} for the specific variation of $\omega_k(t)$ we consider.
In principle, for a given variation of $\omega_{\perp}(t)$, we should first solve Eq.~\eqref{eq:EOM_sigma} to get the evolution of $\sigma(t)$, which we then use as an input to solve Eq.~\eqref{eq:EOM_bogo_phonon}. In practice, in~\cite{Micheli:2022zet}, we made the simplifying assumption that the modulation of  $\omega_{\perp}(t)$ induces a modulation of $\omega_k(t)$ at the same frequency, which we recall is only valid for a slow enough modulation.

\subsection{Parametric amplification}
\label{sec:parametric_amplification}

\subsubsection{Link with preheating}

To exhibit the link between the creation of phonons in the Bose gas and the preheating mechanism described in Sec.~\ref{sec:preheating}, we focus on the case where, for example following a quench, the speed of sound oscillates at a certain frequency
\begin{equation}
\label{def:mod_c}
	c^2(t) = c_0^2 \left[ 1 + a \sin \left( \omega_{m} t \right)\right] \, ,
\end{equation}
so that 
\begin{equation}
	\label{def:mod_omega}
	\omega_k^2(t) = \omega_{k,0}^2 \left[ 1 + \mathcal{A}_k  \sin \left( \omega_{m} t \right) \right] \, ,
\end{equation}
where $\mathcal{A}_k =  a  (1+ k^2 \xi^2/4)^{-1}$, we recall that $\xi$ is the healing length defined below Eq.~\eqref{def:BdG_dispersion}.
This modulation type was considered in~\cite{Busch:2014vda}.
To derive a Mathieu's equation, inspired by~\cite{Busch:2013gna}, we consider the perturbations of the density and phase fields. Using the form~\eqref{def:madelung_form}, after integration over the radial profile, we expand the field in density $\delta \hat{\rho}$ and phase $\delta \hat{\theta}$ perturbations assumed small
\begin{equation}
	\hat{\Psi} = n_0 \sqrt{\hat{\mathds{1}} + \frac{\delta \hat{\rho}}{  n_0}} e^{i \delta \hat{\theta}} \approx  n_0 \left( \mathds{1} + \frac{\delta \hat{\rho}}{2  n_0} + i \delta \hat{\theta} \right) \, .
\end{equation}
Equating with Eq.~\eqref{eq:BdG_longitidunal}, and taking the hermitian and anti-hermitian part, we get
\begin{subequations}
	\begin{align}
		\begin{split}
			\label{eq:density_pert_BdG}
			\delta \hat{\rho}  \left( z , t \right) & = n_0 \left( \delta \hat{\Psi} + \delta \hat{\Psi}^{\dagger} \right) \, ,
		\end{split} 
		\\
		\begin{split}
			\label{eq:phase_pert_BdG}
			\delta \hat{\theta}  \left( z , t \right) & = - \frac{i}{2} \left( \delta \hat{\Psi} - \delta \hat{\Psi}^{\dagger} \right) \, .
		\end{split}
	\end{align}
\end{subequations}
We use the same conventions as~\cite{Micheli:2022tld} for the Fourier transform of these quantities (chosen to make them adimensional)
\begin{subequations}
	\label{def:density_phase_fourier}
	\begin{align}
		\begin{split}
			\delta \hat{\rho} & = \sqrt{\frac{n_0}{L}} \sum_{k \neq 0} e^{i k x}  \hat{\rho}_k \, ,
		\end{split} 
		\\
		\begin{split}
			\delta \hat{\theta}  & = \frac{1}{\sqrt{L n_0}} \sum_{k \neq 0} e^{i k x} \hat{\theta}_k \, ,
		\end{split}
	\end{align}
\end{subequations}
where we recall $L$ is the length of the gas.
In Fourier space, the relation with the perturbations of the atomic field then reads
\begin{subequations}
	\begin{align}
		\begin{split}
			\label{eq:density_pert_BdG_Fourier}
			\hat{\rho}_k & =  \hat{a}_k + \hat{a}^{\dagger}_{-k}  =  \left( u_k + v_k \right) \left( \hat{b}_k + \hat{b}^{\dagger}_{-k} \right)  \, ,
		\end{split} 
		\\
		\begin{split}
			\label{eq:phase_pert_BdG_Fourier}
			\hat{\theta}_k  & = - \frac{i}{2 } \left( \hat{a}_k -  \hat{a}^{\dagger}_{-k} \right) = - \frac{i}{2 } \left( u_k - v_k \right) \left( \hat{b}_{k} - \hat{b}^{\dagger}_{-k} \right) \, ,
		\end{split}
	\end{align}
\end{subequations}
where in the second equality we have written the expressions in terms of phononic operators. 
First, notice that in a thermal state of atoms (or phonons), the above expressions show explicitly that the average density and phase fluctuations vanish $\langle \delta \hat{\theta} \rangle = \langle \delta \hat{\rho} \rangle = 0$. Second, notice that $(i \hbar \delta \hat{\theta} , \delta \hat{\rho})$ form a canonically conjugated pair. Finally, expressing $u_k + v_k =\sqrt{ \hbar k^2 / 2 m \omega_k}$, notice that the normalisation (similar to that of a relativistic scalar field) has a time-dependent piece due to $\omega_k$. Now using Eq.~\eqref{eq:EOM_BdG_timedependent} we can write the equation of motion of the density perturbation modes
\begin{equation}
	\label{eq:Mathieu_eq_rho}
	\ddot{\hat{\rho}}_k + \omega_{k}^2 \hat{\rho}_{k} = 0 \, .
\end{equation}
For a modulation of the frequency as in Eq.~\eqref{def:mod_omega}, Eq.~\eqref{eq:Mathieu_eq_rho} is exactly a Mathieu's equation as that found for a scalar field excited by the oscillations of the inflaton in preheating, see Eq.~\eqref{eq:mathieu_eq}. Notice that Eq.~\eqref{eq:mathieu_eq} was obtained in the regime where the expansion of space is negligible. Having an analogue of the equation in the presence of expansion could maybe be achieved by considering the expansion of the vertical part, which would redshift the modes as in~\cite{Eckel:2017uqx}.
Following the standard analysis of Mathieu's equation~\cite{kovacicMathieuEquationIts2018} we thus expect an exponential creation of density fluctuations and pairs of phonons in some resonant bands.

To conclude this part, we compare two other studies of analogue preheating systems~\cite{Zache:2017dn,Barroso:2022vxg}. 
On the one hand, the setting of the numerical study~\cite{Zache:2017dnz} is very similar to the one of~\cite{Jaskula:2012ab} in that it considers a 1D Bose gas whose longitudinal modes are sourced by initial excitations in the radial modes. The main difference is that while in~\cite{Zache:2017dnz} only the first excited radial mode is populated, in~\cite{Jaskula:2012ab}, the gas is driven to oscillate in a quasi-classical manner in the radial direction. The radial excitation is then treated as a second quantum field in the former, while it is taken to be a classical background in the latter. Consequently, the longitudinal modes are excited via a tachyonic instability triggered by the excited radial mode in ~\cite{Zache:2017dnz}, while they are excited via parametric resonances due to the coherent radial oscillations in~\cite{Jaskula:2012ab}. 
On the other hand, the experiment~\cite{Barroso:2022vxg} is based on a quite different set-up which uses 2D interface waves. 
The system is placed on a moving platform, which oscillates to trigger a parametric amplification of the waves. 
In contrast to~\cite{Zache:2017dnz,Barroso:2022vxg}, the excitations are classical in nature, and the amplification is seeded by environmental noise rather than vacuum or thermal one.
The authors follow the amplification of the modes from their initial exponential growth to its saturation and the redistribution towards other modes of the system, leading to the growth of secondaries. 
In Sec.~\ref{sec:decay_nk}, we consider the effect of damping in a 1D Bose gas due to interaction with a thermal population of quasi-particles. Our analysis focuses on linear damping when the population of the decaying mode is relatively small. In contrast, the system of Ref.~\cite{Zache:2017dnz} features linear damping but not due to interaction with other modes, and the authors focus on damping triggered by the largeness of the amplitude of the resonant mode leading to non-linear damping.

\subsubsection{Growth and decay of phonon numbers and correlation}
\label{sec:growth_nk_ck}

In~\cite{Busch:2014vda}, the authors analysed analytically and numerically the generation of phonons and the correlation between them in the presence of an effective dissipation rate.
Let us first review the results in the absence of dissipation. Assuming that we start from a thermal state (including possibly the vacuum), the initial state is Gaussian, and the evolution via Eq.~\eqref{eq:EOM_bogo_phonon} leads to another Gaussian state.
For the homogeneous system we consider, the covariance matrix for the modes $\pm k$ is then completely characterised by the $2$-point functions
\begin{subequations}
	\begin{align}
		\begin{split}
			\label{def:nk}
			n_{\pm k} & =\left\langle \hat{b}^{\dagger}_{\pm k} \hat{b}_{\pm k} \right \rangle \, ,
		\end{split} 
		\\
		\begin{split}
			\label{def:ck}
			c_{k} & =  \left\langle \hat{b}_{\pm k} \hat{b}_{\mp k} \right \rangle \, ,
		\end{split}
	\end{align}
\end{subequations}
where $n_{\pm k}$ is the number of phonons in the mode $\pm k$ and $c_k$ the correlation in between them. Notice that $c_k$ is a complex quantity and, by definition, $c_k= c_{-k}$. We assume that the system is isotropic $n_{k} = n_{- k}$ for simplicity. Using the
relation~\eqref{eq:in_out_phonon_operator}, we can relate this quantity before and after the variation of the speed of sound. 
We have
\begin{subequations}
	\begin{align}
		\begin{split}
			\label{eq:in_out_nk}
			n_{ k}^{\mathrm{out}} & = \left( \left| \alpha_k \right|^2 + \left| \beta_k \right|^2 \right) n_{ k}^{\mathrm{in}} + \left| \beta_k \right|^2 \, ,
		\end{split} 
		\\
		\begin{split}
			\label{eq:in_out_ck}
			c_{k} & = \alpha_k \beta_k^{\star} \left( 2 n_{k}^{\mathrm{in}} + 1 \right)   \, ,
		\end{split}
	\end{align}
\end{subequations}
where we have used $c_{k}^{\mathrm{in}} = 0$ and $\langle \hat{b}_{ k} \hat{b}^{\dagger}_{- k}  \rangle = 0$ in a homogeneous thermal state.

We see that generically phonons are created by the variation of the sound speed, the extent of which is controlled by the magnitude of the Bogolibuov coefficients and the initial population. There are two types of contributions, the one proportional to the initial population $n_{k}^{\mathrm{in}}$, a stimulated creation, and the spontaneous creation, which would be present even if $n_{k}^{\mathrm{in}} = 0$.
Spontaneous production can occur from an initial vacuum state, and it arises in the computation from the non-commutation of the creation/annihilation operators; it is a genuine quantum process.
One cannot distinguish a phonon created via a stimulated or a spontaneous process, these two channels add phonons in the mode $\pm k$ in a common quantum state. Still~\cite{Busch:2014vda}, a way to check those phonons were created out of the vacuum is to check that their state is entangled, also referred to as `non-separable', see Sec.~\ref{sec:comparing_quantumness_criteria} for details on this notion. For an isotropic and homogeneous Gaussian state, the non-separability condition is equivalent to the inequality~\cite{Busch:2014vda,Campo:2005sy}
\begin{equation}
	\label{def:nonsep_phonons}
	\Delta_k = n_k - \left| c_k \right| < 0 \, .
\end{equation}
After the modulation, we have
\begin{equation}
	\label{eq:delta_out_phonons}
	\Delta_k^{\mathrm{out}} = \frac{ -\left|\beta_k\right| \left(  \left|\alpha_k\right| +  \left|\beta_k\right| \right) + n_k^{\mathrm{in}}  }{ \left( \left|\alpha_k\right| +  \left|\beta_k\right| \right)^2 } \, ,
\end{equation}
which makes clear the effect of the different contributions. If we start from the vacuum, then $\Delta_k^{\mathrm{out}} < 0$, and the phonons are in an entangled state. On the other hand, if we start to increase $n_k^{\mathrm{in}}$, then $\Delta_k$ will also increase, and it can make the state separable if $2 n_k^{\mathrm{in}} + 1 >  ( |\alpha_k| +  |\beta_k | )^2$.
Therefore, \textit{only} creation out of the vacuum via a spontaneous process can generate an entangled state, while stimulated creation will dilute this entanglement.
In addition, with the form~\eqref{eq:delta_out_phonons}, one can check that for a fixed value of $n_k^{\mathrm{in}}$, $\Delta_k^{\mathrm{out}}$ is monotonically decreasing as $| \beta_k |$ increases. For very large $| \beta_k |$, it goes to $\Delta_k^{\mathrm{out}} \to -1/2$. Therefore, the resonant modes of the parametric amplification, which have the largest values of $| \beta_k |$, will exhibit the smallest values of $\Delta_k^{\mathrm{out}}$. 

\begin{figure}
	\centering
	\includegraphics[height=0.85\textheight]{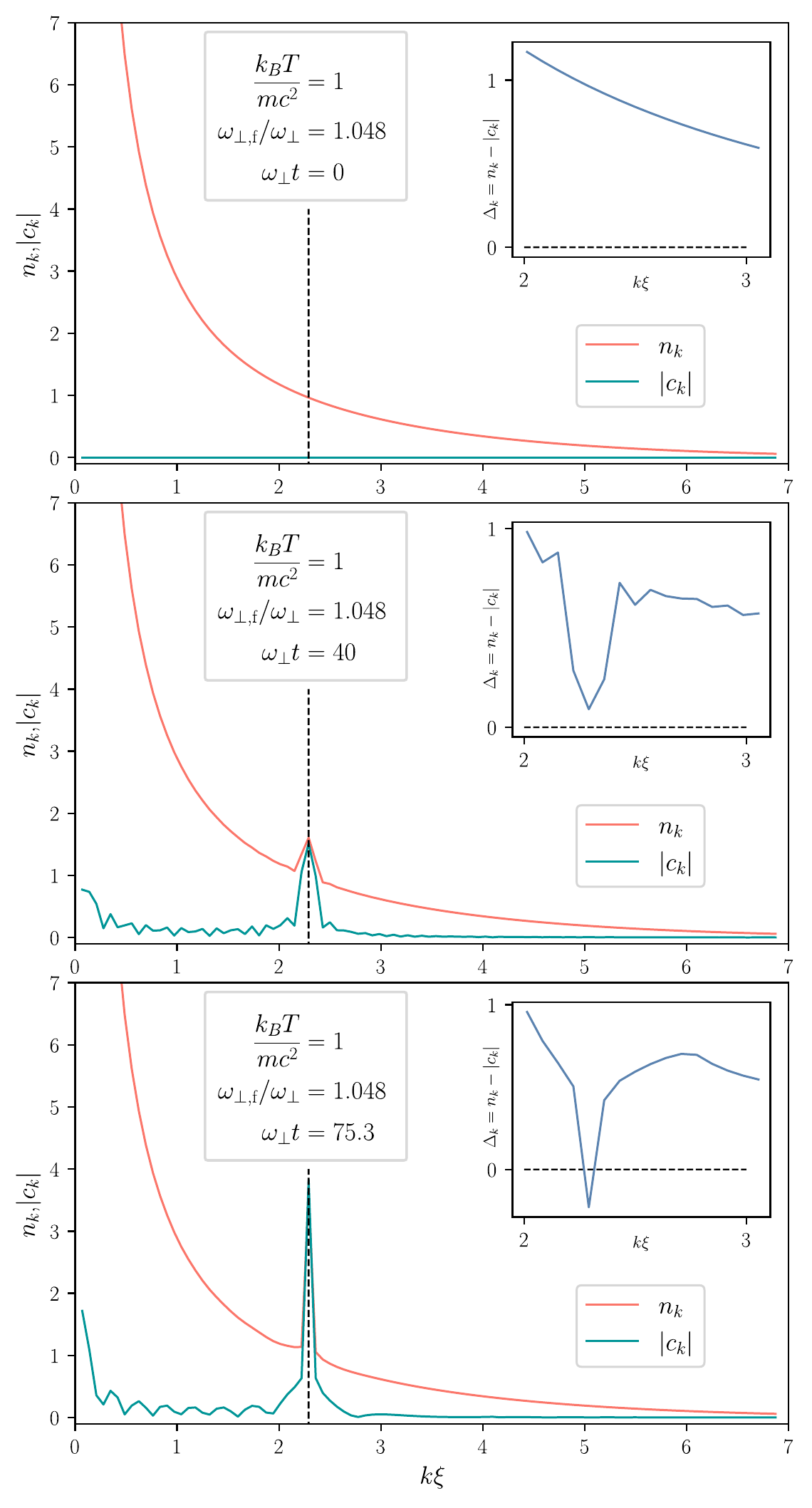} % second figure itself
	\caption{Evolution of the number of quasi-particles $n_k$ and their pair-correlation $|c_k|$, as predicted by Eqs.~(\ref{eq:in_out_nk})-(\ref{eq:in_out_ck}), following a quench of the trap frequency. The vertical dashed line is located at the discrete mode $k_j$ whose frequency is closest to half the final trap frequency $2 \omega_{k_j} = \omega_{\perp,\mathrm{f} }$. We are in the narrow resonance regime where resonant modes are located close to this sub-harmonic mode. We consider a gas of length $L/ a_{\perp}= 256$.  }
	\label{fig:evolution_nk_ck_quench}
\end{figure}

The authors of~\cite{Busch:2014vda} solved Eq.~\eqref{eq:EOM_bogo_phonon} for the modulation of Eq.~\eqref{def:mod_omega} and showed that, as expected from the Mathieu's equation, Eq.~\eqref{eq:Mathieu_eq_rho},  the Bogoliubov coefficients only grows exponentially in some resonant bands and oscillate out of this band. 
They also analysed the evolution of $\Delta_k$ for different modes.
As an illustration we plot in Fig.~\ref{fig:evolution_nk_ck_quench} the evolution of $n_k$ and $|c_k|$ obtained for a small quench of the trapping frequency $\omega_{\perp, \mathrm{f}} /  \omega_{\perp} = 1.048$, that results in a modulation of $\omega_{k}$ at a frequency $2 \omega_{\perp}$. 
The ratio $\omega_{\perp, \mathrm{f}} /  \omega_{\perp}$ determines the amplitude $a$ of the oscillations of $c^2$ in Eq.~\eqref{def:mod_c}, which will be small in this case, putting us in the narrow resonance regime, see Sec.~\ref{sec:preheating}. 
The instantaneous Bogoliubov coefficients of Eq.~\eqref{eq:time_evo_phonons} are numerically computed by first solving Eq.~\eqref{eq:EOM_sigma} and using the result as an input for a numerical resolution of Eq.~\eqref{eq:EOM_BdG_timedependent}. We then compute $n_k$ and $|c_k|$ from Eqs.~(\ref{eq:in_out_nk})-(\ref{eq:in_out_ck}) as a function of time 
for a non-vanishing initial temperature $k_{\mathrm{B}} T / m c^2 = 1$. 
We chose these values to make the entanglement generation clear and visible. We comment on more realistic values below.
Fig.~\ref{fig:evolution_nk_ck_quench} shows the generation of an entangled state for the modes $\pm k$ at late times \ie $|c_k| > n_k$.

\begin{figure}
	\centering
	\includegraphics[height=0.85\textheight]{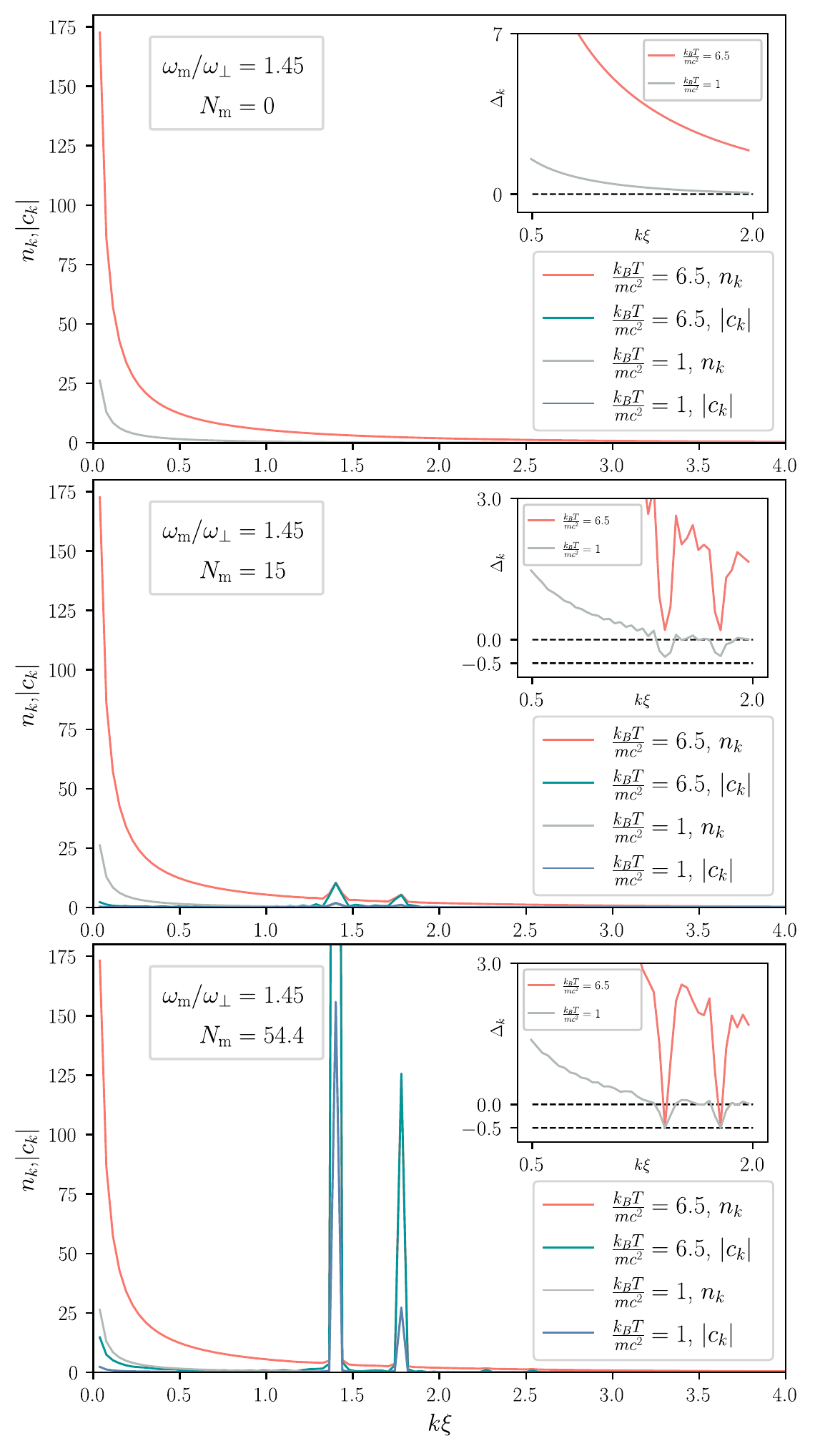} % second figure itself
	\caption{Evolution of the number of quasi-particles $n_k$ and their pair-correlation $|c_k|$ as a function of time during modulation of the trap frequency at $\omega_{\mathrm{m}}$ with amplitude $A=0.1$, as predicted by Eqs.~(\ref{eq:in_out_nk})-(\ref{eq:in_out_ck}). At final time the number of phonons in the first peak is $n_k = 8.3 \times 10^2$, outside of the frame of the figure. 
	We consider a gas of length $L/ a_{\perp}= 256$.}
	\label{fig:evolution_nk_ck_jaskula_modulation}
\end{figure}

In~\cite{Busch:2014vda}, the authors solved Eq.~\eqref{eq:EOM_bogo_phonon} for a modulation lasting $50$ oscillations, close to the value of $N_{\mathrm{m}} \approx 54.4$ used in the second experiment of~\cite{Jaskula:2012ab}. They estimated the initial temperature value $k_{\mathrm{B}} T / \hbar \omega_{\perp} = 1$. They concluded the simple quadratic model overestimates the intensity of correlations, predicting that the state should have been entangled. To illustrate this conclusion, we plot in Fig.~\ref{fig:evolution_nk_ck_jaskula_modulation} the evolution of $n_k$ and $|c_k|$ following a modulation of the trapping frequency. 
We used values estimated from~\cite{Jaskula:2012ab}. We considered a modulation at the benchmark value of $\omega_{\mathrm{m}} / 2 \pi = 2170 \, \mathrm{Hz}$, lasting 
$N_{\mathrm{m}} \approx 54.4$ oscillations.
We used values of the speed of sound and temperature computed from the estimates derived in Sec.~\ref{sec:presentation_exp}, which gives $c \approx 8 \, \mathrm{mm}/ \mathrm{s}$ and $k_{\mathrm{B}} T / m c^2 \approx 6.5$. In this respect, the value of $k_{\mathrm{B}} T / \omega_k = 1$, at resonance, used in~\cite{Busch:2014vda} seems too low. It would correspond to $k_{\mathrm{B}} T / m c^2 \approx 2.3$. Still, it is in the appropriate range for the current run of the experiment where $k_{\mathrm{B}} T / m c^2 \approx 1$.
Another difference with the computations of~\cite{Busch:2014vda} is that, in Fig.~\ref{fig:evolution_nk_ck_jaskula_modulation}, we solved Eq.~\eqref{eq:EOM_sigma} to deduce the change in the speed of sound due to the modulation, rather than using the effective form Eq.~\eqref{def:mod_omega}. 
The oscillation of $\sigma$ is a combination of several oscillations at different frequencies as shown for instance in the upper panel of Fig.~\ref{fig:modulation_and_quench_sigma}. 
In the context of preheating~\cite{Zanchin:1997gf,Zanchin:1998fj,DeCross:2016fdz}, it is known that such non-ideal modulation results in a modified structure of the Mathieu equation resonance bands, for instance by broadening them~\footnote{Quasi-particle interactions can also lead to broadened peak as shown in Sec.~\ref{sec:decay_nk}. However, note that in this case the other modes in the peak are still non-resonant and are fed by the decay of excitations in the resonant modes.}
In our particular case, Fig.~\ref{fig:evolution_nk_ck_jaskula_modulation} shows two resonant peaks, one around $\omega_{k}=\omega_{\perp}$, and one around $\omega_{k}=\omega_{\mathrm{m}}/2$.
As reported in~\cite{Busch:2014vda}, and at odds with the experimental results of~\cite{Jaskula:2012ab}, Fig.~\ref{fig:evolution_nk_ck_jaskula_modulation} shows that the phononic states of the two pairs of modes are entangled at late time.
For comparison, we have plotted the result of the same modulation on the initial state of phonons at a lower temperature $k_{\mathrm{B}} T / mc^2 = 1$, close to current experimental values. We observe that the state becomes entangled at earlier times and that the lower bound $\delta_k = -0.5$ is quickly saturated. Note that despite this saturation, the entanglement will be harder to detect when the number of excitations is very large, see discussion below Eq.~\eqref{def:small_g2}.

The authors of~\cite{Busch:2014vda} suggested as a tentative explanation for the absence of entanglement that some dissipative processes, ignored here, might have weakened the correlations. 
They then introduce a model of dissipation that preserves the Gaussianity and isotropy of the state of the modes so that they are still described by $n_k$ and $c_k$. Therefore, the resulting state is in the same class as the ones considered for the cosmological perturbations in Sec.~\ref{chapt:quantumness_cosmo}. The authors then derived an evolution equation for $n_k$ and $c_k$ parameterised by a dissipation rate $\Gamma$, such that $n_k$ and $c_k$ both decay at a rate $2 \Gamma$.
They show that a large enough $\Gamma$ can prevent the generation of entanglement and, in some cases, even stop the exponential growth of the number of phonons. 
Note that in Fig.~\ref{fig:evolution_nk_ck_jaskula_modulation} the number of phonons in the resonant modes is much larger than that reported in~\cite{Jaskula:2012ab}. Thus, in addition to prevent the generation of entanglement, we do expect dissipation to significantly reduced the number of phonons reduced.
The authors of~\cite{Busch:2014vda} give a threshold $\Gamma / \omega_{k} \geq 4.2 \%$\footnote{We use our convention here where $n_k$ and $c_k$ decay at $\Gamma_k$, while their decay rate is $2 \Gamma_k$ in~\cite{Busch:2014vda}.} above which they estimated the dissipation sufficient to explain the absence of entanglement in~\cite{Jaskula:2012ab}.

However, the treatment of~\cite{Busch:2014vda} was effective. In Sec.~\ref{sec:decay_nk}, by analysing the non-linear evolution of the gas, we exhibit a micro-physically derived dissipation mechanism due to Beliaev-Landau scatterings present in any one-dimensional Bose gas. In Sec.~\ref{sec:decay_ck}, we explain that the same processes are expected to affect correlations in just the same proportion so that due to these scatterings, $n_k$ and $c_k$ decay approximately at the same rate. We then compare the magnitude of this rate with the bounds given in~\cite{Busch:2014vda}.
Notice that we limit our investigations to the effect of the dissipation on the growth of the resonant modes. The loss of entanglement and slowing down of the exponential growth are only the first effects of non-linearities. They will be followed by substantial growth of non-resonant modes and back-reaction on the condensate oscillations, which will be damped. These later stages are very reminiscent of the very non-linear regime following preheating. They were numerically analysed in~\cite{Robertson:2018gwi}.
As preparation for deriving the phonon scattering rate in Sec.~\ref{sec:decay_ck}, we explain in the next sub-section, Sec.~\ref{sec:Madelung}, why the formalism used so far is not adapted for 1D gases and develop a more suitable one.

We close this sub-section by considering how the non-separability criteria of Eq~\eqref{def:nonsep_phonons} can be experimentally accessed.

\subsubsection{Density-density correlation}

In~\cite{Robertson:2016evj,Robertson:2017ysi}, the authors showed that the $2$-point density-density correlation function, a quantity experimentally measured in~\cite{Steinhauer:2015saa} for instance, can be used to demonstrate the non-separability of the phonon pairs. 
We define the connected $2$-point density correlation function at equal time following~\cite{Robertson:2017ysi}
\begin{align}
	\begin{split}
		G^{(2)} \left( z, z^{\prime} ; t \right) & = \left\langle 
		\hat{\rho} \left(z,t \right) \hat{\rho} \left( z^{\prime} , t \right) \right\rangle - \left\langle 
		\hat{\rho} \left(z,t \right) \right\rangle \left\langle  \hat{\rho} \left( z^{\prime} , t \right) \right\rangle \, , \\
		& = \left\langle 
		\delta \hat{\rho} \left(z,t \right) \delta \hat{\rho} \left( z^{\prime} , t \right) \right\rangle \, ,
	\end{split}
\end{align}
where we have used that $\langle \delta \hat{\rho} \rangle =0$ to get the second line. Taking the Fourier transform and expanding in terms of quasi-particles, we get
\begin{align}
	\begin{split}
		\label{eq:G2_k_phonons}
		G^{(2)} \left( k, k^{\prime} ; t \right) & = \int_0^L \int_0^L G^{(2)} e^{-i k z}  e^{-i k^{\prime} z^{\prime} }  \left( z, z^{\prime} ; t \right) \dd z \dd z^{\prime} \, , \\
		& = n_0 L 
		\left\langle 
		\delta \hat{\rho}_k \delta \hat{\rho}_{k^{\prime}} \right\rangle \, , \\
		& = \left( u_k + v_k \right)^2 \left[ 2 n_k + 1 + 2 \Re \left( c_k \right) \right] \delta_{k,-k^{\prime}} \, ,
	\end{split}
\end{align}
where the Kronecker delta comes from the homogeneity of the state.
The only non-trivial information then comes from the anti-diagonal $2$-point function $	G^{(2)} ( k, -k) = 	G^{(2)}_k$. Evaluating the above in the out asymptotic region, where the quasi-particles are evolving freely, we have
\begin{equation}
	\label{eq:g2_out}
	G^{(2)}_k \left( t \right)  = G^{(2) \, \mathrm{out} \, \mathrm{vac}}_k \left[ 2 n_k^{\mathrm{out}} + 1 + 2 \Re \left( c_k^{\mathrm{out}} e^{-i 2 \omega_k^{\mathrm{out}} t} \right) \right] \, ,
\end{equation}
where we have introduced the out vacuum value $G^{(2) \, \mathrm{vac}}_k = (u_k^{\mathrm{out}} + v_k^{\mathrm{out}})^2$.
If at any time $t$, $G^{(2)}_k$ is less than its vacuum value then it implies $ |c_k^{\mathrm{out}}| > n_k^{\mathrm{out}}$ \ie the state of the quasi-particles is non-separable. The density-density correlation can be directly accessed in an experimental set-up allowing for \textit{in situ} measurements, e.g.~\cite{Steinhauer:2015saa}.

In the case of~\cite{Jaskula:2012ab}, the phononic distribution is measured by opening the trap and letting the atomic cloud expand and fall on a measuring device. The velocities are inferred from the expansion of the cloud, a method called Time Of Flight measurement (TOF), and one can show that, in the limit of an adiabatic opening, the state of the phonons is mapped to the atoms~\cite{Robertson:2016evj}, a phenomenon known as phonon evaporation~\cite{Jaskula:2012ab}. 
Using this procedure, the authors were able to measure the following quantity
\begin{equation}
	\label{def:small_g2}
	g^{(2)} = \frac{\left\langle \hat{b}^{\dagger}_{-k} \hat{b}^{\dagger}_k \hat{b}_k \hat{b}_{-k}  \right\rangle}{ \left\langle \hat{b}^{\dagger}_k \hat{b}_k \right\rangle \left\langle \hat{b}^{\dagger}_{-k} \hat{b}_{-k} \right\rangle}  = 1 + \frac{\left| c_k \right|^2}{n_k^2} \, ,
\end{equation}
where we performed Wick contractions to get the second equality. The condition of entanglement~\eqref{def:nonsep_phonons} is now recast as $g^{(2)} > 2$.
Note that measuring $g^{(2)} > 2$, or likewise $G^{(2)}_k < G^{(2) \, \mathrm{vac}}_k$, is not a non-classicality criteria in the sense of Ref.~\cite{Finke:2016wcn}.  Indeed, we could find a classical model of the system giving $G^{(2)}_k < G^{(2) \, \mathrm{vac}}_k$. A Bell inequality would be necessary to rule out a whole class of alternative classical theories. Still, our aim is slightly different here. We already know that the correct description of the system is a quantum one. Adopting this description, we want to show that the system is in a particular type of \textit{quantum} states, the entangled states, which is sufficient to demonstrate vacuum amplification.

We reproduce in Fig.~\ref{fig:measured_small_g2} the measured values in~\cite{Jaskula:2012ab}. We see that across the spectrum, and in particular around the resonant mode, $g^{(2)} < 2$. Therefore, as previously stated, one cannot claim that the phonons produced in the experiment were in an entangled state. Notice that with the values of $n_k$ and $|c_k|$ shown in Fig.~\ref{fig:evolution_nk_ck_jaskula_modulation}, although the state is entangled since $\Delta_k^{\mathrm{out.} }$ is negative, we would have $g^{(2)} \approx 2$ at the resonant modes \ie the threshold value between a separable and an entangled state. 
This convergence to $2$ is due to the normalisation of $g^{(2)}$ by the number of phonons $n_k$. It illustrates the notion of visibility of the entanglement~\cite{Robertson:2017ysi,Robertson:2018gwi}. Even if the state is entangled, it can be difficult to tell if we have to compare two very large numbers, $n_k$ and $|c_k|$, while their difference is bounded from below by $-1/2$. Therefore, the visibility of entanglement in the experiment is optimal when we create a small number of excitations~\cite{Robertson:2018gwi}.
Consequently, as illustrated in Fig.~\ref{fig:evolution_nk_ck_jaskula_modulation}, where entanglement is reached with fewer excitations created for the smallest of the two temperatures, we should use the lowest possible initial temperature for optimal visibility.

Having completed our review of the previous theoretical endeavours to analyse the experiment results, we will present the progress made during this PhD. First, we discuss the Madelung perturbation scheme, which is at the core of the analysis conducted in~\cite{Micheli:2022tld} reproduced in Sec.~\ref{sec:decay_nk}.

\subsection{Madelung approximation of 1D gas}
\label{sec:Madelung}

In this section, we examine more closely the BdG approximation, presented in Sec.~\ref{sec:dynamics_pert_bdg}, in the case of one-dimensional gas.
We show that it is not valid anymore if the length of the gas $L$ is larger than a critical size $r_0$ defined in Eq.~\eqref{def:r0}.
We then present a perturbative scheme based on the Madelung form~\eqref{def:madelung_form}, which remains valid for long one-dimensional quasi-condensates.

\subsubsection{Failure of BdG in 1D gas}

Assuming that we are in a static situation $g = \text{cst}$, the approximate ground state of the system is given by the vacuum of phonons $n_k = c_k  = 0$ for any $k$. The number of depleted atoms in the mode $k$ then reads
\begin{equation}
	\left\langle \hat{a}^{\dagger}_k \hat{a}_k \right\rangle = \left| v_k \right|^2 \sim_{k \to 0 } \frac{g_1 n_0}{2 \hbar c k} \, ,
\end{equation}
where we have taken the small wavenumber limit.
The total number of atoms $N$ given by Eq.~\eqref{eq:nbr_atoms} is always larger than that in the condensate $N_0$, so the total density $n$ always larger than $n_0$.
We want to compute the value of the total density of the gas in the thermodynamic limit $L \to + \infty$ for a condensate of fixed density $n_0$.
The extra contribution is given by the number of depleted atoms normalised by $L$, which by using the formula for Riemann sums, reads
\begin{equation}
	\frac{1}{L} \sum_{k \in 2 \pi \mathbb{Z}^{\star} / L} \left \langle \hat{a}_{k}^{\dagger} \hat{a}_{k} \right \rangle \xrightarrow[L \to + \infty]{} \frac{1}{2 \pi} \int_{- \infty}^{+ \infty} \left \langle \hat{a}_{k}^{\dagger} \hat{a}_{k} \right \rangle \dd k = + \infty  \, . 
\end{equation}
The right-hand side is an IR divergent integral since its integrand behaves as $k^{-1}$.
In the large $L$ limit, the Bogoliubov approximate predicts an unphysical infinite density. This failure of the approximation is directly related to the general Mermin-Wagner-Hohenberg theorem~\cite{Hohenberg:1967zz} stating the absence of long-range order in one and two-dimensional systems. The theorem can be proven using general inequalities on the number of excitation resulting in a divergence of some physical quantities, see Sec.~7.4 of~\cite{pitaevskiiBoseEinsteinCondensation2003}.
In the specific case of Bose gas, it implies that there can be no genuine condensate in reduced dimensions, only \textit{quasi-}condensate. 

The physical reason behind this failure is that one-dimensional Bose gas exhibits large phase fluctuations~\cite{pitaevskiiBoseEinsteinCondensation2003}. One can check that by computing the phase-phase correlation in a thermal state of phonons using Eq.~\eqref{eq:phase_pert_BdG_Fourier}, see Sec.~\ref{sec:phase_density_pert}.
Because of these fluctuations, any pre-existing order will be gradually destroyed.
Thinking in terms of numbers rather than operators, the condition $| \delta \Psi | \ll 1$ of the BdG approximation requires that the atomic field $\Psi$, a complex number, is close to $\Psi_0$. This condition requires that the modulus and the phase of $\Psi$ are close to that of $\Psi_0$ \ie the density \textit{and} the phase perturbations around the condensate are small. The latter condition is not satisfied for a 1D gas of arbitrary length. A finite length $L$ of the gas regulates the unbouded growth of the phase fluctuations, but the $| \delta \Psi |$ term would become more and more problematic as $L$ is increased.
To avoid this problem, we can formulate a perturbation theory for one-dimensional gas in the Madelung form of Eq.~\eqref{def:madelung_form}, where we expand the relative density perturbation, but not the phase one that cannot be assumed small.
We followed this strategy in~\cite{Micheli:2022zet} to define quasi-particles and calculate their interaction rate, and we now give more details on this approach.

\subsubsection{Quantum hydrodynamics and quasi-condensate}
\label{sec:quasi_condensate}

Before making any computations using the density and phase formalism, we have to say a few words about the technical difficulties hidden there.
Indeed, the analysis in density and phase can only describe long enough wavelength fluctuations. As a consequence, as mentioned below the definition Eq.~\eqref{def:madelung_form}, the phase operator cannot be mathematically well-defined. It is physically intuitive that the notion of fluid ceases to make sense when looking at scales shorter than the inter-particle distance: the density is infinite at points where atoms are located and vanishing elsewhere.
The authors of~\cite{moraExtensionBogoliubovTheory2003} showed that in order to make the effective character of the theory manifest and make the mathematical treatment rigorous, it is necessary to discretise space in boxes of size $\dd x$, which provides a UV cut-off to the theory.
Using this discrete theory, we want to describe a quasi-condensed gas which exhibits small density fluctuations \ie the variance of its density is smaller than its average value $\text{Var} (\hat{\rho}) / \langle \hat{\rho} \rangle \ll 1$. Following~\cite{moraExtensionBogoliubovTheory2003}, it is instructive to write the variance in this discrete theory explicitly
\begin{align}
	\begin{split}
		\label{eq:density_fluct_discrete}
		\text{Var} \left(\hat{\rho} \right) & = \left \langle \hat{\rho} ^2 \right \rangle - \left \langle \hat{\rho}  \right \rangle^2 \, \\
		& = \left \langle \hat{\Psi}^{\dagger} \left( \bm{x} \right)  \hat{\Psi}^{\dagger} \left( \bm{x} \right)  \hat{\Psi} \left( \bm{x} \right)  \hat{\Psi} \left( \bm{x} \right) \right \rangle + \frac{\left \langle \hat{\rho} \left( \bm{x} \right) \right \rangle}{\dd x}  - \left \langle \hat{\rho} \left( \bm{x} \right) \right \rangle^2 \, ,
	\end{split}
\end{align}
where we have used the commutation relation of the discrete theory $[ \hat{\Psi} \left( \bm{x} \right)  ,  \hat{\Psi}^{\dagger} \left( \bm{x}^{\prime} \right) ] = \delta_{\bm{x} , \bm{x}^{\prime} } / \dd x$.
We then have
\begin{equation}
	\label{eq:relative_density_fluct_discrete}
	\frac{\text{Var} \left(\hat{\rho} \right)}{\left \langle \hat{\rho}  \right \rangle^2} = \frac{1}{\left \langle \hat{\rho}  \right \rangle \dd x } +
	\frac{\langle : \delta \hat{\rho}(x)^2: \rangle }{\left \langle \hat{\rho}  \right \rangle^2} \, .
\end{equation}
The density fluctuations thus contain two terms. The first one comes from the commutator; it depends on the discretisation and would be infinite in a continuous theory. 
Therefore, to suppress the density fluctuations, we must choose a box size large enough to ensure that
$\left \langle \hat{\rho}  \right \rangle \dd x \gg 1$ \ie
the occupation number in each box is large.
The second term in Eq.~\eqref{eq:relative_density_fluct_discrete} is the normal-ordered (with respect to the atomic field) relative density fluctuations and regular in a well-behaved continuous theory. It encodes information about the state considered rather than the discretisation, and we must consider states where it is small.
Under these first two requirements, we can define an approximate phase operator~\cite{moraExtensionBogoliubovTheory2003}.
Additional requirements are necessary to describe well the physics of long wavelength fluctuations. The size of the boxes must be smaller than the healing length, to capture the departure of excitation from exact sound waves, and smaller than the thermal de Broglie wavelength $\lambda_{\mathrm{T}} = \sqrt{2 \pi \hbar^2 / m k_{\mathrm{B}} T}$, for the energy cut-off $\hbar^2 /m \dd x^2$ to be larger than the typical energy of thermal fluctuations $k_{\mathrm{B}}T$.
In this framework, the authors of~\cite{moraExtensionBogoliubovTheory2003} then develop a perturbation theory around a classical solution of the equation of motion given by $\rho_0(x)$ and $\theta_0(x)$. Assuming that the density fluctuations about this solution are small $| \delta \hat{\rho} (x)| \ll \rho_0$, and that the phase fluctuations are slow $|\partial_{x} \delta \hat{\theta}| \dd x \ll 1$, they compute one and two-body correlation functions and show that no divergence occurs when taking the limit $\dd x \to 0$.
This rigorous procedure is quite technical, and in~\cite{Micheli:2022zet}, we skipped over these aspects when introducing the density and phase representation.
Nonetheless, no divergence appears in the computations, the spectrum of the quadratic theory found does match that derived in~\cite{moraExtensionBogoliubovTheory2003}, and the decay processes that we have identified involve only thermal fluctuations located at very long wavelengths, where the effective density and phase treatment is expected to be valid.
The density and phase representation is also used in the numerical simulations of the one-dimensional gas we performed to confirm our analytical results. There, we represented the system using a grid and chose the space-step $\dd x$ to respect the prescription of~\cite{moraExtensionBogoliubovTheory2003}.
At the end of this section and in the next Sec.~\ref{sec:TWA_simulation_1D_BEC}, we give additional details on the physics of quasi-condensate and the method used to simulate it, that we omitted in~\cite{Micheli:2022zet}.

\subsubsection{Phase and density fluctuations}
\label{sec:phase_density_pert}

We follow Sec.~\ref{sec:decay_nk} and perform the canonical transformation to density and phase using Eq.~\eqref{def:madelung_form} while ignoring previously mentioned technical difficulties.
We first consider a classical theory of fields, where the transformation is indeed canonical, and all quantities are well-defined. We then quantise the density and phase by imposing
\begin{equation}
	\left[ \hat{\rho}\left(x\right) \,,\, \hat{\theta}\left(x^{\prime}\right) \right] = i \, \delta\left(x-x^{\prime}\right) \,.
	\label{eq:rho-theta-CCR}
\end{equation}
We reproduce the Hamiltonian~(3) of~\cite{Micheli:2022zet}
\begin{equation}
	\label{def:H_full_madelung}
	\hat{H} = \int_{0}^{L} \dd x \left[ \frac{\hbar^2}{2m} \frac{\partial \hat{\theta}}{\partial x} \hat{\rho} \frac{\partial \hat{\theta}}{\partial x} +  \frac{\hbar^2}{8m \hat{\rho}} \left( \frac{\partial \hat{\rho}}{\partial x} \right)^2  + \frac{g}{2} \hat{\rho}^2 \right] \,,
\end{equation}
where we dropped infinite commutators coming from the kinetic and interaction term. The latter could be renormalised by an (infinite) chemical potential shift. Notice that we have not expanded the exponential of the phase to obtain the above.
We then assume we are in a state where the density and phase are perturbations around homogeneous and stationary solutions of the classical equations of motion, given by $\rho_0(x)$ and $\theta_0(x)$
\begin{equation}
	\label{eq:madelung_pert}
	\hat{\Psi} = e^{i \theta_0 + i \delta \hat{\theta}} \sqrt{\rho_0 + \delta \hat{\rho}} \, ,
\end{equation}
with small relative density fluctuations $| \delta \hat{\rho} (x)| \rho_0^{-1}$, and slow phase fluctuations \ie $|\partial_{x} \delta \hat{\theta}|$ small. The Hamiltonian Eq.~\eqref{def:H_full_madelung} is then expanded in a perturbation series, see Eq.~(5) of~\cite{Micheli:2022zet}.
The result of this expansion is given in Eq.~(6) of~\cite{Micheli:2022zet}. We diagonalise the quadratic term of this expansion to Eq.~\eqref{eq:time_evo_phonons} by introducing quasi-particle creation/ annihilation operators defined in Eq.~(10) of~\cite{Micheli:2022zet}.
Notice that both the energy spectrum defined by this procedure, see Eq.~(9) of~\cite{Micheli:2022zet}, and the link between quasi-particles and density/phase, see Eq.~(10) of~\cite{Micheli:2022zet}, have the same form as that obtained in the BdG approach, Eq.~\eqref{def:BdG_dispersion} and Eq.~\eqref{eq:phase_pert_BdG_Fourier}\footnote{The published equation reproduced in Sec.~\ref{sec:decay_nk} has a factor $1/2$ too many in the definition of $C_k$ given below Eq.~(10).}.
However, one subtle difference is that $n_0$ in the BdG expressions is replaced by $\rho_0$, the background density, in the Madelung perturbation scheme. Since $\langle \delta \hat{\rho} \rangle =0$, one can check by computing the number of atoms in the gas via
Eq.~\eqref{eq:nbr_atoms} that $\rho_0$ is, in fact, the \textit{total} density of atoms, while $n_0$ was the density of condensed atoms only. The Madelung perturbation scheme is thus, by construction, immune to the appearance of divergences in the total number of atoms. 
Up to this substitution, the two approaches only differ in how the atomic field is reconstructed from the phonons.
In BdG, the density and phase perturbations are treated as additive perturbations of the atomic field, while in the density and phase picture, the reconstruction is non-linear, see Eq.~\eqref{eq:madelung_pert}.

Using these new quasi-particles, we can define an approximate thermal state of the system by assuming a thermal distribution $\langle \hat{b}^{\dagger}_{k} \hat{b}_{k} \rangle = n_k^{\mathrm{th}}$, where
\begin{equation}
\label{def:nk_thermal}
n_k^{\mathrm{th}} =  \frac{1}{e^{\hbar \omega_{k}/k_{\mathrm{B}} T}-1} \, .
\end{equation}
We can then compute the phase and density fluctuations in this state.
We have
\begin{align}
	\begin{split}
		\label{def:phase_fluctuations}
		\chi \left( x-x^{\prime} \right) & = \left \langle \delta \hat{\theta} \left( x \right) \delta \hat{\theta} \left( x^{\prime} \right) \right \rangle = \frac{1}{4  \rho_0 L } \sum_{k \neq 0} e^{i k \left(  x - x^{\prime} \right)} \left( u_k - v_k \right)^2 \left( 2 n_k^{\mathrm{th}} + 1 \right) \, , \\
		& \approx \frac{1}{8 \pi \rho_0} \int_{- \infty}^{+ \infty} e^{i k \left(  x - x^{\prime} \right)} \left( u_k - v_k \right)^2 \left( 2 n_k^{\mathrm{th}} + 1 \right) \dd k \, ,
	\end{split}
\end{align}
where in the second line, we have taken the continuum limit.
Although the integral is taken over the whole range of wavenumbers, there should be a UV cut-off since the hydrodynamical theory is not valid for arbitrarily large wavenumbers, and the integral is divergent in this limit where $u_k - v_k \to 1$. This divergence gives rise to a Dirac delta and we study the finite part of the integral.
We want to extract this integral's physical large wavelength content, which will correspond to the large separation $s = |x - x^{\prime}|$ regime. For that, we expand the integrand in the limit of small $k$. The lowest order term gives 
\begin{equation}
	\label{eq:u_minus_v_low_k}
	\frac{\left( u_k - v_k \right)^2}{4} \left( 2 n_k^{\mathrm{th}} + 1 \right)  \sim_{k \to 0} \frac{k_{\mathrm{B}} T}{m c^2} \frac{1}{k^2 a_{\perp}^2 } \, .
\end{equation}
The phase fluctuations thus diverge for large wavelengths, the limit in which we expect the theory to be valid.
In addition, the term in Eq.~\eqref{eq:u_minus_v_low_k} gives the integral a finite UV contribution, so we do not need to introduce a cut-off to regularise it.
Notice that the dominant fluctuations are of thermal origin, as made manifest by the factor of $T$, while vacuum fluctuations would contribute in $1/k$~\cite{pitaevskiiBoseEinsteinCondensation2003}.
In one dimension, the integral in Eq.~\eqref{def:phase_fluctuations} does not pick up any factor of $k$ from the differential element and therefore exhibits an IR divergence.
The divergence can be renormalised by removing the coincident point fluctuation $\chi(0)$~\cite{pitaevskiiBoseEinsteinCondensation2003}. We then have
\begin{align}
	\begin{split}
		\label{eq:growth_fluctuations} 
		\chi \left( s \right) - \chi \left( 0 \right) & \approx \frac{1}{\pi  \rho_0} \int_{0}^{+ \infty}  \left[ \cos \left( k s \right) - 1  \right] \frac{k_{\mathrm{B}} T}{m c^2} \frac{\dd k }{k^2}  
		=  - \frac{s}{r_0}  \, ,
	\end{split}
\end{align}
where
\begin{equation}
\label{def:r0}
	r_0 = \frac{2 \hbar^2  \rho_0}{k_{\mathrm{B}} T m} \, ,
\end{equation}
gives the characteristic scale for the growth of phase fluctuations.
To compare, we compute the density fluctuations
\begin{align}
	\begin{split}
		\label{def:density_fluctuations}
		\left \langle \delta \hat{\rho} \left( x \right) \delta \hat{\rho} \left( x^{\prime} \right) \right \rangle \approx \frac{\rho_0}{2 \pi L } \int_{- \infty}^{+ \infty} e^{i k \left(  x - x^{\prime} \right)} \left( u_k + v_k \right)^2 \left( 2 n_k^{\mathrm{th}} + 1 \right) \dd k \, .
	\end{split}
\end{align}
We again have a UV divergence giving rise to a Dirac delta as $u_k + v_k \to 1$ for $k \xi \gg 1$. However, in the long wavelength limit, we find
\begin{equation}
	\left( u_k + v_k \right)^2 \left( 2 n_k^{\mathrm{th}} + 1 \right) \sim_{k \to 0} \frac{k_{\mathrm{B}} T}{m c^2} \, ,
\end{equation}
the density fluctuations are not growing in the IR. The integral in Eq.~\eqref{def:density_fluctuations} will be regular in this limit, giving a contribution decaying in $1/s$. In this sense, the density fluctuations are suppressed compared to the phase ones. 

Finally, we use these results to compute the behaviour of the one-body correlation, defined in Eq.~\eqref{def:g1}, for a quasi-condensate at thermal equilibrium.
We will neglect the density fluctuations, and we have~\cite{pitaevskiiBoseEinsteinCondensation2003}
\begin{align}
	\begin{split}
		\label{eq:g1_quasi_condensate}
		g_{1} \left( x,x^{\prime} \right) & = \left \langle \sqrt{ \hat{\rho} \left(x\right) } e^{- i \left[ \hat{\theta} \left( x \right) - \hat{\theta} \left( x^{\prime} \right) \right]} \sqrt{ \hat{\rho} \left(x^{\prime}\right)  } \right \rangle  \, , \\
		& \approx \rho_0 \left \langle e^{- i \left[ \hat{\theta} \left( x \right) - \hat{\theta} \left( x^{\prime} \right) \right]} \right \rangle  \, , \\
		& = \rho_0  e^{ - \frac{\left \langle \left[  \hat{\theta} \left( x \right) - \hat{\theta}\left( x^{\prime} \right) \right]^2  \right \rangle }{2}  }   \, , \\
		& = \rho_0  e^{ \chi \left( s \right) - \chi \left( 0 \right) } = \rho_0 e^{- \frac{s}{r_0}}   \, .
	\end{split}
\end{align}
To compute the expression on the third line, we first expanded the exponential and dropped the odd powers that are vanishing by expansion in creation/annihilation operators
\begin{equation}
\label{eq:expansion_exp_theta}
\left \langle e^{- i \left[ \hat{\theta} \left( x \right) - \hat{\theta} \left( x^{\prime} \right) \right]} \right \rangle = \sum_{j = 0}^{+ \infty} \frac{(-1)^j}{(2 j)! } \left \langle \left[ \hat{\theta} \left( x \right) - \hat{\theta} \left( x^{\prime} \right) \right]^{2j}  \right \rangle \, .
\end{equation}
Each term in the expansion can be computed using Wick contractions and counting the number of pairs. In general
\begin{equation}
\left \langle \hat{X}^{2j} \right \rangle  = \frac{\left( 2 j \right)!}{2^j j !} \left \langle \hat{X}^{2} \right \rangle ^{j} \, .
\end{equation}
Using this relation and the expansion Eq.~\eqref{eq:expansion_exp_theta}, we obtain the third line of Eq.~\eqref{eq:g1_quasi_condensate}.
The above computation is only valid for distances $s$ large compared to the healing length $\xi$ and the thermal length $\lambda_{\mathrm{T}}$. A refined version can be found in~\cite{moraExtensionBogoliubovTheory2003}, which agrees with Eq.~\eqref{eq:g1_quasi_condensate} in the large $s$ limit.
Eq.~\eqref{eq:g1_quasi_condensate} shows that even if the density fluctuations are small, there can be no long-range order in one dimension due to large phase fluctuations.\footnote{This results still holds for zero temperature. The computation is different and lead to a power-law decay~\cite{moraExtensionBogoliubovTheory2003}.}
Notice, however, that if we consider a system of finite size $L$, as we do here, then the Mermin-Wagner theorem does not strictly apply, and the total number of depleted atoms remains finite.
Eq.~\eqref{eq:g1_quasi_condensate} indicates that when $L/r_0 \ll 1$, the effect of the phase fluctuation is not visible, and the gas effectively appears as condensed. Using the estimates computed in Sec.~\ref{sec:condensed_state}, in particular, for the vertical size of the condensate in the Thomas-Fermi regime, and the value of the temperature estimated in Sec.~\ref{sec:dynamics_external_drive}, we compute the value of $L/r_0$ in~\cite{Jaskula:2012ab}. We find $r_0 / L \approx 3 \times 10^{-2}$ \ie we are in the quasi-condensate regime where phase fluctuations are large. 

We reproduce our work~\cite{Micheli:2022zet} in the next section, Sec.~\ref{sec:decay_nk}. In this work, we also use the perturbation theory based on the Madelung form for numerical simulations. More details about the workings of these simulations are given in Sec.~\ref{sec:TWA_simulation_1D_BEC}.

\section{Article: `Phonon decay in one-dimensional atomic Bose quasi-condensates via Beliaev-Landau damping'}
\label{sec:decay_nk}

\includepdf[pages=-]{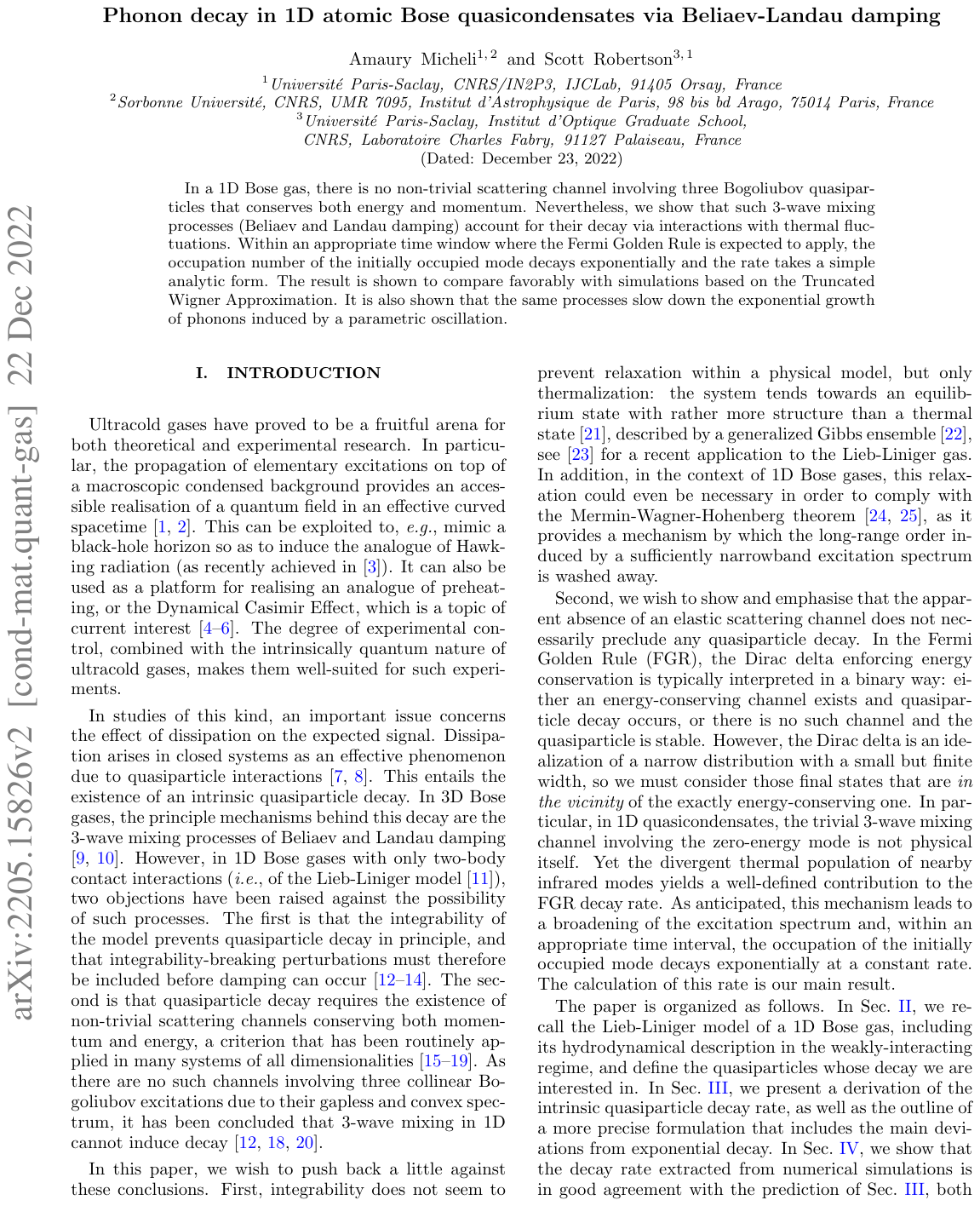}

\section{Simulating 1D Bose gas using TWA}
\label{sec:TWA_simulation_1D_BEC}

\begin{figure}
	\centering
	\includegraphics[width=1.\textwidth]{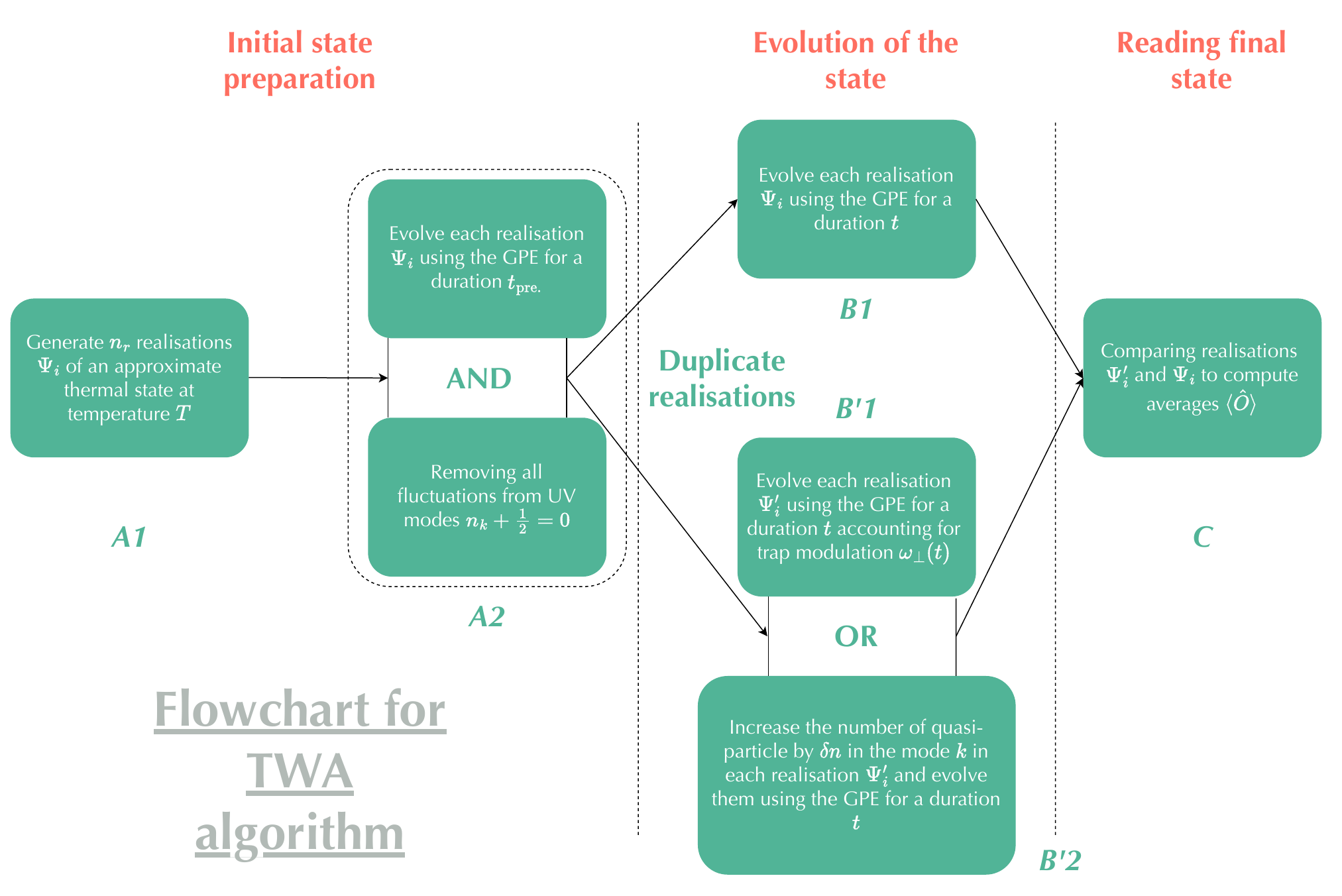}
	\caption{Steps of the TWA algorithm used in~\cite{Micheli:2022zet}. }
	\label{fig:flow_chart}
\end{figure}

The numerical simulations we performed are based on the Truncated Wigner Approximation (TWA)~\cite{steelDynamicalQuantumNoise1998,sinatraTruncatedWignerMethod2002}, the general procedure of which is explained in Appendix C of~\cite{Micheli:2022zet}, reproduced in the previous section.
TWA allows simulating the gas's evolution beyond the BdG approximation where quasi-particles are considered free and there is no backreaction on the gas's condensed part. The strategy is to treat the atomic field $\hat{\Psi}$ as a stochastic variable $\Psi$. At the initial time, we draw realisations of the atomic field  $\Psi_i$, which we evolve using the full Heisenberg equation of motion, corresponding exactly to the GPE for these classical fields. 
Notice that, in the TWA, the non-classical aspects are confined to the stochasticity of the initial state, while the dynamics is entirely classical.
The TWA, therefore, misses part of the quantum dynamics~\cite{Polkovnikov:2009ys}. Still, in~\cite{Micheli:2022zet}, it completely captured the scattering processes we were studying.
Due to the stochasticity of the initial state, the TWA differs from the resolution of the GPE describing the mean atomic field. The procedure we just described can be justified by a truncation of the equation of motion for the full Wigner function, see Appendix C of~\cite{Micheli:2022zet}, and exactly solves the quantum dynamics for a Gaussian state evolved via a quadratic Hamiltonian.
Our goal is to use the TWA to simulate the evolution of a gas initially in thermal equilibrium, which
undergoes a change in interaction constant $g$ leading to the creation of quasi-particles, as in~\cite{Jaskula:2012ab}, or to follow the evolution of a thermal state in which we initially add a few quasi-particles in some modes. These simulations allow us to probe how the gas reacts to the introduction of quasi-particles in a weakly non-linear regime and to follow the propagation of these perturbations to other modes. We summarise the main steps of the algorithm in Fig.~\ref{fig:flow_chart}. We give more details about each of these steps in the dedicated sub-sections below.

\subsection{A - Initial state preparation }

\subsubsection{Space discretisation}
\label{sec:space_discretisation}

We assume that our gas is contained in a one-dimensional box of size $L$. Additionally, to represent the system, we need to discretise space with a space step $\dd x$, and we have $n_x$ sites such that $L = n_x \times \dd x$. This discretisation of real space induces a UV cut-off in Fourier space \ie we do not encode the physics of scales smaller than $|k| \geq \pi  / \dd x$. 
In our case, this UV cut-off is not a limitation since we are mainly interested in the behaviour of long wavelengths. 
As detailed in Sec.~\ref{sec:quasi_condensate}, for the discretisation to still capture the relevant physics, we have to make sure that the space step is small enough \ie $\dd x \ll \xi$ and $\dd x \ll \lambda_{\mathrm{T}} $. We typically work at the limit of validity of these conditions where $\dd x \sim \lambda_{\mathrm{T}}$ and $\dd x \sim \xi / 3$. Decreasing the steps did not noticeably improved the results and require more computation time to run the simulations.
We also have to make sure that the space step is large enough so that the total number of atoms $\rho_0\dd x \gg 1$ in each site is large.
For the smallest density values used in~\cite{Micheli:2022zet} we have roughly $\rho_0\dd x = 10$ atoms per site so that the condition is always satisfied. Combining these two series of conditions gives a range in which we can choose $\dd x$. It thus puts a limit on the number of UV modes we can include in the simulation for a given value of the size of the system $L$ and of the density $n_0$.

The condition of a large occupation in each site is a requirement to build an approximate phase operator and for the density fluctuation to remain small~\cite{moraExtensionBogoliubovTheory2003}, see Eq.~\eqref{eq:relative_density_fluct_discrete}.
This requirement can be understood easily in the case of TWA simulations. The density fluctuations are not \textit{a priori} a normal-ordered combination, and in any state, it contains a sum of commutators encoding the vacuum fluctuations of each mode. 
In principle, this number is infinite, and we have to regularise it by normal ordering.
In TWA simulations, each mode comes with a certain level of vacuum fluctuations included by a stochastic noise, see Sec.~\ref{sec:TWA_thermal_state} below, and these will contribute to every observable.
If we include too many modes, the density fluctuations will grow out of control. 
These vacuum fluctuations are also effectively encoded via a non-zero value of the classical field as any other mode excitation. Therefore, they are not protected against scattering processes affecting the excitation of the system \ie the `vacuum fluctuations' can decay in a TWA code. These processes can lead to a negative occupation number of the highest UV modes and are thus another reason not to include too many UV modes~\cite{vanregemortelSpontaneousBeliaevLandauScattering2017}.

\subsubsection{A1 - Thermal state}
\label{sec:TWA_thermal_state}

We want to use the thermal state of the system as an initial state to mimic the setting of~\cite{Jaskula:2012ab}.
Initialising the atomic field in an actual thermal state of the system would require knowing the exact energy spectrum and the associated eigenfunctions of the system. In general, this is impossible. We instead start from an approximate thermal state using the energy spectrum computed from the quadratic order Hamiltonian $\hat{H}_2$ of Eq.~\eqref{eq:quadratic_H_pert_BdG_phonons}. Notice that at this stage, using BdG, or Madelung, perturbation theory gives the same energy spectrum. Nevertheless, the quasi-particles are related to the atomic field differently in these two schemes. They thus define different approximate thermal states for the system.
As expected, the phase and density approach leads to a more stationary and better-behaved state. We come back to this point in Sec.~\ref{sec:madelung_vs_Bdg}.

A thermal state for the quadratic Hamiltonian is a number state $| \{ n_k^{\mathrm{th}} \} \rangle $ where $n_k^{\mathrm{th}} $ is the thermal state given in Eq.~\eqref{def:nk_thermal}, and $\omega_k$ is the BdG quasi-particles dispersion relation~\eqref{def:BdG_dispersion}.
In the stochastic treatment the operators $\hat{b}_k$ are treated as stochastic c-numbers $b_k$. The probability distribution attached to them is given by the Wigner function of the system's state.
The thermal state is a Gaussian state meaning that its Wigner function is a Gaussian and is entirely characterised by its covariance matrix as detailed in~\cite{Micheli:2022tld} reproduced in Sec.~\ref{sec:quantum_GW}. By analogy with a harmonic oscillator, we can define pseudo-position $\hat{q}_k$, and pseudo-momentum operator $\hat{p}_k$, for each quasi-particle mode related to the hermitian and anti-hermitian part of the operator similar to Eq.~(60) of~\cite{Micheli:2022tld}. This gives us a phase-space representation of the state.
We can then use the stochastic phase-space variables $q_k$ and $p_k$. Their statistics are given by the Wigner function of the modes $\pm k$ \ie a two-dimensional Gaussian probability distribution parameterised by the covariance matrix of the quantum state. For a thermal state, $q_k$ and $p_k$ are independent and they have equal variance $\langle \hat{q}^2_k \rangle = \langle \hat{p}^2_k \rangle = n_k^{\mathrm{th}} +1/2$ see Eq.~(C.3) of~\cite{Micheli:2022zet}. In general, the stochastic phase-space variables satisfy~\cite{gardinerQuantumNoiseHandbook2004}.
\begin{equation}
	\label{eq:Wigner_correspondence}
	\left \langle f \left( q_k , p_k \right) \right \rangle_{W} = \left  \langle f_{\mathrm{sym}} \left( \hat{q}_k , \hat{p}_k \right) \right \rangle \, ,
\end{equation}
where the left-hand side is an average performed with the Wigner function, $f$ is a polynomial and $f_{\mathrm{sym}}$ is a writing of the polynomial totally symmetric in its arguments, e.g. for $f(q,p) = q^2p^2$ we have $f_{\mathrm{sym}}(\hat{q},\hat{p}) = ( \hat{q}^2 \hat{p}^2 + \hat{q} \hat{p}^2 \hat{q} + \hat{q} \hat{p} \hat{q} \hat{p} + \hat{p} \hat{q} \hat{p} \hat{q} + \hat{p} \hat{q}^2 \hat{p} + \hat{p}^2 \hat{q}^2)/6$. 
The left-hand side average in Eq.~\eqref{eq:Wigner_correspondence} can be approximated by drawing a large number $n_r$ of independent realisations of the stochastic phase-space variables $\{ q_{k, i} , p_{k , i} \}_{i \in [1,n_r]}$. Typically we use $n_r = 500$ in the simulations of~\cite{Micheli:2022zet}.
Going back to phonon numbers we have in particular $ \langle b_k^{\star} b_k \rangle_{W} =  n_k^{\mathrm{th}} + 1/2$ see Eqs.~(74)-(75) in Appendix C of~\cite{Micheli:2022zet}. The $1/2$ present in each mode encodes the vacuum fluctuations. To read out the average number of particles in each mode, we must subtract this half. Note then that the total decay of the excitation in a mode $k$, leading to a vanishing value for $b_{k , i}$, would indeed imply a \textit{negative} number of quasi-particles in the mode as explained in Sec.~\ref{sec:space_discretisation}.

Using this stochastic picture based on the Wigner function, we can exactly represent the initial thermal state of quasi-particles.
We only include a finite number of modes in our simulation, but the requirement $\dd x \ll \lambda_{\mathrm{T}} $ ensures that we sample the thermally occupied ones.
Finally, from each realisation of the series $\{ b_{k_j} \}_{j \in [- \pi n_x / L , \pi n_x / L ] \textbackslash \{0\} }$, where $k_j = 2 \pi j / L$, we build a realisation of the atomic field by taking for each point $x_l$ on the grid~\cite{ruostekoskiTruncatedWignerMethod2012}
\begin{subequations}
\begin{align}
\begin{split}
	\delta \rho_i \left( x_l \right) & =  \sqrt{\frac{\rho_0}{L}} \sum_{j \in \left[- \frac{\pi n_x}{L}  , \frac{\pi n_x}{L} \right] \textbackslash \{0 \} } e^{i k_j x_l} \left( u_{k_j} + v_{k_j} \right) \left( b_{k_j , i} + b^{\star}_{k_j , i} \right) \, ,
\end{split} \\
\begin{split}
\delta \theta_i \left( x_l \right) & = \frac{1}{2} \sum_{j \in \left[- \frac{\pi n_x}{L}  , \frac{\pi n_x}{L} \right] \textbackslash \{0 \} } e^{i k_j x_l} \left( u_{k_j} - v_{k_j} \right) \left( b_{k_j , i} - b^{\star}_{k_j , i} \right) \, ,
\end{split} \\
	\begin{split}
		\label{def:realisation_atomic_field_Madelung}
		\Psi_i \left( x_l \right) & =  \sqrt{\rho_0 + \delta \rho_i \left( x_l \right) }  \exp \left[ \delta \theta_i \left( x_l \right) \right] \, .
	\end{split}
\end{align}
\end{subequations}
Notice that with the above, the average density of atoms in each realisation is fixed by construction to $\rho_0$. Eq.~\eqref{def:realisation_atomic_field_Madelung} also shows that the formalism cannot accommodate arbitrarily large density fluctuations, or the number under the square root might become negative.
First, this requires that space discretisation is chosen as described in Sec.~\ref{sec:space_discretisation} to limit the vacuum fluctuations.
Second, we have to consider a physical situation where the gas we model is in the quasi-condensate regime so that the normal-ordered relative density fluctuations remain small~\cite{moraExtensionBogoliubovTheory2003,petrovLowdimensionalTrappedGases2004}. This condition requires a small enough temperature and a large enough density. For the values of parameters used in~\cite{Micheli:2022zet}, the fluctuations are, at maximum, typically an order of magnitude smaller than the background density.

\subsubsection{Thermal state in BdG}

The reconstruction Eq.~\eqref{def:realisation_atomic_field_Madelung} is based on the density and phase perturbation theory that, we have argued in Sec.~\ref{sec:Madelung}, is the most appropriate to use for 1D gas.
In a previous paper~\cite{Robertson:2018gwi} on the analysis of~\cite{Jaskula:2012ab}, the authors had used TWA simulations of the gas but relied on the BdG approximation instead. 
The thermal state of quasi-particles was similarly sampled, but the realisations of the initial state of the atomic field were then built using 
\begin{align}
	\begin{split}
		\label{def:realisation_atomic_field_BdG}
		\Psi_i \left( x_l \right) & =  \frac{n_0}{\sqrt{L}} \sum_{j \in \left[- \frac{\pi n_x}{L}  , \frac{\pi n_x}{L} \right] \textbackslash \{0 \} } e^{i k_j x_l} \left( u_{k_j} b_{k_j , i} + v_{k_j} b^{\star}_{k_j  , i} \right) \, .
	\end{split}
\end{align}
For a given temperature value $T$, interaction constant $g_1$ and background condensed density $n_0$, the perturbation $\delta \Psi$ gives a non-vanishing density of depleted atoms. The state's average density of depleted atoms $\delta n$ is computed from the $n_r$ realisations. Therefore, with the prescription of Eq.~\eqref{def:realisation_atomic_field_BdG}, the average total number of atoms in the system will fluctuate for different values of the parameters. 
In~\cite{Robertson:2018gwi}, in order to keep the total number of atoms fixed in average, irrespective of the parameters chosen, the density of condensed atoms is corrected to $n_0 - \delta n$ and the full field taken to be\begin{equation}
	\Psi_i \left( x_l \right) =  \sqrt{ \frac{n_0 - \delta n}{n_0}}   \delta \Psi_i \left( x_l \right) \, .
\end{equation}
It is clear that this procedure is not entirely consistent with the BdG procedure since we first take the condensate density to be $n_0$ to draw the realisations of the phononic operators $b_{k  , i}$, but we then modify the condensate density to $n_0 - \delta_n$ to build the field.
In addition the procedure can quickly run into problems if the density of depleted atoms $\delta n$ grows too much, as it typically does in one-dimension, see Sec.~\ref{sec:Madelung}.
However, even the safer prescription of Eq.~\eqref{def:realisation_atomic_field_BdG} leads to divergences in the density-density correlation function and to a poorer approximation of the actual thermal state of the gas compared to 
Eq.~\eqref{def:realisation_atomic_field_Madelung}.
To illustrate these aspects, we compared in Sec.~\ref{sec:madelung_vs_Bdg} the evolution of the system's Gaussian entropy for the two constructions when the system is left to evolve without external action.

\subsubsection{A2 - Modifying initial state}
\label{sec:modif_initial_state}

We can apply a couple of modifications to the initial state we have constructed to get around some of the limitations of the simulation.
First, the thermal state of the quasi-particles we use as an initial state is not strictly stationary. Therefore, we expect a certain degree of state evolution even without external perturbation. We can allow this initial state to evolve for some time to limit that effect, hoping it will get closer to a stationary state before perturbing it~\cite{Robertson:2018gwi}. 
%Notice that we should not wait too long because, on long times, the TWA is known~\cite{vanregemortelSpontaneousBeliaevLandauScattering2017} to drive the state towards a \textit{classical} Rayleigh-Jeans distribution of phonons $\hbar \omega_k (n_k +1/2)\sim k_{\mathrm{B}} T $, rather than the correct Bose-Einstein distribution for quantum degrees of freedom. This relaxation will lead to negative occupations for most UV modes, see below.
However, when the initial state is constructed using Eq.~\eqref{def:realisation_atomic_field_Madelung} this initial evolution did not dramatically change the system's response when introducing quasi-particles directly or by a modulation.

Second, we have already mentioned that only a few UV modes should be present in the simulation. The vacuum fluctuations of these modes can decay, leading to \textit{negative} occupation, and even if we do not probe these UV modes, their decay products can be located nearby IR modes of interest for our simulations.
We allow ourselves to remove all fluctuations in the modes with $k$ larger than a certain $k_{\mathrm{CO}}$ to avoid this pollution of the IR modes. How do we pick this cut-off wavenumber?

\begin{figure}
	\centering
	\includegraphics[width=0.9\textwidth]{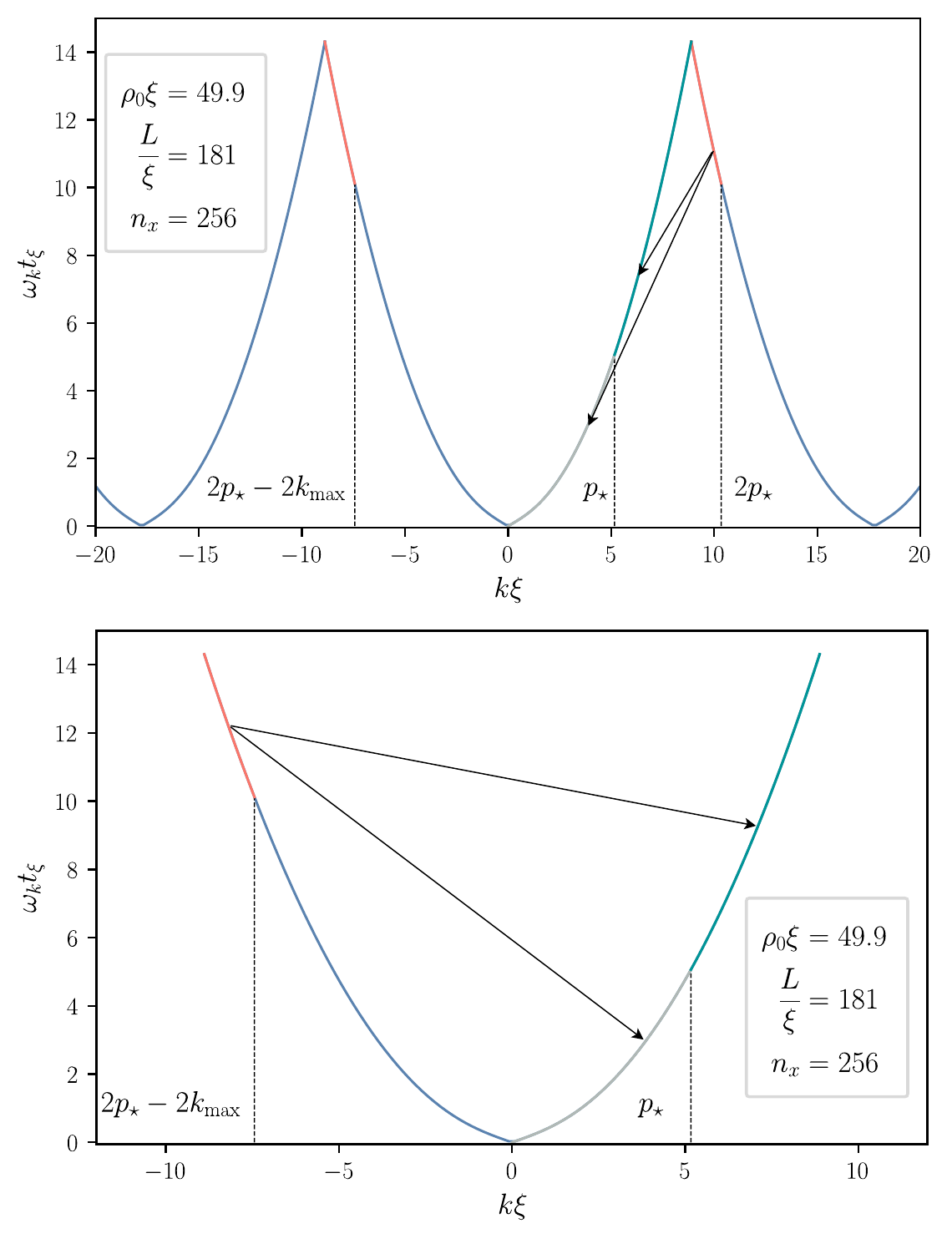}
	\caption{(Top panel) Periodic Bogoliubov-de Gennes dispersion relation given in Eq.~\eqref{def:BdG_dispersion} for a grid of size $n_x=256$. The red region for $k_{\mathrm{max}} < k < 2 p _{\star} $ shows the UV modes decaying towards the green and grey regions, indicated by the two arrows. The second red region for  $- k_{\mathrm{max}} < k < 2 p _{\star} - 2 k_{\mathrm{max}} $ corresponds to the first one translated by $ - 2 k_{\mathrm{max}} $. }
	\label{fig:dispersion_relation_spurious_decay}
\end{figure}

In the preparation of~\cite{Micheli:2022zet}, we identified spurious Beliaev-Landay decay channels for the UV modes. These will be the topic of a future publication, but a brief computation is sufficient to understand their origin.
The energy and momentum conservation for 3-body scattering, such as Beliaev-Landau processes, read
\begin{subequations}
	\begin{align}
		\begin{split}
			\label{eq:momentum_conservation_BL}
			p + q &= k \, ,
		\end{split}\\
		\begin{split}
			\omega_{p} + \omega_{q} &= \omega_{k} \, .
			\label{eq:energy_conservation_BL}
		\end{split}
	\end{align}    
\end{subequations}
The momentum-conservation condition of Eq.~\eqref{eq:momentum_conservation_BL} arises when looking at the third-order Hamiltonian in Fourier space. It comes from the requirement that $e^{i \dd x (p+q - k)}=1$. Since we work with a discrete space, the space step is finite $\dd x = \pi / k_{\mathrm{max}}$. The conservation condition is then actually generalised to $p+q - k = m \times 2 k_{\rm max}$, where $m$ can be any integer \ie Fourier space is also effectively periodic of period $2 k_{\rm max}$. A quick analysis shows that for $(p,q)\in [-k_{\rm max},k_{\rm max}]$ the momentum conservation condition can be satisfied for $m=0$, the usual condition, but also for $m= \pm 1$, leading to additional spurious channels. 
The dispersion relation should now be extended to a periodic one $\omega_{k + 2 m k_{\rm max}} =\omega_k$, see Fig.~\ref{fig:dispersion_relation_spurious_decay}.

Let us now consider the energy conservation for fixed $p>0$. When $p$, $q$ and $p+q$ lie within $[-k_{\rm max},k_{\rm max}]$ the conservation is only satisfied for $q=0$. However, when $p+q > k_{\rm max}$, one can find a solution with non-vanishing $q$. In particular choosing $p=q$ we can always find a solution $p_{\star}> k_{\rm max}/2$ to Eq.~\eqref{eq:energy_conservation_BL}. Since $2 p_{\star}> k_{\rm max}$, if we restrict attention to modes within $[-k_{\rm max},k_{\rm max}]$, $2 p_{\star}$ is mapped to $2 p_{\star} - 2 k_{\rm max}$. The scattering process thus appears as the combination of two quasi-particles with positive momentum $p_{\star}$ to create one with negative, typically UV, momentum $2 p_{\star} - 2 k_{\rm max}$, see Fig.~\ref{fig:dispersion_relation_spurious_decay}. The reverse process also happens, giving a channel for UV modes to decay towards IR modes. The spurious processes will also happen between a quasi-particle with $p > p_{\star}$ combining with one with $q< p_{\star}$. Therefore,
to completely suppress the direct and reverse processes, one can pick the cut-off momentum to be $k_{\mathrm{CO}} = p_{\star}$.
We used this choice in~\cite{Micheli:2022zet} to avoid the decay of quasi-particles injected in the gas by spurious Beliaev-Landau channels, which would deteriorate quantitative comparison with our predictions for the physical channel.

\subsection{B - Evolution}

One can map the Heisenberg evolution equation to an exact evolution equation for the Wigner function. In the TWA, this equation is truncated at its lowest non-trivial order, see Sec.~\ref{sec:decay_nk}, and the resulting equation is solved by evolving the realisation of the atomic field $\Psi_i$ using a space and time discrete version GPE Eq.~\eqref{eq:3D_GPE}.
While we use the density and phase perturbation picture to build the initial realisations, evolving them using the equation of motion over the atomic field given in Eq.~\eqref{eq:EOM_3D_Heinsenberg} proves more convenient. In the absence of external potential $V_{\mathrm{ext}}$, the latter only has two terms which are easy to integrate using a split-step Fourier algorithm.

\subsubsection{Split-step Fourier evolution}
\label{sec:split_step}

We start by describing how we relate a realisation of the atomic field made of the numbers $\Psi_i (x_l ; t)$ to the realisation at time $\Psi_i (x_l ; t+\dd t)$, where $\dd t$ is our time-step.
Since we neglect the external potential $V_{\mathrm{ext}}$, there are two pieces to evaluate on the right-hand side of the GPE to determine the time-derivative at $t$: the kinetic and self-interaction terms.
First, consider the interaction term. It is local in real space and depends only on $\Psi_i(x_l)$. Assuming that any change of $g(t)$ happens over time scales much larger than $\dd t$, and forgetting about the kinetic term, the evolution would be
\begin{equation}
	\label{eq:interaction_term_evo}
	\Psi_i \left(x_l ; t+\dd t \right) = e^{- \frac{i}{\hbar} g(t) \left| \Psi_i \left(x_l ; t \right) \right|^2 \dd t} \Psi_i \left(x_l ; t \right) \, .
\end{equation}
Notice that this evolution takes into account the evolution of the stationary background part $e^{- i g \rho_0\dd t /\hbar }$ since $| \Psi_i |^2$ contains the average density $\rho_0$.
Second, consider the kinetic term. It is not local in real space and would require to use finite differences to be evaluated at $x_l$, which involves the field at nearby points $x_l \pm m \times \dd x$, with $m$ an integer. However, working in Fourier space, this term is `local' because its expression for $\Psi_i (k_j)$ only depends on $\Psi_i (k_j)$. If we temporarily drop the interaction term the evolution can then be solved in Fourier space by
\begin{equation}
	\label{eq:kinetic_term_evo}
	\Psi_i \left(k_j ; t+\dd t \right) = e^{- \frac{i}{\hbar} \frac{\hbar^2 k_j^2}{2 m} \dd t} \Psi_i \left(k_j ; t \right) \, .
\end{equation}
The split-step Fourier algorithm~\cite{agrawalNonlinearFiberOptics2013} we use to evolve the realisations makes use of both these simplifications. We first evaluate the evolution due to the interaction term~\eqref{eq:interaction_term_evo} for half a step $\dd t/2$. We then do a discrete Fourier transform of the resulting $\Psi_i$ and perform the evolution due to the kinetic term for a full step $\dd t$~\eqref{eq:kinetic_term_evo}. Finally, we go back to real space and perform another evolution of half a step for the interaction term $\dd t/2$.
The error of the resulting evolution is expected to be of the third order in the time-step~\cite{agrawalNonlinearFiberOptics2013}.

\subsubsection{Time discretisation}

The time-step $\dd t$ cannot be chosen independently of the space-step $\dd x$. We have to make sure that the highest frequency modes $\omega_{\pm k_{\mathrm{max}}}$ do not evolve too quickly between two time-steps so that we correctly describe their evolution. Approximating $\omega_{\pm k_{\mathrm{max}}}$ as being in the free particle regime we have
\begin{equation}
	\label{eq:time_step_condition}
	\omega_{\pm k_{\mathrm{max}}} \dd t \ll 2 \pi \iff    \frac{\hbar}{m} \frac{\pi \dd t}{4 \dd x^2} \ll 1 \, .
\end{equation}
We have observed that choosing a too large time-step not only leads to a poor description of large frequency modes but can also lead to dramatic resonant behaviour around $\omega_k \approx \pi/ \dd t $, where we witness a spontaneous exponential growth of the number of quasi-particles. The condition \eqref{eq:time_step_condition} ensures that this latter frequency is not present in the system.

\subsubsection{Duplicating realisations}
We want to observe the evolution of a perturbation on top of the initial thermal state, particularly how initial quasi-particles injected in the system in a given mode $k$ redistribute.
In order to isolate precisely the effect of introducing the perturbation, we can compare the evolution from the same initial state
with and without the external perturbation. Recall that our initial thermal state is not exactly stationary, so we expect a small degree of evolution even without perturbation.
The algorithm we use is stochastic in drawing initial realisations of the atomic field but deterministic in its evolution. If we draw $n_r$ realisations $\Psi_i$, compute the evolution without perturbation, then draw $n_r$ more $\Psi_i^{\prime}$, compute the evolution with perturbation and compare $\Psi_i(t_{\mathrm{fin.}})$ and $\Psi_i^{\prime}(t_{\mathrm{fin.}})$ there will always be additional difference owing to the different initial states. A large number of realisation $n_r$ already suppresses these differences. However, we can make the comparison more efficient by duplicating the first $n_r$ realisations $\Psi_i$. We then use one realisation for the evolution with perturbation, corresponding to B'2 or B'1 in Fig.~\ref{fig:flow_chart}, and the other without perturbation,  B in  Fig.~\ref{fig:flow_chart}. This trick allows us to have clean numerical results where we can track precisely the (dis)appearance of quasi-particles in the mode $k$ and their redistribution to other modes even if the transfers are small, see Fig.13 of~\cite{Micheli:2022zet}.

\subsubsection{B'1 - Modulation of transverse trapping frequency }

Modelling the effect of time-dependent trapping frequency $\omega_{\perp}(t)$ is straightforward with the reduction we have described in Sec.~\ref{sec:dynamics_gas}: it is modelled by a time-dependent 1D interaction constant $g(t)$. Although the link between the evolution of the interaction constant and that of the trapping frequency is complicated, see Eq.~\eqref{eq:EOM_sigma}. In~\cite{Micheli:2022zet}, we made the simplifying assumption that modulating the trapping frequency at $\omega_{\mathrm{m}}$ results directly in a modulation of $g$ at the same frequency. This modulation is simply implemented by changing the value of $g(t)$ in the integration of the kinetic term~\eqref{eq:kinetic_term_evo}.

\subsubsection{B'2 - Injection of quasi-particles in a mode $k$}
\label{sec:injection_quasiparticles}

Another series of simulations performed in~\cite{Micheli:2022zet} consists in adding quasi-particles, on top of the thermal population, in a given mode $k$. To do so, we modified the statistics of the state of this mode $k$. In~\cite{Micheli:2022zet}, we added $\delta n$ quasi-particles in the mode $k$ by enhancing the values of the stochastic numbers $b_{k,i}$ that were drawn
\begin{equation}
	b_{k,i}^{\prime} = b_{k,i} \sqrt{1 + \frac{\delta n}{n_k^{\mathrm{th}}  + 0.5} } \, .
\end{equation}
It is straightforward to check that the resulting $b_{k,i}^{\prime}$ still follow a thermal distribution. Nevertheless, the average number of quasi-particles has been enhanced to $n_k^{\mathrm{th}} + \delta n$. A thermal state corresponds to a completely incoherent distribution, so we have added excitations \textit{incoherently} \ie $\langle b_k^2 \rangle_W =0$ before and after the addition. 
This incoherence would not have been achieved if we had added $\sqrt{\delta n}$ to each realisation $b_{k,i} \to b_{k,i} +\sqrt{\delta n}$. The price to pay for this is that the number of injected quasi-particles is also stochastic and only $\delta n$ on average.

Note that the result of parametric amplification of the initial thermal distribution considered in Sec.~\ref{sec:parametric_amplification} would lead to a different state. First, it generates quasi-particles in at least \textit{two} resonant modes $\pm k$. Second, the quasi-particles generated in the two modes are correlated: $c_k = \langle b_k b_{-k} \rangle_W \neq 0$ after the amplification, see Sec.~\ref{sec:dynamics_external_drive}. 
The evolution of the correlation $c_k$ is crucial to understand how entanglement can be lost in~\cite{Jaskula:2012ab}. Preliminary results on this point are presented in Sec.~\ref{sec:decay_ck}.
TWA simulations also guided these results. We injected quasi-particles \textit{coherently} in the initial state by replacing the initial thermal state of the modes $\pm k$ with a Two-Mode Squeezed Vacuum (TMSV) state, see~\cite{Martin:2022kph} definitions. This injection is a proxy for the effect of a parametric amplification on the modes $\pm k$. It is more computationally effective than solving the dynamics in the presence of modulation and also much more straightforward as it leaves the other modes $k^{\prime} \neq k$ unaffected.
A TMSV is a Gaussian state, and we can also faithfully represent it using stochastic variables drawn from a Gaussian probability distribution. The major difference with the initial state is that we deal with a four-dimensional probability distribution for $q_{\pm k}$ and $p_{\pm k}$. An easy way to build realisations of these stochastic numbers is to transform the TMSV into two independent equally distributed One-Mode Squeezed Vacuum (OMSV) state for modes described by $(\hat{q}_{\mathrm{R}/\mathrm{I}},\hat{p}_{\mathrm{R}/\mathrm{I}})$, see Sec.~\ref{sec:comparing_quantumness_criteria}. We draw realisations for the numbers $q_{\mathrm{R}/\mathrm{I}}$ and $p_{\mathrm{R}/\mathrm{I}})$ which are then linearly related to $q_{\pm k}$ and $p_{\pm k}$.
Note that the state after the two-mode squeezing of modes in a thermal state is not strictly speaking a TMSV but a two-mode squeezed thermal state because it is a mixed state, see Sec.~\ref{sec:comparing_quantumness_criteria}. This correction could be implemented in a future version of the code by replacing the OMSV of the fictitious modes with a one-mode squeezed thermal state whose degree of mixedness is fixed by the initial thermal population.

\subsubsection{Independent time-step}

In our simulations, we compute approximate values for quantum operators using its correspondence to stochastic averages computed using the Wigner function, see Eq.~\eqref{eq:Wigner_correspondence}. 
We approximate the stochastic averages by performing averages over a finite number of independent realisations $n_r$. We estimate the error in our approximated average by computing the variance of this quantity in our $n_r$ realisations. The standard deviation then gives the error bars we show in our plots in~\cite{Micheli:2022zet}, e.g. in Fig.~1.

Say we want to follow a quantity in time, for instance, the number of phonons $n_k$ in a mode $k$ and want to compare the evolution of this quantity with our prediction, for instance, an exponential decay of the population $n_k \propto e^{- \Gamma_k t}$. Then we will fit the data points at different times using a curve matching the predicted form with a few fitting parameters. 
The fitting procedure will give an estimated error on the fitting parameters from the error bars given at each point $t_i$.
However, correctly estimating the errors requires knowing how the points at different times $t$ and $t + \dd t$ are correlated, which requires a detailed study for each quantity.
To use the usual fitting procedures, which assume that the data points are independent of each other, we re-draw $n_r$ different realisations and repeat the complete evolution until $t_i$ to evaluate the quantity at this time. It ensures that the data at each time point are independent.

\subsection{C - Reading final state}

The primary quantity that we compute from the realisations of the atomic field $\Psi_{i} (x_l)$ are the values of the phononic creation/annihilation operators $b_{k,i}$. Computing them requires knowing the density and the phase field at every point $x_l$. While computing the density field from the atomic field is straightforward $\rho (x_l) =  |\Psi_{i} (x_l)|^2$, it is not so for the phase field. Indeed, the atomic field only retains the value of the phase up to $2 \pi$. A straightforward reading of the argument of the phase field can then lead to a large jump of the phase field from one point to another, while we need a continuous field to build the quasi-particles. For this reason, we construct, in parallel to the atomic field's values, the phase field's value.
To avoid large phase jumps, we proceed iteratively. First, in the initial state, we set the phase at the site $x_l=0$ to the value of the argument of the atomic field there. Then, assuming that the phase difference between two neighbouring sites $x_l$ and $x_l + \dd x$ is less than $\pi$, we compute the argument $ \Psi (x_l + \dd x)/\Psi (x_l)$ and set $\theta (x_l + \dd x) = \theta (x_l) + \mathrm{arg} [\Psi (x_l + \dd x)/\Psi (x_l)]$.
Having constructed a phase field devoid of large jumps for the initial state, we can repeat the same method for the time evolution.
At any point $t + \dd t$ we set the phase of the field to be $\theta (x_l ; t + \dd t) = \theta (x_l ; t) +  \mathrm{arg} [\Psi (x_l ; t + \dd t)/\Psi (x_l; t)]$.
With this reconstructed phase field, we then build the stochastic number $b_k$ by inverting the transformation 
of Eqs.~(\ref{eq:density_pert_BdG_Fourier})-(\ref{eq:phase_pert_BdG_Fourier}).
From this number, we can compute the average values of the quantities studied in~\cite{Micheli:2022zet}, all constructed from a combination of these operators.

\subsubsection{Following the external perturbation}

Once both the realisations with $\Psi_{i}^{\prime} (x_l)$ and without perturbation $\Psi_{i} (x_l)$ are obtained, we can read out the effect of the perturbation by taking the difference of the two before averaging $\delta \Psi_{i} = \Psi_{i}^{\prime} - \Psi_{i}$, rather than after. The intuition behind this can be formulated as follows. The result of the unperturbed evolution is only a function of the initial background state $\Psi_{i}(t) = f[\Psi_{i}(0);t]$. This piece can be removed from the evolution of the perturbed evolution
\begin{equation}
	\Psi_{i}^{\prime} \left( t \right) = \Psi_{i} \left( t \right) + \delta \left[ \Psi_{i} \left( 0 \right) , \delta \Psi_{i} \left( 0 \right) ;t\right] \, ,
\end{equation}
where due to the non-linear evolution the perturbation $\delta \Psi_{i}$ at time $t$ depends on both its initial value $\delta \Psi_{i} (0)$ and that of the background $\Psi_{i} ( 0 )$. Therefore $\Psi_{i}(t)$ and $\delta \Psi_{i}(t)$ could be correlated $\langle \Psi_{i}(t) \delta \Psi_{i}(t) \rangle \neq 0$ even if initially incoherent $\langle \Psi_{i}(0) \delta \Psi_{i}(0) \rangle = 0$. We are not interested in this correlation between the background and perturbation but solely in the evolution of the perturbation due to the background. For instance, we may ask how many quasi-particles initially contained in $\delta \Psi_i(0)$ were lost at time $t$ by interaction with the background. To extract this number, we first compute $\delta \Psi_i(t)$ and then extract the number of quasi-particles via $\langle \delta \Psi^{\dagger}_i(t) \delta \Psi_i(t) \rangle$, rather than computing $\langle \Psi^{\dagger , \prime}_i(t) \Psi^{\prime}_i(t)  \rangle - \langle \Psi^{\dagger}_i(t) \Psi_i(t) \rangle = \langle \delta \Psi^{\dagger}_i(t) \delta \Psi_i(t) \rangle + \langle \delta \Psi^{\dagger}_i(t) \Psi_i(t) \rangle + \langle \delta \Psi_i(t) \Psi^{\dagger}_i(t) \rangle$.
The additional terms in the above expression need not vanish, and their magnitude \textit{a priori} depends on that of the background, which can be large. Therefore, they could be a noise source, and we favour the first expression where they are absent.
This comparison method is generically used in the plots of~\cite{Micheli:2022zet} unless otherwise specified.

\subsubsection{Madelung vs BdG}
\label{sec:madelung_vs_Bdg}

In this part, we illustrate the superiority of the Madelung perturbation scheme over the BdG one when performing numerical simulations. 
The distinction between the BdG and Madelung formalisms can be relevant at two stages. First, as already explained, when building the initial state: à la BdG as in Eq.~\eqref{def:realisation_atomic_field_BdG}, or using the Madelung scheme as in Eq.~\eqref{def:realisation_atomic_field_Madelung}. We will refer to the first as the BdG state and the second as the Madelung state.
Still, for a given value of the atomic field $\Psi_i$, we can use these expressions the other way around to define BdG quasi-particles and Madelung quasi-particles.
In the following, to be consistent, when we refer to quasi-particles of the BdG state, we mean BdG quasi-particles, and similarly, we only extract Madelung quasi-particles from the Madelung state.
We produced the data presented using a former version of the code that contained no error bars on the data points and where the time steps were not independent. However, the difference between the two schemes is clear enough without needing these refinements.
To compare the two states, we compare the evolution of several quantities without any external perturbation (initial injection of quasi-particles or modulation) computed in the BdG and the Madelung states. We use the same set of realisations to build the two initial states to ensure that they only differ in the method used to construct them.

\begin{figure}
	\centering
	\includegraphics[width=0.9\textwidth]{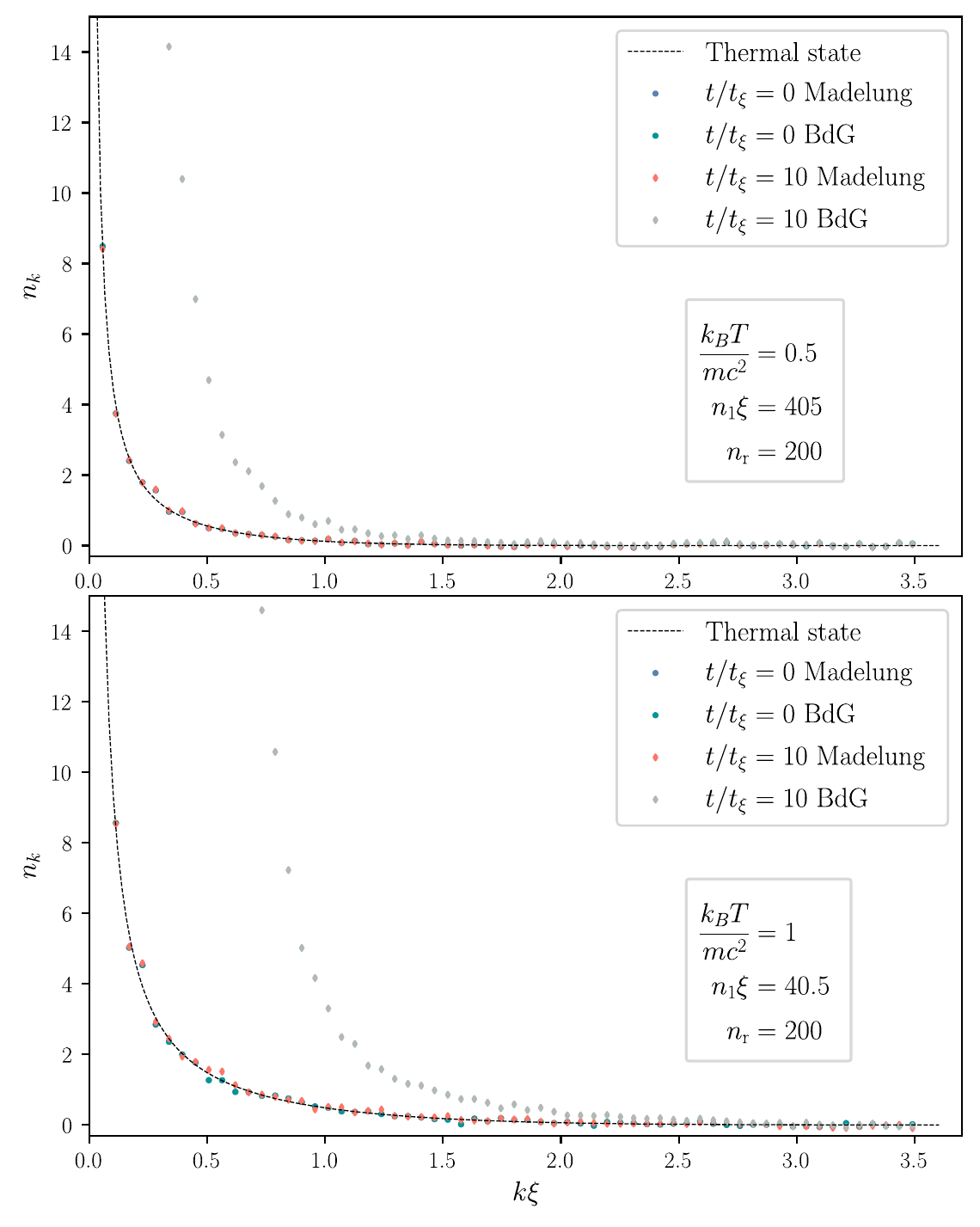}
	\caption{Average number of phonons~\eqref{def:nk} as a function of the wavenumber $k \xi$, for the Madelung and the BdG states, at two different times.  This number is compared to a  thermal distribution~\eqref{def:nk_thermal} of phonons associated with the temperature $T$ shown in dashed lines. The two panels differ by the values of the temperature of the initial phonon distribution $k_{\mathrm{B}}T/m c^2 =0.5$ for the upper panel,  $k_{\mathrm{B}}T/m c^2 =1$ for the lower panel, and the value of the density of the background $\rho_0\xi = 405$ for the upper, and $\rho_0\xi = 40.5$ for the lower panel. Recall that lower densities imply larger interactions.  }
	\label{fig:nbr_phonons_two_values_interaction}
\end{figure}

First, in Fig.~\ref{fig:nbr_phonons_two_values_interaction}, we compare the number of quasi-particles extracted from the BdG and Madelung states at two different times.
At the initial time, both states give a spectrum matching the prediction for an exact Bose-Einstein distribution at temperature $T$ given in Eq.~\eqref{def:nk_thermal}. This has to be the case by construction.
The Madelung state stays close to the prediction as a function of time for both series of values. On the other hand,
Fig.~\ref{fig:nbr_phonons_two_values_interaction} shows that, in the BdG state, $n_k$ suffers from a significant degree of non-stationarity. 
For larger interactions, lower panel, the non-stationarity of the BdG state becomes more dramatic and propagates towards more UV modes.
Note that additional snapshots would reveal oscillations rather than a continuous growth of these modes.
We point out that the non-stationarity could be partially cured at lower interaction/temperature by `reading' the BdG state with Madelung phonons, but the non-stationarity still appears at larger interaction/temperature.
The fact that the non-stationarity persists shows that it is not only due to how we extract the quasi-particles from the BdG state; the BdG state itself is plagued with non-stationarity.

\begin{figure}
	\centering
	\includegraphics[width=0.9\textwidth]{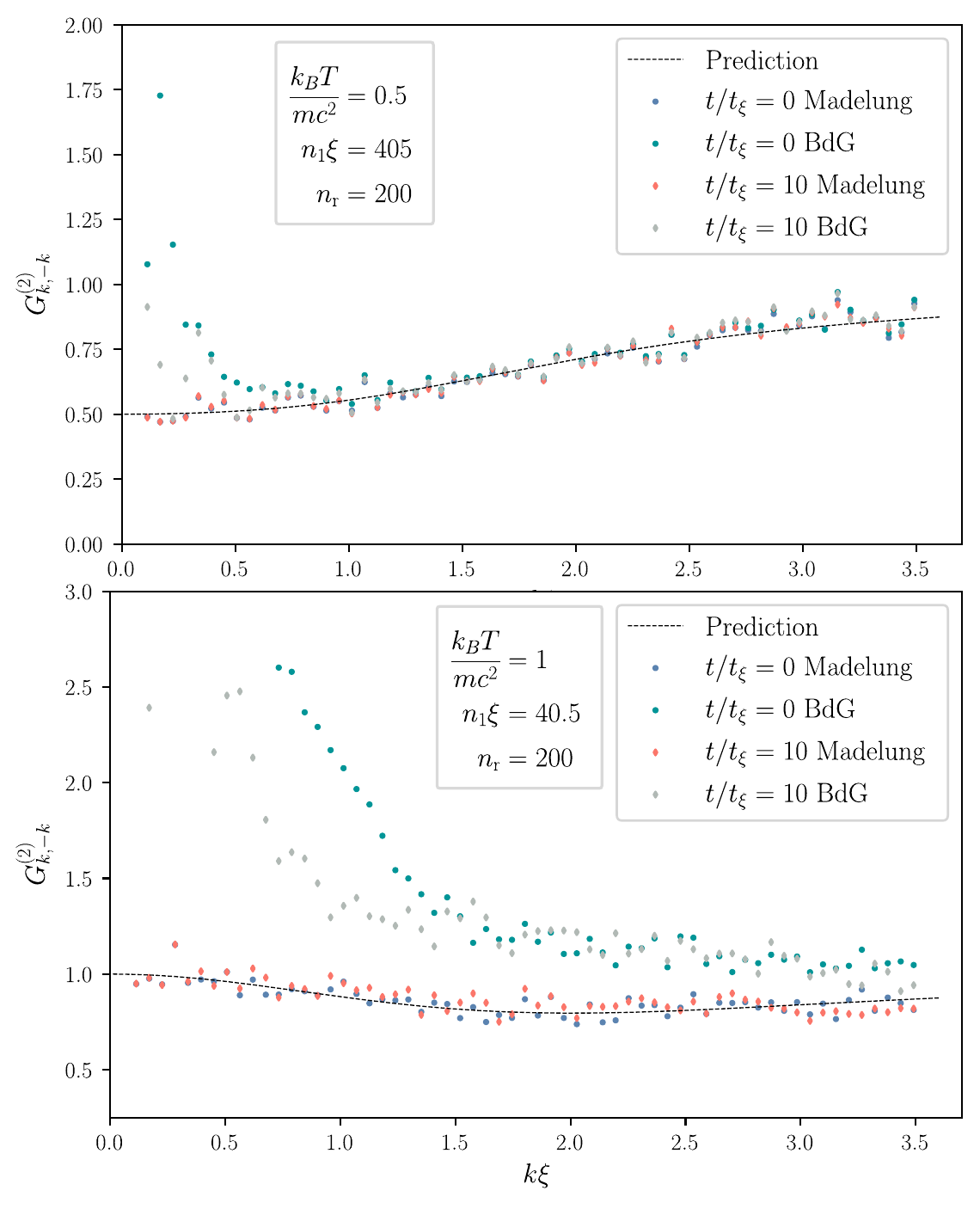}
	\caption{
	Density-density correlation~\eqref{eq:G2_k_phonons} as a function of the wavenumber $k \xi$, for the Madelung and the BdG states, at two different times.  The results of the numerical simulations are compared to the prediction~\eqref{eq:g2_out} for a thermal state of phonons, where $n_k = n_k^{\mathrm{th}}$ and $c_k=0$. As in Fig. ~\ref{fig:nbr_phonons_two_values_interaction}, the two panels differ by the values of the temperature and density. }
	\label{fig:G2_two_values_interaction}
\end{figure}

In the current version of the algorithm, the primary quantity used in the computation are the phononic amplitudes computed from the Madelung state $b_{k,i}$. Previously, the primary quantity was the atomic field $\Psi_i$ itself. The number of quasi-particles $n_k$, and their pair-correlation $c_k$, were deduced from evaluating the density-density correlation function assuming Gaussianity using the formula~\eqref{eq:G2_k_phonons}.  
This procedure mimics an experimental `in-situ' measurement of these quantities. Given the importance of $G^{(2)}_{k,-k}$, we compared its values computed in the BdG and the Madelung states. The results are shown in Fig.~\ref{fig:G2_two_values_interaction}.
We compare the values estimated in the simulations to the predicted values of $G^{(2)}_{k,-k}$ for a thermal state of Madelung quasi-particles at temperature $T$. It is given by Eq.~\eqref{eq:g2_out}, where $n_k = n_k^{\mathrm{th}}$ is a thermal distribution given in Eq.~\eqref{def:nk_thermal}, while $c_k=0$.
In the upper panel of Fig.~\ref{fig:G2_two_values_interaction},
corresponding to lower interaction/temperature, $G^{(2)}_{k,k}$ for the BdG state is manifestly non-stationary in the IR part of the spectrum.
In the lower panel corresponding to larger interaction and temperature, already at the initial time, almost no point of the BdG state lies on the prediction of Eq.~\eqref{eq:g2_out}. 
At later times, its IR values decrease but are still far off the expected values.
On the other hand, the Madelung state is again observed to be stationary around the expected value for both series of parameters.

\begin{figure}
	\centering
	\includegraphics[width=0.9\textwidth]{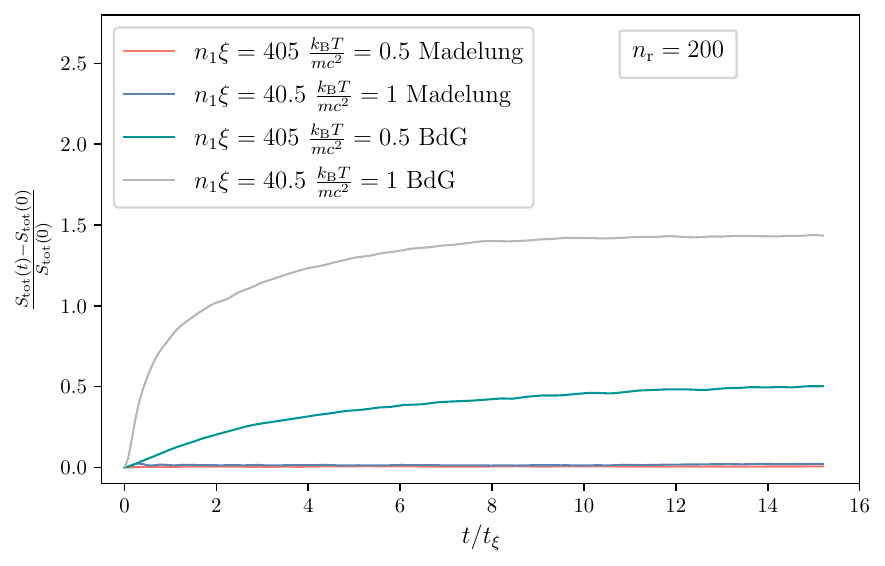}
	\caption{Gaussian approximation of the total entropy, given in Eq.~\eqref{eq:von_neumann_entropy_tot}, computed for the Madelung and the BdG states, as a function of time $t /t_{\xi}$.   }
	\label{fig:entropy_atoms_two_values_interaction}
\end{figure}

Finally, to quantify the level of non-stationarity, we can compute a Gaussian approximate value of the entropy of the state. This corresponds to the best estimate of the entropy based on $2$-point functions~\cite{Campo:2008ju}, and would match the exact entropy when the state is Gaussian.
This entropy will be computed using the state of the quasi-particles and we will assume each pairs of modes $\pm k \neq 0$ to be independent of each other. The von Neumann entropy for one such pair reads~\cite{Robertson:2018gwi}
\begin{equation}
	\label{def:von_neumann_entropy_pm_k}
	S_{\bm{k}} = \left( n^{\mathrm{eff}}_k + 1 \right) \ln \left( n^{\mathrm{eff}}_k + 1 \right)  - n^{\mathrm{eff}}_k \ln \left( n^{\mathrm{eff}}_k  \right) \, ,
\end{equation}
where we defined the quantity
\begin{equation}
	\label{def:neff}
	\left( n^{\mathrm{eff}}_k + \frac{1}{2} \right)^2 = \left( n_k + \frac{1}{2} \right)^2 - \left| c_k \right|^2 \, .
\end{equation}
We can then compute the total entropy of the state by summing this over all modes in the simulation
\begin{equation}
	\label{eq:von_neumann_entropy_tot}
	S_{\mathrm{tot}} = \sum_{k > 0 } S_{\bm{k}} \, .
\end{equation}
In Fig.~\ref{fig:entropy_atoms_two_values_interaction}, we show the evolution of the total entropy.~\eqref{eq:von_neumann_entropy_tot} as a function of time for the two initial states.
As expected from the evolution of the number of quasi-particles in Fig.~\ref{fig:nbr_phonons_two_values_interaction}, since the BdG state is more evidently non-stationary for lower background density and larger temperature, its entropy grows to even larger values. On the other hand, the entropy of the Madelung state is constant for both series of temperature and background density.

We conclude that to make reliable predictions in numerical simulations, one should relinquish the BdG state and use the Madelung state instead.

\subsection{Improvements compared to previous versions}

Since a similar algorithm was used in~\cite{Robertson:2018gwi} to study the evolution of the same 1D Bose gas numerically, we quickly summarise the improvements made over the code used then.
The primary improvement is using the density and phase perturbations in the Madelung form rather than the BdG approximation. It made the code more stable and reliable, especially at higher interactions, see Sec.~\ref{sec:madelung_vs_Bdg}. 
Second, we have also implemented the possibility of directly extracting the value of $b_{k,i}$, and compute the number of quasi-particles $n_k$ and their correlation $c_k$. 
Thanks to this, we can directly manipulate the number of phonons in the system, allowing for the simulation series where we inject phonons in a precisely controlled manner. 
Third, we have implemented the duplication of realisations to visualise the effect of an external perturbation better.
Finally, we eliminated several noise sources by identifying and removing spurious decay of UV modes.

\section{Absence of entanglement during parametric resonance}
\label{sec:decay_ck}

In the analysis of the evolution of entanglement made in~\cite{Busch:2014vda}, the authors assumed that the pair-correlation $c_k$, and the particle numbers $n_{\pm k}$ decay at the same rate. Note that under this assumption, starting in a state where $|c_k| > n_k$, in the absence of an external drive, entanglement would \textit{persist}. No pure decoherence process affects the correlation without changing the number of particles.
Still, the authors found that, in the context of parametric resonance, a large enough rate $\Gamma_k / \omega_k \approx 4.2 \%$ would prevent the appearance of an entangled state starting from a thermal state. 
Due to the simultaneity of parametric resonance and dissipation processes, there seems to be a non-trivial effect preventing entanglement formation. Therefore, the growth rates of $n_k$ and $c_k$ are not simply reduced by the decay rate, as suggested by the form used in~\cite{Micheli:2022zet} when studying $n_k$ alone.
In~\cite{Busch:2014vda}, the authors assumed $\rho_0 \xi =0.6$ and $k_{\mathrm{B}} T / \hbar \omega_k = 1$ and in the conclusion of~\cite{Micheli:2022zet} we computed for these values that $\Gamma_k / \omega_k \approx 5\%$. We thus concluded that the Beliaev-Landau processes are sufficient to explain the absence of entanglement~\cite{Jaskula:2012ab}.
To draw this conclusion, we implicitly assumed that $c_k$ would also decay at the rate $\Gamma_k$. 
The evolution of pair correlation is the topic of current investigations, and we close this chapter by briefly reporting on our latest findings.

\subsection{Decay of correlations}

In~\cite{Micheli:2022zet}, we analysed the effects of Beliaev-Landau scattering processes on the number of quasi-particles $n_k = \langle \hat{b}^{\dagger}_k \hat{b}_{k} \rangle$. 
We derived an equation of motion for a small perturbation of the number of particles $\delta n_k$ in the mode $k$. The strategy is to start from the full Heisenberg equation of motion of $\hat{b}_k$, given by Eq.~(A1) of~\cite{Micheli:2022zet}, and compute an equation of motion for $n_k$ with and without perturbation. Under the assumption that the only relevant non-Gaussian connected correlation functions are $C^{(3)}_{p,q} = \langle \hat{b}_{p}^{\dagger} \hat{b}_{q}^{\dagger} \hat{b}_{p+q} \rangle$ we can write a system of differential equations over $n_k$ and $C^{(3)}_{p,q}$. 
We re-expressed the latter as an integro-differential equation for $\delta n_k$, Eq.~(19) of~\cite{Micheli:2022zet}, which involves the perturbation of all modes $\delta n_q$.
Finally, neglecting the inverse processes that partially restore the initial $\delta n_k$ \ie the terms in $\delta n_q$ for $q \neq k$, we checked that our equation predicts the same decay rate as the Fermi Golden rule taken between two number states of quasi-particles.

Assume now that we excite quasi-particles in the modes $\pm k$ by an amount $\delta n_{\pm k}$, in a correlated fashion such that $\delta c_k = \langle \hat{b}_{-k} \hat{b}_{k} \rangle \neq 0$. We can repeat the same steps to derive equations of motion for $\delta n_{\pm k}$ and $\delta c_k$. The equation for $\delta c_k$ features an extra non-Gaussian correlation function $D^{(3)}_{p,q} = \langle \hat{b}_{p} \hat{b}_{q} \hat{b}_{-p-q} \rangle$.
Neglecting all other connected correlation functions, we find a system of coupled integro-differential equations over $\delta n_k$ and $\delta c_k$. These equations also \textit{a priori} features the whole spectrum of $\delta n_q$ and $\delta c_q$. Nevertheless, an analysis of the magnitude of each term shows that those mixing $\delta n_k$ and $\delta c_k$ are sub-dominant; the dynamics of the number of phonons and their correlation are independent.
$\delta n_k$ still obeys Eq.~(19) of~\cite{Micheli:2022zet}, while $\delta c_k$ obeys an equation of the same form with a different response function. Neglecting finite size effects and reverse processes, $\delta n_k$ still decays at a rate $\Gamma_k$ given by Eq.~(17) of~\cite{Micheli:2022zet} that we repeat here
\begin{equation}
	\label{eq:gamma_k_repetition}
	\Gamma_{k} t_{\xi} = \frac{k_{\mathrm{B}} T}{m c^2} \frac{1}{\rho_0 \xi} \left[  f_{-}\left( k \xi \right)  +  f_{+}\left( k \xi \right)\right] \, .
\end{equation}
Finally, under the same assumptions, we find that $|\delta c_k|$ decays at the same rate $\Gamma_k$.

\begin{figure}
	\centering
	\includegraphics[width=0.9\textwidth]{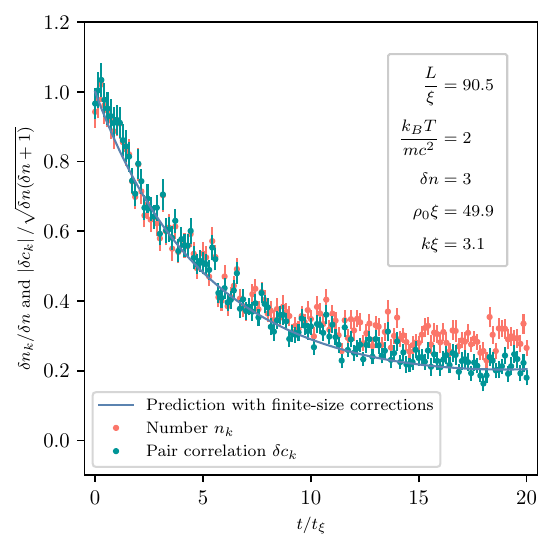}
	\caption{Evolution of the number of quasi-particles $n_k$ in the modes $\pm k$ and of their pair correlation $|c_k|$, as a function of time. At the initial time, the modes' state is considered a TMSV with $\delta n$ quasi-particles on average. The coloured dots shows the result of TWA simulations. The blue shows an exponential decay at a rate given by  Eq.~\eqref{eq:gamma_k_repetition} (corrected for finite size-effect see Eq.~(20) of~\cite{Micheli:2022zet} ).  }
	\label{fig:decay_ck_and_nk}
\end{figure}

In Fig.~\ref{fig:decay_ck_and_nk}, we show the evolution of $|\delta c_k|$ and $\delta n_k$ from an initial TMSV. We normalised their values by their expected initial values $\delta n_k(0) = \delta n$ and $|\delta c_k|(0) = \sqrt{ \delta n (\delta n  +1)}$. These two series are compared to the prediction of an exponential decay at rate $\Gamma_k$ given by  Eq.~\eqref{eq:gamma_k_repetition} (corrected for finite size-effect, see Eq.~(20) of~\cite{Micheli:2022zet} ) and we observe an excellent agreement, at least at early-times.
The fact that correlation and number of particles decay at least at the same rate is physically reasonable: when a quasi-particle is lost in one of the modes $\pm k$, the pair-correlation $c_k$ is lost. One could expect other pure decoherence processes to affect pair-correlation leaving the number of quasi-particles $n_k$ unaffected. Indeed, we see signs of such processes at late times in Fig.~\ref{fig:decay_ck_and_nk}. However, they are sub-dominant, and would only speed up the decoherence process. They are thus not expected to change our conclusion that the Beliaev-Landau dissipative processes are sufficient to explain the absence of entanglement.
 
In conclusion, our latest numerical and analytical analyses support the minimal assumptions of an equal dissipation for quasi-particle numbers and pair correlation made in~\cite{Busch:2014vda}. With this extra piece of the puzzle, the claim of~\cite{Micheli:2022zet} that Beliaev-Landau scatterings are sufficient to explain the absence of entanglement in~\cite{Jaskula:2012ab} is then adequately justified.

%%%%%%%
%To add the bibliography when compiling the subfile only
% \bibliographystyle{plain}
% \bibliography{bibmanuscript}

\chapter{Conclusion and perspectives}
\label{chapt:conclusion}

This PhD was dedicated to time-dependent situations relevant to cosmology that are described using quantum field theory in curved spacetimes. 
We studied the appearance and disappearance of quantum features in these contexts.

In the first part of this manuscript, we reported on our study of such quantum aspects in the cosmological perturbations, which are expected to have emerged from the quantum vacuum.
Our goal was to assess the robustness of quantum features in the state of the perturbations to decoherence phenomena.
To do so, we focused on the non-classicality of bi-partite correlation in the state of the cosmological perturbations, arguably the simplest manifestation of its quantum nature.
We have measured the efficiency of decoherence on the cosmological perturbations using the quantum discord~\cite{Martin:2021znx} and a GMKR-type Bell inequality~\cite{Martin:2022kph}. We showed that these measures reveal a competition between decoherence and squeezing to determine the character (quantum or classical) of the bi-partite correlations between opposite Fourier modes. This phenomenon is already known for other criteria, e.g. separability or another Bell inequality~\cite{Campo:2005sv,Adamek:2013vw}. We systematically compared three such quantumness criteria in a context broader than that of cosmological perturbations~\cite{Martin:2022kph}. Although they all qualitatively exhibit the competition between squeezing and decoherence, the precise threshold of classicality depends on the chosen measure.
Finally, we tried to clarify and condense the discussions on this question in a review~\cite{Micheli:2022tld}.
The precise decoherence level in the early Universe is still being determined. Therefore, we cannot yet assess definitively whether the correlations were already classical at the end of inflation. Nevertheless, a lower bound was recently obtained in~\cite{Burgess:2022nwu}. 

\begin{figure}
	\centering
	\includegraphics[width=0.9\textwidth]{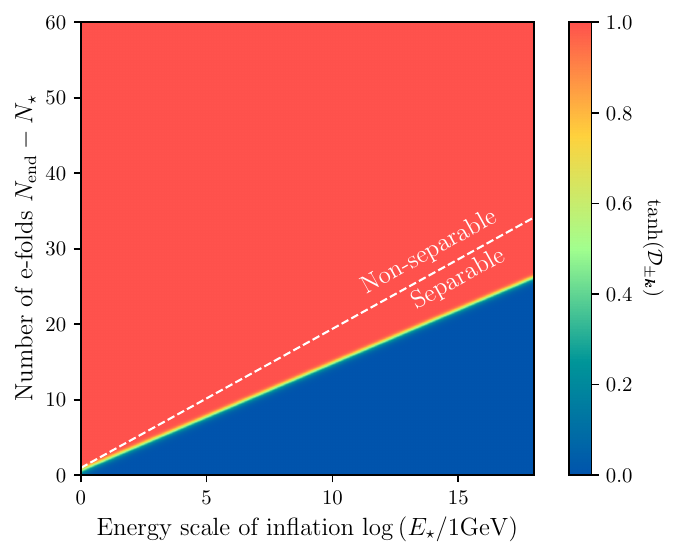} % second figure itself
	\caption{Hyperbolic tangent of the quantum discord $\tanh ( \mathcal{D}_{\pm \bm{k}} )$ of the correlation between modes $\pm \bm{k}$ as a function of the energy scale of inflation $E_{\star}$ and the number of e-folds from the time the modes crossed out the Hubble radius to the end of inflation $N-N_{\star}$. The white dashed line shows the separability threshold. We refer to~\cite{Martin:2022kph} reproduced in Sec.~\ref{sec:comparing_quantumness_criteria} for details on these quantities. The quantum discord given in Eq.~(2.28) of~\cite{Martin:2021znx} for $\theta = - \pi /4$, depends on the squeezing parameter $r$ and the purity $p$. The squeezing parameter is estimated using the de Sitter behaviour given in Eq.~(5.6) of~\cite{Martin:2021znx}, while the purity is estimated using Eq.~\eqref{eq:max_purity_mode_k}.  }
	\label{fig:sep_discord_min_deco}
\end{figure}

As a continuation of the above programme and a direct application of the PhD results, we give a quick evaluation of the expected level of quantumness left at the end of inflation.
The authors of~\cite{Burgess:2022nwu} found that the purity of a mode $\bm{k}$ that crossed out the Hubble radius $N-N_{\star}$ e-folds before the end of inflation is given by
\begin{equation}
	\label{eq:max_purity_mode_k}
	p_k = \frac{1}{\sqrt{1+ \Xi_k }} \, ,
\end{equation}
where
\begin{equation}
	\label{eq:xi_mode_k}
	\Xi_{k} = 1.6 \times 10^4 \times \left(  \frac{A_{\mathrm{S}}}{2.2 \times 10^{-9}} \right)^(-1) \left( \frac{E_{\star}}{M_{\mathrm{Pl}}c^2} \right)^8 e^{3 (N-N_{\star})} \, .
\end{equation}
Although the exact number depends on the details of reheating and the energy scale of inflation, we expect that for modes corresponding to cosmological scales today $N-N_{\star} \geq 30$, so that the exponentially is typically dominant in the above expression.
The purity then scales as $p_k \propto  e^{- 3 (N-N_{\star})/2}$. To assess the entanglement of the state, we need to compare this number to the exponential of twice the squeezing parameter $r$~\cite{Martin:2022kph}. Using the conventions of~\cite{Martin:2021znx} for the conjugated momentum and taking the de Sitter limit, we have $r_k \sim 2 (N - N_{\star})$. 
Taking the ratio with the square root of the purity, we have $e^{2 r}/ \sqrt{p} \sim e^{13 (N - N_{\star})/4}$, which grows with $N - N_{\star}$. Thus, irrespective of the energy scale of inflation, if the mode crosses the Hubble radius early enough during inflation, the state of the pair $\pm \bm{k}$ will still be entangled at the end of inflation! 
A more precise illustration is given in Fig.~\ref{fig:sep_discord_min_deco}, where we plotted the value of the quantum discord and the threshold of separability as a function of $N-N_{\star}$ and the energy scale of inflation $E_{\star}$. Fig.~\ref{fig:sep_discord_min_deco} shows that modes $\pm \bm{k}$ such that $N-N_{\star}\geq 30$, in particular cosmological scales, are still entangled at the end of inflation and have a large quantum discord. We thus observe that entanglement can persist despite an exponential suppression of the purity in the number of e-folds. This suppression led the authors of~\cite{Burgess:2022nwu} to conclude that modes were all `decohered' on cosmological scales. Nevertheless, the above shows the necessity to consider the competition between squeezing and decoherence to conclude on the classicalisation of the perturbations.

It is thus not obvious that by the end of inflation, the cosmological perturbations are in an entirely classical state.
Since the quantum character of the perturbations might have persisted until the end of inflation, one may then want to ask whether it can be experimentally demonstrated.
Except for the conclusion of Sec.~\ref{sec:quantum_GW}, we did not deal with this question in the manuscript, and we say a few words about it now.

We believe that, in the context of single field slow-roll inflation, with a Gaussian state, the prospects of experimental demonstrating the quantum origin of the perturbations are scarce.
First, in this scenario, there are only two fields that carry all information on the quantum state of the Universe, the Mukhanov-Sasaki field $\hat{v}$ and the tensor perturbations $\hat{h}$. Let us first discuss the scalar perturbations.
In the single field scenario, where $\hat{v}$ is not substantially perturbed by other fields that would bear traces of its state, we only have access to $n$-point functions of $\hat{v}$ which are imprinted on CMB's photons via the Sachs-Wolfe effect~\cite{Sachs:1967er}. However, in the Gaussian case, all correlation functions of $\hat{v}$, and its conjugated field (which we do not measure), can be reproduced to an excellent approximation by a classical stochastic process~\cite{Micheli:2022tld}. They are thus insufficient to demonstrate the quantum origin of the state.
On the other hand, tensor perturbations have yet to be detected. 
Some authors have analysed the effect of a gravitational wave impinging on a gravitational interferometer~\cite{Kanno:2020usf,Parikh:2020fhy}. They showed that a squeezed state of the perturbations would induce noise in the detector akin to that of a Gaussian stochastic process. Therefore, a naive measure of the amplitude of these waves can only lead to an indirect proof of their quantum origin, as for scalar perturbations, by measuring the $k$-dependence of their power-spectrum, see Sec.~\ref{sec:quantum_IC}.
Still, if we were to measure the noise of primordial gravitational waves in interferometers, then we would be measuring directly waves produced during inflation, as opposed to measuring CMB's photons, produced much later, in the case of scalar perturbations or $B$-mode detection, see Sec.~\ref{sec:quantum_GW}.
In principle, we can then access the complete state of these waves, and one could potentially design some intelligent measurement protocol to reveal their quantum nature, e.g.~\cite{Kanno:2021gpt}.
Unfortunately, this remaining window of opportunity might also be closed due to the smallness of the amplitude of the primordial gravitational waves. Even if the tensor to scalar ratio, see Sec.~\ref{sec:quantum_IC}, is close to the current upper bound 
$r < 0.036$ at $95 \%$~\cite{BICEP:2021xfz}, the waves would not be in the sensitivity of Big-Bang Observer (BBO), the most futuristic interferometer planned~\cite{Caprini:2018mtu}.

Before discussing possibilities to go beyond the current modelling, note that, in this work, we performed computations solely in Fourier space. Since each pair of modes $\pm \bm{k}$ is independent of the others, the only correlations to consider in Fourier space are between the two members of the pair.
Nevertheless, it could be objected, see Sec.~\ref{sec:quantum_GW}, that checking for a Bell inequality violation for non-local degrees of freedom is meaningless since the objects are not well separated.
In addition, since the correlations in real space potentially depend on the state of \textit{all} Fourier modes, one may hope to escape the no-go theorem on $n$-points functions of $\hat{v}$ and its conjugate field.
Some authors have analysed how correlations in Fourier space manifest when correlating regions in real space~\cite {Martin:2021qkg,Martin:2021xml,Espinosa-Portales:2022yok}. Unfortunately, they found that, already in the absence of decoherence, Bell inequalities are never violated and that the quantum discord was very small, albeit non-vanishing.

With the work done in this PhD, we made a point that the presence of decoherence does not straightforwardly make the cosmological perturbations classical at the end of inflation. 
However, as long as the quantum features are concealed in operators for which no clear measurement protocol, even for a tabletop system, can be given, it could be seen as a moot point.
Therefore, we believe that in the next studies, more attention should be given to operational approaches describing what could \textit{actually} be measured. For instance, in the spirit of~\cite{Green:2020whw}, it would be worth investigating whether some of the possible non-Gaussian signals can, under reasonable assumptions, not be attributed to a classical theory and how multi-partite~\cite{Horodecki:2009zz} entanglement could manifest there.

In the other part of this PhD, we progressed on analysing the analogue preheating experiment~\cite{Jaskula:2012ab}. 
We demonstrated the existence of scattering processes for quasi-particles of 1D quasi-condensate~\cite{Micheli:2022zet}, addressing some claims to the contrary in parts of the literature.
We improved a TWA simulation algorithm to study these processes, making it more precise and reliable.

A first continuation of this PhD work would be to quantitatively explain the spectrum of phonons reported in~\cite{Jaskula:2012ab}. Comparing these experimental results with the dissipation-less prediction of Fig.~\ref{fig:evolution_nk_ck_jaskula_modulation}, we see a few differences. First, we predict two narrow and distinct peaks, while only a broad one was observed experimentally. Several things can be brought up to explain it. It could be that the two peaks are too close to be well resolved in the experiment. 
Additionally, although the modulation in our model is already not simply the sum of two oscillations, one associated with each peak, any supplementary noise in this modulation could affect the peak structure~\cite{Zanchin:1997gf,Zanchin:1998fj,DeCross:2016fdz}.
Finally, the peaks will also broaden due to the dissipation processes we identified. Therefore, they could appear as a single broad peak in the experimental data.
Second, the number of phonons observed is much less than predicted. Again, the dissipation processes could provide a sufficient reduction to explain it. As a first estimation, we can compare the growth $G_k$ and decay rate $\Gamma_k$ expected for our resonant modes. Based on~\cite{Busch:2014vda}, and using the decay rate given in Eq.~(17) of~\cite{Micheli:2022zet}, for the values $c = 8$ mm/s, $k_{\mathrm{B}} T / m c^2 \approx 6.5$ and $n_{\mathrm{1D}} a_s \approx 0.19$  of~\cite{Jaskula:2012ab} used in Fig.~\ref{fig:evolution_nk_ck_jaskula_modulation}, we find $\Gamma_k/G_k \sim 15.6$ for the first resonant mode. In this regime, we expect the growth of the resonant modes to saturate and the final number of phonons to be strongly reduced with respect to that shown in Fig.~\ref{fig:evolution_nk_ck_jaskula_modulation}.
Nevertheless, the comparison should be done precisely at the level of the full spectrum.

A natural next step in the analysis is to use the TWA simulations to simulate the current experimental set-up's precise conditions and test whether our equations accounting for the dissipative effects, Eq.~(19) of~\cite{Micheli:2022zet}, allow us to predict the evolution of entanglement correctly. These equations should also be compared with the effective equations of motion used in~\cite{Busch:2014vda} to understand the fine details of decoherence during the parametric amplification process.
In this respect, the latest results shown in Fig.~\ref{fig:decay_ck_and_nk} indicate that decoherence channels are present in the system, even without parametric resonance. Note that, in this figure, the pair-correlation follows quite closely the predicted exponential decay, while the decay of the number of phons $n_k$ seems to slow down.
Our current understanding is that this slowing down is due to inverse processes, \ie decay products recombining to the initial excitation. These inverse processes appear to be more efficient in regenerating the number of excitations $n_k$ in the two modes $\pm k$ than their correlation $c_k$.
This result still requires a finer analysis, but it would constitute a nice illustration of the fragility of quantum correlations compared to other quantities, such as the number of excitations. 
From a broader perspective, it would be interesting to investigate how general this decoherence mechanism via differentiated inverse process rate is. Is it a specific feature of the unusual decay process we studied, which involves many modes as decay products, or a general one?

In addition, the model still requires some improvements. We should consider the trapping of the gas that breaks periodic boundary conditions so that the nature of the eigenmodes is altered. In particular, the notion of direction of the mode $\pm \bm{k}$ loses its meaning as it is reflected back and forth. It should then be analysed how our non-separability witnesses are affected by these changes in the mode structure. For instance, the quantity $\langle \hat{b}_{\bm{k}} \hat{b}_{-\bm{k}}^{\dagger}  \rangle$, which encodes the mixes in between the two directions $\pm \bm{k}$ and was neglected in previous works, should be taken into account.

Finally, as a more long-term goal, following-up on~\cite{Robertson:2018gwi}, we would like to extend the analysis of the experiment to later times when it enters a very non-linear regime.
Several other groups have studied, numerically and experimentally, this regime in similar set-ups, e.g.~\cite{Butera:2022kwi,Barroso:2022vxg,Zache:2017dnz}. 
First, as done in~\cite{Barroso:2022vxg,Zache:2017dnz}, we should be able to predict and experimentally observe the generation of secondaries in our system.
Second, the authors~\cite{Butera:2022kwi} report the numerical observation of fragmentation of the initially coherent condensate in different incoherent regions of space due to the back-reaction of the amplified quantum fluctuations. It is unclear, but worth studying, if there are signatures of this fragmentation in the distribution of excitations in Fourier space beyond the mere decrease of the condensate occupation. If so, we might observe it using the TOF method used in the current set-up, making it possible to experimentally study the back-reaction from a quantum field on a classical one expected in preheating.
A last question we wish to investigate is whether the initial bi-partite $\pm \bm{k}$ entanglement that is distilled to other degrees of freedom in the system can still be captured experimentally by analysing well-chosen quantities in this regime. As a first example, at early times, since the relevant Beliaev-Landau scattering processes only involve nearby modes, it may seem natural to believe that part of the entanglement is distilled to them. It is then an experimentally relevant question to know whether this distilled entanglement can be recovered in the Fourier analysis by simply correlating larger boxes \ie larger than $\delta k \sim 2 \pi /L$.
More generically, it would be interesting to follow the distillation of entanglement from the two resonant modes to more, in a similar way that the number of excitations initially injected in the resonant modes can be tracked down.
Nevertheless, the problem appears much more challenging. Indeed, while following which modes are excited or not only requires computing the $2$-point function $n_k$, measuring the formation of muti-partite entanglement requires accessing higher order $n$-point function since $c_k$ only encodes pair-correlation.

\chapter*{Appendices}
\addcontentsline{toc}{chapter}{\protect\numberline{}Appendices}%

%To get the numbering of appendices to be letters
%https://tex.stackexchange.com/questions/174621/numbering-appendices-by-letter-instead-of-number
\renewcommand{\thesection}{\Alph{section}}

\section{Geodesics and time dilation in FLRW}

\subsection{Geodesics in an expanding Universe}
\label{app:geodesics_FLRW}

The Christoffel symbols of the metric~\eqref{def:FLRW} in spherical co-moving coordinates read
\begin{align*}
	\label{eq:christo_FLRW}
	\Gamma^{t}_{rr} & = \frac{\dot{a}a}{1 - \mathcal{K}r^2}, && \Gamma^{t}_{\theta \theta} = \dot{a}a r^2,  &&& \Gamma^{t}_{\phi \phi} = \dot{a}a r^2 \sin^2 \theta, \\
	\Gamma^{r}_{tr} & = \Gamma^{\theta}_{t \theta} = \Gamma^{\phi}_{t \phi} = H, && \Gamma^{r}_{rr}  = \frac{\mathcal{K}r}{1-\mathcal{K}r^2},  &&& \Gamma^{r}_{\theta \theta} = - r \left( 1 - \mathcal{K}r^2 \right), \\
	\Gamma^{r}_{\phi\phi} & = - r \left( 1 - \mathcal{K}r^2 \right) \sin^2 \theta , && \Gamma^{\theta}_{\theta r} =  \Gamma^{\phi}_{\phi r} = \frac{1}{r},  &&& \Gamma^{\theta}_{\phi \phi} = - \sin \theta \cos \theta, \quad \Gamma^{\phi}_{\phi \theta} = - \frac{\sin \theta}{\cos \theta}.
\end{align*}

We give the solution of the geodesic equation over the energy-momentum vector~\eqref{eq:geodesic_p} for massive and massless particles. We consider geodesics that are purely radial at initial time $P^{\theta} = P^{\phi} = 0$, which can always be arranged by rotating the system of coordinates. The geodesic then remains radial at any time, and we find that as a function of cosmic time, its components read
\begin{align}
	P^{t} \left( t \right) & = \frac{E \left( t \right)}{c} =  m c \sqrt{1 + \left[ \frac{a_0}{a\left( t \right)} \right]^22 \left( \frac{p_0}{mc} \right)^2 } \, , \\
	P^{r} \left( t \right) & = \left[ \frac{a_0}{a\left( t \right)} \right]^2 \frac{p_0}{a_0} \sqrt{1 - \mathcal{K}r^2\left( t \right) } \, ,
\end{align}
where $p_0$ is the physical 3-momentum at the present time. At an earlier time $t$ we have $p = [a_0 / a(t)] p_0 $ \ie the physical momentum decays with the expansion as $a^{-1}$. Notice that test bodies at rest in the co-moving coordinates, \ie $p = 0$ at any time, follow spacetime's geodesics; they are said to be \textit{co-moving}. For massless particles, in particular photons, we similarly have
\begin{align}
	P^{t} \left( t \right) & = \frac{E \left( t \right)}{c} =  \frac{a_0}{a} p_0 \, , \\
	P^{r} \left( t \right) & = \left[ \frac{a_0}{a\left( t \right)} \right]^2 \sqrt{1 - \mathcal{K}r^2\left( t \right) } \frac{p_0}{a_0} \, .
\end{align}
In this case, the physical momentum and the energy evolve as $a^{-1}$. For photons, this translates into a redshift of their frequencies, see Sec.~\ref{sec:physics_FLRW}.

\subsection{Time dilation in an expanding Universe}

In a spacetime described by the metric~\eqref{def:FLRW}, time duration as measured by observers at different times will differ. 
Consider an emitter $\mathrm{E}$ sending signals to a receptor $\mathrm{R}$. We assume that $\mathrm{E}$ and $\mathrm{R}$ are co-moving observers $\underline{U}^{\mathrm{E}/\mathrm{R}} = (c, 0,0,0)$. By redefining the co-moving coordinates, one can always arrange for $\mathrm{E}$ to be located at the coordinates' origin and $\mathrm{R}$ to be located at $(r_{\mathrm{R}},0,0)$. Assume that $\mathrm{E}$ sends a first light signal from $(c t_{\mathrm{E}} , 0,0,0)$ received by $\mathrm{R}$ at $(c t_{\mathrm{R}} , r_{\mathrm{R}},0,0)$, and then a second signal just after at $(c t_{\mathrm{E}} + c \delta t_{\mathrm{E}}, 0,0,0)$
which is received at $(c t_{\mathrm{R}} + c \delta t_{\mathrm{R}}, 0,0,0)$. The relation between the arrival time and the distance covered by these light signals is found by setting $\dd^2 s = 0$ in Eq.~\eqref{def:FLRW}. We find 
\begin{equation}
	\label{eq:photon_FLRW}
	r_{\mathrm{R}} = c \int_{t_{\mathrm{E}}}^{t_{\mathrm{R}}} \frac{\dd t^{\prime}}{a (t^{\prime})} = c \int_{t_{\mathrm{E}} + \delta t_{\mathrm{E}}}^{t_{\mathrm{R}} + \delta t_{\mathrm{R}}} \frac{\dd t^{\prime}}{a (t^{\prime})} \, .
\end{equation}
By taking the difference between these last two relations and exploiting the infinitesimal character of the time intervals, we find
\begin{equation}
	\delta t_{\mathrm{R}} = \frac{a \left( t_{\mathrm{R}} \right)}{a \left(t_{\mathrm{E}} \right)} \delta t_{\mathrm{E}} \, ,
\end{equation}
assuming that the scale factor increases with cosmic time, there is a time dilation $\delta t_{\mathrm{R}} > \delta t_{\mathrm{E}}$. In particular, assuming that $\mathrm{E}$ sends a continuous monochromatic light signal to $\mathrm{R}$ and taking $\delta t_{\mathrm{E}}$ to be the interval in between two crests of the light signals \ie the period of the signal, then the above relation gives the same relation as Eq.~\eqref{def:redshift_photons} for the redshift of the signal. This time dilation can then be understood as the physical reason behind the redshift derived from the machinery of null geodesics.

\section{Stress-energy tensor and $R^2$ inflation}

\subsection{Stress-energy tensor of a scalar field}
\label{app:tensor}

We first derive Jacobi's formula for the variation of the determinant of
a matrix with respect to one of its coordinates. We have
\begin{equation}
	\label{eq:variation_g}
	g = \sum_{\mu_{1}, ... , \mu_{n}} \epsilon\indices{_{\mu_{1}, ... , \mu_{n}}} g \indices{_{1 \mu_{1}}} ... g \indices{_{n \mu_{n}}} \, ,
\end{equation}
where $\epsilon\indices{_{\mu_{1}, ... , \mu_{n}}}$ is the Levi-Civita tensor in $n$-dimensions. Notice that one cannot forget about the sum by putting the indices up. It would imply having the coefficient of the inverse of metric, which would be wrong. Using the formula~\eqref{eq:variation_g}, we have 
\begin{equation}
	\omega\indices{^{\alpha \beta} } = \frac{\delta g}{\delta g_{\alpha \beta}} 
	= \sum_{s} \sum_{\mu_{1}, ... , \mu_{s-1}, \mu_{s+1}, ... , \mu_{n} } \epsilon\indices{_{\mu_{1}, ... , \mu_{n}}} g \indices{_{1 \mu_{1}}} ... g \indices{_{s-1 \mu_{s-1}}} \delta\indices{^{\alpha}_{s}} \delta\indices{^{\beta}_{\mu_{s}}} g \indices{_{s+1 \mu_{s+1}}} ... g \indices{_{n \mu_{n}}}
\end{equation}
Then one can check that $   \omega\indices{^{\alpha \beta} }  g \indices{_{\beta  \gamma} } = g \delta\indices{^{\alpha}_{\gamma}}$ and we get Jacobi's formula
\begin{equation}
	\frac{\delta g}{\delta g_{\alpha \beta}} g g \indices{^{\alpha \beta}} \, .
\end{equation}
Now using that $g \indices{^{\alpha \beta}} g \indices{_{\beta \gamma}} = \delta \indices{^{\alpha}_{\gamma}}$ we get
\begin{equation}
	\delta g \indices{_{\mu \nu}} = - g \indices{_{\mu \alpha}} g \indices{_{\nu \beta}} \delta g \indices{^{\alpha \beta}} \, .
\end{equation}
Combining the above
\begin{equation}
	T \indices{_{\mu \nu}} = L_{\mathrm{m}} g \indices{_{\mu \nu}} - 2 \frac{\delta  L_{\mathrm{m}}}{\delta g \indices{^{\mu \nu}}} \, .
\end{equation}
For the scalar field action, this gives Eq.~\eqref{eq:Tmunu_scalar_field}.

\subsection{Slow-roll at first-order in $R^2$-inflation}
\label{app:R2inflation}

In this appendix, we analyse briefly the case of slow-roll $R^2$ inflation.

\subsubsection{A potential for $R^2$ inflation.}

This model was originally introduced by Starobinsky in~\cite{Starobinsky:1980te} by considering a modification to Einstein-Hilbert action due to quantum corrections.
The relevant piece reads
\begin{equation}
	\label{def:R2_action}
	S_{R^2} = \frac{1}{2 \kappa c} \int \sqrt{-g} \left( R + \frac{\hbar^2}{6 m^2 c^2} R^2 \right) \dd^4 \bm{x} \, ,
\end{equation}
where $m$ is a mass scale. The usual Einstein-Hilbert action only has two degrees of freedom; the two polarisations of the gravitons, see Sec.~\ref{sec:quantum_GW}. 
The theory described by the action~\eqref{def:R2_action} has, in fact, an extra scalar degree of freedom. This action can be recast as Einstein-Hilbert action plus an action for a scalar field $\phi$~\cite{Sotiriou:2006hs}. We first introduce an auxiliary field $\chi$ with the action
\begin{equation}
	S_{R^2}^{\prime} = \frac{1}{2 \kappa c} \int \sqrt{-g} \left( R + \frac{\hbar^2}{3 m^2 c^2} \chi R - \frac{\hbar^2}{6 m^2 c^2} \chi^2\right)  \dd^4 \bm{x} \, .
\end{equation}
The Euler-Lagrange equation for $\chi$ gives $\chi = R$ so that the equations of motion are the same as that of Eq.~\eqref{def:R2_action}. By re-parameterising the scalar field via $\varphi = 1 + \frac{\hbar^2}{3 m^2 c^2} \chi$ the action reads
\begin{equation}
	\label{eq:R2_brans_dicke}
	S_{R^2}^{\prime} = \frac{1}{2 \kappa c} \int \sqrt{-g} \left[ \varphi R - \frac{3 m^2 c^2}{2 \hbar^2} \left( \varphi - 1 \right)^2 \right]  \dd^4 \bm{x} \, .
\end{equation}
This action is a sub-case of Brans-Dicke's theory of gravity. 
Under a  conformal transformation of the metric $g = \Omega^2 \tilde{g}$ the action transforms to
\begin{equation}
	S_{R^2}^{\prime} = \frac{1}{2 \kappa c} \int \sqrt{-\tilde{g}} \left[ \Omega^2 \varphi \left( \tilde{R} - 6 \Omega^{-1} \tilde\Box \Omega \right) - \Omega^4 \frac{3 m^2 c^2}{2 \hbar^2} \left( \varphi - 1 \right)^2 \right]  \dd^4 \bm{x} \, .
\end{equation}
To cast the gravitational part of the action in the Einstein-Hilbert form, we pick
$\Omega^2 \varphi = 1$, or $\Omega = \varphi^{-1/2}$, and reparameterise $\varphi = \exp \left( \sqrt{ \frac{2}{3}} \sqrt{\kappa} \phi \right)$. Then
\begin{equation}
	6 \Omega^{-1} \tilde\Box \Omega = 2 \kappa \frac{1}{2} \partial_{\mu} \phi \partial_{\nu} \phi - \sqrt{\frac{6 \kappa}{2}} \partial_{\mu} \partial_{\nu} \phi \, ,
\end{equation}
and dropping the boundary term in the action, we finally get
\begin{equation}
	\label{def:R2_action_potential}
	S_{R^2}^{\prime} = \frac{1}{2 \kappa c} \int \sqrt{-\tilde{g}} \tilde{R}  \dd^4 \bm{x}
	- \frac{1}{c} \int \sqrt{-\tilde{g}} \left[ \frac{1}{2} \partial_{\mu} \phi \partial_{\nu} \phi + V_{\mathrm{S}} \left( \phi \right) \right]  \dd^4 \bm{x} \, ,
\end{equation}
where the potential $V_{\mathrm{S}}$ is defined in Eq.~\eqref{def:R2potential}.
The actions given by Eq.~\eqref{def:R2_action} and Eq.~\eqref{def:R2_action_potential} are equivalent. The latter form is that of a scalar field minimally coupled to general relativity, which corresponds to the form used in Sec.~\ref{sec:SSR}. This form makes manifest the presence of an extra scalar degree of freedom compared to the Einstein-Hilbert action.

\subsubsection{Slow-roll}

We now derive the equation of evolution of the field in the first-order in slow-roll parameters. It is useful first to rewrite the equations of motion using $N$, the number of e-folds, as a time variable rather than cosmic time. The Klein-Gordon equation can be recast has
\begin{equation}
	\frac{\dd^2 \phi}{\dd N^2} + \left( 3 - \epsilon_1 \right) \frac{\dd \phi}{\dd N} + \frac{V^{\prime} (\phi) c^2}{H^2} = 0 \, .
\end{equation}
The second slow-roll parameter then assumes the simple form
\begin{equation}
	\epsilon_2 = 2 \left( \frac{\dd \phi}{\dd N} \right)^{-1} \frac{\dd^2 \phi}{\dd N^2} \, ,
\end{equation}
which in turn allows us to rewrite the Klein-Gordon equation as
\begin{equation}
	\left( 3 - \epsilon_1 + \frac{\epsilon_2}{2} \right) \frac{\dd \phi}{\dd N} + \frac{V^{\prime} (\phi) c^2}{H^2} = 0 \, .
\end{equation}
Under this form, it is clear that neglecting the two first slow-roll parameters, we have
\begin{equation}
	\frac{\dd \phi}{\dd N} \approx - \frac{V^{\prime} (\phi) c^2}{3 H^2} \approx - \frac{1}{\kappa} \frac{V^{\prime}}{V} \, ,
\end{equation}
where we used Eq.~\eqref{eq:friedmann_SR} in the second equality.
The expression of the first flow function in the slow-roll regime~\eqref{eq:SR_potential_1} shows that the first slow-roll
condition $\epsilon \ll 1$, imposes to be in the large-field regime $\sqrt{k} \phi$.
In this limit, defining the number of e-folds to be $0$ at the start of inflation $N = \ln \left( a_{\mathrm{in}} / a \right)$ and the initial value
of the field to be $\phi_{\mathrm{in}}$, the equation of motion is easily integrated to Eq.~\eqref{eq:field_N_R2inflation}.
We can then express $\dot{\phi}$, $\epsilon_1$ and $H$ as a function of the number of e-folds
\begin{subequations}
	\begin{equation}
		\dot{\phi} \left( N \right) = - \frac{ m c^2 \sqrt{\kappa}}{4 \hbar} \sqrt{ \frac{3}{8} } \frac{1}{N_e - N} \, ,
	\end{equation}
	\begin{equation}
		\epsilon_{1} \left( N \right) = \frac{3}{4} \frac{1}{\left( N_e - N - \frac{3}{4} \right)^2} \, ,
	\end{equation}
	\begin{equation}
		H \left( N \right) = \frac{m c^2}{2 \hbar} \left( 1 - \frac{3}{4} \frac{1}{N_e - N}\right) \, .
	\end{equation}
\end{subequations}

Recasting the equation for $\dot{\phi}$ we get the expression of the slow-roll trajectory in the $\left( \phi, \dot{\phi} \right)$-plane
\begin{equation}
	\label{eq:SR_trajectory_R2inflation}
	\dot{\phi} = - \frac{m c^2 \sqrt{\kappa}}{\hbar} \sqrt{\frac{2}{3}} e^{- \sqrt{\frac{2}{3}} \sqrt{\kappa} \phi } \, ,
\end{equation}
which is plotted in Fig.~\ref{fig:trajectories_R2inflation}.
Solving for $\epsilon_{1} \left( N \right) = 1$ we find that inflation lasts roughly $N_e - N - \frac{3}{4} - \sqrt{\frac{3}{4}}$.

\section{Perturbations of a 1D weakly interacting Bose gas}

\subsection{Canonical transformation for relative perturbations}
\label{app:canonical_transfo_BdG}

For an Hamiltonian system with Hamiltonian $H$ described by a conjugated pair $(p,q)$
\begin{equation}
	\left\{ p , q \right\} = 1 \, ,
\end{equation}
a generating function is a function that generates a canonical transformation to another conjugated pair of variables $(P,Q)$ :
\begin{equation}
	\left\{ P , Q \right\} = 1 \, .
\end{equation}
Hamilton's equations also describe the time evolution of the new pair
but with respect to a possibly new Hamiltonian $K$ computed from the generating function. If the transformation is time-independent, then $H=K$. A type-2 generating function $F_2$ depends on the old position $q$ and the new momentum $P$ and gives the transformation laws
\begin{align}
	Q & = \frac{\partial F_2}{\partial P} \, , \\
	p & = \frac{\partial F_2}{\partial q} \, ,
\end{align}
and
\begin{equation}
	K = H + \frac{\partial F_2}{\partial t} \, .
\end{equation}

We want to use this formalism to describe the relative perturbations defined in sub-section Sec.~\ref{sec:dynamics_pert_bdg}. We consider a classical version of the system, where the atomic field $\Psi$ is a complex function. We want to write a generating function for the transformation $\psi  \to e^{- i \mu t / \hbar } \psi$, where we defined $\psi = \sqrt{n_0} \delta \hat{\Psi}$ to shorten the notations in this appendix. The canonical pair used to describe the system is initially $( \psi , i \hbar  \psi^{\star})$ (the additional factor of $i \hbar$ appears in the classical counterpart theory), and the final one $( \psi^{\prime} , i \hbar \psi^{\star \, \prime}) = ( e^{- i \mu t / \hbar } \psi , i \hbar  e^{ i \mu t / \hbar } \psi^{\star})$. Notice that we have one degree of freedom at each position $\bm{x}$, so implicitly, we perform the canonical transformation separately for each point. If we define the following type-2 generating function
\begin{equation}
	\label{eq:F2_gen_function}
	F_2 \left( \psi , \psi^{\star \, \prime } , t \right)  = e^{- i \mu t / \hbar } i \hbar  \psi^{\star \, \prime } \psi \, ,
\end{equation}
the associated equations of transformation are 
\begin{align}
	\psi^{\prime} & = \frac{\partial F_2}{\partial i \hbar  \psi^{\dagger \, \prime }} = e^{ i \mu t} \psi  \, , \\
	i \hbar  \Psi^{\dagger } & = \frac{\partial F_2}{\partial \psi} = e^{ i \mu t} i \hbar  \psi^{\dagger \, \prime } \, ,
\end{align}
which are equivalent to a $ \psi \to e^{- i \mu t / \hbar } \psi$.
Therefore, $F_2$ given by Eq.~\eqref{eq:F2_gen_function} is a generating function for the transformation we want to perform
and the Hamiltonian density for the new variable is given by
\begin{equation}
	\label{eq:can_transfor_H}
	K(x) = H(x) + \frac{\partial F_2}{\partial t} =  H(x) +  e^{ i \mu t / \hbar } \frac{\partial  e^{- i \mu t / \hbar }}{\partial t} i \hbar \psi^{\star} \psi  \, .
\end{equation}
Starting from Eq.~\eqref{eq:EOM_3D_Heinsenberg} and expanding at second order in $\delta \hat{\Psi}$ we get
\begin{align}
		\hat{H}^{(2)} &= \int \dd {\bm{x} }   \left[  \left| \Psi_0 \right|^2 \frac{\hbar^2}{2 m} \frac{\partial \delta \hat{\Psi}^{\dagger}}{\partial z} \frac{\partial \delta \hat{\Psi}}{\partial z} + \left( \frac{\hbar^2}{2 m}  \left| \frac{\partial \Psi_0}{\partial r}\right|^2  + \frac{1}{2} \omega_{\perp}^2 r^2  \left| \Psi_0 \right|^2 + g \left| \Psi_0 \right|^4  \right) \delta \hat{\Psi}^{\dagger} \delta \hat{\Psi} \right. \nonumber \\ 
		& \left. \phantom{\left( \frac{\hbar^2}{2 m}  \left| \frac{\partial \Psi_0}{\partial r}\right|^2 \right)} + g  \left| \Psi_0 \right|^4 \hat{\Psi}^{\dagger} \delta \hat{\Psi} + \frac{g}{2} \left| \Psi_0 \right|^4  \left( \delta \hat{\Psi}^{\dagger \, 2} + \delta \hat{\Psi}^2 \right)   \right] \, , 
\end{align}
where, in the second term, we have isolated a combination of terms corresponding to the right-hand side of the GPE, Eq.~\eqref{eq:3D_GPE}.
Performing the transformation, Eq.~\eqref{eq:can_transfor_H} cancels this term. The remaining Hamiltonian is integrated over $r$ and $\theta$ to give Eq.~\eqref{eq:quadratic_H_pert_BdG_longitidunal}, where we have reused the name $\hat{H}^{(2)}$ for this modified Hamiltonian in the main text.

\bibliographystyle{unsrt_2023}
\bibliography{bibmanuscript}

\end{document}